\documentclass[onecolumn,romanappendices]{IEEEtran}
\usepackage[utf8]{inputenc}
\usepackage[T1]{fontenc}
\usepackage{url}
\usepackage{cite}
\usepackage{longtable}
\usepackage[cmex10]{amsmath} 
\usepackage{amsmath}
\usepackage{amsfonts}
\usepackage{amssymb}
\usepackage{enumerate}
\usepackage{enumitem}
\usepackage{bm}
\usepackage{graphicx}
\usepackage{color}
\usepackage[bottom]{footmisc}
\usepackage{dsfont}
\usepackage{multirow} 
\usepackage{arydshln}
\usepackage[caption=false]{subfig}
\usepackage{enumitem}

\allowdisplaybreaks
\normalsize  

\newtheorem{Theorem}{Theorem}

\newtheorem{Lemma}{Lemma}

\begin{document}	
	\title{On Noncoherent Multiple-Antenna Rayleigh Block-Fading Channels at Finite Blocklength}
	
	\author{Chao Qi and Tobias Koch, \emph{Senior Member, IEEE}
		\thanks{Tobias Koch has received funding from the European Research Council (ERC) under the European Union's Horizon 2020 research and innovation programme (Grant No.~714161). Tobias Koch has further received funding from the Spanish Ministerio de Ciencia e Innovación under Grants RYC-2014-16332, TEC2016-78434-C3-3-R (AEI/FEDER, EU), and PID2020-116683GB-C21~/~AEI~/~10.13039/501100011033, and from the Comunidad de Madrid under Grant IDEA-CM (TEC-2024/COM-89). The material in this paper was presented in part at the IEEE International Symposium on Information Theory, Los Angeles, CA, June 2020.}
		\thanks{Chao Qi is with School of Computer and Software Engineering, Xihua University, 610039, Chengdu, China (email: chaoqi@mail.xhu.edu.cn).}
		\thanks{Tobias Koch is with the Department of Signal Theory and Communications, Universidad Carlos III de Madrid, 28911, Legan\'es, Spain and also with the Gregorio Mara\~n\'on Health Research Institute, 28007, Madrid, Spain (email: tkoch@ing.uc3m.es).}
	}
	\maketitle
	
	\begin{abstract}
	This paper investigates the maximum coding rate at which data can be transmitted over a noncoherent, multiple-input, multiple-output (MIMO) Rayleigh block-fading channel using an error-correcting code of a given blocklength with a block-error probability not exceeding a given value. A high-SNR normal approximation is derived that becomes accurate as the signal-to-noise ratio (SNR) and the number of coherence intervals over which we code tend to infinity. The obtained normal approximation complements the nonasymptotic bounds that have appeared in the literature, but whose evaluation is computationally demanding. It further lays the theoretical foundation for an analytical analysis of the fundamental tradeoff between diversity, multiplexing, and channel-estimation cost at finite blocklength and finite SNR.
	\end{abstract}
	

	\section{Introduction}\label{Sec_Introduction} 
	
	There exists an increasing interest in the problem of transmitting short packets in wireless communications \cite{durisi2016toward}. For example, the vast majority of wireless connections in the next generations of cellular systems will most likely be originated by autonomous machines and devices, which predominantly exchange short packets. It is also expected that enhanced mobile-broadband services will be complemented by new services that target systems requiring reliable real-time communication with stringent requirements on latency and reliability. While \emph{capacity} and \emph{outage capacity} provide accurate benchmarks for the throughput achievable in wireless communication systems when the package length is not restricted, for short-package wireless communications, a more refined analysis of the maximum coding rate as a function of the blocklength is needed. Such an analysis is provided in this paper. 
	
	Let $R^{\ast}(n, \epsilon)$ denote the maximum coding rate at which data can be transmitted using an error-correcting code of blocklength $n$ with a block-error probability no larger than $\epsilon$. Hayashi \cite{hayashi09} and Polyanskiy, Poor, and Verd\'u \cite{polyanskiy2010channel} showed that, for various channels with a positive capacity $C$, $R^{\ast}(n, \epsilon)$ can be tightly approximated as 
	\begin{eqnarray}
		R^{\ast}(n, \epsilon)=C-\sqrt{\frac{V}{n}}Q^{-1}(\epsilon)+\mathcal{O}\left(\frac{\log n}{n}\right) \label{Equ_Yury_R}
	\end{eqnarray}
	where $V$ is the so-called channel dispersion; $Q^{-1}(\epsilon)$ denotes the inverse of the Q-function
	\begin{equation}
		Q(x)\triangleq \frac{1}{\sqrt{2\pi}} \int_{x}^{\infty} e^{-t^2/2}\mathrm{d}t
	\end{equation}
	and $\mathcal{O}(\log n/n)$ comprises terms that decay no slower than $\log n/n$.  
	The approximation that follows by ignoring the $\mathcal{O}(\log n/n)$ term is sometimes referred to as \emph{normal approximation}. The normal approximation has been established as a benchmark for short error-correcting codes; see, e.g., \cite{Shirvanimoghaddam19,coskun19}. It further serves as a proxy for the maximum coding rate in the analysis and optimization of communication systems that exchange short packets and has appeared in numerous papers on short-packet wireless communications; see, e.g., \cite{Makki15,Makki16,Zhang18,Li19,Mahmood21,Munari21,Feng23,Xie23,Zheng24,Li25}. However, many of these works consider the normal approximation of the Gaussian channel, which may fail to capture the effects of key parameters in wireless communication systems, such as coherence time, diversity, or multiplexing gain.

	\begin{table}[!t]
	\centering
	\caption{Summary of finite-blocklength results and refined asympotics for wireless communication channels}
	\label{summary_table}
	\begin{tabular}{|p{3.8cm}|p{4.6cm}|p{3.4cm}|p{4cm}|}
	\hline
	\textbf{Paper} & \textbf{Fading Model and Setting} & \textbf{Channel-State Information\textsuperscript{\textdagger}} & \textbf{Result} \\\hline\hline
	Polyanskiy \& Verdú 2011 \cite{polyanskiy2011scalar} &\vspace{-1.2\topsep}\begin{itemize}[leftmargin=*] \item stationary \& ergodic fading\item single antenna \end{itemize} & CSIR & normal approximation\newline (available in closed form) \\\hline
	Collins \& Polyanskiy 2014 \cite{collins2014orthogonal} & \vspace{-1.2\topsep}\begin{itemize}[leftmargin=*] \item block-fading\item MISO \end{itemize}& CSIR & achievable normal approximation\newline (available in closed form)\\\hline
 	Collins \& Polyanskiy 2016 \cite{collins2016dispersion}, \newline Collins \& Polyanskiy 2019 \cite{Collins2019coherent} & \vspace{-1.2\topsep}\begin{itemize}[leftmargin=*] \item block-fading\item MIMO \end{itemize} & CSIR & normal approximation \newline (available in closed form)\\\hline
	Hoydis, Couillet \& Piantanida 2015 \cite{hoydis2015second} & \vspace{-1.2\topsep}\begin{itemize}[leftmargin=*] \item quasi-static Rayleigh fading\item MIMO \item number of transmit and receive antennas tends to infinity with blocklength\end{itemize} & CSIR & bounds on optimal error probability for given second-order coding rate\newline (available in closed form)\\\hline
	Yang \emph{et al.} 2015 \cite{Yang15} & \vspace{-1.2\topsep}\begin{itemize}[leftmargin=*] \item quasi-static fading\item single antenna \item long-term power constraint \end{itemize} & CSIRT & normal approximation \newline (available in closed form) \\\hline
	Yang \emph{et al.} 2014 \cite{yang2014quasi} & \vspace{-1.2\topsep}\begin{itemize}[leftmargin=*] \item quasi-static fading\item MIMO \end{itemize} & no-CSI, CSIT, CSIR, CSIRT & normal approximation \newline (available in closed form) \\\hline
	Yang \emph{et al.} 2012 \cite{yang2012diversity} & \vspace{-1.2\topsep}\begin{itemize}[leftmargin=*] \item Rayleigh block-fading\item single antenna \end{itemize} & no-CSI & nonasymptotic bounds \newline (must be evaluated numerically) \\\hline
	Durisi \emph{et al.} 2016 \cite{durisi2015short} & \vspace{-1.2\topsep}\begin{itemize}[leftmargin=*] \item Rayleigh block-fading\item MIMO \end{itemize} & no-CSI & nonasymptotic bounds \newline (must be evaluated numerically)\\\hline
	Lancho \emph{et al.} 2019 \cite{lancho2019saddlepoint},\newline Lancho \emph{et al.} 2020 \cite{LanchoSaddlepoint2020} & \vspace{-1.2\topsep}\begin{itemize}[leftmargin=*] \item Rayleigh block-fading\item single antenna \end{itemize} & no-CSI & saddlepoint approximations \newline (must be evaluated numerically) \\\hline
	Lancho, Koch \& Durisi 2019 \cite{lancho2019single} & \vspace{-1.2\topsep}\begin{itemize}[leftmargin=*] \item Rayleigh block-fading\item single antenna \end{itemize}& no-CSI & high-SNR normal approximation \newline (available in closed form)\\\hline
	\end{tabular}
	\begin{flushleft}
	\textsuperscript{\textdagger} no-CSI: noncoherent setting (only statistics of fading coefficients known); CSIT: fading coefficients available at transmitter; CSIR: fading coefficients available at receiver; CSIRT: fading coefficients available at transmitter and receiver.
	\end{flushleft}
	\end{table}
	
To address this shortcoming, the work of Polyanskiy \textit{et al.} has been generalized to several wireless communication channels \cite{polyanskiy2011scalar,collins2014orthogonal,collins2016dispersion,Collins2019coherent,hoydis2015second,Yang15,yang2014quasi,yang2012diversity,durisi2015short,lancho2019saddlepoint,LanchoSaddlepoint2020,lancho2019single};  see Table~\ref{summary_table} for a summary.
	In particular, the channel dispersion of coherent fading channels, where the receiver has perfect knowledge of the realizations of the fading coefficients, was obtained by Polyanskiy and Verd\'{u} for the single-antenna case \cite{polyanskiy2011scalar}, and by Collins and Polyanskiy for the multiple-input single-output (MISO) \cite{collins2014orthogonal} and the multiple-input multiple-output (MIMO) case \cite{collins2016dispersion,Collins2019coherent}. The case where the number of transmit and receive antennas grows with the blocklength was considered in \cite{hoydis2015second}.
	When both the transmitter and the receiver have perfect knowledge of the realization of the fading coefficients and the transmitter satisfies a long-term power constraint, the channel dispersion of single-antenna quasi-static fading channels was obtained by Yang \emph{et al.}
	\cite{Yang15}.  
	In the noncoherent setting, the channel dispersion is only known in the quasi-static case,  where it is zero \cite{yang2014quasi}.
	For general noncoherent Rayleigh block-fading channels, nonasymptotic bounds on the maximum coding rate were presented in \cite{yang2012diversity} and \cite{durisi2015short}. 
	Saddlepoint approximations that accurately approximate these bounds in the single-antenna case with a negligible computational cost were given in \cite{lancho2019saddlepoint,LanchoSaddlepoint2020}. 
	However, a closed-form expression of the channel dispersion for general noncoherent Rayleigh block-fading channels is still unknown. 
	Obtaining such an expression is difficult because the capacity-achieving input distribution is in general unknown. 
	Thus, the standard approach of obtaining expressions of the  form \eqref{Equ_Yury_R}, which consists of first evaluating nonasymptotic upper and lower bounds on $R^{\ast}(n, \epsilon)$ for the capacity-achieving input and output distributions and then analyzing these bounds in the limit as $n\to \infty$, cannot be followed.
	Fortunately, the asymptotic behavior of the capacity of such channels at high signal-to-noise ratio (SNR) is well understood \cite{zheng2002communication,yang2013capacity}. 
	This fact was exploited in \cite{lancho2019single} to derive a high-SNR normal approximation of $R^{\ast}(n, \epsilon)$ for noncoherent single-antenna Rayleigh block-fading channels.
	
	In this paper, we generalize \cite{lancho2019single} to the MIMO case. In particular, we present an expression of $R^{\ast}(n, \epsilon)$ similar to \eqref{Equ_Yury_R} for noncoherent MIMO Rayleigh block-fading channels. 
	By deriving asymptotically-tight approximations of the capacity and the channel dispersion at high SNR, we obtain a high-SNR normal approximation of $R^{\ast}(n, \epsilon)$, which complements the existing nonasymptotic bounds.
	
	The rest of the paper is organized as follows. Section~\ref{Sec_Model} introduces the system model. Section~\ref{Sec_MainResult} presents and discusses the main result of this paper: a high-SNR normal approximation for noncoherent MIMO block-fading channels that becomes accurate as the SNR and the number of coherence intervals over which we code tend to infinity. The proof of the main result is given in Section~\ref{Sec_Proof}. The paper concludes with a summary and discussion of our results in Section~\ref{Sec_Concl}. Some of the proofs are deferred to the appendices.
	
	\subsection*{Notation}
	Upper-case letters such as $X$ denote scalar random variables and their realizations are written in lower case, e.g., $x$. 
	We use boldface upper-case letters to denote random matrices, e.g., $\mathbf{X}$, and upper-case letters of a special font for their realizations, e.g., $\mathsf{X}$. The distribution of a circularly-symmetric complex Gaussian random variable with mean $\mu$ and variance $\sigma^2$ is denoted by $\mathcal{CN}(\mu,\sigma^2)$. The Gamma distribution of shape parameter $\alpha$ and rate parameter $\lambda$ is denoted by $\Gamma(\alpha,\lambda)$. We use $\mathsf{E}[\cdot]$ and $\mathrm{Var}(\cdot)$ to denote expectation and variance, respectively. The symbol ``$\stackrel{d}{=}$" indicates equivalence in distribution.
	
	We write $(\cdot)^{\mathsf{H}}$, $(\cdot)^\ast$, and $(\cdot)^{\mathsf{T}}$ to denote Hermitian transposition, complex conjugation, and transpose (without conjugation), respectively, and $\text{tr}(\cdot)$ and $\det(\cdot)$ to denote the trace and the determinant, respectively.  The identity matrix of size $a\times a$ is written as $\mathsf{I}_a$, and $\text{diag}\{d_1,\ldots,d_a\}$ denotes an $(a\times a)$-dimensional diagonal matrix with entries $d_1,\ldots,d_a$. The diagonal submatrix that is composed of the first $i$ columns and rows of a matrix $\mathsf{X}$ is denoted as $[\mathsf{X}]_{(1:i)^2}$. For any matrix $\mathsf{X}$, $\|\mathsf{X}\|_{\textnormal{F}}$ denotes the Frobenius norm.
		
	Throughout the paper, $\log(\cdot)$ denotes the natural logarithm function, $[a]^+$ stands for $\max\{a, 0\}$, and  $\mathds{1}\{\cdot \}$ denotes the indicator function. We shall further use the following Gamma and digamma functions:
	\begin{subequations}
	\begin{IEEEeqnarray}{rCll}
	\Gamma(x) & \triangleq & \int_0^{\infty} t^{x-1}e^{-t}\textnormal{d} t, \quad & \textnormal{Gamma function} \\
	\Gamma_m(x) & \triangleq & \pi^{m(m-1)/2}\prod_{k=1}^m \Gamma(x-k+1), \quad & \textnormal{complex multivariate Gamma function} \\
	\varPsi(x) & \triangleq & \frac{\textnormal{d}}{\textnormal{d} x} \log \Gamma(x), \quad & \textnormal{Euler's digamma function} \\
	\varPsi'(x) & \triangleq & \frac{\textnormal{d}}{\textnormal{d} x} \varPsi(x), \quad & \textnormal{derivative of $\varPsi(\cdot)$.}
	\end{IEEEeqnarray}
	\end{subequations}
	Last but not least, we denote by $\varlimsup$ the \emph{limit superior} and by $\varliminf$ the \emph{limit inferior}.
	
	\section{System Model}\label{Sec_Model} 
	We consider a Rayleigh block-fading channel with $n_t$ transmit antennas, $n_r$ receive antennas, and coherence interval $T$. 
	For this channel, within the $\ell$-th coherence interval, the channel input-output relation is given by
	\begin{eqnarray}
		\mathbf{Y}_{\ell}=\mathbf{X}_{\ell}\mathbf{H}_{\ell}+\mathbf{W}_{\ell} \label{Equ_Channel}
	\end{eqnarray}
	where $\mathbf{X}_{\ell}$ is the complex-valued, $(T\times n_t)$-dimensional, transmitted matrix; $\mathbf{Y}_{\ell}$ is the complex-valued, \mbox{$(T\times n_r)$-dimensional}, received matrix; $\mathbf{H}_{\ell}$ is the complex-valued, $(n_t \times n_r)$-dimensional fading matrix with independent and identically distributed (i.i.d.) $\mathcal{CN}(0,1)$ entries; $\mathbf{W}_{\ell}$ is the complex-valued, $(T\times n_r)$-dimensional, additive noise matrix at the receiver with i.i.d. $\mathcal{CN}(0,1)$ entries. 
	We assume that $\mathbf{H}_{\ell}$ and $\mathbf{W}_{\ell}$ are independent and take on independent realizations over successive coherence intervals. 
	We further assume that the joint law of $\mathbf{H}_{\ell}$ and $\mathbf{W}_{\ell}$ does not depend on $\mathbf{X}_{\ell}$.
	We consider a noncoherent setting where transmitter and receiver only know the statistics of the fading matrix $\mathbf{H}_{\ell}$ but do not have \emph{a priori} knowledge of its realization.
	
	We assume that $n_t\leq n_r$ and $T\geq n_t + n_r$.  The assumption $n_t\leq n_r$ incurs no loss in capacity at high SNR \cite{zheng2002communication}. More specifically, Zheng and Tse \cite{zheng2002communication} showed that, at high SNR, the capacity of the noncoherent Rayleigh block-fading channel can be expressed as
	\begin{equation}
	\label{eq:capacity_ZhenTse}
		C(\rho)=n_{\star}\left(1-\frac{n_{\star}}{T}\right)\log(\rho)+\mathcal{O}_{\rho}(1)
	\end{equation}
	where $n_{\star}\triangleq\min\{n_t,n_r, \lfloor T/2\rfloor\}$ and $\mathcal{O}_{\rho}(1)$ summarizes terms that are bounded in the SNR $\rho$. 
	This implies that, for a given coherence time $T$ and number of receive antennas $n_r$, the capacity pre-log is maximized by using \mbox{$n_t=\min\{n_r,\lfloor T/2\rfloor\}$} transmit antennas. Thus, using more than $n_r$ transmit antennas does not increase the high-SNR capacity. The assumption $T\geq n_t + n_r$ is reasonable for slow-fading channels when the number of antennas is moderate. Under this assumption, an input distribution referred to as \emph{unitary space-time modulation (USTM)} achieves a lower bound on the capacity that is asymptotically tight in the limit as the SNR tends to infinity \cite{zheng2002communication,yang2013capacity}. When $T < n_t + n_r$, USTM inputs are no longer optimal and an input distribution called \emph{beta-variate space-time modulation (BSTM)} should be considered instead \cite{yang2013capacity}. Analyzing the maximum coding rate for this input distribution requires a different analysis that is beyond the scope of this paper.
	
	For simplicity, we shall restrict ourselves to codes whose blocklength is $n = LT$, where $L$ denotes the number of blocks of coherence interval $T$ a codeword spans.  An $(L,T,M,\epsilon,\rho)$ code consists of:
	\begin{itemize} 
		\item[(1)] An encoder $f$: $\{1,\cdots,M\} \rightarrow (\mathbb{C}^{T\times n_t})^L$ that maps a message $\Phi$, which is uniformly distributed on $\{1,\cdots,M\}$, to a codeword $\mathbf{X}^L =[\mathbf{X}_1,\cdots,\mathbf{X}_L]$. The codewords are assumed to satisfy the power constraint\footnote{In the information theory literature, it is more common to impose a power constraint per codeword $\mathbf{X}^L$. Such a constraint would allow for a more flexible power allocation across coherence intervals and may give rise to a larger maximum coding rate. However, practical systems typically require a per-coherence-interval constraint to avoid large peak-to-average-power ratios. In any case, allocating equal power to each coherence interval is optimal with respect to channel capacity due to the concavity of mutual information in the input distribution. In this case, the per-codeword power constraint specializes to the per-coherence-interval constraint. It therefore seems plausible that both power constraints give rise to the same maximum coding rate for $L$ sufficiently large, where the channel-capacity term dominates the channel-dispersion term.}   
		\begin{equation}
			\label{eq:power}
			\|\mathbf{X}_{\ell}\|_{\textnormal{F}}^2\leq T\rho, \quad \ell=1,\cdots,L, \, \textnormal{with probability one.}
		\end{equation}
		Since the variances of $\mathbf{H}_{\ell}$ and $\mathbf{W}_{\ell}$ are normalized, $\rho$ can be interpreted as the average SNR at the receiver. 
		\item[(2)] A decoder $g$: $(\mathbb{C}^{T\times n_t})^L\rightarrow \{1,\cdots,M\}$ satisfying the maximum error probability constraint
		\begin{equation}
			\max_{1\leq \phi\leq M} \Pr\big[g(\mathbf{Y}^L)\neq \Phi|\Phi=\phi\big]\leq \epsilon
		\end{equation}
		where $\mathbf{Y}^L =[\mathbf{Y}_1,\cdots,\mathbf{Y}_L]$ is the channel output induced by the codeword $\mathbf{X}^L= f(\phi)$, according to \eqref{Equ_Channel}.
	\end{itemize}
	The maximum coding rate is defined as 
	\begin{equation}
		R^{\ast}(L,T, \epsilon,\rho)\triangleq \sup\left\{\frac{\log M}{LT}: \exists(L,T,M,\epsilon,\rho) \text{ code}\right\}. \label{eq:Rstar}
	\end{equation} 
	
	\section{Main Results} \label{Sec_MainResult} 	
	
	The main result of this paper is a high-SNR normal approximation of $R^{\ast}(L,T, \epsilon,\rho)$ presented in Section~\ref{sub:mainresults}. In Section~\ref{Sec_MainResultB}, we discuss the accuracy of this approximation by means of numerical results. Possible applications are discussed in Section~\ref{Sec_EngineeringWisdom}. 
	
	\subsection{High-SNR Normal Approximation}
	\label{sub:mainresults}
	\begin{Theorem} \label{Thm_R}
		Assume that $T\geq n_r+n_t$, $n_r\geq n_t$, and $0<\epsilon<\frac{1}{2}$. Then, at high SNR, 
		\begin{IEEEeqnarray}{lCl}
			R^{\ast}(L, T, \epsilon, \rho) & = & \tilde{I}(T,\rho)+K_{\tilde{I}}(T,\rho)-\sqrt{\frac{\tilde{V}(T)+K_{\tilde{V}}(T,\rho)}{L}} Q^{-1}(\epsilon) +K_L(L,T,\rho) \label{Equ_Thm_R}
		\end{IEEEeqnarray}
		where
		\begin{IEEEeqnarray}{rCl}
			\tilde{I}(T,\rho) & \triangleq& n_t \left(1-\frac{n_t}{T}\right) \log\left(\frac{\rho}{n_t}\right) + n_t \left(1-\frac{n_t}{T}\right) \log\left(\frac{T}{e}\right) 
			+\frac{1}{T}\log\frac{\Gamma_{n_t}(n_t)}{\Gamma_{n_t}(T)} +\left(1-\frac{n_t}{T}\right)\sum_{i=0}^{n_t-1} \varPsi(n_r-i) \label{Def_It}\\
			\tilde{V}(T) & \triangleq &   \frac{n_t}{T}\left(1-\frac{n_t}{T}\right)+\left(1-\frac{n_t}{T}\right)^2  \sum_{i=0}^{n_t-1} \varPsi'(n_r-i). \label{Def_Vt}
		\end{IEEEeqnarray} 
		In \eqref{Equ_Thm_R},  $K_{\tilde{I}}(T,\rho)$ and $K_{\tilde{V}}(T,\rho)$ are functions of $T$ and $\rho$ that satisfy
		\begin{equation}
			\lim_{\rho\to\infty}K_{\xi}(T,\rho)=0, \quad \xi \in\{\tilde{I},\tilde{V}\} \label{eq:thm_little_o}
		\end{equation}
		and $K_L(L,T,\rho)$ is a function of $L$, $T$, and $\rho$ that satisfies 
		\begin{equation}
			\sup_{\rho\geq \rho_0} \big|K_L(L,T,\rho)\big| \leq A\frac{\log L}{L}, \quad L\geq L_0 \label{eq:thm_big_O}
		\end{equation}
		for some $A$, $L_0$, and $\rho_0$ independent of $L$ and $\rho$.
		
	\end{Theorem}
	\begin{IEEEproof}
		See Sec. \ref{Sec_Proof}.
	\end{IEEEproof} 
	
	The quantity $\tilde{I}(T,\rho)$ is an asymptotically-tight lower bound on the capacity $C(\rho)$ of the noncoherent MIMO Rayleigh block-fading channel \cite{yang2013capacity}. Similarly, $\tilde{V}(T)$ can be viewed as a high-SNR approximation of the channel dispersion. By ignoring the error terms $K_{\tilde{I}}(T,\rho)$, $K_{\tilde{V}}(T,\rho)$, and $K_L(L,T,\rho)$ in \eqref{Equ_Thm_R}, we obtain the high-SNR normal approximation
	\begin{equation}
		\label{eq:highSNR_NA}
		R^{\ast}(L, T, \epsilon, \rho) \approx  \tilde{I}(T,\rho)-\sqrt{\frac{\tilde{V}(T)}{L}} Q^{-1}(\epsilon).
	\end{equation}
	
	Observe that $\tilde{I}(T,\rho)$ and $\tilde{V}(T)$ depend on $n_r$ only via the digamma function $\varPsi(n_r-i)$ and its derivative $\varPsi'(n_r-i)$, respectively. On the domain of positive integers, the digamma function is monotonically increasing, and its derivative is monotonically decreasing. As a consequence, the approximation \eqref{eq:highSNR_NA} is monotonically increasing in $n_r$. As we shall observe in Section~\ref{Sec_MainResultB}, the dependence of \eqref{eq:highSNR_NA} on $n_t$ is more intricate. Intuitively, increasing the number of transmit antennas achieves a higher multiplexing gain $\min\{n_t,n_r\}$, but it also requires the estimation of more channel coefficients.
		
	For comparison, at high SNR, the capacity $C_{c}(\rho)$ of the coherent MIMO Rayleigh block-fading channel satisfies \cite{telatar1999capacity}
	\begin{equation}
		\lim_{\rho\to\infty} \left\{C_c(\rho) -  n_t\log \left(\frac{\rho}{n_t}\right) - \mathsf{E}\left[\log\det(\mathbf{H}_{\ell}\mathbf{H}_{\ell}^\mathsf{H})\right] \right\} = 0.
	\end{equation}
	Furthermore, the channel dispersion $V_c(\rho)$ converges to \cite{Collins2019coherent}
	\begin{equation}
		\lim_{\rho\to\infty} V_c(\rho) = \frac{n_t}{T}+\mathrm{Var}\bigl( \log \det (\mathbf{H}_{\ell}\mathbf{H}_{\ell}^\mathsf{H})\bigr).
	\end{equation}
	Applying \cite[Lemma~A.2]{grant2002rayleigh} (see also Lemma~\ref{lem:grant_wishart} below), the sum of the Euler digamma functions $\varPsi(n_r-i)$ in \eqref{Def_It} can be identified as $\mathsf{E}\left[\log\det(\mathbf{H}_{\ell}\mathbf{H}_{\ell}^\mathsf{H})\right]$, and the sum of the derivatives of these functions in \eqref{Def_Vt} can be identified as $\mathrm{Var}\bigl( \log \det (\mathbf{H}_{\ell}\mathbf{H}_{\ell}^\mathsf{H})\bigr)$. The high-SNR capacity and dispersion of the noncoherent block-fading channel can thus be written as
	\begin{IEEEeqnarray}{lCl}
		\tilde{I}(T,\rho) & = & n_t\left(1-\frac{n_t}{T}\right)\log\left(\frac{\rho}{n_t}\right) + \left(1-\frac{n_t}{T}\right) \mathsf{E}\left[\log\det(\mathbf{H}_{\ell}\mathbf{H}_{\ell}^\mathsf{H})\right] + n_t \left(1-\frac{n_t}{T}\right) \log\left(\frac{T}{e}\right) 
		+\frac{1}{T}\log\frac{\Gamma_{n_t}(n_t)}{\Gamma_{n_t}(T)}
	\end{IEEEeqnarray}
	and
	\begin{equation}
		\tilde{V}(T) = \frac{n_t}{T}\left(1-\frac{n_t}{T}\right) + \left(1-\frac{n_t}{T}\right)^2 \mathrm{Var}\bigl( \log \det (\mathbf{H}_{\ell}\mathbf{H}_{\ell}^\mathsf{H})\bigr).
	\end{equation}
	Observe that, up to SNR-independent terms, $\tilde{I}(T,\rho)$ is given by $(1-n_t/T)$ times the high-SNR approximation of $C_{c}(\rho)$. Similarly, it can be shown that $\tilde{V}(T)$ corresponds to the high-SNR channel dispersion one obtains by transmitting in each coherence block one pilot symbol per transmit antenna to estimate the fading coefficient, and by then transmitting  $T-n_t$ symbols over a coherent fading channel. This suggests the heuristic that, at high SNR, one pilot symbol per transmit antenna should be transmitted in each coherence block followed by coherent transmission. However, this heuristic may be misleading, since the SNR-independent terms of $\tilde{I}(T,\rho)$ may be significant, and since it is \emph{prima facie} unclear whether one pilot symbol per transmit antenna and coherence block suffices to obtain a fading estimate of sufficient accuracy. For the single-antenna case, a more refined analysis of the maximum coding rate achievable with pilot-assisted transmission can be found in \cite{ostman2019}. 
	
	Theorem~\ref{Thm_R} provides an asymptotic analysis of the maximum coding rate $R^{\ast}(L, T, \epsilon, \rho)$ as $L$ tends to infinity for a fixed coherence interval $T$. Indeed, the error terms in \eqref{Equ_Thm_R} vanish as $L \to \infty$ and $\rho \to \infty$, provided that $T$ is held fixed. This corresponds to the ergodic case where codewords span independent coherence intervals. The complementary case where $T\to\infty$ as $L$ is held fixed was considered in \cite{yang2014quasi}. Specifically, \cite{yang2014quasi} considers a quasi-static multiple-antenna fading channel and analyzes the convergence of the maximum coding rate to the outage capacity in the limit as $T\to\infty$. By replacing the fading matrix $\mathbb{H}$ in \cite{yang2014quasi} by a block-diagonal matrix, one obtains a channel model that is equivalent to the block-fading channel \eqref{Equ_Channel}. The corresponding results in \cite[Ths.~3 \& 9]{yang2014quasi} demonstrate that the $\epsilon$-dispersion is zero, irrespective of whether channel-state information is available at the transmitter or receiver. This implies that the maximum coding rate converges quickly to the outage capacity as $T\to\infty$. Consequently, in this regime, the outage capacity is a good proxy for the maximum coding rate. In general, both analyses (ergodic and quasi-static) characterize the maximum coding rate in different regimes and complement each other well. For the single-antenna case, a comparison between the high-SNR normal approximation \eqref{Equ_Thm_R} and a normal approximation based on \cite[Eq.~(95)]{yang2014quasi} can be found in \cite[Sec.~V-B]{lancho2019single}. As expected, the high-SNR normal approximation \eqref{Equ_Thm_R} developed for the ergodic case is accurate when $L$ is large, whereas the normal approximation \cite[Eq.~(95)]{yang2014quasi} developed for the quasi-static case is accurate when $T$ is large.

	\subsection{Numerical Results} \label{Sec_MainResultB} 
	We next provide numerical examples that illustrate the accuracy of the high-SNR normal approximation in Theorem~\ref{Thm_R}.  
	In the following figures, we depict the high-SNR normal approximation \eqref{Equ_Thm_R}, the normal approximation of the coherent MIMO Rayleigh block-fading channel obtained in \cite{Collins2019coherent}, 
	a nonasymptotic (in $\rho$ and $L$) lower bound on $R^{\ast}(L, T, \epsilon, \rho)$ that is based on the dependence testing (DT) bound \cite[Th. 22]{polyanskiy2010channel}, and a nonasymptotic upper bound  that is based on the meta converse (MC) bound \cite[Th.~31]{polyanskiy2010channel}, both specialized to the noncoherent MIMO Rayleigh block-fading channel in \cite{durisi2015short}. The nonasymptotic bounds are evaluated using the communication toolbox SPECTRE \cite{collins2016spectre}. While these bounds are very accurate, their computational complexity grows exponentially in the blocklength and can be substantial. For the noncoherent single-antenna block-fading channel, a discussion of the computational complexity of these bounds and approximations thereof can be found in \cite[Sec.~VIII-A]{LanchoSaddlepoint2020}. In the figures, the shaded area indicates the area where $R^{\ast}(L,T, \epsilon, \rho)$ lies. 
	
	\begin{figure}[!htpb]
		\centering
		\subfloat[$n=168$, $\epsilon=10^{-3}$, $\rho=15$~dB, and $n_t=n_r=2$.\label{Fig_TR_S15}]{
			\includegraphics[width=0.42\textwidth]{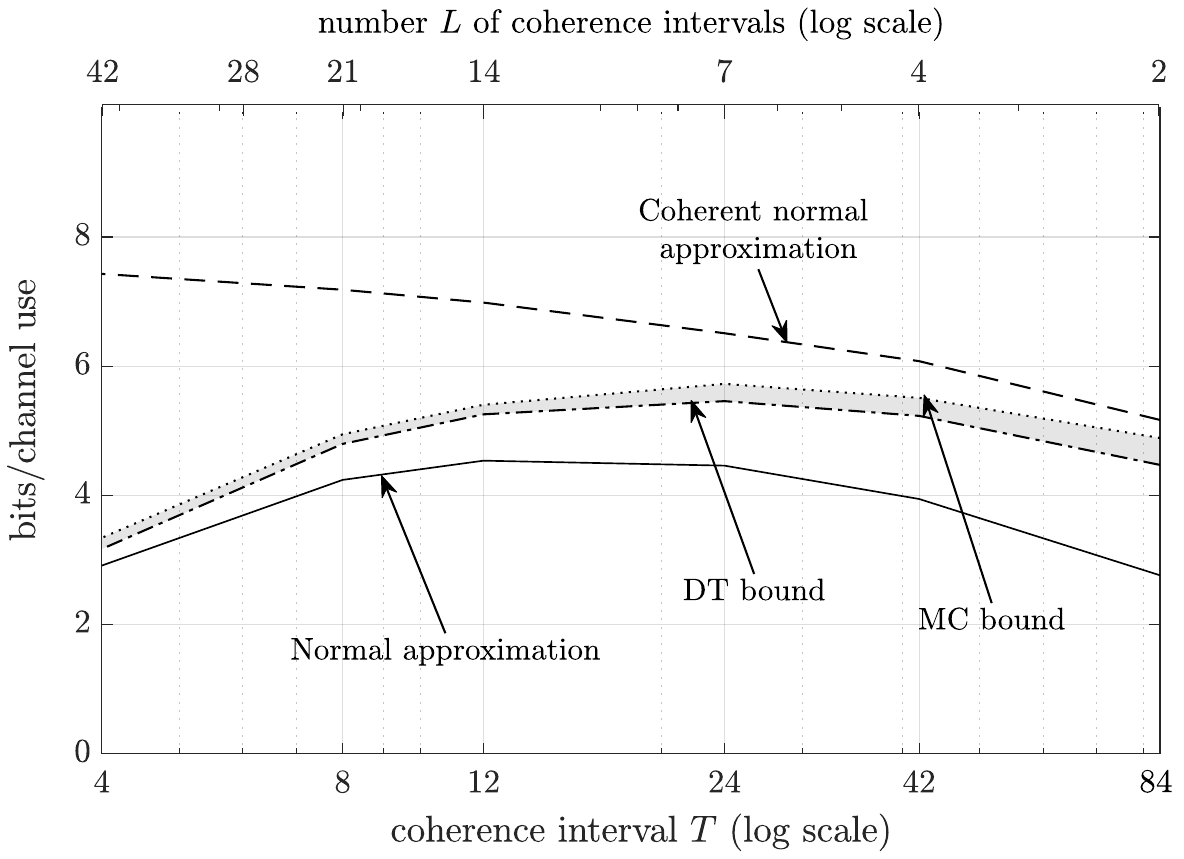}
		}
		\hspace{2em}
		\subfloat[$n=168$, $\epsilon=10^{-3}$, $\rho=25$~dB, and $n_t=n_r=2$.\label{Fig_TR_S25}]{
			\includegraphics[width=0.42\textwidth]{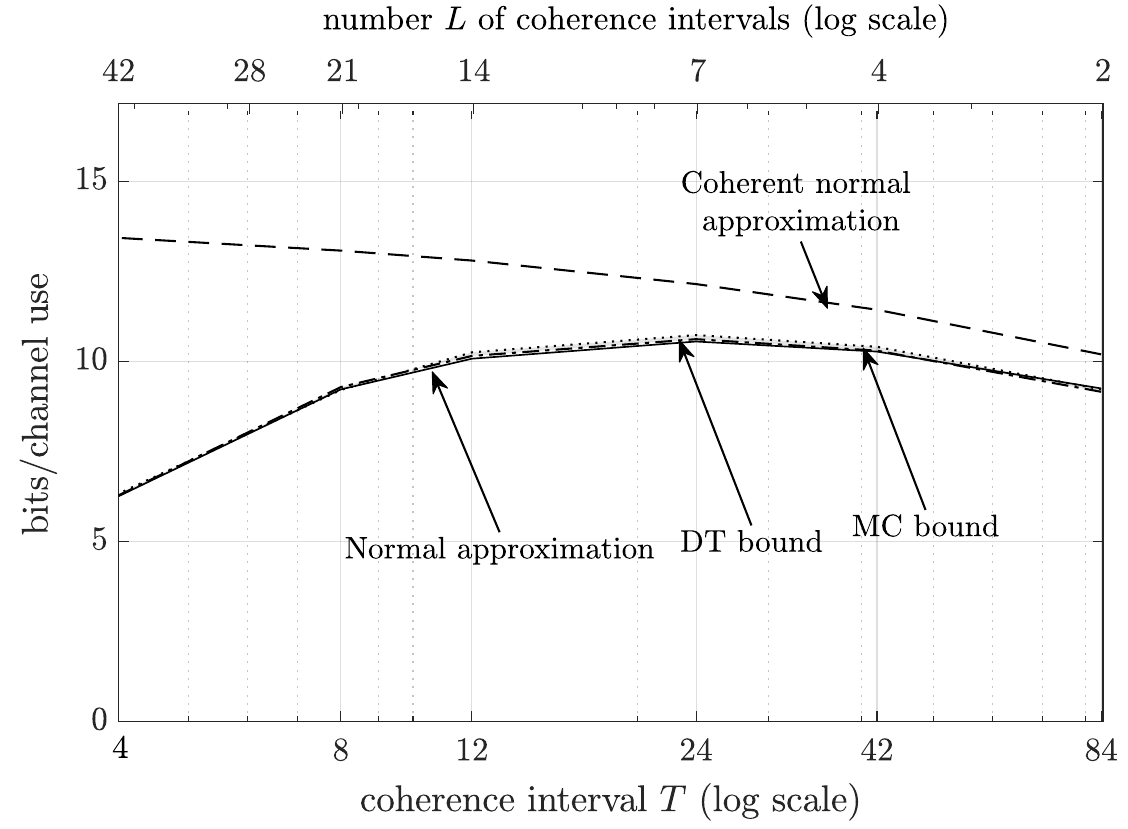}
		}
		\caption{$R^{*}(L,T,\epsilon,\rho)$ as a function of $T$ for a fixed $n=TL$.  \label{fig:1}}
	\end{figure}

	In Figs.~\ref{Fig_TR_S15} and \ref{Fig_TR_S25}, we show $R^{\ast}(L,T, \epsilon, \rho)$ as a function of  $T$ for a fixed blocklength\footnote{Thus, $L$ is inversely proportional to $T$.} $n=LT=168$ and SNR values $\rho=15$~dB and $\rho=25$~dB. Observe that the high-SNR normal approximation \eqref{Equ_Thm_R} accurately describes the maximum coding rate for $\rho=25$~dB but is loose for $\rho=15$~dB unless $L$ is large. Furthermore, as expected, the normal approximation of the coherent setting is strictly larger than that of the noncoherent setting, and it becomes more accurate as $T$ increases, indicating that the cost for estimating the channel decreases with $T$.
	
	\begin{figure}[!htpb]
		\centering  
		\subfloat[$T=24$, $\epsilon=10^{-3}$, $n_t =2, n_r = 2$, and $\rho=15$~dB. \label{Fig2_LR_N22}]{\includegraphics[width=0.43\textwidth]{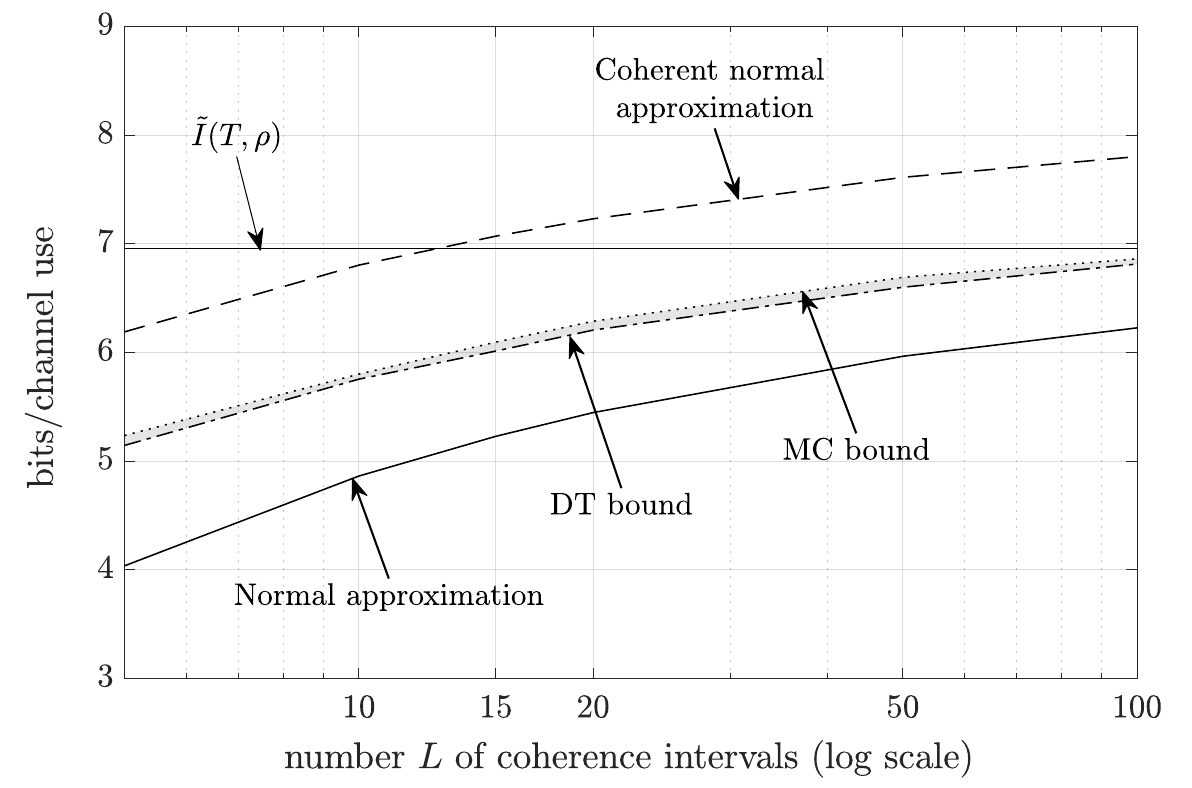}}	
		\hspace{2em}
		\subfloat[$T = 24$, $\epsilon=10^{-3}$, $n_t =2, n_r = 4$, and $\rho=15$~dB. \label{Fig2_LR_N24}]{\includegraphics[width=0.43\textwidth]{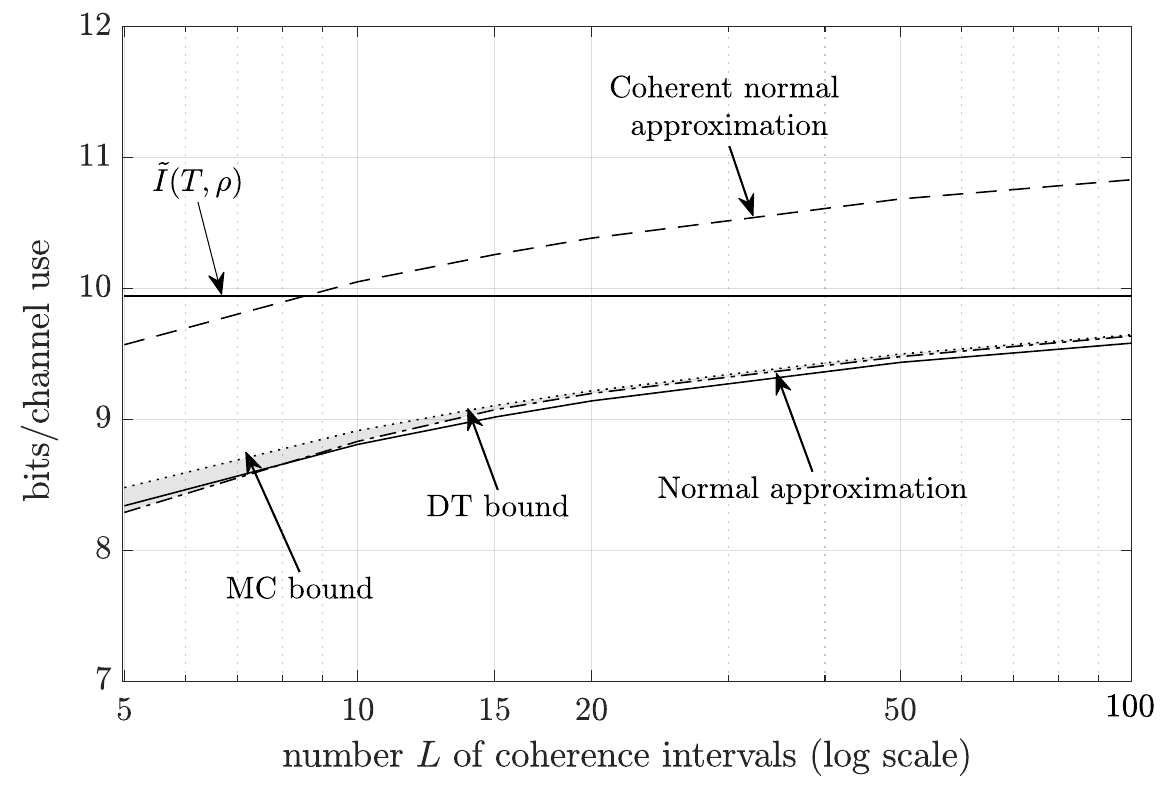}}
		\caption{$R^{\ast}(L, T,\epsilon, \rho)$ as a function of $n$ for a fixed $T$. }
		\label{fig:2}
	\end{figure}
	
	In Figs. \ref{Fig2_LR_N22} and \ref{Fig2_LR_N24}, we show $R^{\ast}(L, T,\epsilon, \rho)$ as a function of the blocklength $n=LT$ for $T=24$ and $\rho=15$~dB, and for  $n_t=n_r=2$, and $n_t=2, n_r=4$, respectively. For comparison, we further depict $\tilde{I}(T,\rho)$ \eqref{Def_It}, which is an asymptotically tight lower bound on capacity. 
	Observe that the normal approximation \eqref{Equ_Thm_R} becomes more accurate as $L$ and $n_r$ increase, and for $n_r=4$, it is very close to the nonasymptotic bounds for the considered values of $L$. Moreover, the gap between the coherent and the noncoherent normal approximations appears to be independent of $L$. This agrees with the intuition that the cost for estimating the channel is determined by the coherence interval $T$.
	
	\begin{figure}[!htpb]
		\centering  
		\subfloat[$n_t=2$, $n_r=2$, $T=24$, and $L=7$. \label{Fig3_ErrR_N22}]{\includegraphics[width=0.42\textwidth]{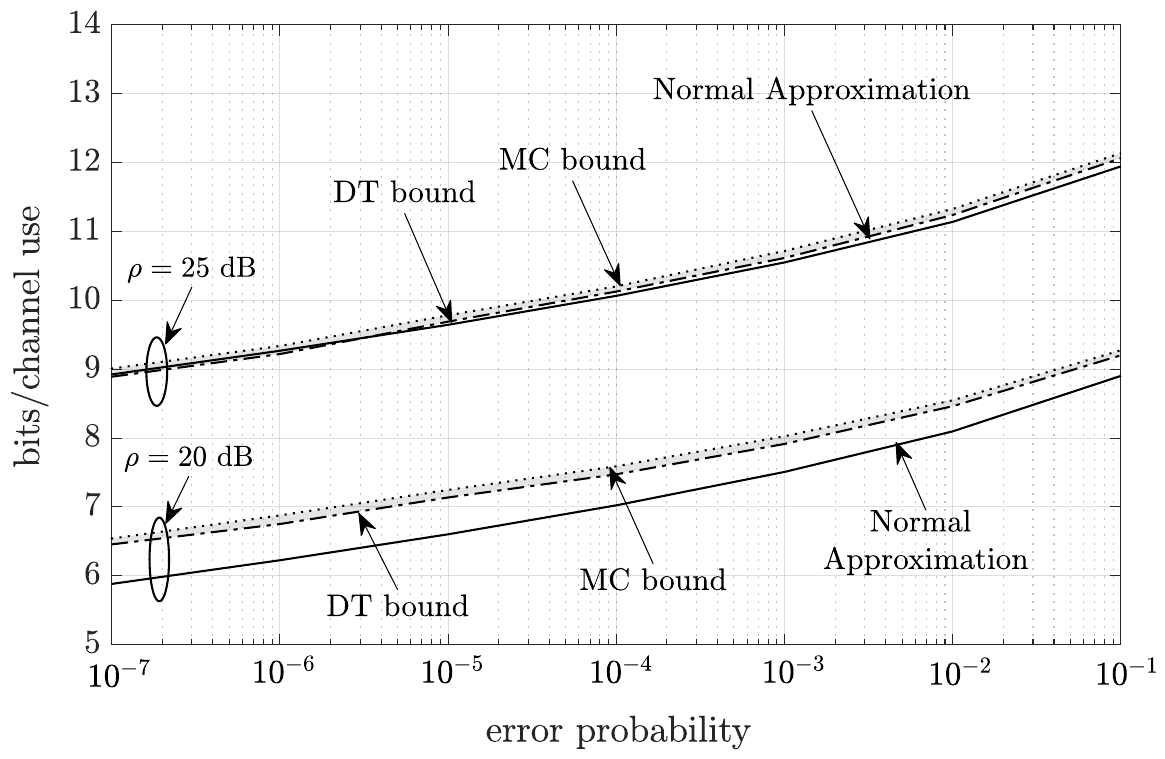}} 
		\hspace{2em}
		\subfloat[$\rho=25$dB, $T=24$, and $L=7$. \label{Fig3_ErrR_N12} ]{\includegraphics[width=0.42\textwidth]{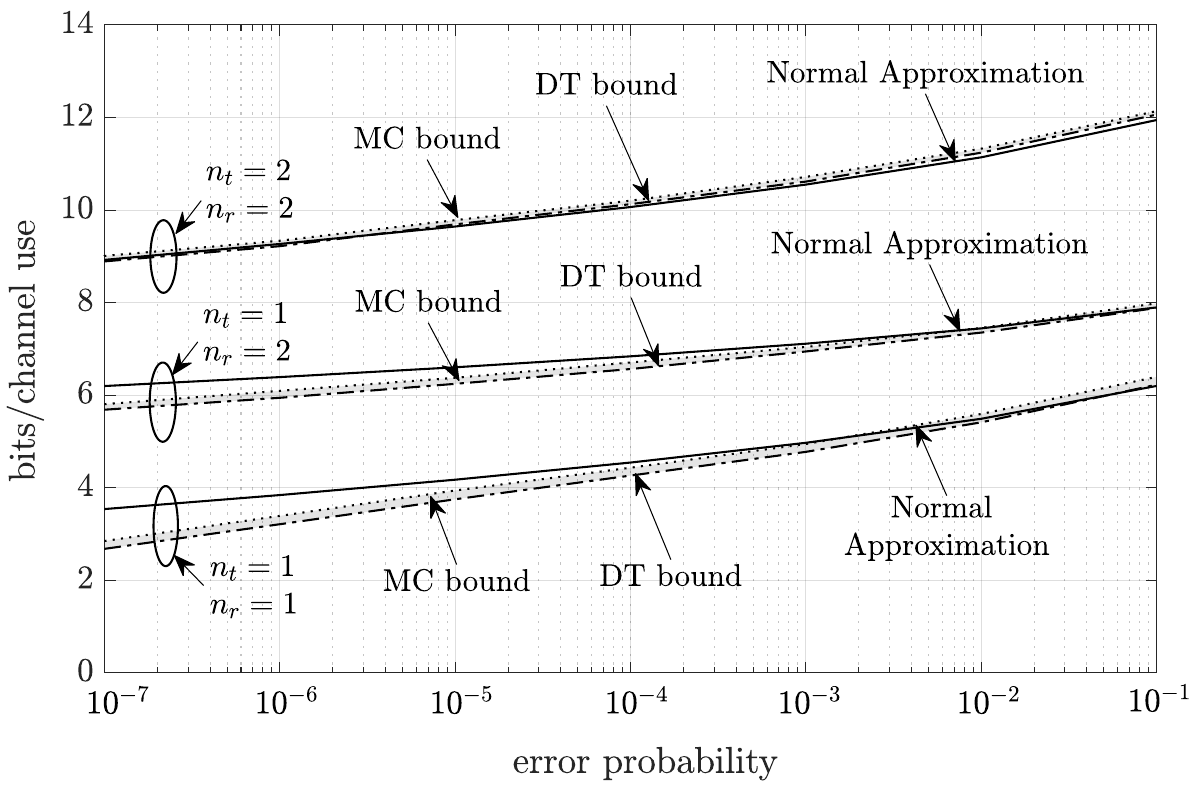}} 
		\caption{$R^{\ast}(L, T,\epsilon, \rho)$ as a function of $\epsilon$.  \label{fig:3}}
	\end{figure}

	In Figs.~\ref{Fig3_ErrR_N22} and \ref{Fig3_ErrR_N12}, we study $R^{\ast}(L, T,\epsilon, \rho)$ as a function of $\epsilon$ for $T=24$ and $L=7$. Specifically,
	Fig.~\ref{Fig3_ErrR_N22} plots $R^{\ast}(L, T,\epsilon, \rho)$ for $n_t=2, n_r=2$, and two different SNR values $20$~dB and $25$~dB. Fig.~\ref{Fig3_ErrR_N12} plots $R^{\ast}(L, T,\epsilon, \rho)$ for $25$~dB and three different numbers of antennas $(n_t,n_r)=(1,1)$, $(1,2)$, and $(2,2)$. Observe that the accuracy of normal approximation \eqref{Equ_Thm_R} increases as the SNR value and the number of antennas become larger, and for $\rho = 25$~dB and $(n_t,n_r) = (2,2)$ it is very close to the nonasymptotic bounds over the entire range of error probabilities considered. For smaller numbers of antennas, the normal approximation loses accuracy at small error probabilities.
	
	
	\begin{figure}[!htpb]
		\centering  
		\subfloat[$T=24$, $L=7$, $\epsilon=10^{-5}$, $n_t=1$, and $n_r=2$. \label{Fig4_rhoR_N12}]{\includegraphics[width=0.42\textwidth]{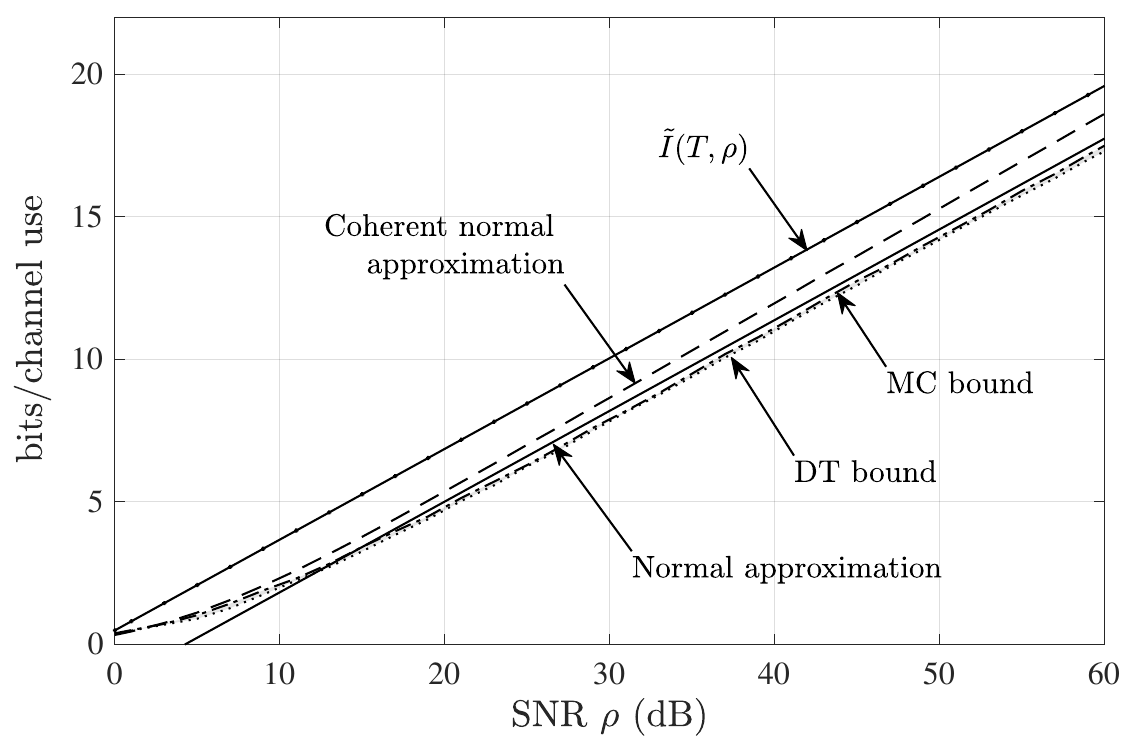}} 
		\hspace{2em}
		\subfloat[$T=24$, $L=7$, $\epsilon=10^{-5}$, $n_t=2$, and $n_r=2$. \label{Fig4_rhoR_N22} ]{\includegraphics[width=0.42\textwidth]{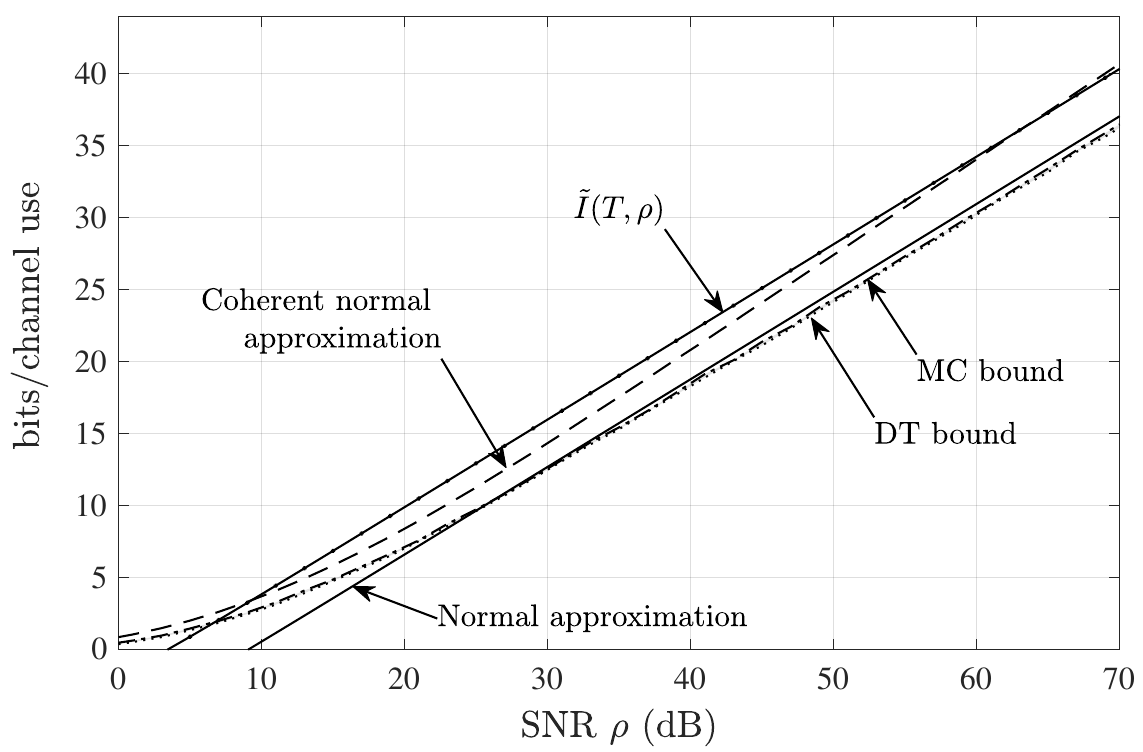}}  			 	
		\caption{$R^{\ast}(L, T,\epsilon, \rho)$ as a function of $\rho$.  \label{fig:4}}
	\end{figure}  
	
	In Figs.~\ref{Fig4_rhoR_N12} and \ref{Fig4_rhoR_N22}, we study $R^{\ast}(L, T,\epsilon, \rho)$ as a function of the SNR for $T=24$, $L=7$,  $\epsilon=10^{-5}$, and $(n_t,n_r)=(1,2)$ and $(2,2)$, respectively. For comparison, we also show the high-SNR approximation $\tilde{I}(T,\rho)$ of channel capacity. First observe that the DT lower bound on $R^{\ast}(L, T,\epsilon, \rho)$ (which is based on USTM channel inputs) is close to the MC upper bound (which is valid for any input distribution satisfying the power constraint \eqref{eq:power}). Thus, USTM channel inputs are nearly capacity-achieving for all SNR values considered.
	Further observe that the SNR range over which the normal approximation \eqref{Equ_Thm_R} is accurate depends on the number of antennas. Specifically, when $n_t=1$ and $n_r=2$, the normal approximation is accurate for SNR values above $\rho=12$dB, whereas when $n_t=2$ and $n_r=2$, it is accurate for SNR values above $\rho=23$dB. As expected, the normal approximation of the coherent channel is strictly larger than both the high-SNR normal approximation of the noncoherent channel and the nonasymptotic bounds. However, its gap to the nonasymptotic bounds diminishes as $\rho$ becomes small. Intuitively, this is because, as $\rho$ decreases, knowledge of the fading coefficients becomes less important.
	Furthermore, it can be observed from Fig.~\ref{fig:4} that, at high SNR, there is a constant gap between the normal approximation and the asymptotic bounds. This is consistent with the characterization \eqref{eq:thm_big_O} of the error term $K_L(L, T, \rho)$, which is only guaranteed to vanish as $L$ tends to infinity but may be bounded away for a fixed $L$.
	
	\begin{figure}[!htpb]
		\centering  
		\includegraphics[width=0.55\textwidth]{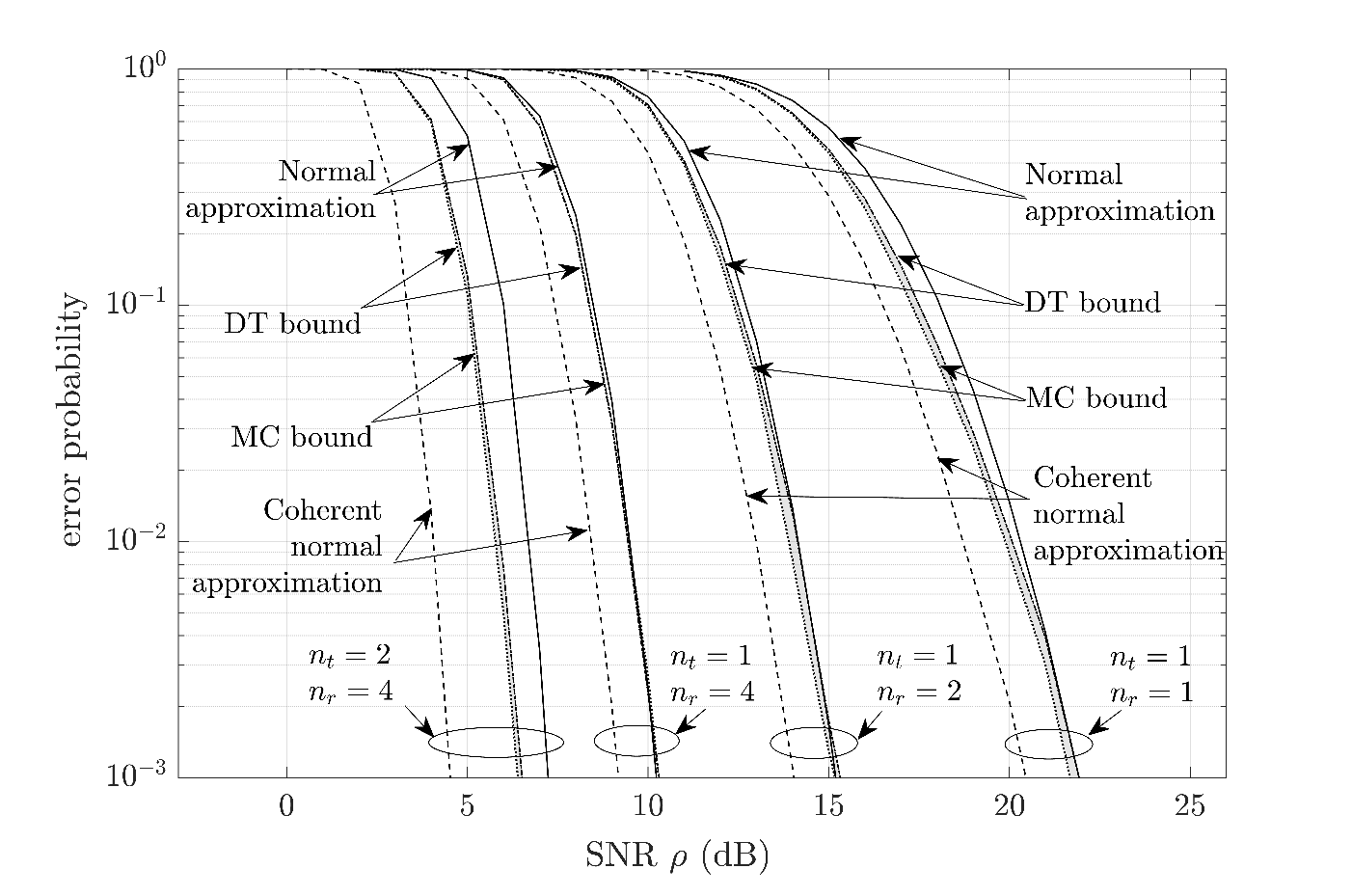}
		\caption{The minimum error probability $\epsilon$ as a function of the SNR for $R=4$, $T=24$, and $L=7$. }
		\label{Fig5_ErrRho_Nr124} 
	\end{figure}  
		
	In Fig. \ref{Fig5_ErrRho_Nr124}, we plot the error probability as a function of the 
	SNR for a fixed rate $R=4$ and for $T=24$, $L=7$, and the numbers of antennas $(n_t,n_r) = (1,1)$, $(1,2)$, $(1,4)$, and $(2,4)$. 
	As in the previous figures, the plotted DT bounds are based on USTM channel inputs, whereas the plotted MC bounds indicate the smallest probability of error that can be achieved by channel inputs satisfying the power constraint \eqref{eq:power}.
	Observe that the error probability curves become steeper as the number of transmit and receive antennas increases. This is consistent with the diversity-multiplexing tradeoff (DMT) proposed by Zheng and Tse \cite{zhengtse2003DMT}. Indeed, let the \emph{multiplexing gain} $r$ and the \emph{diversity gain} $d$ be defined as
	\begin{equation}
	r \triangleq \lim_{\rho\to\infty} \frac{R^{\ast}(L, T,\epsilon, \rho)}{\log \rho}
	\end{equation}
	and
	\begin{equation}
	d \triangleq - \lim_{\rho\to\infty} \frac{\log\epsilon}{\log\rho}
	\end{equation}
	respectively. Intuitively, if a system has a diversity gain of $d$, then the corresponding error probability decays as $\rho^{-d}$ with the SNR $\rho$. A steeper error probability curve thus corresponds to a higher diversity gain. For the noncoherent Rayleigh block-fading channel, the DMT is given by the piecewise linear curve joining the points \cite{zhengtse2002DMT_noncoherent}
		\begin{equation}
		\label{eq:DMT}
		\left(r=\left(1-\frac{n_{\star}}{T}\right) k,\, d=L(n_t - k)(n_r - k)\right),\quad k = 0, \ldots, \min(n_t, n_r)
		\end{equation}
		where $n_{\star}$ is as in \eqref{eq:capacity_ZhenTse}. If the multiplexing gain $r$ (or, equivalently, $k$) is held fixed, then \eqref{eq:DMT} implies that the diversity gain grows with the number of transmit and receive antennas. This is the setting considered in Fig.~\ref{Fig5_ErrRho_Nr124}, where we assume a fixed rate, and hence also a fixed multiplexing gain. Our observation that the slope of $\epsilon$ becomes steeper as the number of transmit and receive antennas increases is thus supported theoretically by the DMT.
	Further observe that the normal approximation \eqref{Equ_Thm_R} is accurate when the number of transmit antennas is $n_t=1$, but it is overly-pessimistic when $n_t=2$. In contrast, the coherent normal approximation is overly-optimistic for all parameters considered in this figure. Last but not least, observe that the error probability decreases significantly as the number of antennas increases. This demonstrates the benefit of multiple antennas at the transmitter and receiver at short blocklengths.
	
	\begin{figure}[!htpb]
		\centering  
		\subfloat[$R^{\ast}(L, T,\epsilon, \rho)$ as a function of $T$ for a fixed $TL=168$, $\epsilon=10^{-3}$, and $\rho=25$~dB. \label{Fig7_TR_N124}]{\includegraphics[width=0.42\textwidth]{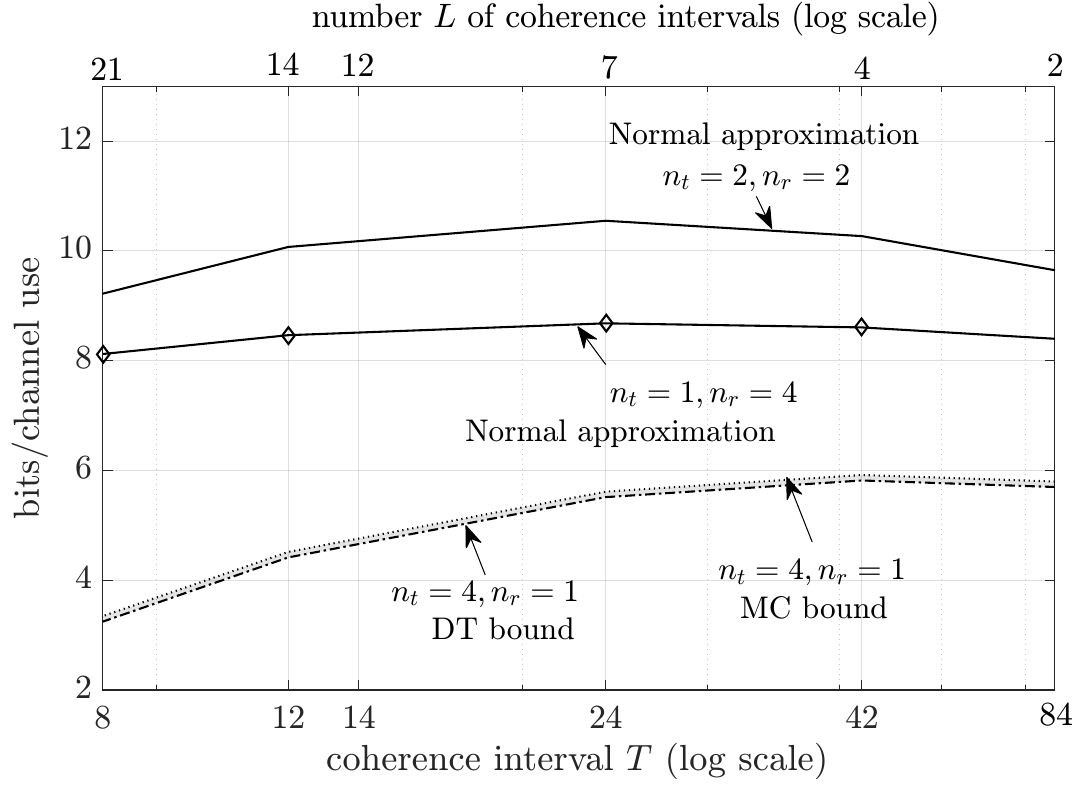}} 
		\hspace{2em}
		\subfloat[Error probability $\epsilon$ as a function of the SNR for $R=4$, $T=24$, and $L=7$. \label{Fig7_RhoErr_N124}]{\includegraphics[width=0.43\textwidth]{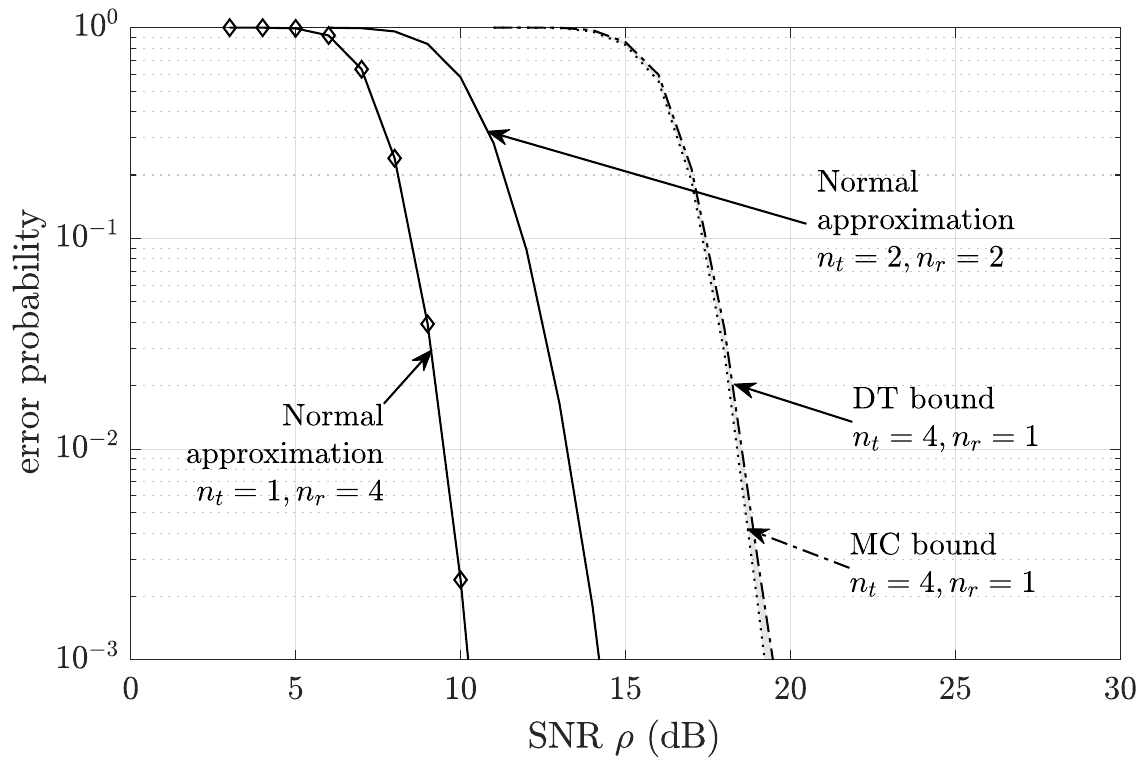}} 
		\caption{Maximum coding rate and minimum error probability for $(n_t,n_r)=(1,4)$, $(2,2)$, and $(4,1)$. \label{fig:6}}
	\end{figure}
	
	Finally, we study the impact of antenna allocation on the maximum coding rate and the error probability when the total number of antennas is equal to $4$. To this end, we plot in Fig.~\ref{Fig7_TR_N124} the maximum coding rate $R^{\ast}(L, T,\epsilon, \rho)$ as a function of $T$ for a fixed blocklength $n=168$ for  $\rho=25$~dB, error probability $\epsilon=10^{-3}$, and the numbers of antennas $(n_t,n_r)=(1,4)$, $(2,2)$, and $(4,1)$. Similarly, we plot in Fig.~\ref{Fig7_RhoErr_N124} the minimum error probability $\epsilon$ as a function of the SNR for $R=4$, $T=24$, $L=7$, and the same numbers of antennas. For the cases $(n_t,n_r)=(1,4)$ and $(n_t,n_r) = (2,2)$, we plot the normal approximation \eqref{Equ_Thm_R}, whereas for the case $(n_t,n_r)=(4,1)$ we plot the DT and the MC bounds, as the case $n_t>n_r$ is not covered by Theorem~\ref{Thm_R}. Observe that, for a fixed error probability, the coding rate is maximized by allocating the same number of antennas to the transmitter and receiver. This corresponds to the case where the transmit and receive antennas are used to maximize spatial multiplexing. In contrast, for a fixed rate, the error probability is minimized by maximizing the number of receive antennas. Intuitively, all cases exhibit the same diversity order $n_t n_r$, but the smallest number of transmit antennas results in the smallest cost for estimating the fading matrix $\mathbf{H}_{\ell}$. For example, estimating $\mathbf{H}_{\ell}$ by means of pilot symbols would require one pilot symbol per coherence interval and transmit antenna. This agrees with the observation made in \cite{zheng2002communication} that, at high SNR, using more transmit than receive antennas does not increase channel capacity.

It is well-known that normal approximations are accurate for sufficiently large blocklengths and moderate error probabilities; see also the discussion in \cite[Sec.~VIII-B]{LanchoSaddlepoint2020}. The same applies to the high-SNR normal approximation presented in Theorem~\ref{Thm_R}. Specifically, comparing the high-SNR normal approximation \eqref{Equ_Thm_R} against the nonasymptotic bounds, we observe that, for $T=24$ and $L\geq 7$, the normal approximation deviates from the true value of $R^{\ast}(L, T,\epsilon, \rho)$ by less than $5$\% for SNR values $\rho\geq 25$~dB and error probabilities $\epsilon \geq 10^{-5}$, and for SNR values $\rho\geq 21$~dB and error probabilities $\epsilon \geq 10^{-3}$, irrespective of the number of transmit and receive antennas.

\subsection{Engineering Wisdom: Optimal Number of Active Transmit Antennas} \label{Sec_EngineeringWisdom}
In \cite{zheng2002communication}, Zheng and Tse showed that, at high SNR, the channel capacity of the noncoherent Rayleigh block-fading channel behaves as
		\begin{equation}
			C(\rho)=n_{\star}\left(1-\frac{n_{\star}}{T}\right)\log(\rho)+\mathcal{O}_{\rho}(1) \label{Equ_High_Capcity}
		\end{equation}
		where $n_{\star}$ is as in \eqref{eq:capacity_ZhenTse}.
		If $T\geq n_t+n_r$ and $n_t\leq n_r$, then $n_{\star}=n_t$ and the pre-log factor $n_t (1-n_t/T)$ is monotonically increasing in $n_{t}$. In this case, it is optimal to use all available transmit antennas. However, it is \emph{prima facie} unclear whether the same is true at finite blocklength. Intuitively, increasing the number of transmit antennas achieves a higher multiplexing gain $\min\{n_t,n_r\}$, but it also requires the estimation of more channel coefficients. To gain some insights on this question, we analyze the behavior of the high-SNR normal approximation \eqref{Equ_Thm_R} as a function of the number of transmit antennas $n_t$ for a given number of receive antennas $n_r$.

	\begin{figure}[!htpb]
	\centering  
	\subfloat[$n=168$, $\epsilon=10^{-3}$, $\rho=25$~dB, and $n_r=6$. The crossing points are $T_{4,6}=14.9$, and $T_{5,6}=22.9$.  \label{Fig_7a} ]{\includegraphics[width=0.41\textwidth]{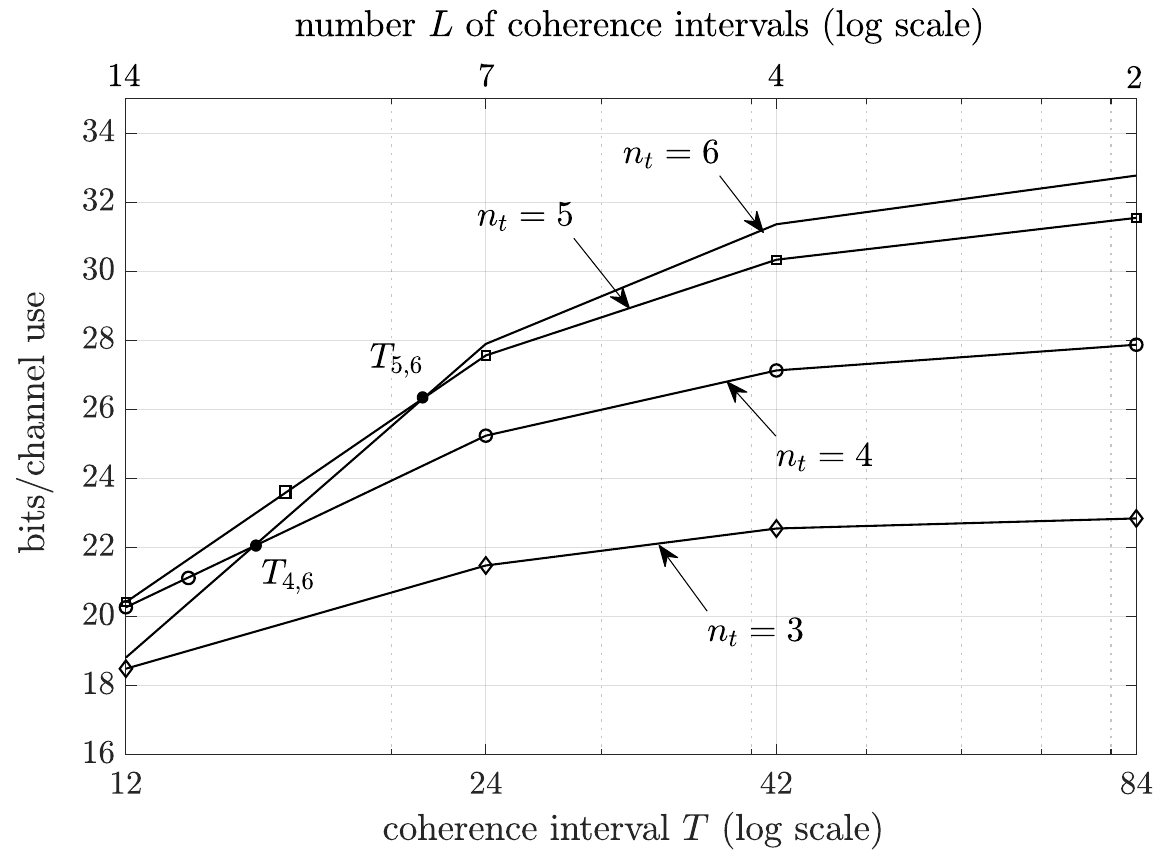}}
	\hspace{2em}
	\subfloat[$n=512$, $\epsilon=10^{-5}$, $\rho=25$~dB, and $n_r=8$.  The crossing points are $T_{6,7}=17.9$, $T_{5,8}=18.3$, $T_{6,8}=22.3$, and $T_{7,8}=32.5$.   \label{Fig_7b}]{\includegraphics[width=0.41\textwidth]{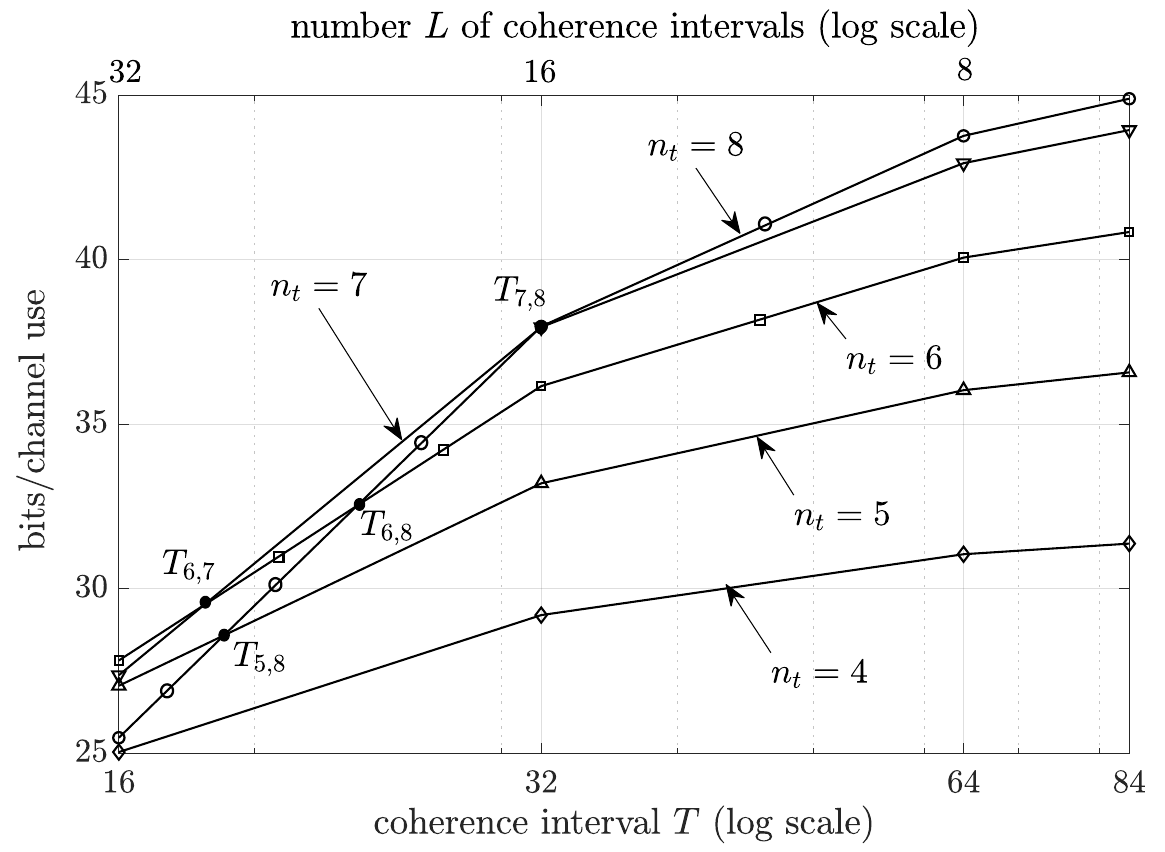} }
	\hspace{0.6em}			
	\caption{High-SNR normal approximation as a function of $T$ for a fixed $n=TL$, a given number of receive antennas, and varying numbers of transmit antennas. } \label{fig:7}
\end{figure} 

In Fig.~\ref{fig:7}, we plot the high-SNR normal approximation as a function of  $T$ for a fixed blocklength $n=LT$, a given number of receive antennas, and varying numbers of transmit antennas. We consider the cases $\rho=25~\textnormal{dB}$, $\epsilon=10^{-3}$, $n=168$, and $n_r=6$ (Fig.~\ref{Fig_7a}), and $\rho=25~\textnormal{dB}$, $\epsilon=10^{-5}$, $n=512$, and $n_r=8$ (Fig.~\ref{Fig_7b}). Comparing the high-SNR normal approximation against the nonasymptotic bounds, it can be observed that, for the considered parameters, the normal approximation deviates from the true value of $R^{\ast}(L, T,\epsilon, \rho)$ by less than $5.2$\%, hence it is a good proxy for the maximum coding rate. We further indicate the values of the coherence interval where two lines cross. Specifically, we use the notation $T_{n_1,n_2}$ to indicate the crossing point of the curves for $n_t=n_1$ and $n_t=n_2$. Observe that, in contrast to the pre-log factor of the high-SNR asymptotic capacity, the number of transmit antennas that maximizes the high-SNR normal approximation is not necessarily equal to the maximum value $n_t=n_r$ and depends on the coherence interval $T$. 		
	For example, according to Fig.~\ref{Fig_7a}, when $n=168$ and $n_r=6$, the maximum number of available transmit antennas should only be used when $T\geq 23$. Furthermore, when $n=512$ and $n_r=8$, using all available transmit antennas is actually suboptimal when $T< 32$; cf.~Fig.~\ref{Fig_7b}.  			
	Finally, the optimal number of active transmit antennas is not necessarily monotonically increasing in $T$. As can be observed from Fig.~\ref{Fig_7b}, when $n=512$ and $n_r=8$, using $n_t=7$ transmit antennas is optimal when $18 \leq T \leq 32$. 
	
	In summary, while the high-SNR asymptotic capacity suggests that, at large blocklengths and high SNR, it is optimal to use all available transmit antennas, in general the maximum coding rate has a more intricate dependence on the number of active transmit antennas. The high-SNR normal approximation presented in Theorem 1 allows us to unveil this dependence without having to resort to nonasymptotic bounds that need to be evaluated numerically at a high computation cost.

\section{Proof of Theorem~\ref{Thm_R}} \label{Sec_Proof}

To prove Theorem~\ref{Thm_R}, we derive in Section~\ref{Sec_DT_Proof} a lower bound on $R^*(L,T,\epsilon,\rho)$ and in Section~\ref{Sec_MC_Proof} an upper bound on $R^*(L,T,\epsilon,\rho)$.  Since both bounds are equal to \eqref{Equ_Thm_R} up to error terms that have the same behavior as $K_{\tilde{I}}(T,\rho)$, $K_{\tilde{V}}(T,\rho)$, and $K_L(L,T,\rho)$ (cf.~\eqref{Equ_R_Inner} and \eqref{eq:MC_FINAL}), we conclude that $R^*(L,T,\epsilon,\rho)$ is given by \eqref{Equ_Thm_R}. 
							
\subsection{Lower Bound on $R^{\ast}(L, T, \epsilon, \rho)$} \label{Sec_DT_Proof}
To derive a lower bound on $R^{\ast}(L, T, \epsilon, \rho)$, we evaluate the DT lower bound \cite[Th. 22]{polyanskiy2010channel} for an USTM input distribution.
Such an input can be written as $\mathbf{X}^L=\sqrt{T\rho/n_t}\mathbf{U}^L$, where $\mathbf{U}^L=[\mathbf{U}_1, \ldots, \mathbf{U}_L]$  is a sequence of i.i.d., isotropically distributed, random matrices satisfying $\mathbf{U}_{\ell}^\mathsf{H}\mathbf{U}_{\ell}=\mathsf{I}_{n_t}$, $\ell=1,\ldots,L$. 
We define the \emph{information density} between the random vectors $\mathbf{X}^L$ and $\mathbf{Y}^L$ as 
\begin{equation}
	i(\mathbf{X}^L;\mathbf{Y}^L) \triangleq \log\left( \frac{f_{\mathbf{Y}^L|\mathbf{X}^L}(\mathbf{Y}^L|\mathbf{X}^L)}{f_{\mathbf{Y}^L}(\mathbf{Y}^L)}  \right)
\end{equation}
where $f_{\mathbf{Y}^L|\mathbf{X}^L}$ denotes the conditional probability density function (pdf) of the channel outputs $\mathbf{Y}^L$ of the channel \eqref{Equ_Channel} given the channel inputs $\mathbf{X}^L$, and $f_{\mathbf{Y}^L}$ is the output pdf induced by the input distribution and the channel law.
When the input distribution is USTM, the channel outputs $\mathbf{Y}^L=[\mathbf{Y}_1, \ldots, \mathbf{Y}_L]$ are i.i.d., $(T\times n_r)$-dimensional random matrices whose joint pdf is given by
\begin{equation}
	\label{eq:UMTS_output}
	f_{\mathbf{Y}^L}(\mathsf{Y}^L)=\prod_{\ell=1}^{L} f_{\mathbf{Y}}^{(U)}(\mathsf{Y}_{\ell}), \quad \mathsf{Y}^L=[\mathsf{Y}_1,\ldots,\mathsf{Y}_L]
\end{equation}
where $f_{\mathbf{Y}}^{(U)}$ denotes the pdf of the outputs of the channel \eqref{Equ_Channel} in each coherence interval induced by USTM channel inputs. Since the channel is blockwise memoryless, the information density can then be expressed as
\begin{equation}
\label{eq:USTM_density}
	i(\mathbf{X}^L;\mathbf{Y}^L) = \sum_{\ell=1}^{L}  i_{\ell}^{(U)}(\mathbf{X}_{\ell};\mathbf{Y}_{\ell})
\end{equation}
where the \emph{USTM information density} $i_{\ell}^{(U)}(\mathbf{X}_{\ell};\mathbf{Y}_{\ell})$ is defined as
\begin{equation}
 i_{\ell}^{(U)}(\mathbf{X}_{\ell};\mathbf{Y}_{\ell})\triangleq \log \frac{f_{\mathbf{Y}|\mathbf{X}}(\mathbf{Y}_{\ell}|\mathbf{X}_{\ell})}{f_{\mathbf{Y}}^{(U)}(\mathbf{Y}_{\ell})} 
\end{equation}
and $f_{\mathbf{Y}|\mathbf{X}}$ denotes  the conditional pdf of the channel output $\mathbf{Y}_{\ell}$ in coherence interval $\ell$ given the corresponding channel input $\mathbf{X}_{\ell}$. 

To apply the DT bound, we first note that the cumulative distribution function $\Pr\big[i(\mathbf{x}^L;\mathbf{Y}^L) \leq \alpha\big]$ for \eqref{eq:USTM_density} and $\mathbf{Y}^L$ distributed according to \eqref{eq:UMTS_output} does not depend on $\mathbf{x}^L$. This follows from the isotropy of USTM inputs combined with the unitary invariance of the output distribution; see also \cite[App.~A]{durisi2015short}. Furthermore, the USTM input distribution satisfies the power constraint \eqref{eq:power} with probability one. Under the above assumptions, the DT bound can be expressed as follows:
 							
\begin{Lemma}[Dependence-Testing Bound]\label{lem:DT_bound}
Fix an arbitrary distribution on $\mathbf{X}^L$. Assume that the cumulative distribution function $\Pr\big[i(\mathbf{x}^L;\mathbf{Y}^L) \leq \alpha\big]$ does not depend on $\mathbf{x}^L$ when $\mathbf{Y}^L$ is distributed according to $f_{\mathbf{Y}^L}$,  and that $\mathbf{X}^L$ satisfies the power constraint \eqref{eq:power} with probability one. Then, there exists a code of blocklength $n=LT$ with $M$ codewords whose maximum error probability satisfies
	\begin{equation}
	\label{eq:lem_DT_bound}
		\epsilon \leq \mathsf{E} \left[ \exp \left\{ -[i(\mathbf{X}^L;\mathbf{Y}^L) - \log(M-1)]^+ \right\} \right].
	\end{equation}
\end{Lemma}
\begin{IEEEproof}
The lemma follows by optimizing the bound in \cite[Th.~23]{polyanskiy2010channel} for the case where $\Pr\big[i(\mathbf{x}^L;\mathbf{Y}^L) \leq \alpha\big]$ does not depend on $\mathbf{x}^L$, as in the proof of \cite[Th.~22]{polyanskiy2010channel}.
\end{IEEEproof}

Distinguishing between the cases where $i(\mathbf{X}^L;\mathbf{Y}^L) > \log (M-1)$ and $i(\mathbf{X}^L;\mathbf{Y}^L) \leq \log (M-1)$, the inequality \eqref{eq:lem_DT_bound} in Lemma~\ref{lem:DT_bound} can be written as
	\begin{equation}
		\epsilon \leq  (M-1) \mathsf{E}\left[ e^{-i(\mathbf{X}^L;\mathbf{Y}^L) } \mathds{1}\left\{i(\mathbf{X}^L;\mathbf{Y}^L) > \log (M-1) \right\}\right] 
		+\Pr\left[i(\mathbf{X}^L;\mathbf{Y}^L) \leq \log (M-1) \right]. \label{Ine_DT}
\end{equation}
Consequently, if we can find an $M$ such that the right-hand side (RHS) of \eqref{Ine_DT} is upper-bounded by $\epsilon$, then there exists an $(L,T,M,\epsilon,\rho)$ code with these parameters. By the definition of the maximum coding rate \eqref{eq:Rstar}, we then obtain the lower bound
\begin{equation}
	R^*(L,T,\epsilon,\rho) \geq \frac{\log M}{LT}.
\end{equation}
To find such an $M$, we shall closely follow \cite[Eqs. (258)--(267)]{polyanskiy2010channel} (see also \cite[Eqs. (69)--(85)]{lancho2019single}). To this end, we will need the following auxiliary results. In the following, and throughout the paper, we shall omit the subscript $\ell$ where it is immaterial. We further define $I(T,\rho)\triangleq\mathsf{E}[i^{(U)}(\mathbf{X};\mathbf{Y})]$ and $\underline{U}(T, \rho)\triangleq \mathsf{E}\bigl[\bigl(i^{(U)}(\mathbf{X};\mathbf{Y})-I(T,\rho)\bigl)^2\bigl]$. Recall that these quantities, as well as the third moment $\mathsf{E}[|i^{(U)}(\mathbf{X};\mathbf{Y})-I(T,\rho)\big|^3]$, are computed for USTM channel inputs.

\begin{Lemma} \label{DT_Lem_1}
	At high SNR, $\underline{U}(T, \rho)$ can be approximated as
	\begin{equation}
		\underline{U}(T, \rho)=T^2\tilde{V}(T)+K_{\underline{U}}(T,\rho)
	\end{equation}
	where $\tilde{V}(T)$ is defined in \eqref{Def_Vt} 
	and $K_{\underline{U}}(T,\rho)$ is a function of $T$ and $\rho$ that satisfies
	\begin{equation}
	\lim\limits_{\rho\to\infty}K_{\underline{U}}(T,\rho)=0.
	\end{equation}
	\begin{IEEEproof}
		See Appendix \ref{Sec_Pf_DTLem1}.
	\end{IEEEproof} 
\end{Lemma}

\begin{Lemma}\label{DT_Lem_23order}
	There exists a sufficiently large $\rho_0$ (that only depends on $T$) such that 
	\begin{IEEEeqnarray}{rCl}
		\sup_{\rho\geq \rho_0} \underline{U}(T, \rho)&< & \infty\\
		\sup_{\rho\geq \rho_0}\mathsf{E}\big[\big|i^{(U)}(\mathbf{X};\mathbf{Y})-I(T,\rho)\big|^3\big] & < & \infty.
	\end{IEEEeqnarray}
	\begin{IEEEproof}
		See Appendix \ref{Sec_Pf_DTLem2}.
	\end{IEEEproof} 
\end{Lemma}
							
Lemma~\ref{DT_Lem_1} implies that there exists a sufficiently large $\rho_0$ (that only depends on $T$) such that
\begin{equation}
	\label{eq:U_LB_obvious}
	\underline{U}(T, \rho)  \geq   \frac{T^2}{2}\tilde{V}(T), \quad \rho\geq\rho_0.
\end{equation}
Furthermore, by Lemma~\ref{DT_Lem_23order}, there exist a sufficiently large $\rho_0$ and an $\underline{S}(T)$ that is independent of $\rho$ such that
\begin{equation}
	\mathsf{E}\big[\big|i^{(U)}(\mathbf{X};\mathbf{Y})-I(T,\rho)\big|^3\big] \leq \underline{S}(T), \qquad \rho\geq \rho_0. \label{Equ_Srho0}
\end{equation}
Combining \eqref{eq:U_LB_obvious} and \eqref{Equ_Srho0}, it then follows that
\begin{equation}
	\frac{6\mathsf{E}\big[ \big|i^{(U)}(\mathbf{X};\mathbf{Y})-I(T,\rho)\big|^3\big]}{\underline{U}(T, \rho)^{3/2}}
	\leq \frac{6\underline{S}(T)}{\left[ \frac{T^2}{2}\tilde{V}(T) \right]^{3/2}}
	\triangleq \underline{B}(T), \quad \rho\geq \rho_0. \label{DT_B}
\end{equation}
The first term on the RHS of \eqref{Ine_DT} can be upper-bounded using the following lemma.

\begin{Lemma}[Polyanskiy-Poor-Verd\'u---Lemma~47] \label{Lem_47}  
	Let $Z_1,Z_2,\ldots,Z_L$  be independent random variables, $\sigma^{2} =\sum_{\ell=1}^{L}\mathrm{Var}(Z_{\ell})$  be nonzero, and $T=\sum_{\ell=1}^{L}\mathsf{E}\left[|Z_{\ell}-\mathsf{E}[Z_{\ell}]|^{3}\right]<\infty$. Then, for any $A$,
	\begin{equation}
		\mathbb{E}\left[\exp\left\{-\sum_{\ell=1}^{L} Z_{\ell}\right\}\mathds{1}\left\{\sum_{\ell=1}^L Z_{\ell}>A\right\}\right] 
		\leq 2\left(\frac{\log2}{\sqrt{2\pi}}+\frac{12T}{\sigma^2}\right)\frac{1}{\sigma}\exp\{-A\}.
	\end{equation}
\end{Lemma}
\begin{IEEEproof}
See \cite[Lemma~47]{polyanskiy2010channel}.
\end{IEEEproof}

Applying Lemma~\ref{Lem_47} to the first term on the RHS of \eqref{Ine_DT}, we obtain that					
\begin{equation}
						(M-1) \mathsf{E}\left[ e^{-i(\mathbf{X}^L;\mathbf{Y}^L) }\mathds{1}\left\{i(\mathbf{X}^L;\mathbf{Y}^L) >  \log(M-1)\right\}\right]
						\leq 2\left(\frac{\log 2}{\sqrt{2\pi}} + 2\underline{B}(T)\right)\frac{1}{\sqrt{L}}. \label{DT_Pr2}
					\end{equation}
					We next use \eqref{DT_B} together with the Berry-Esseen theorem  to upper-bound the second term on the RHS of \eqref{Ine_DT}:
					
	\begin{Lemma}[Berry-Esseen Theorem] \label{Lem_BerryEsseen}  
	Let $Z_1,\ldots,Z_{L}$ be independent with $\mu_{\ell}=\mathsf{E}[Z_{\ell}]$, $\sigma_{\ell}^{2}=\mathrm{Var}[Z_{\ell}]$, $t_{\ell}=\mathsf{E}\left[|Z_{\ell}-\mu_{\ell}|^{3}\right]$, $\sigma^{2}=\sum_{\ell=1}^{L}\sigma_{\ell}^{2}$, and $T=\sum_{\ell=1}^{L}t_{\ell}$. Then, for any $\lambda\in\mathbb{R}$,
	\begin{equation}
		\left|\mathsf{P}\left[\sum_{\ell=1}^{L}(Z_{\ell}-\mu_{\ell})\geq\lambda\sigma\right]-Q(\lambda)\right|\leq\frac{6T}{\sigma^3}.\label{Equ_BerryEsseen}
	\end{equation} 
\end{Lemma}
\begin{IEEEproof}
See \cite[Ch.~XVI.5]{feller71}.
\end{IEEEproof}

Let
\begin{equation}
	\tau=Q^{-1}\left(\epsilon-\left(\frac{2\log 2}{\sqrt{2\pi}} + 5\underline{B}(T)\right)\frac{1}{\sqrt{L}}  \right)
\end{equation}
and set
\begin{equation}
	\log(M-1)=LI(T,\rho)-\tau\sqrt{L\underline{U}(T, \rho)}. \label{eq:Mchoice}
\end{equation}
Then, by setting in \eqref{Equ_BerryEsseen} $Z_{\ell} =  i^{(U)}(\mathbf{X}_{\ell}; \mathbf{Y}_{\ell})$, $\lambda \triangleq \tau$, $T \triangleq L\mathsf{E}\big[\big| i^{(U)}(\mathbf{X};\mathbf{Y}) -I(T,\rho) \big|^3 \big]$, and $\sigma\triangleq \sqrt{L\underline{U}(T, \rho)}$,\footnote{Note that $T$ and $\sigma$ defined in \eqref{Equ_BerryEsseen} are different from the definitions in this paper.} the Berry-Esseen theorem applied to the second term on the RHS of \eqref{Ine_DT} yields that
\begin{IEEEeqnarray}{lCl} 
	\Pr\left[i(\mathbf{X}^L;\mathbf{Y}^L) \leq \log(M-1)\right] & \leq & Q(\tau) + \frac{6\mathsf{E}\big[\big| i^{(U)}(\mathbf{X};\mathbf{Y}) -I(T,\rho) \big|^3 \big]}{\sqrt{L}\big( \underline{U}(T, \rho)\big)^{3/2}}  \nonumber\\
	& \leq &\epsilon-\left(\frac{2\log 2}{\sqrt{2\pi}} + 5\underline{B}(T)\right)\frac{1}{\sqrt{L}} + \frac{\underline{B}(T)}{\sqrt{L}} \nonumber\\
	& =& \epsilon-2\left(\frac{\log 2}{\sqrt{2\pi}} + 2\underline{B}(T)\right)\frac{1}{\sqrt{L}} \label{DT_Pr1}
\end{IEEEeqnarray}
where the second step follows from \eqref{DT_B} and the definition of $\tau$.
It can be checked that the sum of the RHSs of \eqref{DT_Pr1} and \eqref{DT_Pr2} is equal to $\epsilon$. Consequently, there exists an $(L,T,M,\epsilon,\rho)$ code with $M$ given in \eqref{eq:Mchoice} and we have
\begin{IEEEeqnarray}{lCl}
	R^{\ast}(L, T, \epsilon, \rho)
	& \geq & \frac{\log(M-1)}{LT} \nonumber\\
	& = & \frac{I(T,\rho)}{T}-\sqrt{\frac{\underline{U}(T, \rho)}{LT^2}}Q^{-1}\left(\epsilon-\left(\frac{2\log 2}{\sqrt{2\pi}} + 5\underline{B}(T)\right)\frac{1}{\sqrt{L}}\right). \label{Equ_R_DT1}
\end{IEEEeqnarray}
A Taylor-series expansion of $Q^{-1}(\cdot)$ around $\epsilon$ yields that
\begin{align}
	\left| Q^{-1}\left(\epsilon-\left(\frac{2\log 2}{\sqrt{2\pi}} + 5\underline{B}(T)\right)\frac{1}{\sqrt{L}}\right)-Q^{-1}(\epsilon)\right| \leq \frac{A}{\sqrt{L}}, \quad L\geq L_0 \label{DT_Qfun}
\end{align}
for some constants $A$ and $L_0$ that are independent of $L$ and $\rho$. 
Furthermore, Lemma 2 shows that $\underline{U}(T,\rho)$ is uniformly bounded in $\rho\geq\rho_0$.
Thus, combining \eqref{Equ_R_DT1} and \eqref{DT_Qfun}, we obtain
\begin{equation}
	R^{\ast}(L, T, \epsilon, \rho)
	\geq \frac{I(T,\rho)}{T}-\sqrt{\frac{\underline{U}(T, \rho)}{LT^2}}Q^{-1}(\epsilon) + K_{\textnormal{DT}}(L,T,\rho) \label{Equ_R_DT2}
\end{equation}  
where $K_{\textnormal{DT}}(L,T,\rho)$ is a function of $L$, $T$, and $\rho$ that satisfies 
\begin{align}
	\sup_{\rho\geq \rho_0} \big| K_{\textnormal{DT}}(L,T,\rho)\big| \leq \frac{A}{L}, \quad L\geq L_0 \label{Equ_KDT_1}
\end{align}
for some constants $A$, $L_0$, and $\rho_0$ that are independent of $L$ and $\rho$.

The expressions $I(T,\rho)/T$ and $\underline{U}(T, \rho)/T^2$ can be approximated at high SNR as follows:
\begin{IEEEeqnarray}{lCl}
	\frac{I(T,\rho)}{T} & = & \tilde{I}(T,\rho)+\tilde{K}_{{I}}(T,\rho) \label{Def_I_HS} \\
	\frac{\underline{U}(T, \rho)}{T^2} & = & \tilde{V}(T)+\tilde{K}_{\underline{U}}(T,\rho)\label{Def_U_HS}
\end{IEEEeqnarray}
where $\tilde{I}(T,\rho)$ and $\tilde{V}(T)$ are given in \eqref{Def_It} and \eqref{Def_Vt}, and where $\tilde{K}_{\xi}(T,\rho)$, $\xi\in\{{I},\underline{U}\}$ are functions of $T$ and $\rho$ that both satisfy $\lim\limits_{\rho\to\infty}\tilde{K}_{\xi}(T,\rho)=0$.
Here, \eqref{Def_I_HS} follows from \cite[Eq. (49)]{yang2013capacity}, which studied the channel capacity of noncoherent MIMO Rayleigh block-fading channels and demonstrated that, at high SNR, $I(T,\rho)/T$ can be approximated by \eqref{Def_It};  \eqref{Def_U_HS} follows from Lemma~\ref{DT_Lem_1}. Combining \eqref{Def_I_HS} and \eqref{Def_U_HS} with \eqref{Equ_R_DT2}, we obtain that
\begin{equation}
	R^{\ast}(L, T, \epsilon, \rho) \geq \tilde{I}(T,\rho) +\tilde{K}_{{I}}(T,\rho) -\sqrt{\frac{\tilde{V}(T)+\tilde{K}_{\underline{U}}(T,\rho)}{L}} Q^{-1}(\epsilon)+K_{\textnormal{DT}}(L,T,\rho). \label{Equ_R_Inner}
\end{equation}
Note that $\tilde{K}_{\xi}(T,\rho)$, $\xi\in\{I,\underline{U}\}$ satisfy \eqref{eq:thm_little_o} and $K_{\textnormal{DT}}(L,T,\rho)$ satisfies \eqref{eq:thm_big_O}.

\subsection{Upper Bound on $R^{\ast}(L, T, \epsilon, \rho)$} \label{Sec_MC_Proof}
Our upper bound on $R^{\ast}(L, T, \epsilon, \rho)$ is based on the MC bound:

\begin{Lemma}[Meta-Converse Bound]
The cardinality of a codebook $\mathcal{C}$ with codewords $(\mathsf{X}_1,\ldots,\mathsf{X}_L)$ belonging to a set $\mathcal{F}$ and maximum error probability not exceeding $\epsilon$ satisfies
\begin{equation}
	\label{eq:MC}
	\log|\mathcal{C}| \leq \sup_{\mathsf{X}^L\in\mathcal{F}} \log\frac{1}{\beta_{1-\epsilon}(\mathsf{X}^L,q_{\mathbf{Y}^L})}
\end{equation}
where $\beta_{1-\epsilon}(\mathsf{X}^L,q_{\mathbf{Y}^L})$ denotes the minimum probability of error of a binary hypothesis test under hypothesis $q_{\mathbf{Y}^L}$ if the probability of error under hypothesis $f_{\mathbf{Y}^L|\mathbf{X}^L}$ does not exceed $\epsilon$ \cite[Eq.~(100)]{polyanskiy2010channel}, and $q_{\mathbf{Y}^L}$ is an auxiliary pdf on the output alphabet. 
\end{Lemma}
\begin{IEEEproof}
This lemma follows by considering a binary hypothesis test between  $f_{\mathbf{Y}^L|\mathbf{X}^L}$ and $q_{\mathbf{Y}^L}$.  Intuitively, testing whether the channel outputs $\mathbf{Y}^L$ are generated by transmitting a given codeword $\mathbf{x}^L$ over the channel  $f_{\mathbf{Y}^L|\mathbf{X}^L}$ can be viewed as a hypothesis test that decides between two hypotheses:
\begin{itemize}
\item $H_0$: the channel outputs $\mathbf{Y}^L$ are distributed according to the conditional distribution of the channel outputs given the inputs $\mathbf{x}^L$;
\item $H_1$: the channel outputs $\mathbf{Y}^L$ are independent of the inputs $\mathbf{x}^L$.
\end{itemize}
The meta-converse bound follows then because the best decoder cannot outperform the best hypothesis test. For a proof, see \cite[Th.~31]{polyanskiy2010channel}.
\end{IEEEproof}

The RHS of \eqref{eq:MC} cannot be evaluated in closed form, since characterizing $\beta_{1-\epsilon}(\mathsf{X}^L,q_{\mathbf{Y}^L})$ asymptotically as $L\to\infty$ and $\rho\to\infty$ is intractable. To sidestep this problem, we use \cite[Eq.~(106)]{polyanskiy2010channel} to lower-bound 
\begin{equation} 
	\beta_{1-\epsilon}(\mathsf{X}^L,q_{\mathbf{Y}^L})\geq\frac{1}{\xi(\mathsf{X}^L)}\left(1-\epsilon-\Pr\biggl[\frac{f_{{\mathbf{Y}^L}|\mathbf{X}^L}(\mathbf{Y}^L|\mathbf{X}^L)}{q_{\mathbf{Y}^L}(\mathbf{Y}^L)}\geq\xi(\mathsf{X}^L)\biggm| \mathbf{X}^L = \mathsf{X}^L\biggr]\right)
\end{equation}							
for arbitrary $\xi(\mathsf{X}^L)>0$ which may depend on $\mathsf{X}^L$. Together with \eqref{eq:MC}, this gives rise to the weakened MC bound							
\begin{equation} \label{eq:MC_weak}
	M\leq\sup_{\mathsf{X}^L\in\mathcal{F}}\left\{\log\xi(\mathsf{X}^L)-\log\left(1-\epsilon-\Pr\Bigl[j(\mathbf{X}^L;\mathbf{Y}^L)\geq\log\xi(\mathsf{X}^L)\Bigm| \mathbf{X}^L=\mathsf{X}^L\Bigr]\right)\right\}
\end{equation}
where 
\begin{equation} \label{eq:jL}
	j(\mathbf{X}^L;\mathbf{Y}^L) \triangleq  \sum_{\ell=1}^{L} \log\left( \frac{f_{\mathbf{Y}|\mathbf{X}}(\mathbf{Y}_{\ell}|\mathbf{X}_{\ell})}{q_{\mathbf{Y}}(\mathbf{Y}_{\ell})}  \right)
\end{equation}
is the so-called \emph{mismatched information density}\footnote{We use the word ``mismatched" to indicate that the output distribution $q_{\mathbf{Y}}$ is not the one induced by the input distribution and the channel.}. 

In \eqref{eq:MC}, we choose the auxiliary pdf $q_{\mathbf{Y}}$ that was chosen in \cite[Sec.~IV-A]{yang2013capacity} to derive an upper bound on the channel capacity of noncoherent MIMO Rayleigh block-fading channels. That is, $q_{\mathbf{Y}^L}=q_{\mathbf{Y}}\times\ldots \times q_{\mathbf{Y}}$ is a product distribution with 
\begin{equation}
	q_{\mathbf{Y}}(\mathsf{Y})=\frac{1}{\mu^{n_tn_r}\pi^{n_rT}}\frac{\Gamma_{n_t}(T)}{\Gamma_{n_t}(n_t)} 
	e^{-\sum_{i=1}^{n_t}\frac{\sigma_i^2}{\mu}} e^{-\sum_{i=n_t+1}^{n_r}\sigma_i^2}
	\frac{1}{\big(\prod_{i=1}^{n_t}\sigma_i^2\big)^{T-2n_r+n_t}} 
	\frac{1}{\prod_{i=1}^{n_t}\prod_{j=n_t+1}^{n_r}(\sigma_i^2-\sigma_j^2)^2} \label{Equ_fY_outer} 
\end{equation}
where $\sigma_1 \geq \ldots \geq \sigma_{n_r}$ are the ordered nonzero singular values of $\mathsf{Y}$ and $\mu\triangleq T\rho/n_t$.

We next note that, for every $T\times T$ unitary matrix $\mathsf{A}$ and every $\mathsf{X}$ and $\mathsf{Y}$, \cite[Eqs.~(54) \& (55)]{durisi2015short}
\begin{subequations}
\begin{IEEEeqnarray}{rCl}  
	f_{\mathbf{Y}|\mathbf{X}}(\mathsf{Y}_{\ell}|\mathsf{A}^{\mathsf{H}}\mathsf{X}_{\ell}) & = &  f_{\mathbf{Y}|\mathbf{X}}(\mathsf{A}\mathsf{Y}_{\ell}|\mathsf{X}_{\ell}) \\
	q_{\mathbf{Y}}(\mathsf{A}\mathsf{Y}) & = & q_{\mathbf{Y}}(\mathsf{Y}).
\end{IEEEeqnarray}
\end{subequations}
It then follows from \cite[Prop.~19]{polyanskiy2013saddlepoint} that $\beta_{1-\epsilon}(\mathsf{X}^L,q_{\mathbf{Y}^L})$ does not change if we multiply the channel input $\mathsf{X}_{\ell}$ by a unitary matrix $\mathsf{A}_{\ell}$. We further note that
\begin{equation} \label{Equ_fY_x} 
	f_{\mathbf{Y}|\mathbf{X}}(\mathsf{Y}_{\ell}|\mathsf{X}_{\ell}) = \frac{e^{-\text{tr}\left(\mathsf{Y}_{\ell}^{\mathsf{H}}(\mathsf{I}_{T}+ \mathsf{X}_{\ell}\mathsf{X}_{\ell}^{\mathsf{H}})^{-1}\mathsf{Y}_{\ell}\right)}}{\pi^{n_r T} \det(\mathsf{I}_{T}+\mathsf{X}_{\ell}\mathsf{X}_{\ell}^{\mathsf{H}})^{n_r}}
\end{equation}  
and, hence, also $\beta_{1-\epsilon}(\mathsf{X}^L,q_{\mathbf{Y}^L})$ depends on $\mathsf{X}_{\ell}$ only via the product $\mathsf{X}_{\ell} \mathsf{X}_{\ell}^{\mathsf{H}}$. By expressing $\mathsf{X}_{\ell}$ in terms of its singular value decomposition (SVD)
\begin{equation}
	\mathsf{X}_{\ell} = \mathsf{U}_{\ell} \Sigma_{\ell} \mathsf{V}_{\ell}^{\mathsf{H}} \label{equ_Xl}
\end{equation}
(where $\mathsf{U}_{\ell}$ and $\mathsf{V}_{\ell}$ are unitary matrices of dimensions $T\times T$ and $n_t\times n_t$, respectively, and $\Sigma_{\ell}$ is a $T\times n_t$ rectangular diagonal matrix), and by setting $\mathsf{A}_\ell = \mathsf{U}_\ell$, we thus obtain that we can restrict ourselves without loss of optimality to channel inputs $\mathsf{X}_{\ell}$ that are given by a $T\times n_t$ rectangular diagonal matrix with diagonal entries $d_{1,\ell},\ldots,d_{n_t,\ell}$. Intuitively, $d_{i,\ell}^2$ denotes the power at transmit antenna $i=1,\ldots,n_t$ and coherence interval $\ell=1,\ldots,L$. Let $\mathsf{D}_{\ell} \triangleq  \text{diag}\bigl\{d_{1,\ell},\ldots,d_{n_t,\ell}\bigr\}$ and $\alpha_\ell\triangleq \text{tr}(\mathsf{D}_{\ell}^2)/T$. Thus, $\mathsf{D}_{\ell}$ is the diagonal square matrix that is composed of the first $n_t$ columns and rows of $\mathsf{X}_{\ell}$, i.e., $\mathsf{D}_{\ell}=[\mathsf{X}_{\ell}]_{(1:n_t)^2}$ (where we use the notation $[\mathsf{X}_{\ell}]_{(1:n_t)^2}$ to denote the diagonal submatrix that is composed of the first $n_t$ columns and rows of $\mathsf{X}_{\ell}$; see also the notation subsection at the end of Section~\ref{Sec_Introduction}). By the power constraint \eqref{eq:power}, we have for every $\ell$ that $0\leq \alpha_{\ell} \leq \rho$.

Since directly maximizing the weakened MC bound over $\mathsf{X}^L$ is challenging, we wish to first derive an asymptotic expansion (in the limit as $L\to\infty$) of the conditional cumulative distribution function of $j(\mathbf{X}^L;\mathbf{Y}^L)$ by resorting to the Berry-Esseen theorem (Lemma~\ref{Lem_BerryEsseen}). Maximizing the resulting expansion over $\mathsf{X}^L$ is then feasible. However, the Berry-Esseen theorem only yields a meaningful asymptotic expansion if the absolute third central moment of $j(\mathbf{X}^L;\mathbf{Y}^L)$ divided by its variance to the power of $3/2$ is bounded in $\mathsf{X}^L\in\mathcal{F}$. Unfortunately, this is not the case if the norm of $\mathsf{X}^L$ is small, since the variance of $j(\mathbf{X}^L;\mathbf{Y}^L)$ vanishes as $\mathsf{X}^L$ tends to zero. To avoid this issue, we follow the idea in \cite{Collins2019coherent} and separate the codebook $\mathcal{C}$ into two sub-codebooks $\mathcal{C}_1$ and $\mathcal{C}_2$, where $\mathcal{C}_1$ contains all the codewords for which $d_{i,\ell}^2>\bar{\delta} \rho$, $i=1,\cdots,n_t$ (with $\bar{\delta}$ defined in \eqref{eq:delta_bar}) in at least half of the coherence intervals, and where $\mathcal{C}_2$ contains the remaining codewords. Clearly, if the maximum error probabilities of $\mathcal{C}_1$ and $\mathcal{C}_2$ are $\epsilon$, then the maximum error probability of $\mathcal{C}$ cannot be smaller than $\epsilon$. For codewords in $\mathcal{C}_1$, the variance of $j(\mathbf{X}^L;\mathbf{Y}^L)$ is bounded away from zero, so the Berry-Esseen theorem can be applied to obtain a meaningful asymptotic expansion. More precisely, the Berry-Esseen ratio $\bar{B}(\mathsf{D}^L,T,\rho)$ (see \eqref{Equ_converse_BEratio} below)  can be upper-bounded by a positive value. For codewords in $\mathcal{C}_2$, we apply Chebyshev's inequality, which is less precise but applies to all values of $\mathsf{X}^L$. We then show that, as $L\to\infty$, the cardinality of the entire codebook $\mathcal{C}$ is dominated by the cardinality of $\mathcal{C}_1$, so $R^{\ast}(L, T, \epsilon, \rho)$ is asymptotically upper-bounded by the upper bound on the maximum coding rate of $\mathcal{C}_1$ and a less precise analysis for codewords in $\mathcal{C}_2$ is unproblematic.
	 													
In the following, we provide a detailed proof of the upper bound on $R^{\ast}(L, T, \epsilon, \rho)$. We begin by dividing the codebook $\mathcal{C}$ into the sub-codebooks $\mathcal{C}_1$ and $\mathcal{C}_2$ and upper-bound the cardinalities of both codebooks using the weakened MC bound \eqref{eq:MC_weak}
\begin{equation}
	\log |\mathcal{C}_i| 
	\leq  \sup_{\mathsf{X}^L\in\mathcal{C}_i}  \bigg\{ \log \xi(\mathsf{D}^L) 
	-\log\bigg(1-\epsilon-\Pr\Bigl[j(\mathbf{X}^L;\mathbf{Y}^L)  \geq \log \xi(\mathsf{D}^L) \Bigm| \mathbf{X}^L = \mathsf{X}^L\Bigr] \bigg)  \bigg\}, \quad i=1,2  \label{eq:weakened_MC}
\end{equation}
where (with a slight abuse of notation) we replaced $\xi(\mathsf{X}^L)$ in \eqref{eq:MC_weak} by $\xi(\mathsf{D}^L)$. Thus, $\xi(\mathsf{D}^L)>0$ is an arbitrary threshold which may depend on $\mathsf{D}^L$ and which we shall define later. We conclude by showing that, when both $\rho$ and $L$ tend to infinity, the cardinality of the entire codebook $\mathcal{C}$ can be approximated by the cardinality of $\mathcal{C}_1$. Consequently, $R^{\ast}(L, T, \epsilon, \rho)$ is asymptotically upper-bounded by the upper bound on the maximum coding rate of $\mathcal{C}_1$.

To define $\mathcal{C}_1$ and $\mathcal{C}_2$ mathematically, we first introduce the sets 
\begin{subequations}
\begin{IEEEeqnarray}{lCl} 
	\mathcal{D}_1 & \triangleq & \big\{\mathsf{D}\in \mathcal{D}_{n_t}:  d_{i}^2 > \bar{\delta} \rho, ~i = 1,\ldots,n_t\big\}, \label{Def_D1}\\
	\mathcal{D}_2 & \triangleq & \mathcal{D}_{n_t} \setminus \mathcal{D}_1, \label{Def_D2}
\end{IEEEeqnarray}
\end{subequations}
where $\mathcal{D}_{n_t}$ denotes the set of $(n_t\times n_t)$-dimensional diagonal matrices $\mathsf{D}$ with non-negative, real-valued entries that satisfy $\text{tr}(\mathsf{D}^2) \leq T\rho$; and $d_{i}$ denotes the $i$-th diagonal element of $\mathsf{D}$. Further let
\begin{equation}
	\label{eq:delta_bar}
	\bar{\delta}\triangleq \frac{T}{n_t}-\frac{T}{2n_r \sqrt{n_t}\sqrt{\mathsf{E}\big[\log \det( \mathbf{H}_{\ell} \mathbf{H}_{\ell}^\mathsf{H})^2\big]}+1}.
\end{equation}		
This choice of $\bar{\delta}$ satisfies the power constraint $n_t \bar{\delta}\rho\leq T\rho$ and ensures that the Berry-Esseen ratio $\bar{B}(\mathsf{D}^L,T,\rho)$ (see \eqref{Equ_converse_BEratio} below) is bounded. We further argue that $\bar{\delta}>0$ by evaluating $\mathsf{E}[( \log\det(\mathbf{H}_{\ell} \mathbf{H}_{\ell}^\mathsf{H}))^2]$ using the following lemma.

\begin{Lemma}[Wishart matrices]\label{lem:grant_wishart}
Let $\mathbf{W}$ be a  Wishart matrix with distribution $\mathcal{W}(m, n)$, $m\leq n$, i.e., $\mathbf{W}=\mathbf{Z}\mathbf{Z}^\mathsf{H}$, where $\mathbf{Z}$ is an $(m\times n)$-dimensional matrix with i.i.d.\ $\mathcal{CN}(0,1)$ entries. Then  
\begin{IEEEeqnarray}{rCl}
	\mathsf{E}[\log \det \mathbf{W}] &=& \sum_{i=0}^{m-1} \varPsi(n-i)  \\
	\mathrm{Var}\big(\log \det \mathbf{W}  \big) & = &\sum_{i=0}^{m-1} \varPsi'(n-i).
\end{IEEEeqnarray}
\end{Lemma}
\begin{IEEEproof}
See \cite[Lemma~A.2]{grant2002rayleigh}.
\end{IEEEproof}

Lemma~\ref{lem:grant_wishart} yields that
\begin{equation}
	\mathsf{E}\bigl[\bigl( \log\det(\mathbf{H}_{\ell} \mathbf{H}_{\ell}^\mathsf{H})\bigr)^2\bigr]=\mathsf{E}\bigl[\log \det ( \mathbf{H}_{\ell} \mathbf{H}_{\ell}^\mathsf{H})\bigr]^2+\mathrm{Var}\bigl(\log \det ( \mathbf{H}_{\ell} \mathbf{H}_{\ell}^\mathsf{H})\bigr)=\left(  \sum_{i=0}^{n_t-1} \varPsi(n_t-i)\right)^2+\sum_{i=0}^{n_t-1} \varPsi'(n_t-i). \label{eq:ElogdetHH}
\end{equation}
Since $n_t\leq n_r$ (by assumption) and $\varPsi'(1)=\pi^2/6$ \cite[Th. 2.11]{tulino2004random}, it follows that $\bar{\delta}>0$.

For a sequence of matrices $\mathsf{D}^L=(\mathsf{D}_1,\ldots,\mathsf{D}_L)$, let
\begin{equation}
	L_{\mathcal{D}_1}(\mathsf{D}^L) \triangleq\sum_{\ell=1}^L \mathds{1}\{\mathsf{D}_{\ell}\in \mathcal{D}_1 \}.
\end{equation}
Then, we define the two sub-codebooks $\mathcal{C}_1$ and $\mathcal{C}_2$ as
\begin{subequations}
	\begin{IEEEeqnarray}{lCl}
		\mathcal{C}_1 & \triangleq & \big\{\mathsf{X}^L\in \mathcal{C}\colon L_{\mathcal{D}_1}(\mathsf{D}^L)\geq L/2, \mathsf{D}_{\ell} = [\mathsf{X}_{\ell}]_{(1:n_t)^2} \big\}  \\
		\mathcal{C}_2 & \triangleq &\big\{\mathsf{X}^L\in \mathcal{C}\colon L_{\mathcal{D}_1}(\mathsf{D}^L)< L/2,\mathsf{D}_{\ell} = [\mathsf{X}_{\ell}]_{(1:n_t)^2} \big\}.
	\end{IEEEeqnarray}
\end{subequations}
In words, the sub-codebook $\mathcal{C}_1$ contains all codewords $\mathsf{X}^L$ for which the diagonal entries satisfy $d^2_{i,\ell} > \bar{\delta}\rho$ for all $i=1,\ldots,n_t$ and for at least half of the time instants $\ell=1,\ldots,L$.

We next derive upper bounds on the cardinalities of $\mathcal{C}_1$ and $\mathcal{C}_2$.

\subsubsection{Upper bound on the cardinality of $\mathcal{C}_1$}
An upper bound on $\log |\mathcal{C}_1|$ follows from the weakened MC bound \eqref{eq:weakened_MC}. To obtain a tractable expression, we further upper-bound \eqref{eq:weakened_MC} by upper-bounding the probability
\begin{equation}
	\Pr\biggl[j(\mathbf{X}^L;\mathbf{Y}^L)  \geq \log \xi(\mathsf{D}^L) \biggm| \mathbf{X}^L = \mathsf{X}^L\biggr].
\end{equation}
Indeed, as argued above, we can assume without loss of optimality that $\mathsf{X}_{\ell}$ is given by a $T\times n_t$ rectangular diagonal matrix with diagonal entries $d_{1,\ell},\ldots,d_{n_t,\ell}$. By substituting $q_{\mathbf{Y}}(\mathsf{Y})$ in \eqref{Equ_fY_outer} and $f_{\mathbf{Y}|\mathbf{X}}(\mathsf{Y}_{\ell}|\mathsf{X}_{\ell})$ in \eqref{Equ_fY_x} into \eqref{eq:jL}, we then obtain that
\begin{equation}
	j(\mathbf{X}^L;\mathbf{Y}^L) = \sum_{\ell=1}^{L}  j_{\ell}(\mathbf{X}_{\ell};\mathbf{Y}_{\ell})
\end{equation}
where
\begin{IEEEeqnarray}{lCl}
	j_{\ell}(\mathbf{X}_{\ell};\mathbf{Y}_{\ell})  & = & n_t n_r\log \left(\frac{T\rho}{n_t}\right)+\log \left( \frac{\Gamma_{n_t}(n_t)}{\Gamma_{n_t}(T)} \right) 	
	+ \frac{n_t}{T\rho} \sum_{i=1}^{n_t} \boldsymbol{\sigma}_{i,\ell}^2 + \sum_{i=n_t+1}^{n_r}\boldsymbol{\sigma}_{i,\ell}^2 \nonumber\\
	& & {} - \text{tr}\left(\mathbf{Y}_{\ell}^{\mathsf{H}}\left(\mathsf{I}_{T}+\mathbf{X}_{\ell}\mathbf{X}_{\ell}^{\mathsf{H}}\right)^{-1}\mathbf{Y}_{\ell}\right) - n_r \log\det\big(\mathsf{I}_{n_t}+\mathbf{X}_{\ell}\mathbf{X}_{\ell}^{\mathsf{H}} \big) \nonumber\\
	& & {} + (T-2n_r+n_t)\log \left(\prod_{i=1}^{n_t}{\boldsymbol{\sigma}}_{i,\ell}^2\right) + 2\log\left( \prod_{i=1}^{n_t}\prod_{j=n_t+1}^{n_r}({\boldsymbol{\sigma}}_{i,\ell}^2-{\boldsymbol{\sigma}}_{j,\ell}^2)\right). \label{eq:def_j(D,s)}
\end{IEEEeqnarray}
In \eqref{eq:def_j(D,s)}, $\boldsymbol{\sigma}_{\ell}=(\boldsymbol{\sigma}_{1,\ell},\ldots,\boldsymbol{\sigma}_{n_r,\ell})$ are the ordered singular values of $\mathbf{Y}_{\ell}$.

We show in Appendix~\ref{Sec_UB_j} that  $j_{\ell}(\mathbf{X}_{\ell};\mathbf{Y}_{\ell})$ can be upper-bounded by a term that, conditioned on $\mathbf{X}_{\ell}=\mathsf{X}_{\ell}$, has the same distribution as  
\begin{IEEEeqnarray}{lCl}
	 \bar{j}(\mathsf{D}_{\ell}, \mathbf{Z}_{\ell}',\mathbf{Z}_{\ell}'',\mathbf{H}_{\ell},\mathbf{Q}_{\ell})
	& \triangleq & n_tn_r\log \left(\frac{T\rho}{n_t}\right) 
	+\log \left( \frac{\Gamma_{n_t}(n_t)}{\Gamma_{n_t}(T)} \right)
	+\sum_{i=1}^{n_t} \left( \frac{n_t + n_t d_{i,\ell}^2}{T\rho} -1 \right)Z'_{i, \ell}  -  \frac{T\rho-n_t}{T\rho}\sum_{i=1}^{n_t}Z_{i,\ell}'' \nonumber\\
	& & {} 
	+(T-n_t-n_r) \log\det\bigl(\mathsf{I}_{n_t}+ \mathsf{D}^2_{\ell}\bigr)  +(T-n_t)\log \det \Bigl( \mathbf{H}_{\ell} \mathbf{H}_{\ell}^\mathsf{H}+  \lambda_1(\mathbf{Q}_{\ell}^\mathsf{H}\mathbf{Q}_{\ell})\bigl(\mathsf{I}_{n_t}+\mathsf{D}^2_{\ell}\bigr)^{-1}  \Bigr) \IEEEeqnarraynumspace \label{MC_Def_j_up} 
\end{IEEEeqnarray} 
 where $\mathbf{Z}_{\ell}' = \{Z_{1,\ell}',\ldots, Z_{n_t,\ell}'\}$ and $\mathbf{Z}_{\ell}''=\{Z_{1,\ell}'',\ldots, Z_{n_t,\ell}''\}$ are both sequences of i.i.d. random variables, the former distributed according to the gamma distribution $\Gamma(n_r,1)$ and satisfying $\sum_{i=1}^{n_t} Z_{i,\ell}' = \textnormal{tr}(\mathbf{H}_{\ell}^{\mathsf{H}}\mathbf{H}_{\ell})$, and the latter distributed according to the gamma distribution $\Gamma(T-n_t,1)$; $\mathbf{Q}_{\ell}$ is an $(T-n_t)\times n_r$ random matrices with i.i.d. $\mathcal{CN}(0,1)$ entries that are independent of $\mathbf{H}_{\ell}$; and $\lambda_1(\mathbf{Q}_{\ell}^\mathsf{H}\mathbf{Q}_{\ell})$ denotes the largest eigenvalue of $\mathbf{Q}_{\ell}^\mathsf{H}\mathbf{Q}_{\ell}$.  Furthermore, the sequences $\{\mathbf{Z}_1',\ldots,\mathbf{Z}_L'\}$, $\{\mathbf{Z}_1'',\ldots,\mathbf{Z}_L''\}$, and $\{\mathbf{Q}_1,\ldots,\mathbf{Q}_L\}$ are all i.i.d., and $\{\mathbf{Z}_{1}'',\ldots, \mathbf{Z}_{L}''\}$ and $\{\mathbf{H}^L,\mathbf{Z}_{1}',\ldots, \mathbf{Z}_{L}'\}$ are independent of each other. Since $\bar{j}_{\ell}(\mathsf{D}_{\ell}, \mathbf{Z}_{\ell}',\mathbf{Z}_{\ell}'',\mathbf{H}_{\ell},\mathbf{Q}_{\ell})$ depends on $\mathsf{X}^L$ only through $\mathsf{D}^L$, it then follows that
\begin{equation}
	\label{eq:Prob_UB_jbar}
	\Pr\Biggl[ \sum_{\ell=1}^{L} j_{\ell}(\mathbf{X}_{\ell};\mathbf{Y}_{\ell}) \geq \log \xi(\mathsf{D}^L) \Biggm| \mathbf{X}^L = \mathsf{X}^L\Biggr] \leq \Pr\left[\sum_{\ell=1}^{L} \bar{j}(\mathsf{D}_{\ell}, \mathbf{Z}_{\ell}',\mathbf{Z}_{\ell}'',\mathbf{H}_{\ell},\mathbf{Q}_{\ell})\geq \log\xi(\mathsf{D}^L)\right].
\end{equation}
For future reference, the expected value of $\bar{j}(\mathsf{D}_{\ell}, \mathbf{Z}_{\ell}',\mathbf{Z}_{\ell}'',\mathbf{H}_{\ell},\mathbf{Q}_{\ell})$ is given by
\begin{IEEEeqnarray}{lCl} 
\bar{J}(\mathsf{D}_{\ell},T,\rho)
& =& n_tn_r\log \left(\frac{T\rho}{n_t}\right)
+\log \left(\frac{\Gamma_{n_t}(n_t)}{\Gamma_{n_t}(T)} \right)
+\sum_{i=1}^{n_t} \left( \frac{n_t+n_t d_{i,\ell}^2 }{T\rho} -1 \right)n_r    -\frac{T\rho-n_t}{T\rho}n_t(T-n_t) \nonumber\\
& & {}
+(T-n_t-n_r)\log\det\left(\mathsf{I}_{n_t}+  \mathsf{D}^2_{\ell}\right) +(T-n_t)\mathsf{E}\left[ \log \det \Bigl( \mathbf{H}_{\ell} \mathbf{H}_{\ell}^\mathsf{H}+  \lambda_1(\mathbf{Q}_{\ell}^\mathsf{H}\mathbf{Q}_{\ell})\bigl(\mathsf{I}_{n_t}+ \mathsf{D}^2_{\ell}\bigr)^{-1}  \Bigr) \right]\label{MC_Def_J_up}
\end{IEEEeqnarray}
since $Z_{i,\ell}'$ and $Z_{i,\ell}''$ are Gamma distributed with means $\mathsf{E}[Z_{i,\ell}'] = n_r$ and $\mathsf{E}[Z_{i,\ell}'']=T-n_t$, respectively.

Combining \eqref{eq:Prob_UB_jbar} with \eqref{eq:weakened_MC}, we can upper-bound the cardinality of the sub-codebook $\mathcal{C}_1$ as
\begin{equation}
\log |\mathcal{C}_1| 
\leq  \sup_{\mathsf{X}^L\in\mathcal{C}_1}  \left\{ \log \xi(\mathsf{D}^L)
-\log\left(1-\epsilon-\Pr\left[\sum_{\ell=1}^{L} \bar{j}(\mathsf{D}_{\ell}, \mathbf{Z}_{\ell}',\mathbf{Z}_{\ell}'',\mathbf{H}_{\ell},\mathbf{Q}_{\ell})\geq \log \xi(\mathsf{D}^L) \right] \right)  \right\}  \label{Equ_R_MC} 
\end{equation} 
for every $\xi: \mathcal{D}_{n_t}^L\rightarrow (0,\infty)$. We next perform an asymptotic analysis of \eqref{Equ_R_MC} based on the Berry-Esseen theorem. To this end, we need the following auxiliary lemmas:

\begin{Lemma}\label{MC_Lemma4}
Let $\bar{U}(\mathsf{D},T,\rho) \triangleq \mathsf{E}\bigl[\bigl( \bar{j}(\mathsf{D}_{\ell}, \mathbf{Z}_{\ell}',\mathbf{Z}_{\ell}'',\mathbf{H}_{\ell},\mathbf{Q}_{\ell})-\bar{J}(\mathsf{D},T,\rho)\bigr)^2\bigr]$ and $\bar{S}(\mathsf{D},T,\rho) \triangleq \mathsf{E}\bigl[\bigl| \bar{j}(\mathsf{D}_{\ell}, \mathbf{Z}_{\ell}',\mathbf{Z}_{\ell}'',\mathbf{H}_{\ell},\mathbf{Q}_{\ell})-\bar{J}(\mathsf{D},T,\rho)\bigr|^3\bigr]$ for $\mathsf{D} \in\mathcal{D}_{n_t}$. Then, we have for every $T\geq n_r + n_t$ and any positive $\rho_0$
\begin{IEEEeqnarray}{rCl}
	\sup_{\rho \geq\rho_0} \sup_{\mathsf{D}\in \mathcal{D}_{n_t}} \bar{U}(\mathsf{D},T,\rho)&< & \infty \label{eq:Lemma3_first}\\
	\sup_{\rho \geq\rho_0} \sup_{\mathsf{D}\in \mathcal{D}_{n_t}} \bar{S}(\mathsf{D},T,\rho) & < & \infty. \label{eq:Lemma3_second}
\end{IEEEeqnarray}
\end{Lemma}	
\begin{IEEEproof}
See Appendix \ref{Appx_Pf_MC_Lem4}.
\end{IEEEproof}

\begin{Lemma}\label{MC_Lemma5}
For every $\mathsf{D} \in \mathcal{D}_1$, we have
\begin{equation}
	\bar{U}(\mathsf{D},T,\rho) \geq  (T-n_t) n_t + K_{\bar{U}}(\mathsf{D},T,\rho) 
\end{equation}
where $K_{\bar{U}}(\mathsf{D},T,\rho)$ is a function of $\mathsf{D}$, $T$, and $\rho$ that satisfies
\begin{equation*}
	\lim_{\rho\to\infty} \sup_{\mathsf{D}\in\mathcal{D}_1} \big|K_{\bar{U}}(\mathsf{D},T,\rho) \big|= 0.
\end{equation*}
\end{Lemma}
\begin{IEEEproof}
See Appendix \ref{Appx_Pf_MC_Lem5}.
\end{IEEEproof}
Lemma \ref{MC_Lemma4} implies that there exists an $\bar{S}(T)$ that only depends on $T$ and satisfies
\begin{equation}
\bar{S}(\mathsf{D},T,\rho) \leq \bar{S}(T), \quad \mathsf{D} \in \mathcal{D}_{n_t},\, \rho> \rho_0.\label{Equ_upSrho0}
\end{equation}
Furthermore, for $\rho\geq\rho_0$ and sufficiently large $\rho_0$, we have $K_{\bar{U}}(\mathsf{D},T,\rho)\leq (T-n_t) n_t/2$, $\mathsf{D}\in\mathcal{D}_1$, by the definition of $K_{\bar{U}}(\mathsf{D},T,\rho)$. It thus follows from Lemma~\ref{MC_Lemma5} that, for $\rho\geq\rho_0$ and $\mathsf{X}^L \in \mathcal{C}_1$,
\begin{IEEEeqnarray}{lCl}
\sum_{\ell=1}^{L} \bar{U}(\mathsf{D}_{\ell},T,\rho) & \geq & L_{\mathcal{D}_1}(\mathsf{D}^L) \frac{(T-n_t) n_t}{2} \nonumber\\
& \geq & \frac{L}{2} \frac{(T-n_t) n_t}{2}
\end{IEEEeqnarray}
since all codewords in $\mathcal{C}_1$ satisfy $L_{\mathcal{D}_1}(\mathsf{D}^L)\geq L/2$. Combining these two results, we can upper-bound the ratio
\begin{equation} \label{Equ_converse_BEratio}
\bar{B}(\mathsf{D}^L,T,\rho) \triangleq \frac{6\sum_{\ell=1}^{L} \bar{S}(\mathsf{D},T,\rho)}{\left(\sum_{\ell=1}^{L} \bar{U}(\mathsf{D}_{\ell},T,\rho)\right)^{3/2}} 
\end{equation}
for $\mathsf{X}^L\in\mathcal{C}_1$ as
\begin{IEEEeqnarray}{lCl}
\bar{B}(\mathsf{D}^L,T,\rho) &\leq & \frac{6L \bar{S}(T)}{\left(\frac{L(T-n_t) n_t}{4}\right)^{3/2}} \nonumber\\
& \triangleq & \frac{\bar{B}(T)}{\sqrt{L}}, \quad \mathsf{X}^L\in\mathcal{C}_1,\, \mathsf{D}_{\ell} = [\mathsf{X}_{\ell}]_{(1:n_t)^2} \label{MC_Def_B_a}
\end{IEEEeqnarray}
where $\bar{B}(T)$ is a function of $T$.

With this, we are ready to apply the Berry-Esseen theorem (Lemma~\ref{Lem_BerryEsseen}) to the upper bound \eqref{Equ_R_MC}. Let
\begin{subequations}
\label{eq:MC_Defs}
\begin{IEEEeqnarray}{rCl}
	\log\xi(\mathsf{D}^L) & = & \sum_{\ell=1}^{L} \bar{J}(\mathsf{D}_{\ell},T,\rho)-\lambda \sqrt{\sum_{\ell=1}^{L} \bar{U}(\mathsf{D}_{\ell},T,\rho)} \label{MC_Def_Xi} \\
	\lambda & = & Q^{-1}\left(\epsilon+\frac{2\bar{B}(T)}{\sqrt{L}}\right),\quad 0<\epsilon<1/2. \label{MC_Def_lambda} 
\end{IEEEeqnarray}
\end{subequations}
It then follows from the Berry-Esseen theorem that 
\begin{align}
\Pr \left[ \sum_{\ell=1}^{L} \bar{j}(\mathsf{D}_{\ell}, \mathbf{Z}_{\ell}',\mathbf{Z}_{\ell}'',\mathbf{H}_{\ell},\mathbf{Q}_{\ell}) \geq \log\xi(\mathsf{D}^L)\right]
\leq 1- \epsilon-\frac{\bar{B}(T)}{\sqrt{L}}. \label{MC_Equ_BEa}
\end{align}
Substituting \eqref{eq:MC_Defs} and \eqref{MC_Equ_BEa} into the upper bound \eqref{Equ_R_MC}, we obtain for $L\geq L_0$ and a sufficiently large $L_0$ that
\begin{IEEEeqnarray}{lCl}
\frac{\log |\mathcal{C}_1| }{L} 
& \leq &  \frac{1}{L}  \sup_{\mathsf{X}^L\in\mathcal{C}_1}   \left\{ \log \xi(\mathsf{D}^L) 
-\log\left(1-\epsilon-\Pr\left[\sum_{\ell=1}^{L}\bar{j}(\mathsf{D}_{\ell}, \mathbf{Z}_{\ell}',\mathbf{Z}_{\ell}'',\mathbf{H}_{\ell},\mathbf{Q}_{\ell})\geq \log \xi(\mathsf{D}^L) \right] \right)   \right\} \nonumber\\
& \leq & \frac{1}{L} \sup_{\mathsf{X}^L\in\mathcal{C}_1}    \left\{ \sum_{\ell=1}^{L} \bar{J}(\mathsf{D}_{\ell},T,\rho)
-\sqrt{\sum_{\ell=1}^{L} \bar{U}(\mathsf{D}_{\ell},T,\rho)} Q^{-1}\left(\epsilon+\frac{2\bar{B}(T)}{\sqrt{L}}\right)\right\}-\frac{\log \bar{B}(T) }{L}+ \frac{\log L}{2L} \nonumber\\
& \leq & \frac{1}{L} \sup_{\mathsf{X}^L\in\mathcal{C}_1}   \left\{ \sum_{\ell=1}^{L} \bar{J}(\mathsf{D}_{\ell},T,\rho)
-\frac{\sum_{\ell=1}^{L}  \sqrt{\bar{U}(\mathsf{D}_{\ell},T,\rho)}}{\sqrt{L}} Q^{-1}\left(\epsilon+\frac{2\bar{B}(T)}{\sqrt{L}}\right)\right\}-\frac{\log \bar{B}(T) }{L}+ \frac{\log L}{2L} \nonumber  \\
& = & \frac{1}{L} \sup_{\mathsf{X}^L\in\mathcal{C}_1}   \sum_{\ell=1}^{L}\left\{ \bar{J}(\mathsf{D}_{\ell},T,\rho) 
-\sqrt{\frac{\bar{U}(\mathsf{D}_{\ell},T,\rho)}{L}}Q^{-1}\left(\epsilon+\frac{2\bar{B}(T)}{\sqrt{L}}\right)\right\}-\frac{\log \bar{B}(T) }{L}+ \frac{\log L}{2L}\label{MC_Equ_Rub_meta2} 
\end{IEEEeqnarray}  
where the third inequality follows because, by the assumption $0<\epsilon<1/2$, the inverse Q-function is positive for sufficiently large $L$, and by applying Jensen’s inequality to the square-root function.

Performing a Taylor-series expansion of $Q^{-1}\Big(\epsilon+2\bar{B}(T)/\sqrt{L}\Big)$ around $\epsilon$, we obtain 
\begin{equation}
\left| Q^{-1}\left(\epsilon+\frac{2\bar{B}(T)}{\sqrt{L}}\right)-Q^{-1}(\epsilon)\right| \leq \frac{A}{\sqrt{L}}, \quad L\geq L_0 
\end{equation}
for some constants $A$, $L_0$, and $\rho_0$ that are independent of $L$ and $\rho$. Further using that, by Lemma~\ref{MC_Lemma4}, $\bar{U}(\mathsf{D}_{\ell},T,\rho)$ is bounded in $\mathsf{D}_{\ell}\in\mathbb{D}_{n_t}$ and $\rho\geq\rho_0$, and collecting terms of order $\log L/L$, we can then rewrite the upper bound \eqref{MC_Equ_Rub_meta2} as
\begin{equation}
\frac{\log |\mathcal{C}_1| }{L} 
\leq  \frac{1}{L} \sup_{\mathsf{X}^L\in\mathcal{C}_1}   \sum_{\ell=1}^{L}\left\{ \bar{J}(\mathsf{D}_{\ell},T,\rho) 
-\sqrt{\frac{\bar{U}(\mathsf{D}_{\ell},T,\rho)}{L}}Q^{-1}(\epsilon)\right\}+K_{1}(L,T,\rho)  \label{MC_Equ_Rub_meta3}
\end{equation}  
where $K_{1}(L,T,\rho)$ is a function of $L$, $T$, and $\rho$ that satisfies 
\begin{equation}
\sup_{\rho\geq \rho_0} \big| K_1(L,T,\rho)\big| \leq A\frac{\log L}{L}, \quad L\geq L_0
\end{equation}
for some constants $A$, $L_0$, and $\rho_0$ that are independent of $L$ and $\rho$.

Since $\mathsf{X}_{\ell}$ satisfies the power constraint $\|\mathsf{X}_{\ell}\|^2\leq T\rho$ in each coherence block, and since each summand in \eqref{MC_Equ_Rub_meta2} only depends on one $\mathsf{X}_{\ell}$, the supremum on the RHS of \eqref{MC_Equ_Rub_meta2} can be written as
\begin{IEEEeqnarray}{lCl}
\IEEEeqnarraymulticol{3}{l}{\sum_{\ell\colon \mathsf{D}_{\ell}\in\mathcal{D}_1} \sup_{\mathsf{D}\in \mathcal{D}_1}
	\left\{\bar{J}(\mathsf{D},T,\rho) 
	-\sqrt{\frac{\bar{U}(\mathsf{D},T,\rho)}{L}}Q^{-1}(\epsilon)\right\} 
	+\sum_{\ell\colon \mathsf{D}_{\ell}\in\mathcal{D}_2} \sup_{\mathsf{D}\in \mathcal{D}_2}
	\left\{\bar{J}(\mathsf{D},T,\rho) 
	-\sqrt{\frac{\bar{U}(\mathsf{D},T,\rho)}{L}}Q^{-1}(\epsilon)\right\}} \nonumber\\
\quad & = & L_{\mathcal{D}_1}(\mathsf{D}^L) \sup_{\mathsf{D}\in \mathcal{D}_1}
\left\{\bar{J}(\mathsf{D},T,\rho) 
-\sqrt{\frac{\bar{U}(\mathsf{D},T,\rho)}{L}}Q^{-1}(\epsilon)\right\} \nonumber\\
& & {} + \bigl(L- L_{\mathcal{D}_1}(\mathsf{D}^L)\bigr)  \sup_{\mathsf{D}\in \mathcal{D}_2}
\left\{\bar{J}(\mathsf{D},T,\rho) 
-\sqrt{\frac{\bar{U}(\mathsf{D},T,\rho)}{L}}Q^{-1}(\epsilon)\right\}\label{MC_Equ_C1_RX2B}
\end{IEEEeqnarray}
maximized over $L_{\mathcal{D}_1}(\mathsf{D}^L)\geq L/2$. We next upper-bound \eqref{MC_Equ_C1_RX2B} using the following lemmas:
\begin{Lemma}\label{MC_Lemma6}  
For every $\mathsf{D}\in\mathcal{D}_{n_t}$ satisfying $\text{tr}(\mathsf{D}^2)\geq T\rho(1-\delta)$ for some $0 < \delta \leq \frac{1}{2}$, we have
\begin{equation}
	\bar{U}(\mathsf{D},T,\rho) \geq  \bar{U}\left(\sqrt{\frac{\rho}{\alpha}} \mathsf{D},T,\rho\right)-\Upsilon(T) \delta, \quad  \rho \geq \rho_0 \label{MC_Equ_Low_Va}  
\end{equation} 
where $\Upsilon(T)$ is a positive constant that only depends on $T$ and $\rho_0$ is an arbitrary positive constant.
\end{Lemma}
\begin{IEEEproof}
See Appendix \ref{Appx_Pf_MC_Lem6}.
\end{IEEEproof}

\begin{Lemma}\label{MC_Lemma7b}
Assume that $T\geq n_t+n_r$.  For $\rho\geq \rho_0$ and $L\geq L_0$ and sufficiently large $\rho_0$ and $L_0$ (that only depend on $T$ and $\epsilon$),
\begin{IEEEeqnarray}{lCl}
	\IEEEeqnarraymulticol{3}{l}{\sup_{\mathsf{D}\in\mathcal{D}_1} \left\{\bar{J}(\mathsf{D},T,\rho) 
		-\sqrt{\frac{\bar{U}(\mathsf{D},T,\rho)}{L}}Q^{-1}(\epsilon)\right\}} \nonumber\\
	\qquad & \leq & \sup_{\mathsf{D}\in\mathcal{D}_{n_t}\colon \textnormal{tr}(\mathsf{D}^2) \geq T\rho\left(1-\frac{K(T)}{L}\right)} \left\{\bar{J}^{\ast}(\mathsf{D},T,\rho)
	-\sqrt{\frac{\bar{U}(\mathsf{D},T,\rho)}{L}}Q^{-1}(\epsilon)\right\} + K_{\mathcal{D}_1}(T,\rho)
\end{IEEEeqnarray}
for some nonnegative constant $K(T)$ that only depends on $T$. Here,
\begin{IEEEeqnarray}{lCl} 
	\bar{J}^{\ast}(\mathsf{D},T,\rho)
	& \triangleq &  n_tn_r\log \left(\frac{T\rho}{n_t}\right)
	+\log \left(\frac{\Gamma_{n_t}(n_t)}{\Gamma_{n_t}(T)} \right)
	+\left(\frac{T\alpha-T\rho}{T\rho}\right)n_rn_t-n_t(T-n_t) \nonumber \\
	& & {} +(T-n_t-n_r)\log\det\big(\mathsf{I}_{n_t}+  \mathsf{D}^2\big) +(T-n_t)\mathsf{E}\left[ \log \det \left( \mathbf{H}_{\ell} \mathbf{H}_{\ell}^\mathsf{H} \right) \right] \label{MC_Ap3_Def_Jast}
\end{IEEEeqnarray}
and $K_{\mathcal{D}_1}(T,\rho)$ is a nonnegative constant that only depends on $T$ and $\rho$ and satisfies $\lim_{\rho\to \infty} K_{\mathcal{D}_1}(T,\rho)= 0$.

\end{Lemma}
\begin{IEEEproof}
See Appendix \ref{Appx_Pf_MC_Lem7b}.
\end{IEEEproof}

\begin{Lemma}\label{MC_Lemma7}
Maximized over $\mathsf{D} \in \mathcal{D}_1$, the term $\bar{J}^{\ast}(\mathsf{D},T,\rho)$ is given by
\begin{equation}
	\sup_{\mathsf{D}\in \mathcal{D}_1}  \bar{J}^{\ast}(\mathsf{D},T,\rho) =  \bar{J}_{\mathcal{D}_1}(T,\rho), \quad \rho> 0\label{MC_C1_Equ_RX1}
\end{equation}
where 
\begin{IEEEeqnarray}{lCl}
	\bar{J}_{\mathcal{D}_1}(T,\rho) & \triangleq &n_tn_r\log \left(\frac{T\rho}{n_t}\right)
	+\log \left(\frac{\Gamma_{n_t}(n_t)}{\Gamma_{n_t}(T)} \right)
	-n_t(T-n_t) 
	+(T-n_t-n_r)\log \left(1+\frac{T\rho}{n_t}\right)^{n_t} \nonumber \\
	& & {} +(T-n_t)\mathsf{E}\big[ \log \det \big( \mathbf{H}_{\ell} \mathbf{H}_{\ell}^\mathsf{H}\big) \big] \label{Def_tildeJ}
\end{IEEEeqnarray} 
\end{Lemma}
\begin{IEEEproof}
See Appendix \ref{Appx_Pf_MC_Lem7}.
\end{IEEEproof}

\begin{Lemma}\label{MC_Lem_Vhigh}
For every $\mathsf{D} \in \mathcal{D}_1$, we have 
\begin{equation}
	\bar{U}\left(\sqrt{\frac{\rho}{\alpha}}\mathsf{D},T,\rho\right) \geq T^2\tilde{V}(T)+ K_{\bar{U}}(\mathsf{D},T,\rho) \label{Equ_L5_UA}
\end{equation}
where $\tilde{V}(T)$ was defined in \eqref{Def_Vt} and $K_{\bar{U}}(\mathsf{D},T,\rho)$ is a function of $\mathsf{D}$, $T$, and $\rho$ which satisfies
\begin{equation}
	\lim_{\rho\to \infty} \sup_{{\mathsf{D}}\in {\mathcal{D}}_1} \left|K_{\bar{U}}(\mathsf{D},T,\rho)\right|=0. \label{eq:vanishing_KU}
\end{equation}
\end{Lemma}	    
\begin{IEEEproof}
See Appendix \ref{Appx_Pf_MC_Lem8}.
\end{IEEEproof}

\begin{Lemma}\label{MC_Lem8} 
Maximized over $\mathsf{D}\in\mathcal{D}_2$, the term $\bar{J}(\mathsf{D},T,\rho)$ can be bounded as
\begin{equation}
	\bar{J}_{\mathcal{D}_2}(T,\rho) \leq \sup_{\mathsf{D}\in \mathcal{D}_2}  \bar{J}(\mathsf{D},T,\rho) \leq  \bar{J}_{\mathcal{D}_2}(T,\rho) +K_{\mathcal{D}_2}(T,\rho), \quad \rho \geq \rho_0
\end{equation}
for a sufficiently large $\rho_0>0$ (that only depends on $T$), where
\begin{IEEEeqnarray}{lCl}
	\bar{J}_{\mathcal{D}_2}(T,\rho) & \triangleq & n_tn_r\log \left(\frac{T\rho}{n_t}\right)
	+\log \left(\frac{\Gamma_{n_t}(n_t)}{\Gamma_{n_t}(T)} \right) -n_t(T-n_t)  
	+(T-n_t)\mathsf{E}\left[ \log \det \big( \mathbf{H}_{\ell} \mathbf{H}_{\ell}^\mathsf{H} \big) \right] \nonumber \\
	& & {} +(T-n_t-n_r)\log \left((1+\bar{\delta}\rho)\left(1+\frac{T-\bar{\delta}}{n_t-1}\rho\right)^{n_t-1}\right) \label{Equ_JD2} 
\end{IEEEeqnarray}
and $K_{\mathcal{D}_2}(T,\rho)$ is a nonnegative correction term that only depends on $T$ and $\rho$ and satisfies
\begin{equation}
	\lim_{\rho\to\infty} K_{\mathcal{D}_2}(T,\rho) = 0. \label{eq:vanishing_KD2}
\end{equation}
\end{Lemma}
\begin{IEEEproof}
See Appendix \ref{Appx_Pf_MC_Lem8b}.
\end{IEEEproof}

We are now ready to upper-bound \eqref{MC_Equ_C1_RX2B}. Indeed, the first supremum on the RHS of \eqref{MC_Equ_C1_RX2B} can be upper-bounded as
\begin{IEEEeqnarray}{lCl}
\IEEEeqnarraymulticol{3}{l}{\sup_{\mathsf{D}\in \mathcal{D}_1}
	\left\{\bar{J}(\mathsf{D},T,\rho) 
	-\sqrt{\frac{\bar{U}(\mathsf{D},T,\rho)}{L}}Q^{-1}(\epsilon)\right\}}\nonumber\\
\quad & \leq & \sup_{\mathsf{D}\in\mathcal{D}_{n_t}\colon \textnormal{tr}(\mathsf{D}^2) \geq T\rho\left(1-\frac{K(T)}{L}\right)} \left\{\bar{J}^{\ast}(\mathsf{D},T,\rho)
-\sqrt{\frac{\bar{U}(\mathsf{D},T,\rho)}{L}}Q^{-1}(\epsilon)\right\} + K_{\mathcal{D}_1}(T,\rho) \nonumber\\
& \leq & \max\left\{ \sup_{\mathsf{D}\in\mathcal{D}_{1}\colon \textnormal{tr}(\mathsf{D}^2) \geq T\rho\left(1-\frac{K(T)}{L}\right)} \left\{\bar{J}^{\ast}(\mathsf{D},T,\rho)
-\sqrt{\frac{\bar{U}(\mathsf{D},T,\rho)}{L}}Q^{-1}(\epsilon)\right\}, \sup_{\mathsf{D}\in\mathcal{D}_2} \bar{J}(\mathsf{D},T,\rho) \right\} + K_{\mathcal{D}_1}(T,\rho) \label{Equ_C1_JU4_a}
\end{IEEEeqnarray}
where the first inequality follows from Lemma~\ref{MC_Lemma7b}; the second inequality follows by writing the supremum over $\mathcal{D}_{n_t}$ as the maximum of the supremum over $\mathcal{D}_1$ and the supremum over $\mathcal{D}_2$ (because $\mathcal{D}_{n_t}=\mathcal{D}_1\cup \mathcal{D}_2$ and $\mathcal{D}_1\cap \mathsf{D}_2 = \varnothing$), and by upper-bounding the latter supremum by removing the trace constraint and by lower-bounding $\bar{U}(\mathsf{D},T,\rho)\geq 0$.
								
The supremum over $\mathcal{D}_1$ in \eqref{Equ_C1_JU4_a} can be upper-bounded as
	\begin{IEEEeqnarray}{lCl}
	\IEEEeqnarraymulticol{3}{l}{\sup_{\mathsf{D}\in\mathcal{D}_{1}\colon \textnormal{tr}(\mathsf{D}^2) \geq T\rho\left(1-\frac{K(T)}{L}\right)} \left\{\bar{J}^{\ast}(\mathsf{D},T,\rho)
	-\sqrt{\frac{\bar{U}(\mathsf{D},T,\rho)}{L}}Q^{-1}(\epsilon)\right\}} \nonumber\\
	\quad & \leq &  \sup_{\mathsf{D}\in\mathcal{D}_{1}\colon \textnormal{tr}(\mathsf{D}^2) \geq T\rho\left(1-\frac{K(T)}{L}\right)} \left\{\bar{J}^{\ast}(\mathsf{D},T,\rho)
	-\sqrt{\frac{\bar{U}\left(\sqrt{\frac{\rho}{\alpha}} \mathsf{D},T,\rho\right)-\Upsilon(T) \frac{K(T)}{L}}{L}}Q^{-1}(\epsilon)\right\} \nonumber\\
	& \leq & \sup_{\mathsf{D}\in \mathcal{D}_1}  \bar{J}^{\ast}(\mathsf{D},T,\rho) - \sqrt{\frac{T^2 \tilde{V}(T) - K_{\bar{U}}(T,\rho) -\Upsilon(T) \frac{K(T)}{L}}{L}}Q^{-1}(\epsilon) \nonumber\\
	& \leq &  \bar{J}_{\mathcal{D}_1}(T,\rho) - \sqrt{\frac{T^2 \tilde{V}(T) - K_{\bar{U}}(T,\rho)}{L}}Q^{-1}(\epsilon) + K_{\mathcal{D}_1}(L,T) \nonumber\\
	& = & \bar{R}_{\mathcal{D}_1}(L,T,\rho) + K_{\mathcal{D}_1}(L,T) \label{eq:Equ_C1_JUhu}
\end{IEEEeqnarray}
where
\begin{equation}
\bar{R}_{\mathcal{D}_1}(L,T,\rho) \triangleq \bar{J}_{\mathcal{D}_1}(T,\rho)- \sqrt{\frac{T^2 \tilde{V}(T) - K_{\bar{U}}(T,\rho)}{L}}Q^{-1}(\epsilon) 
\end{equation}
and
\begin{subequations}
\begin{IEEEeqnarray}{rCl}
	K_{\bar{U}}(T,\rho) & \triangleq & \sup_{\mathsf{D}\in\mathcal{D}_1} |K_{\bar{U}}(\mathsf{D},T,\rho)|\\
	K_{\mathcal{D}_1}(L,T) & \triangleq & \frac{\sqrt{\Upsilon(T)K(T)}}{L}Q^{-1}(\epsilon).
\end{IEEEeqnarray}
\end{subequations}
In \eqref{eq:Equ_C1_JUhu}, the first inequality follows from Lemma~\ref{MC_Lemma6}; the second inequality follows from Lemma~\ref{MC_Lem_Vhigh} and by optimizing $\bar{J}^{\ast}(\mathsf{D},T,\rho)$ and the constant $K_{\bar{U}}(\mathsf{D},T,\rho)$ in \eqref{Equ_L5_UA} over $\mathsf{D}\in\mathcal{D}_1$; the third inequality follows from Lemma~\ref{MC_Lemma7} and the inequality $\sqrt{x-y} \geq \sqrt{x} - \sqrt{y}$, $x \geq y$.

We next use Lemma~\ref{MC_Lem8} to upper-bound the supremum over $\mathcal{D}_2$ in \eqref{Equ_C1_JU4_a} as
\begin{equation}
\sup_{\mathsf{D}\in \mathcal{D}_2}  \bar{J}(\mathsf{D},T,\rho) \leq  \bar{J}_{\mathcal{D}_2}(T,\rho) +K_{\mathcal{D}_2}(T,\rho), \quad \rho \geq \rho_0
\end{equation}
for a sufficiently large $\rho_0>0$. Comparing $\bar{J}_{\mathcal{D}_2}(T,\rho)$ with $\bar{J}_{\mathcal{D}_1}(T,\rho)$, we obtain that
\begin{equation}
	\bar{J}_{\mathcal{D}_1}(T,\rho) - \bar{J}_{\mathcal{D}_2}(T,\rho) = (T-n_t - n_r)\log\left(\frac{\left(1+\frac{T\rho}{n_t}\right)^{n_t}}{(1+\bar{\delta}\rho)\left(1+\frac{T\rho-\bar{\delta}\rho}{n_t-1}\right)^{n_t-1}}\right) \triangleq \Theta(T,\rho).  \label{Equ_MC_JD2diff}
\end{equation}
In the limit as $\rho$ tends to infinity, $\Theta(T,\rho)$ tends to
\begin{equation}
	\lim_{\rho\to\infty} \Theta(T,\rho) = (T-n_t-n_r) \log\left(\frac{\left(\frac{T}{n_t}\right)^{n_t}}{\bar{\delta} \left(\frac{T-\bar{\delta}}{n_t-1}\right)^{n_t-1}}\right) \triangleq \Theta(T) \label{eq:MC_JD2diff_asymp}
\end{equation}
which is strictly positive since, by the geometric-arithmetic mean inequality,
\begin{equation}
	\bar{\delta}\left(\frac{T-\bar{\delta}}{n_t-1}\right)^{n_t-1} < \left(\frac{\bar{\delta} + \frac{T-\bar{\delta}}{n_t-1} + \ldots + \frac{T-\bar{\delta}}{n_t-1}}{n_t}\right)^{n_t} = \left(\frac{T}{n_t}\right)^{n_t}.
\end{equation}
By the definitions of $\bar{R}_{\mathcal{D}_1}(L,T,\rho)$ and $\Theta(T,\rho)$, we have that
\begin{equation}
\sup_{\mathsf{D}\in \mathcal{D}_2}  \bar{J}(\mathsf{D},T,\rho) - \bar{R}_{\mathcal{D}_1}(L,T,\rho) - K_{\mathcal{D}_1}(T,\rho) \leq  K_{\mathcal{D}_2}(T,\rho) -  \Theta(T,\rho)+ \sqrt{\frac{T^2 \tilde{V}(T) - K_{\bar{U}}(T,\rho)}{L}}Q^{-1}(\epsilon)- K_{\mathcal{D}_1}(T,\rho).\label{eq:Equ_C1_JU4_b}
\end{equation}
Since the first, second, and last terms on the RHS of \eqref{eq:Equ_C1_JU4_b} only depend on $\rho$, and the second-to-last term vanishes as $L\to\infty$ uniformly in $\rho$, we conclude that the double limit of the RHS of \eqref{eq:Equ_C1_JU4_b} as $\rho\to\infty$ and $L\to\infty$ exists and is equal to $-\Theta(T)<0$. It follows that there exist $\rho_0$ and $L_0$ such that
\begin{equation}
\label{eq:Equ_C1_JU4_ba}
\sup_{\mathsf{D}\in \mathcal{D}_2}  \bar{J}(\mathsf{D},T,\rho) < \bar{R}_{\mathcal{D}_1}(L,T,\rho) + K_{\mathcal{D}_1}(T,\rho)
\end{equation}
for $\rho\geq\rho_0$ and $L\geq L_0$. Combining \eqref{eq:Equ_C1_JU4_ba} and \eqref{eq:Equ_C1_JUhu} with \eqref{Equ_C1_JU4_a}, we obtain that
\begin{IEEEeqnarray}{lCl}
\sup_{\mathsf{D}\in \mathcal{D}_1}\left\{\bar{J}(\mathsf{D},T,\rho) 
				-\sqrt{\frac{\bar{U}(\mathsf{D},T,\rho)}{L}}Q^{-1}(\epsilon)\right\} & \leq & \bar{R}_{\mathcal{D}_1}(L,T,\rho) + K_{\mathcal{D}_1}(T,\rho) + K_{\mathcal{D}_1}(L,T)\label{Equ_C1_JU4}
\end{IEEEeqnarray}
for $\rho\geq\rho_0$ and $L\geq L_0$ and sufficiently large $\rho_0$ and $L_0$. 
							
We next show that the second supremum on the RHS of \eqref{MC_Equ_C1_RX2B} does not exceed the RHS of \eqref{Equ_C1_JU4}. It follows then that our upper bound on \eqref{MC_Equ_C1_RX2B} is maximized for $ L_{\mathcal{D}_1}(\mathsf{D}^L)=L$. To prove this claim, we use again Lemma~\ref{MC_Lem8} together with the nonnegativity of $\bar{U}(\mathsf{D},T,\rho)$ to upper-bound
\begin{equation}
	\sup_{\mathsf{D}\in \mathcal{D}_2}
	\left\{\bar{J}(\mathsf{D},T,\rho) 
	-\sqrt{\frac{\bar{U}(\mathsf{D},T,\rho)}{L}}Q^{-1}(\epsilon)\right\} \leq  \bar{J}_{\mathcal{D}_2}(T,\rho) +K_{\mathcal{D}_2}(T,\rho). \label{eq:UB_lemma9}
\end{equation}

							It then follows from \eqref{eq:UB_lemma9} and \eqref{Equ_MC_JD2diff} that
\begin{IEEEeqnarray}{lCl}
	\IEEEeqnarraymulticol{3}{l}{\bar{R}_{\mathcal{D}_1}(L,T,\rho)  + K_{\mathcal{D}_1}(T,\rho) + K_{\mathcal{D}_1}(L,T) -  \sup_{\mathsf{D}\in \mathcal{D}_2}
		\left\{\bar{J}(\mathsf{D},T,\rho) 
		-\sqrt{\frac{\bar{U}(\mathsf{D},T,\rho)}{L}}Q^{-1}(\epsilon)\right\}} \nonumber\\
	\quad & \geq & \Theta(T,\rho) - \sqrt{\frac{T^2 \tilde{V}(T) - K_{\bar{U}}(T,\rho)}{L}}Q^{-1}(\epsilon) + K_{\mathcal{D}_1}(T,\rho)  + K_{\mathcal{D}_1}(L,T) - K_{\mathcal{D}_2}(T,\rho). \label{eq:MC_intstep_D2}
\end{IEEEeqnarray}
Using similar arguments as the ones leading to \eqref{eq:Equ_C1_JU4_ba}, we can show that the double limit of the RHS of \eqref{eq:MC_intstep_D2} as $\rho\to\infty$ and $L\to\infty$ exists and is equal to $\Theta(T)$. Consequently, there exist $\rho_0$ and $L_0$ such that, for $\rho\geq\rho_0$ and $L\geq L_0$, the RHS of \eqref{eq:MC_intstep_D2} is strictly positive. This implies that, for $\rho\geq\rho_0$ and $L\geq L_0$,
\begin{equation}
	\sup_{\mathsf{D}\in \mathcal{D}_2}
	\left\{\bar{J}(\mathsf{D},T,\rho) 
	-\sqrt{\frac{\bar{U}(\mathsf{D},T,\rho)}{L}}Q^{-1}(\epsilon)\right\} 
		< \bar{R}_{\mathcal{D}_1}(L,T,\rho) + K_{\mathcal{D}_1}(T,\rho) + K_{\mathcal{D}_1}(L,T). \label{eq:Equ_C1_X2U}
\end{equation}

Combining \eqref{Equ_C1_JU4} and \eqref{eq:Equ_C1_X2U} with \eqref{MC_Equ_Rub_meta2} and \eqref{MC_Equ_C1_RX2B}, we obtain that
\begin{IEEEeqnarray}{lCl}
	\frac{\log |\mathcal{C}_1| }{L} & \leq &\bar{R}_{\mathcal{D}_1}(L,T,\rho) + K_{\mathcal{D}_1}(T,\rho) + K_{\mathcal{D}_1}(L,T) -\frac{\log \bar{B}(T) }{L}+ \frac{\log L}{2L} \nonumber\\
	& = & \bar{J}_{\mathcal{D}_1}(T,\rho)- \sqrt{\frac{T^2 \tilde{V}(T) - K_{\bar{U}}(T,\rho)}{L}}Q^{-1}(\epsilon)  + K_{\mathcal{D}_1}(T,\rho) + K_{\mathcal{C}_1}(L,T) \label{Equ_Car_C1}
\end{IEEEeqnarray}
where
\begin{equation}
	K_{\mathcal{C}_1}(L,T) \triangleq K_{\mathcal{D}_1}(L,T) -\frac{\log \bar{B}(T) }{L}+ \frac{\log L}{2L}
\end{equation}
is a function of $L$ and $T$ that satisfies
\begin{equation}
	\left|K_{\mathcal{C}_1}(L,T)\right| \leq A \frac{\log L}{L}, \quad L\geq L_0 \label{eq:asymp_KC1}
\end{equation}
for some constants $A$ and $L_0$ that are independent of $L$ and $\rho$. Recall that $\lim_{\to\infty} K_{\mathcal{D}_1}(T,\rho)=\lim_{\rho\to\infty} K_{\bar{U}}(T,\rho)=0$.

\subsubsection{Upper bound on the cardinality of $\mathcal{C}_2$}\label{MC_Sec: C2}

An upper bound on $\log|\mathcal{C}_2|$ follows from the weakened MC bound \eqref{eq:weakened_MC} and from \eqref{eq:Prob_UB_jbar}:
\begin{equation}
	\log |\mathcal{C}_2| 
	\leq  \sup_{\mathsf{X}^L\in\mathcal{C}_2}  \left\{ \log \xi(\mathsf{D}^L)  
	-\log\left(1-\epsilon-\Pr\left[\sum_{\ell=1}^{L} \bar{j}(\mathsf{D}_{\ell}, \mathbf{Z}_{\ell}',\mathbf{Z}_{\ell}'',\mathbf{H}_{\ell},\mathbf{Q}_{\ell})\geq \log \xi(\mathsf{D}^L)  \right] \right)  \right\}. \label{eq:MC_C2}
\end{equation} 
We next note that, by Lemma~\ref{MC_Lemma4}, there exists an $\bar{U}(T)$ that only depends on $T$ and satisfies
\begin{equation}
	\bar{U}(\mathsf{D},T,\rho) \leq \bar{U}(T), \quad \mathsf{D} \in\mathcal{D}_{n_t}, \, \rho > \rho_0.
\end{equation}
If we choose
\begin{equation}
	\log\xi(\mathsf{D}^L)  = \sum_{\ell=1}^{L} \bar{J}_{\ell}(\mathsf{D}_{\ell},T,\rho) + \sqrt{\frac{2 L \bar{U}(T)}{1-\epsilon}}
\end{equation}
then Chebyshev's inequality yields that
\begin{equation}
	\Pr\left[\sum_{\ell=1}^{L} \bar{j}(\mathsf{D}_{\ell}, \mathbf{Z}_{\ell}',\mathbf{Z}_{\ell}'',\mathbf{H}_{\ell},\mathbf{Q}_{\ell})\geq \log \xi(\mathsf{D}^L)  \right] \leq \frac{1-\epsilon}{2}
\end{equation}
and we can upper-bound \eqref{eq:MC_C2} as
\begin{equation}
	\log |\mathcal{C}_2| 
	\leq  \sup_{\mathsf{X}^L\in\mathcal{C}_2} \sum_{\ell=1}^{L} \bar{J}_{\ell}(\mathsf{D}_{\ell},T,\rho) + \sqrt{\frac{2 L \bar{U}(T)}{1-\epsilon}} - \log\left(\frac{1-\epsilon}{2}\right). \label{MC_C2_Ori} 
\end{equation}

It remains to upper-bound the supremum on the RHS of \eqref{MC_C2_Ori}. Dividing the sum into indices for which $\mathsf{D}_{\ell} \in \mathcal{D}_1$ and indices for which $\mathsf{D}_{\ell}\in\mathcal{D}_2$, we obtain that
\begin{IEEEeqnarray}{lCl}
	\sup_{\mathsf{X}^L\in\mathcal{C}_2} \sum_{\ell=1}^{L} \bar{J}_{\ell}(\mathsf{D}_{\ell},T,\rho) & = & \sup_{\mathsf{D}^L\colon L_{\mathcal{D}_1}(\mathsf{D}^L) < L/2} \left\{ \sum_{\ell\colon \mathsf{D}_{\ell}\in\mathcal{D}_1} \sup_{\mathsf{D}\in \mathcal{D}_1} \bar{J}_{\ell}(\mathsf{D}_{\ell},T,\rho) + \sum_{\ell\colon \mathsf{D}_{\ell}\in\mathcal{D}_2} \sup_{\mathsf{D}\in \mathcal{D}_2}\bar{J}_{\ell}(\mathsf{D}_{\ell},T,\rho) \right\} \nonumber\\
	& \leq & \sup_{\mathsf{D}^L\colon L_{\mathcal{D}_1}(\mathsf{D}^L) < L/2} \bigl\{ L_{\mathcal{D}_1}(\mathsf{D}^L)\left(\bar{J}_{\mathcal{D}_1}(T,\rho)  + K_{\mathcal{D}_1}(T,\rho)\right) \nonumber\\
	& & \qquad\qquad\qquad\qquad {} + \left(L-L_{\mathcal{D}_1}(\mathsf{D}^L)\right)\left(\bar{J}_{\mathcal{D}_2}(T,\rho) +K_{\mathcal{D}_2}(T,\rho)\right) \bigr\} \nonumber\\
	& \leq & L \left(\bar{J}_{\mathcal{D}_1}(T,\rho)  + K_{\mathcal{D}_1}(T,\rho) + K_{\mathcal{D}_2}(T,\rho)\right) - \frac{L}{2} \Theta(T,\rho) \label{Equ_MC_C2sup}
\end{IEEEeqnarray}
where $K_{\mathcal{D}_1}(T,\rho)$ and $K_{\mathcal{D}_2}(T,\rho)$ are functions of $T$ and $\rho$ that vanish as $\rho\to\infty$. Here, the first inequality follows from \eqref{eq:SUI_sucks} in the proof of Lemma~\ref{MC_Lemma7b} and from Lemmas~\ref{MC_Lemma7} and \ref{MC_Lem8}; the second inequality follows from the definition of $\Theta(T,\rho)$ in \eqref{Equ_MC_JD2diff} and because $0\leq L_{\mathcal{D}_1}(\mathsf{D}^L) < L/2$.

Substituting \eqref{Equ_MC_C2sup} into \eqref{MC_C2_Ori}, we obtain the desired upper bound
\begin{equation}
	\frac{\log |\mathcal{C}_2|}{L} \leq \bar{J}_{\mathcal{D}_1}(T,\rho)  - \frac{1}{2} \Theta(T,\rho) +K_{\mathcal{D}_1}(T,\rho) + K_{\mathcal{D}_2}(T,\rho) + K_{\mathcal{C}_2}(L,T) \label{Equ_Car_C2}
\end{equation}
where
\begin{equation}
	K_{\mathcal{C}_2}(L,T) \triangleq \sqrt{\frac{2 \bar{U}(T)}{L(1-\epsilon)}} - \frac{1}{L}\log\left(\frac{1-\epsilon}{2}\right)
\end{equation}
is a function of $L$ and $T$ that satisfies
\begin{equation}
	\left|K_{\mathcal{C}_2}(L,T)\right| \leq \frac{A}{L}, \quad L\geq L_0 \label{eq:asymp_KC2}
\end{equation}
for some constants $A$ and $L_0$ that are independent of $L$ and $\rho$.

\subsubsection{Upper bound on the cardinality of $\mathcal{C}$}\label{MC_Sec: C12}
As shown in the previous subsections, the cardinalities of $\mathcal{C}_1$ and $\mathcal{C}_2$ can be upper-bounded by (cf.\ \eqref{Equ_Car_C1} and \eqref{Equ_Car_C2})
\begin{subequations}
	\begin{IEEEeqnarray}{lClCl}
		\frac{\log |\mathcal{C}_1|}{L} & \leq & \bar{J}_{\mathcal{D}_1}(T,\rho)- \sqrt{\frac{T^2 \tilde{V}(T) - K_{\bar{U}}(T,\rho)}{L}}Q^{-1}(\epsilon)  + K_{\mathcal{D}_1}(T,\rho) + K_{\mathcal{C}_1}(L,T) & \triangleq & \bar{R}_{\mathcal{C}_1}(L,T,\rho) \label{eq:barR1}\\
		\frac{\log |\mathcal{C}_2|}{L}
		& \leq & \bar{J}_{\mathcal{D}_1}(T,\rho)  - \frac{1}{2} \Theta(T,\rho) +K_{\mathcal{D}_1}(T,\rho) + K_{\mathcal{D}_2}(T,\rho)+ K_{\mathcal{C}_2}(L,T) 
		& \triangleq & \bar{R}_{\mathcal{C}_2}(L,T,\rho). \label{eq:barR2} 
	\end{IEEEeqnarray}
\end{subequations}
Therefore, the cardinality of $\mathcal{C}$, i.e., $|\mathcal{C}|=M$, can be upper-bounded as
\begin{IEEEeqnarray}{lCl}
	\log M & \leq & \log\left(e^{L \bar{R}_{\mathcal{C}_1}(L,T,\rho)} + e^{L\bar{R}_{\mathcal{C}_2}(L,T,\rho)} \right) \nonumber\\
	& = & L \bar{R}_{\mathcal{C}_1}(L,T,\rho) + \log\left(1+e^{L\left(\bar{R}_{\mathcal{C}_2}(L,T,\rho) - \bar{R}_{\mathcal{C}_1}(L,T,\rho)\right)}\right).\label{eq:bla3} 
\end{IEEEeqnarray}

By the asymptotic behaviors of $\Theta(T,\rho)$, $K_{\bar{U}}(T,\rho)$, and $K_{\mathcal{D}_2}(T,\rho)$ (cf.~\eqref{eq:MC_JD2diff_asymp}, \eqref{eq:vanishing_KU}, and \eqref{eq:vanishing_KD2}), we have that  $\Theta(T,\rho)\geq \Theta(T)/2$, $T^2 \tilde{V}(T) - K_{\bar{U}}(T,\rho)\leq 2T^2 \tilde{V}(T)$, and $K_{\mathcal{D}_2}(T,\rho) \leq \Theta(T)/8$ for $\rho\geq\rho_0$ and a sufficiently large $\rho_0$ (that is independent of $L$). Thus, for $\rho\geq\rho_0$,
\begin{IEEEeqnarray}{lCl}
	\IEEEeqnarraymulticol{3}{l}{\bar{R}_{\mathcal{C}_2}(L,T,\rho) - \bar{R}_{\mathcal{C}_1}(L,T,\rho)} \nonumber\\
	\quad & = & - \frac{1}{2} \Theta(T,\rho) + \sqrt{\frac{T^2 \tilde{V}(T) - K_{\bar{U}}(T,\rho)}{L}}Q^{-1}(\epsilon) + K_{\mathcal{D}_2}(T,\rho) + K_{\mathcal{C}_2}(L,T) - K_{\mathcal{C}_1}(L,T) \nonumber\\
	& \leq & -\frac{\Theta(T)}{8} + \sqrt{\frac{2T^2 \tilde{V}(T)}{L}}Q^{-1}(\epsilon) + K_{\mathcal{C}_2}(L,T) - K_{\mathcal{C}_1}(L,T). 
\end{IEEEeqnarray}
Similarly, by the asymptotic behaviors of $K_{\mathcal{C}_1}(L,T)$ and $K_{\mathcal{C}_2}(L,T)$ (cf.~\eqref{eq:asymp_KC1} and \eqref{eq:asymp_KC2}), we have that 
\begin{equation}
	\sqrt{\frac{2T^2 \tilde{V}(T)}{L}}Q^{-1}(\epsilon) + K_{\mathcal{C}_2}(L,T) - K_{\mathcal{C}_1}(L,T) \leq \frac{\Theta(T)}{16}
\end{equation}
for $L\geq L_0$ and a sufficiently large $L_0$ that is independent of $\rho$. Consequently,
\begin{equation}
	\bar{R}_{\mathcal{C}_2}(L,T,\rho) - \bar{R}_{\mathcal{C}_1}(L,T,\rho) \leq - \frac{\Theta(T)}{16}, \quad \rho\geq\rho_0,\,L\geq L_0.
\end{equation}
Together with \eqref{eq:bla3}, this allows us to upper-bound the cardinality of $\mathcal{C}$ as
\begin{equation}
	\label{eq:MC_almost}
	\frac{\log  M}{L} \leq \bar{R}_{\mathcal{C}_1}(L,T,\rho)  + \frac{1}{L}\log\left(1+e^{-L \frac{\Theta(T)}{16}}\right), \quad \rho\geq\rho_0,\,L\geq L_0.
\end{equation}
The second term on the RHS of \eqref{eq:MC_almost} decays super-exponentially in $L$. We thus conclude from \eqref{eq:barR1} and \eqref{eq:MC_almost} that
\begin{equation}
	R^{\ast}(L, T, \epsilon, \rho) = \frac{\log  M }{LT} \leq \frac{\bar{J}_{\mathcal{D}_1}(T,\rho)}{T} - \sqrt{\frac{\tilde{V}(T) - K_{\bar{U}}(T,\rho)}{L}}Q^{-1}(\epsilon)  + K_{\mathcal{D}_1}(T,\rho) + K_{\textnormal{MC}}(L,T) \label{Equ_R_upper}
\end{equation}
where
\begin{equation}\label{Equ_KMC_1}
	K_{\textnormal{MC}}(L,T) \triangleq K_{\mathcal{C}_1}(L,T)  +  \frac{1}{L}\log\left(1+e^{-L \frac{\Theta(T)}{16}}\right)
\end{equation}
is a function of $L$ and $T$ which satisfies
\begin{equation} \label{Equ_KMC_2}
	\left|K_{\textnormal{MC}}(L,T)\right| \leq A\frac{\log L}{L}, \quad L\geq L_0
\end{equation}
for some constants $A$ and $L_0$ that are independent of $L$ and $\rho$.

Comparing \eqref{Def_tildeJ} with \eqref{Def_It}, we note that
\begin{equation}
	\frac{\bar{J}_{\mathcal{D}_1}(T,\rho)}{T} = \tilde{I}(T,\rho) + K_{\bar{J}}(T,\rho)
\end{equation}
where $K_{\bar{J}}(T,\rho)$ is a function of $T$ and $\rho$ that vanishes as $\rho\to\infty$. We can thus write \eqref{Equ_R_upper} as
\begin{equation}
	\label{eq:MC_FINAL}
	R^{\ast}(L, T, \epsilon, \rho) \leq  \tilde{I}(T,\rho) + K_{\textnormal{MC}}(T,\rho) - \sqrt{\frac{\tilde{V}(T) - K_{\bar{U}}(T,\rho)}{L}}Q^{-1}(\epsilon) + K_{\textnormal{MC}}(L,T)
\end{equation}
where $K_{\textnormal{MC}}(T,\rho)\triangleq K_{\mathcal{D}_1}(T,\rho) + K_{\bar{J}}(T,\rho)$. Note that $\tilde{K}_{\xi}(T,\rho)$, $\xi\in\{\textnormal{MC},\bar{U}\}$ satisfy \eqref{eq:thm_little_o} and $K_{\textnormal{MC}}(L,T)$ satisfies \eqref{eq:thm_big_O}.

\section{Conclusion}\label{Sec_Concl}
We presented a high-SNR normal approximation for the maximum coding rate $R^{\ast}(L, T, \epsilon, \rho)$ achievable over noncoherent MIMO Rayleigh block-fading channels using an error-correcting code that spans $L$ coherence intervals of length $T$, satisfies a per-block power constraint $\rho$, and achieves a probability of error not larger than $\epsilon$. While the approximation was derived under the assumption that the number of coherence intervals $L$ and the SNR $\rho$ tend to infinity, numerical analyses suggest that it becomes accurate already for moderate SNR values and coherence intervals. For example, for SNR values of $25$~dB, the high-SNR normal approximation is accurate for arbitrary numbers of transmit and receive antennas and 5 coherence intervals or more. For $2$ transmit antennas and $4$ receive antennas, it is even accurate for SNR values of $15$~dB.

The obtained normal approximation complements the nonasymptotic bounds presented in \cite{yang2012diversity} and \cite{durisi2015short}, whose evaluation is computationally demanding. Furthermore, it lays a theoretical foundation for analytical analyses that study the behavior of the maximum coding rate as a function of system parameters such as SNR, number of coherence intervals, number of transmit and receive antennas, or blocklength. \emph{Inter alia}, it enables an analysis of the fundamental tradeoff between diversity, multiplexing, and channel-estimation cost at finite blocklength and finite SNR. In Section~\ref{Sec_EngineeringWisdom}, we applied the normal approximation to determine how many of the $n_t$ available transmit antennas should be employed in order to optimize the maximal coding rate at finite blocklength. Indeed, there is a tradeoff between multiplexing gain and cost of estimating the fading coefficients, which both grow with the number of active transmit antennas. While this tradeoff is trivial with respect to channel capacity, the maximum coding rate has a more intricate dependence on the number of active transmit antennas at finite blocklength. Last but not least, our normal approximation may serve as a proxy for the maximum coding rate in the analysis of communication protocols for short-packet wireless communication systems, similarly to the normal approximations used in \cite{Makki15,Makki16,Zhang18,Li19,Mahmood21,Munari21,Feng23,Xie23,Zheng24,Li25}.
				
\appendices





\section{{Proof of Lemma \ref{DT_Lem_1}}} \label{Sec_Pf_DTLem1} 
To prove Lemma~\ref{DT_Lem_1}, we analyze in Appendix~\ref{Sec_pf_Lem1_A} the  USTM information density $i^{(U)}(\mathbf{X};\mathbf{Y})$ and express it in terms of the singular values of $\mathbf{Y}$.\footnote{Throughout this appendix, the subscript $\ell$ is immaterial and therefore omitted.} The asymptotic behavior of these singular values in the limit as $\rho$ tends to infinity is studied in Appendix~\ref{Sec_pf_Lem1_B}. The convergence of the channel dispersion, and hence Lemma~\ref{DT_Lem_1}, is proved in Appendix~\ref{Sec_pf_Lem1_C}. Some auxiliary derivations are deferred to Appendix~\ref{sub:(107)}.

\subsection{The USTM Information Density $i^{(U)}(\mathbf{X};\mathbf{Y})$} \label{Sec_pf_Lem1_A}
As argued in \cite[App.~A]{durisi2015short}, the conditional distribution of the USTM information density $i^{(U)}(\mathbf{X};\mathbf{Y})$ given $\mathbf{X}=\sqrt{\frac{T\rho}{n_t}}\mathsf{U}$ does not depend on $\mathsf{U}$. Without loss of generality, we can thus set in the following $\mathbf{X}=\sqrt{\frac{T\rho}{n_t}}\mathsf{U}$ with
\begin{equation}
	\mathsf{U} =
	\begin{bmatrix}
		\mathsf{I}_{n_t} \\
		\mathbf{0}_{(T-n_t)\times n_t}
	\end{bmatrix}.  \label{Equ_U}
\end{equation}
Consider the SVD of $\mathbf{H}$, i.e., $\mathbf{H}=\mathbf{U}_{\mathbf{H}}\mathbf{S}_{\mathbf{H}} \mathbf{V}_{\mathbf{H}}^\mathsf{H}$, where $\mathbf{U}_{\mathbf{H}}$ and $\mathbf{V}_{\mathbf{H}}$ are unitary matrices of dimensions $n_t\times n_t$ and $n_r\times n_r$, respectively, and $\mathbf{S}_{\mathbf{H}}$ is an $(n_t \times n_r)$-dimensional rectangular diagonal matrix containing the $n_t$ singular values $\sigma_{\mathbf{H},1},\ldots,\sigma_{\mathbf{H},n_t}$ of $\mathbf{H}$. It follows that $\mathbf{X}\mathbf{H}$ can be expressed as \mbox{$\mathbf{X}\mathbf{H}
=a \mathsf{U} \mathbf{U}_{\mathbf{H}}\mathbf{S}_{\mathbf{H}} \mathbf{V}_{\mathbf{H}}^\mathsf{H}$}, where $a\triangleq\sqrt{T\rho/n_t}$. Using the unitary invariant property \cite[Ch.~5.2]{horn1990matrix} and the circular symmetry of the complex Gaussian matrix $\mathbf{W}$ \cite[App.~A]{Tse_Viswanath_2005}, \cite[Lemma 16]{zheng2002communication}, conditioned on $\mathbf{X}$,  the random matrix $\mathbf{Y}=\mathbf{X}\mathbf{H}+\mathbf{W}$  has the same conditional distribution as $\tilde{\mathbf{Y}}=a\mathsf{U}\mathbf{U}_{\mathbf{H}}\mathbf{\Sigma}_{\mathbf{H}} + \mathbf{W}$, where $\mathbf{\Sigma}_{\mathbf{H}}$ is an $(n_t\times n_t)$-dimensional diagonal matrix containing the singular values $\sigma_{\mathbf{H},1},\ldots,\sigma_{\mathbf{H},n_t}$ \cite[Lemma~16]{zheng2002communication}.
Hence, by \eqref{Equ_U}, we obtain that
\begin{equation}
	\tilde{\mathbf{Y}}=\left[  \begin{array}{c c}
		a\tilde{\mathbf{\Sigma}}_{\mathbf{H}}& \mathbf{0} \\
		\mathbf{0} & \mathbf{0}
	\end{array}
	\right]
	+\left[  \begin{array}{c c}
		\mathbf{W}_{11} & \mathbf{W}_{12}  \\
		\mathbf{W}_{21} & \mathbf{W}_{22} 
	\end{array}
	\right] \label{Euq_transD_Y}
\end{equation}
where $\tilde{\mathbf{\Sigma}}_{\mathbf{H}}\triangleq \mathbf{U}_{\mathbf{H}} \mathbf{\Sigma}_{\mathbf{H}}$, and where $\mathbf{W}_{11},\mathbf{W}_{12},\mathbf{W}_{21}$, and $\mathbf{W}_{22}$ are submatrices of dimensions $n_t \times n_t$, $n_t \times (n_r-n_t)$, $(T-n_t) \times n_t$, and $(T-n_t) \times (n_r-n_t)$, respectively, with i.i.d.\ $\mathcal{CN}(0,1)$ entries which are independent of $\mathbf{H}$. Because $\det(\tilde{\mathbf{\Sigma}}_{\mathbf{H}})=\det(\mathbf{U}_{\mathbf{H}})\det(\mathbf{\Sigma}_{\mathbf{H}} )>0$, we have that $\tilde{\mathbf{\Sigma}}_{\mathbf{H}}$ is invertible and has non-zero eigenvalues.

We next use that, for $\mathbf{X}=a\mathsf{U}$,
\begin{IEEEeqnarray}{lCl}
	\text{tr}(\tilde{\mathbf{Y}}^\mathbf{H}(\mathsf{I}_{T}+ \mathbf{X}\mathbf{X}^\mathsf{H})^{-1}\tilde{\mathbf{Y}}) & = & \frac{a^2}{1+a^2} \text{tr}(\mathbf{H}^\mathsf{H}\mathbf{H}) + \text{tr}(\mathbf{W}_{21}^\mathsf{H}\mathbf{W}_{21})+ \text{tr}(\mathbf{W}_{22}^\mathsf{H}\mathbf{W}_{22}) \nonumber\\
	& & {} + \frac{2a}{1+a^2} \text{Re}\left(\text{tr}(\tilde{\mathbf{\Sigma}}_{\mathbf{H}}^{\mathsf{H}}\mathbf{W}_{11})\right) + \frac{1}{1+a^2} \Bigl(\text{tr}(\mathbf{W}_{11}^\mathsf{H}\mathbf{W}_{11})+ \text{tr}(\mathbf{W}_{12}^\mathsf{H}\mathbf{W}_{12})  \Bigr). \label{Equ_fyx_tr}
\end{IEEEeqnarray}
where $\text{Re}(\cdot)$ represents the imaginary part.
Consequently, the conditional pdf of $\mathbf{Y}$ given $\mathbf{X}=a\mathsf{U}$ satisfies 
\begin{IEEEeqnarray}{lCl}
	f_{\mathbf{Y}|\mathbf{X}}(\tilde{\mathbf{Y}}|a\mathsf{U})
	& = & \dfrac{e^{-\text{tr}\{\tilde{\mathbf{Y}}^\mathsf{H}(\mathsf{I}_T+ a^2\mathsf{U}\mathsf{U}^\mathsf{H})^{-1}\tilde{\mathbf{Y}}\}}}{ \pi^{Tn_r} \det(\mathsf{I}_T+ a^2\mathsf{U}\mathsf{U}^\mathsf{H})^{n_r}} \nonumber \\
	& = & \dfrac{e^{-\frac{a^2}{1+a^2} \text{tr}(\mathbf{H}^\mathsf{H}\mathbf{H}) - \text{tr}(\mathbf{W}_{21}^\mathsf{H}\mathbf{W}_{21}) - \text{tr}(\mathbf{W}_{22}^\mathsf{H}\mathbf{W}_{22}) - \frac{2a}{1+a^2}  \text{Re}\left(\text{tr}(\tilde{\mathbf{\Sigma}}_{\mathbf{H}}^{\mathsf{H}}\mathbf{W}_{11})\right) - \frac{1}{1+a^2} \left(\text{tr}(\mathbf{W}_{11}^\mathsf{H}\mathbf{W}_{11})+ \text{tr}(\mathbf{W}_{12}^\mathsf{H}\mathbf{W}_{12}) \right)} }{ \pi^{Tn_r}(1+a^2)^{n_tn_r}}. \IEEEeqnarraynumspace\label{Equ_fy_x}
\end{IEEEeqnarray}

We next consider the output pdf $f_{\mathbf{Y}}^{(U)}$ induced by USTM channel inputs and the channel \eqref{Equ_Channel}. To this end, we denote the singular values of $\mathbf{Y}$ arranged in decreasing order by $\boldsymbol{\sigma}_1,\ldots,\boldsymbol\sigma_{n_r}$. Then, by expressing $\mathbf{Y}$ in terms of its SVD $\mathbf{Y}=\mathbf{U}_{\mathbf{Y}}\mathbf{\Sigma}_{\mathbf{Y}} \mathbf{V}_{\mathbf{Y}}^\mathsf{H}$, where $\mathbf{U}_{\mathbf{Y}}$ and $\mathbf{V}_{\mathbf{Y}}$ are (truncated) unitary matrices and $\mathbf{\Sigma}_{\mathbf{Y}}$ is a diagonal matrix containing the singular values $\boldsymbol\sigma_1, \ldots, \boldsymbol\sigma_{n_r}$, the pdf $f_{\mathbf{Y}}^{({\footnotesize U})}$ can be written as \cite{yang2013capacity}
\begin{equation}
	f_{\mathbf{Y}}^{({\footnotesize U})}(\mathsf{Y})
	=\frac{f_{\footnotesize \mathbf{U}_{\mathbf{Y}}}(\mathsf{U}_{\mathsf{Y}})f_{\mathbf{\Sigma}_{\mathbf{Y}}}(\sigma_1, \ldots, \sigma_{n_r})f_{\footnotesize \mathbf{V}_{\mathbf{Y}}}(\mathsf{V}_{\mathsf{Y}})}{J(T,\Sigma_{\mathsf{Y}})}, \quad \mathsf{Y} = \mathsf{U}_{\mathsf{Y}}\Sigma_{\mathsf{Y}} \mathsf{V}_{\mathsf{Y}}^{\mathsf{H}} \label{Equ_fy_ori}
\end{equation}
where $\Sigma_{\mathsf{Y}}$ is a diagonal matrix containing the singular values of $\mathsf{Y}$. In \eqref{Equ_fy_ori}, $J(T,\Sigma_{\mathsf{Y}})$ is the Jacobian of the SVD, i.e.,
\begin{equation}
	J(T,\Sigma_{\mathsf{Y}})=\prod_{1\leq i<j\leq n_r}(\sigma_i^2-\sigma_j^2)^2 \cdot \prod_{i=1}^{n_r}\sigma_i^{2(T-n_r)+1}
\end{equation}
and the pdfs of $\mathbf{U}_{\mathbf{Y}}$ and $\mathbf{V}_{\mathbf{Y}}$ are given by $f_{\footnotesize \mathbf{U}_{\mathbf{Y}}}(\mathsf{U}_{\mathsf{Y}})=\frac{\Gamma_{n_r}(T)}{\pi^{n_r(T-1)}}$ and $f_{\footnotesize \mathbf{V}_{\mathbf{Y}}}(\mathsf{V}_{\mathsf{Y}})=\frac{\Gamma_{n_r}(n_r)}{2^{n_r} \pi^{n_r^2}}$. We next define
\begin{equation}  
	\bar{\boldsymbol\sigma}_i \triangleq \begin{cases} \sqrt{\frac{n_t}{T\rho}}\boldsymbol\sigma_i, \quad & i=1,\ldots,n_t\\\boldsymbol\sigma_i, \quad & i=n_t+1,\ldots,n_r.\end{cases} \label{Def_tsigma}
\end{equation}
It follows that $f_{\bm{\Sigma}_{\mathbf{Y}}}(\sigma_1, \ldots, \sigma_{n_r})= \left(\frac{T\rho}{n_t}\right)^{n_t/2}f_{\bar{\boldsymbol\sigma}}(\bar{\sigma}_1,\ldots,\bar{\sigma}_{n_r})$ and $J(T,\Sigma_{\mathsf{Y}})= \bar{J}(T,\bar\Sigma_{\mathsf{Y}}) \left(\frac{n_t}{T\rho}\right)^{n_t(T+n_r-n_t-1/2)}$, where $f_{\bar{\boldsymbol\sigma}}$ denotes the joint pdf of $\bar{\boldsymbol\sigma}_1,\ldots,\bar{\boldsymbol\sigma}_{n_r}$ and
\begin{IEEEeqnarray}{lCl}
	\bar{J}(T,\bar\Sigma_{\mathsf{Y}}) & \triangleq & \left(\prod_{i=1}^{n_r} \bar{\sigma}_i^{2(T-n_r)+1}\right) \left(\prod_{1\leq i<j\leq n_t}  (\bar{\sigma}_i^2-\bar{\sigma}_j^2)^2\right) \times \nonumber\\
	& & {} \times \left(\prod_{n_t+1\leq i<j\leq n_r}(\bar{\sigma}_i^2-\bar{\sigma}_j^2)^2 \right) 
	 \left(\prod_{i=1}^{n_t}\prod_{j=n_t+1}^{n_r}\left(\bar{\sigma}_i^2-\frac{n_t}{T\rho}\bar{\sigma}_j^2\right)^2\right). \label{Def_J_sigma}
\end{IEEEeqnarray}

Finally, we denote by $\tilde{\boldsymbol\sigma}_i$, $i=1,\ldots,n_r$ the random variables that are obtained by replacing in \eqref{Def_tsigma} the singular values $\boldsymbol\sigma_i$ by the singular values of $\tilde{\mathbf{Y}}$. Since $\tilde{\mathbf{Y}}$ has the same distribution as $\mathbf{Y}$, it follows that $(\tilde{\boldsymbol\sigma}_1,\ldots,\tilde{\boldsymbol\sigma}_{n_r})$ have the same joint distribution as $(\bar{\boldsymbol\sigma}_1,\ldots,\bar{\boldsymbol\sigma}_{n_r})$, and the joint pdf of $(\tilde{\boldsymbol\sigma}_1,\ldots,\tilde{\boldsymbol\sigma}_{n_r})$, denoted by $f_{\tilde{\boldsymbol\sigma}}$, is equal to $f_{\bar{\boldsymbol\sigma}}$.

Combining \eqref{Equ_fy_x}--\eqref{Def_J_sigma}, we obtain that the USTM information density $i^{(U)}(\mathbf{X};\mathbf{Y})$ satisfies
\begin{IEEEeqnarray}{lCl}
 i^{(U)}(\mathbf{X};\mathbf{Y}) & \stackrel{d}=&\log \frac{f_{\mathbf{Y}|\mathbf{X}}(\tilde{\mathbf{Y}}|a\mathsf{U})}{f_{\mathbf{Y}}^{(U)}(\tilde{\mathbf{Y}})} \nonumber\\
& =& n_t(T-n_t)\log \left(\frac{T\rho}{n_t}\right) - \log f_{\tilde{\boldsymbol\sigma}}(\tilde{\boldsymbol\Sigma}_{\tilde{\mathbf{Y}}})+\log \bar{J}(T,\tilde{\boldsymbol\Sigma}_{\tilde{\mathbf{Y}}}) - \Xi(\mathbf{H},\mathbf{W}) -  \Delta_\Xi(\rho,\mathbf{H},\mathbf{W})
		\label{Def_i}
\end{IEEEeqnarray}
where
\begin{subequations}
\begin{IEEEeqnarray}{rCl}
	 \Xi(\mathbf{H},\mathbf{W}) & \triangleq & \text{tr}(\mathbf{H}^\mathsf{H}\mathbf{H})+\text{tr}(\mathbf{W}_{21}^\mathsf{H}\mathbf{W}_{21})+\text{tr}(\mathbf{W}_{22}^\mathsf{H}\mathbf{W}_{22})
	+Tn_r\log \pi 
	+\log \left(\frac{\Gamma_{n_r}(T)\Gamma_{n_r}(n_r)}{2^{n_r}\pi^{n_r(T-1)+n_r^2}}\right) \\
	\Delta_\Xi(\rho,\mathbf{H},\mathbf{W}) & \triangleq &  \frac{1}{1+a^2} \left(\text{tr}(\mathbf{W}_{11}^\mathsf{H}\mathbf{W}_{11})+ \text{tr}(\mathbf{W}_{12}^\mathsf{H}\mathbf{W}_{12}) - \text{tr}(\mathbf{H}^\mathsf{H}\mathbf{H}) \right)\nonumber\\
	& & {} + \frac{2a}{1+a^2} \text{Re}\left(\text{tr}(\tilde{\mathbf{\Sigma}}_{\mathbf{H}}^{\mathsf{H}}\mathbf{W}_{11})\right) + n_tn_r \log\left(1+\frac{1}{a^2}\right)
\end{IEEEeqnarray}
\end{subequations}
and where, with a slight abuse of notation, we denote by $\tilde{\boldsymbol\Sigma}_{\tilde{\mathbf{Y}}}$ both the random variables $(\tilde{\boldsymbol\sigma}_1,\ldots,\tilde{\boldsymbol\sigma}_{n_r})$ and the diagonal matrix that contains these random variables.
 
 In order to characterize the convergence of the channel dispersion, we next analyze the asymptotic behavior of the RHS of \eqref{Def_i}. This, in turn, requires an asymptotic analysis of $\tilde{\boldsymbol\Sigma}_{\tilde{\mathbf{Y}}}$ and of the pdf $f_{\tilde{\boldsymbol\sigma}}$, which we shall carry out in the next subsection.
  
\subsection{Asymptotics of $\tilde{\boldsymbol\Sigma}_{\tilde{\mathbf{Y}}}$ and $f_{\tilde{\boldsymbol\sigma}}$} \label{Sec_pf_Lem1_B}

Let $\boldsymbol\varsigma_1,\ldots,\boldsymbol\varsigma_{n_t}$ be the ordered nonzero singular values of $\mathbf{H}$, and let $\boldsymbol\varsigma_{n_t+1},\ldots,\boldsymbol\varsigma_{n_r}$ be the ordered nonzero singular values of $\mathbf{W}_{22}$. The following lemmas concern the convergence of $(\tilde{\boldsymbol\sigma}_1,\ldots,\tilde{\boldsymbol\sigma}_{n_r})$ to $(\boldsymbol\varsigma_1,\ldots,\boldsymbol\varsigma_{n_r})$ and of $f_{\bar{\boldsymbol\sigma}}$ to $f_{\boldsymbol\varsigma}$ (where $f_{\boldsymbol\varsigma}$ denotes the joint pdf of $(\boldsymbol\varsigma_1,\ldots,\boldsymbol\varsigma_{n_r})$).
  
\begin{Lemma}[$\!\!$\protect{\cite[Lemmas~12 \& 13]{yang2013capacity}}] \label{Lem_13}  
	The singular values $(\tilde{\boldsymbol\sigma}_1,\ldots,\tilde{\boldsymbol\sigma}_{n_r})$ and $(\boldsymbol\varsigma_1,\ldots,\boldsymbol\varsigma_{n_r})$, and their corresponding pdfs, satisfy the following: 
	\begin{enumerate} 
		\item\label{Lem_13_1} $\displaystyle\sum_{i=1}^{n_r} \mathsf{E}\left[\tilde{\bm\sigma}_i^2\right] \leq \mathsf{K}_1$, where $\mathsf{K}_1$ is a finite constant which is independent of $\rho$.
		\item\label{Lem_13_2)}  For any subset $\tilde{\boldsymbol\Sigma} \subseteq \{\tilde{\boldsymbol\sigma}_1,\ldots,\tilde{\boldsymbol\sigma}_{n_r}\}$, the pdf of $\tilde{\boldsymbol\Sigma}$ satisfies $\displaystyle \left|f_{\tilde{\boldsymbol\Sigma}}(\tilde{\Sigma})\right| \leq \mathsf{K}_2$ for every $\tilde{\Sigma}$ and $\rho\geq\rho_0$, where $\mathsf{K}_2$ and $\rho_0$ are finite constants which are independent of $\rho$.
		\item\label{Lem_13_3)} $f_{\tilde{\boldsymbol\sigma}}$ converges pointwise to $f_{\boldsymbol\varsigma}$ as $\rho\to\infty$. 
	\end{enumerate}
\end{Lemma}
\begin{IEEEproof}
	The boundedness of $\sum_{i=1}^{n_r} \mathsf{E}\left[\tilde{\bm\sigma}_i^2\right]$ and $f_{\tilde{\boldsymbol\sigma}}$ (Parts~\ref{Lem_13_1}) and \ref{Lem_13_2)})) was shown in \cite[App.~B]{yang2013capacity}. The boundedness of $f_{\tilde{\boldsymbol\Sigma}}$ follows by marginalizing the pdf $f_{\tilde{\boldsymbol\sigma}}$, characterized in \cite[App.~A]{yang2013capacity}, over $\{\tilde{\boldsymbol\sigma}_1,\ldots,\tilde{\boldsymbol\sigma}_{n_r}\}\setminus\tilde{\Sigma}$ and by bounding the corresponding integrals. These integrals can be bounded by integrals of the form $\int_0^{\infty} e^{-x^2}x^m\text{d} x$, which are bounded. Part~\ref{Lem_13_3)}) is \cite[Lemma~12]{yang2013capacity}. Note that, in \cite{yang2013capacity}, $(\boldsymbol\varsigma_{n_t+1},\ldots,\boldsymbol\varsigma_{n_r})$ are defined as the nonzero singular values of an $(T-n_t)\times (n_r-n_t)$  random matrix with i.i.d.\ $\mathcal{CN}(0,1)$ entries which are independent of $\mathbf{H}$. Since $\mathbf{W}_{22}$ has the same distribution as such matrix, the singular values $(\boldsymbol\varsigma_{n_t+1},\ldots,\boldsymbol\varsigma_{n_r})$ have the same distribution as the ones considered in \cite{yang2013capacity}. 
\end{IEEEproof}

\begin{Lemma}[$\!\!$\protect{\cite[Lemma~16]{zheng2002communication}}] \label{Lem_singularY}  
	As $\rho$ tends to infinity, the singular values $(\tilde{\boldsymbol\sigma}_1,\ldots,\tilde{\boldsymbol\sigma}_{n_r})$ converge almost surely to $(\boldsymbol\varsigma_1,\ldots,\boldsymbol\varsigma_{n_r})$.
\end{Lemma}
\begin{IEEEproof}
The proof of Lemma~\ref{Lem_singularY} follows the steps outlined in \cite[App.~F]{zheng2002communication}. Indeed, the convergence of $(\tilde{\boldsymbol\sigma}_1,\ldots,\tilde{\boldsymbol\sigma}_{n_r})$ to $(\boldsymbol\varsigma_1,\ldots,\boldsymbol\varsigma_{n_r})$ is determined by the asymptotic behavior of the eigenvalues $(\tilde{\boldsymbol\lambda}_1,\ldots,\tilde{\boldsymbol{\lambda}}_{n_r})$ of $\tilde{\mathbf{Y}}^{\mathsf{H}}\tilde{\mathbf{Y}}$.

We shall first consider the eigenvalues that are bounded in $\rho$. So let $\tilde{\boldsymbol{\lambda}}$ be an eigenvalue of $\tilde{\mathbf{Y}}^{\mathsf{H}}\tilde{\mathbf{Y}}$ that satisfies $\varlimsup_{\rho\to\infty} \tilde{\boldsymbol{\lambda}} < \infty$. It follows that the characteristic polynomial
\begin{equation}
f(\tilde{\boldsymbol\lambda}) \triangleq \det( \tilde{\boldsymbol\lambda} \mathsf{I}_{n_r}-\tilde{\mathbf{Y}}^\mathsf{H}\tilde{\mathbf{Y}})
\end{equation}
is zero. Using \eqref{Euq_transD_Y}, $f(\tilde{\boldsymbol\lambda})$ can be written as
\begin{equation}
f(\tilde{\boldsymbol\lambda}) = \det \left(\left[  \begin{array}{c c}
		\mathbf{A} & \mathbf{B} \\
		\mathbf{B}^{\mathsf{H}} & \mathbf{D}
	\end{array}
	\right] \right)
\end{equation}
where
\begin{subequations}
	\begin{IEEEeqnarray}{lCl}
	\mathbf{A} & \triangleq &	\tilde{\boldsymbol\lambda} \mathsf{I}_{n_t} 
	-\left( \bigl(a\tilde{\mathbf{\Sigma}}_{\mathbf{H}}+ \mathbf{W}_{11}\bigr)^\mathsf{H}\bigl(a\tilde{\mathbf{\Sigma}}_{\mathbf{H}}+ \mathbf{W}_{11}\bigr)+ \mathbf{W}_{21}^\mathsf{H} \mathbf{W}_{21}\right) \label{Equ_MatrixA} \\
	\mathbf{B} & \triangleq &- \bigl(a \tilde{\mathbf{\Sigma}}_{\mathbf{H}}^\mathsf{H} \mathbf{W}_{12}
	+ \mathbf{W}_{11}^\mathsf{H} \mathbf{W}_{12}
	+\mathbf{W}_{21}^\mathsf{H} \mathbf{W}_{22}
	\bigr)\\
	\mathbf{D} & \triangleq & \tilde{\boldsymbol\lambda} \mathsf{I}_{n_r-n_t} 
	-\bigl(\mathbf{W}_{12}^\mathsf{H}\mathbf{W}_{12}+\mathbf{W}_{22}^\mathsf{H}\mathbf{W}_{22}\bigr).
\end{IEEEeqnarray} 
\end{subequations}
Next note that, since $\tilde{\boldsymbol\lambda}$ is bounded, $\frac{1}{a^2}\mathbf{A}$ converges almost surely to $-\tilde{\mathbf{\Sigma}}_{\mathbf{H}}^{\mathsf{H}}\tilde{\mathbf{\Sigma}}_{\mathbf{H}}$ as $\rho$ (and hence also $a$) tends to infinity. Since $\det(\tilde{\mathbf{\Sigma}}_{\mathbf{H}})>0$, it follows that $\varliminf_{\rho\to\infty} |\det(\mathbf{A})| > 0$. We can therefore use the Schur complement to express the characteristic polynomial as
\begin{equation}
f(\tilde{\boldsymbol\lambda}) = \det(\mathbf{A}) \det(\mathbf{D}-\mathbf{B}^{\mathsf{H}}\mathbf{A}^{-1}\mathbf{B}). 
\end{equation}
Furthermore, $f(\tilde{\boldsymbol\lambda})=0$ implies that $ \det(\mathbf{D}-\mathbf{B}^{\mathsf{H}}\mathbf{A}^{-1}\mathbf{B}) = 0$. Through algebraic manipulations, it can be shown that, as $\rho\to\infty$,
\begin{equation}
	\mathbf{B}^{\mathsf{H}}\mathbf{A}^{-1}\mathbf{B} \to - \mathbf{W}_{12}^{\mathsf{H}} \mathbf{W}_{12}, \quad \textnormal{almost surely.}
\end{equation}
It thus follows from the continuity of the determinant that $\det(\mathbf{D}-\mathbf{B}^{\mathsf{H}}\mathbf{A}^{-1}\mathbf{B})$ converges almost surely to $\det(\tilde{\boldsymbol\lambda}\mathsf{I}_{n_r-n_t}-\mathbf{W}_{22}^{\mathsf{H}}\mathbf{W}_{22})$. Consequently, as $\rho\to\infty$, the eigenvalue $\tilde{\boldsymbol\lambda}$ converges almost surely to an eigenvalue of $\mathbf{W}_{22}^{\mathsf{H}}\mathbf{W}_{22}$. We conclude that $\tilde{\mathbf{Y}}^{\mathsf{H}}\tilde{\mathbf{Y}}$ has $n_r-n_t$ bounded eigenvalues that converge almost surely to the eigenvalues of $\mathbf{W}_{22}^{\mathsf{H}}\mathbf{W}_{22}$. Clearly, these are the $n_r-n_t$ smallest eigenvalues and correspond to the singular values $(\tilde{\boldsymbol\sigma}_{n_t+1},\ldots,\tilde{\boldsymbol\sigma}_{n_r})$, since the remaining eigenvalues are unbounded. It follows that $(\tilde{\boldsymbol\sigma}_{n_t+1},\ldots,\tilde{\boldsymbol\sigma}_{n_r})$ converge almost surely to the $n_r-n_t$ nonzero singular values of $\mathbf{W}_{22}$.

We next consider the unbounded eigenvalues of $\tilde{\mathbf{Y}}^{\mathsf{H}}\tilde{\mathbf{Y}}$. Let $\tilde{\boldsymbol\lambda}=a^2\tilde{\boldsymbol\sigma}^2$ for some $\tilde{\boldsymbol\sigma}>0$. Since $\tilde{\boldsymbol\lambda}$ is unbounded, it follows that $\varliminf_{\rho\to\infty} |\det(\mathbf{D})| > 0$. We can therefore use again the Schur complement to express the characteristic polynomial as
\begin{equation}
f(\tilde{\boldsymbol\lambda}) = \det(\mathbf{D})\det(\mathbf{A}-\mathbf{B}\mathbf{D}^{-1}\mathbf{B}^{\mathsf{H}}).
\end{equation}
Furthermore, $f(\tilde{\boldsymbol\lambda})=0$ implies that $\det(\mathbf{A}-\mathbf{B}\mathbf{A}^{-1}\mathbf{B}^{\mathsf{H}})=0$. Through algebraic manipulations, it can be shown that, as $\rho\to\infty$,
\begin{subequations}
\begin{IEEEeqnarray}{rCll}
\frac{1}{a^2} \mathbf{A} & \to & \tilde{\boldsymbol\sigma}\mathsf{I}_{n_t} - \tilde{\boldsymbol\Sigma}_{\mathbf{H}}^{\mathsf{H}}\tilde{\boldsymbol\Sigma}_{\mathbf{H}}, \quad & \textnormal{almost surely}\\
\frac{1}{a} \mathbf{B} & \to & - \tilde{\boldsymbol\Sigma}_{\mathbf{H}}^{\mathsf{H}}\mathbf{W}_{12}, \quad & \textnormal{almost surely} \\
\mathbf{D}^{-1} & \to & 0, \quad & \textnormal{almost surely.}
\end{IEEEeqnarray}
\end{subequations}
It thus follows from the continuity of the determinant that $\det(\frac{1}{a^2}\mathbf{A}-\frac{1}{a^2}\mathbf{B}\mathbf{A}^{-1}\mathbf{B}^{\mathsf{H}})$ converges almost surely to $\det(\tilde{\boldsymbol\sigma}^2\mathsf{I}_{n_t}-\tilde{\boldsymbol\Sigma}_{\mathbf{H}}^{\mathsf{H}}\tilde{\boldsymbol\Sigma}_{\mathbf{H}})$. Consequently, as $\rho\to\infty$, the scaled eigenvalue $\tilde{\boldsymbol\sigma}^2$ converges almost surely to an eigenvalue of $\tilde{\boldsymbol\Sigma}_{\mathbf{H}}^{\mathsf{H}}\tilde{\boldsymbol\Sigma}_{\mathbf{H}}$. It follows that $(\tilde{\boldsymbol\sigma}_1,\ldots,\tilde{\boldsymbol\sigma}_{n_t})$ converge almost surely to the $n_t$ nonzero singular values of $\mathbf{H}$.
\end{IEEEproof}

\subsection{The Convergence of the Channel Dispersion} \label{Sec_pf_Lem1_C}

In Appendix~\ref{sub:(107)}, we shall show that
\begin{equation}
\lim_{\rho\to\infty} \mathsf{E}\left[\bigl( i^{(U)}(\mathbf{X};\mathbf{Y})-I(T,\rho)\bigr)^2\right] = \mathsf{E}\left[\bigl( i^*(\rho,\boldsymbol\varsigma,\mathbf{H},\mathbf{W}) - I^*(T,\rho)\bigr)^2\right] \label{eq:i_to_i*}
\end{equation}
where
\begin{subequations}
\begin{IEEEeqnarray}{rCl}
 i^*(\rho,\boldsymbol\varsigma,\mathbf{H},\mathbf{W}) & \triangleq & n_t(T-n_t)\log \left(\frac{T\rho}{n_t}\right)-\log f_{\boldsymbol{\varsigma}}(\tilde{\boldsymbol\Sigma}_{\boldsymbol\varsigma}) +\log \tilde{J}(T,\tilde{\boldsymbol\Sigma}_{\boldsymbol\varsigma})-\Xi(\mathbf{H}, \mathbf{W}) \label{Def_i_ast}\\
	I^*(T,\rho)
		& \triangleq & \mathsf{E}[ i^*(\rho,\boldsymbol\varsigma,\mathbf{H},\mathbf{W})] \nonumber \\
		& = & n_t(T-n_t)\log \left(\frac{\rho}{n_t}\right)+n_t(T-n_t)\log \left(\frac{T}{e}\right)
		+\log\frac{\Gamma_{n_t}(n_t)}{\Gamma_{n_t}(T)} 
		+(T-n_t)\sum_{i=0}^{n_t-1} \varPsi(n_r-i) \IEEEeqnarraynumspace\label{Def_I_ast} 
	\end{IEEEeqnarray}
\end{subequations}
and
\begin{equation}
\tilde{J}(T,\tilde{\boldsymbol\Sigma}_{\boldsymbol\varsigma}) \triangleq \left(\prod_{i=1}^{n_r}\boldsymbol{\varsigma}_i^{2(T-n_r)+1}\right) \left(\prod_{1\leq i<j\leq n_t}(\boldsymbol{\varsigma}_i^2-\boldsymbol{\varsigma}_j^2)^2\right)\left(\prod_{n_t+1\leq i<j\leq n_r}(\boldsymbol{\varsigma}_i^2-\boldsymbol{\varsigma}_j^2)^2\right) \left(\prod_{i=1}^{n_t}\boldsymbol{\varsigma}_i^{4(n_r-n_t)}\right).  \label{Def_logJu}
\end{equation}
Similar to \eqref{Def_i}, and with a slight abuse of notation, we denote by $\tilde{\boldsymbol\Sigma}_{\boldsymbol\varsigma}$ both the random variables $(\boldsymbol\varsigma_1,\ldots,\boldsymbol\varsigma_{n_r})$ and a diagonal matrix containing these random variables. To compute the expected value in \eqref{Def_I_ast}, we use Lemma~\ref{Lem_singularY} together with a change of variable to express the pdfs of $(\boldsymbol\varsigma_1,\ldots,\boldsymbol\varsigma_{n_t})$ and $(\boldsymbol\varsigma_{n_t+1},\ldots,\boldsymbol\varsigma_{n_r})$ in terms of the pdfs of $\mathbf{H}$ and $\mathbf{W}_{22}$, respectively. After some algebraic manipulations, we obtain
\begin{equation}
\log \tilde{J}(T,\tilde{\boldsymbol\Sigma}_{\boldsymbol\varsigma}) - \log f_{\boldsymbol{\varsigma}}(\tilde{\boldsymbol\Sigma}_{\boldsymbol\varsigma}) -  \Xi(\mathbf{H},\mathbf{W}) = (T-n_t) \log\det(\mathbf{H}^{\mathsf{H}}\mathbf{H}) + \log\frac{\Gamma_{n_t}(n_t)}{\Gamma_{n_t}(T)} - \text{tr}(\mathbf{W}_{21}^{\mathsf{H}}\mathbf{W}_{21}) \label{eq:logJ-logf-Xi}
\end{equation}
whose expected value can be evaluated using that $\sum_{i=1}^{n_t}\mathsf{E}[\log(\boldsymbol\varsigma_i^2)]=\sum_{i=0}^{n_t-1} \varPsi(n_r-i)$ (Lemma~\ref{lem:grant_wishart}) and $\mathsf{E}[\text{tr}(\mathbf{W}_{21}^{\mathsf{H}}\mathbf{W}_{21})]=n_t(T-n_t)$. We next note that
\begin{IEEEeqnarray}{lCl}
\IEEEeqnarraymulticol{3}{l}{\mathsf{E}\left[\big(i^*(\rho,\boldsymbol\varsigma,\mathbf{H},\mathbf{W})-I^*(T,\rho) \big)^2\right]} \nonumber\\ 
\quad &= &\mathsf{E}\left[ \left(
	-\text{tr}(\mathbf{W}_{21}^\mathsf{H}\mathbf{W}_{21})+n_t(T-n_t) 
	+(T-n_t)\log \det(\mathbf{H}^{\mathsf{H}}\mathbf{H}) -(T-n_t)\sum_{i=0}^{n_t-1} \varPsi(n_r-i)  \right)^2 \right] \nonumber \\
	& =& \mathrm{Var}\left(\text{tr}(\mathbf{W}_{21}^\mathsf{H}\mathbf{W}_{21})\right)+(T-n_t)^2 \mathrm{Var}\left(\log\det(\mathbf{H}^{\mathsf{H}}\mathbf{H})\right) \nonumber\\
	& = & n_t(T-n_t) + (T-n_r)^2 \sum_{i=0}^{n_t-1} \varPsi'(n_r-i) \label{eq:Var_i*}
\end{IEEEeqnarray} 
where in the second step, we used that $\mathbf{H}$ and $\mathbf{W}_{21}$ are independent; and in the last step we used that $\mathrm{Var}\left(\log\det(\mathbf{H}^{\mathsf{H}}\mathbf{H}) \right)=\sum_{i=0}^{n_t-1} \varPsi'(n_t-i)$ (Lemma~\ref{lem:grant_wishart}). Lemma~\ref{DT_Lem_1} follows therefore directly from \eqref{eq:i_to_i*} and by comparing the RHS of \eqref{eq:Var_i*} with \eqref{Def_Vt}.

\subsection{Proof of \eqref{eq:i_to_i*}}
\label{sub:(107)}
By \eqref{Def_i}, we have that
\begin{IEEEeqnarray}{lCl}
\underline{U}(T,\rho) & \triangleq & \mathsf{E}\left[\bigl( i^{(U)}(\mathbf{X};\mathbf{Y})-I(T,\rho)\bigr)^2\right] \nonumber\\
 & = & \mathsf{E}\left[\left(n_t(T-n_t)\log \left(\frac{T\rho}{n_t}\right)-\log f_{\tilde{\boldsymbol\sigma}}(\tilde{\boldsymbol\Sigma}_{\tilde{\mathbf{Y}}})+\log\bar{J}(T,\tilde{\boldsymbol\Sigma}_{\tilde{\mathbf{Y}}})- \Xi(\mathbf{H},\mathbf{W})- \Delta_{\Xi}(\rho,\mathbf{H},\mathbf{W})-I(T,\rho)\right)^2\right] \IEEEeqnarraynumspace
\end{IEEEeqnarray}
where $I(T,\rho)\triangleq \mathsf{E}[i^{(U)}(\mathbf{X};\mathbf{Y})]$. We next note that
\begin{equation}
\lim_{\rho\to\infty} \mathsf{E}\left[\bigl(\Delta_{\Xi}(\rho,\mathbf{H},\mathbf{W})\bigr)^2\right] = 0 \label{eq:DeltaXi_lim}
\end{equation}
from which we obtain that, for every $0<\varepsilon<1$, there exists a sufficiently large $\rho_0$ such that
\begin{equation}
\mathsf{E}\left[\bigl(\Delta_{\Xi}(\rho,\mathbf{H},\mathbf{W})\bigr)^2\right] \leq \varepsilon, \quad \rho\geq\rho_0. \label{eq:approx_DeltaXi}
\end{equation}
Indeed, the absolute value of $\Delta_{\Xi}(\rho,\mathbf{H},\mathbf{W})$ converges almost surely to zero as $\rho\to\infty$ and is bounded by
\begin{equation*}
\left|\text{tr}(\mathbf{W}_{11}^\mathsf{H}\mathbf{W}_{11})\right|+ \left|\text{tr}(\mathbf{W}_{12}^\mathsf{H}\mathbf{W}_{12})\right| + \left|\text{tr}(\mathbf{H}^\mathsf{H}\mathbf{H}) \right| +2\left|\text{Re}\left(\text{tr}(\tilde{\mathbf{\Sigma}}_{\mathbf{H}}^{\mathsf{H}}\mathbf{W}_{11})\right)\right| + n_tn_r \log\left(1+\frac{n_t}{T\rho_0}\right), \quad \rho\geq\rho_0
\end{equation*}
which has a finite second moment. The claim \eqref{eq:DeltaXi_lim} thus follows from the dominated convergence theorem. We next note that, by \eqref{Def_I_HS}, there exists a sufficiently large $\rho_0>0$ such that
\begin{equation}
\left|I(T,\rho) - I^{*}(T,\rho)\right| \leq \varepsilon, \quad \rho\geq\rho_0. \label{eq:approx_I}
\end{equation}
Defining
\begin{equation}
\hat{\underline{U}}(T,\rho) \triangleq \mathsf{E}\left[\left(n_t(T-n_t)\log \left(\frac{T\rho}{n_t}\right)-\log f_{\tilde{\boldsymbol\sigma}}(\tilde{\boldsymbol\Sigma}_{\tilde{\mathbf{Y}}})+\log\bar{J}(T,\tilde{\boldsymbol\Sigma}_{\tilde{\mathbf{Y}}})-\Xi(\mathbf{H}, \mathbf{W})-I^*(T,\rho)\right)^2\right] 
\end{equation}
we can then use the Cauchy-Schwarz inequality to approximate $\underline{U}(T,\rho)$ by $\hat{\underline{U}}(T,\rho)$ as
\begin{equation}
\hat{\underline{U}}(T,\rho)\left(1-4\varepsilon^2\right) - 4\varepsilon^2 \leq \underline{U}(T,\rho) \leq \hat{\underline{U}}(T,\rho)\left(1+4\varepsilon^2\right) + 4\varepsilon^2. \label{eq:Uhat_approx}
\end{equation}
Consequently, the limit of $\underline{U}(T,\rho)$ as $\rho\to\infty$ follows from the limit of $\hat{\underline{U}}(T,\rho)$ upon letting $\varepsilon$ tend to zero from above.

To analyze $\hat{\underline{U}}(T,\rho)$, we next define the set
\begin{equation}
\mathcal{B}_{\varepsilon} \triangleq \left\{\Sigma\in \mathcal{D}_{n_r}^{+}\colon \varepsilon \leq \Sigma_{ii} \leq \frac{1}{\varepsilon}, i=1,\ldots,n_r;\, f_{\tilde{\boldsymbol\sigma}}(\Sigma) \geq \varepsilon; \textnormal{ and } \bar{J}(T,\Sigma) \geq \varepsilon\right\}
\end{equation}
where $\mathcal{D}_{n_r}^{+}$ denotes the set of $(n_r\times n_r)$-dimensional diagonal matrices with non-negative, real-valued entries, and where $\Sigma_{ii}$ denotes the $i$-th diagonal element of $\Sigma$. We shall study the asymptotic behavior of $\hat{\underline{U}}(T,\rho)$ as $\rho\to\infty$ by dividing the expected value into two parts, depending on whether $\tilde{\boldsymbol\Sigma}_{\tilde{\mathbf{Y}}}$ lies in $\mathcal{B}_{\varepsilon}$, and by letting then $\varepsilon$ tend to zero from above. To this end, we define
\begin{IEEEeqnarray}{lCl}
\hat{\underline{U}}(T,\rho,\mathcal{S}) & \triangleq & \mathsf{E}\left[\left(n_t(T-n_t)\log \left(\frac{T\rho}{n_t}\right)-\log f_{\tilde{\boldsymbol\sigma}}(\tilde{\boldsymbol\Sigma}_{\tilde{\mathbf{Y}}})+\log\bar{J}(T,\tilde{\boldsymbol\Sigma}_{\tilde{\mathbf{Y}}})-\Xi(\mathbf{H}, \mathbf{W})-I^*(T,\rho)\right)^2\mathds{1}\left\{\tilde{\boldsymbol\Sigma}_{\tilde{\mathbf{Y}}}\in\mathcal{S}\right\}\right]\nonumber\\
\end{IEEEeqnarray}
and
\begin{IEEEeqnarray}{lCl}
\tilde{\underline{U}}(T,\rho,\mathcal{S})  & \triangleq & \mathsf{E}\left[\left(n_t(T-n_t)\log \left(\frac{T\rho}{n_t}\right)-\log f_{\boldsymbol\varsigma}(\tilde{\boldsymbol\Sigma}_{\tilde{\mathbf{Y}}})+\log\tilde{J}(T,\tilde{\boldsymbol\Sigma}_{\tilde{\mathbf{Y}}})-\Xi(\mathbf{H},\mathbf{W})-I^*(T,\rho)\right)^2\mathds{1}\left\{\tilde{\boldsymbol\Sigma}_{\tilde{\mathbf{Y}}}\in\mathcal{S}\right\}\right] \nonumber\\\label{eq:Ubartilde}
\end{IEEEeqnarray}
for $\mathcal{S} \subseteq \mathcal{D}_{n_r}^{+}$. Consequently, $\hat{\underline{U}}(T,\rho)$ can be written as
\begin{equation}
\hat{\underline{U}}(T,\rho) = \hat{\underline{U}}(T,\rho,\mathcal{B}_{\varepsilon}) + \hat{\underline{U}}(T,\rho,\mathcal{B}_{\varepsilon}^c). \label{eq:U-splitup}
\end{equation}
In the following subsections, we analyze the terms on the RHS of \eqref{eq:U-splitup} separately.

\subsubsection{Limit of $\hat{\underline{U}}(T,\rho,\mathcal{B}_{\varepsilon})$}\label{sub:UB}
By Lemma~\ref{Lem_13}, $f_{\tilde{\boldsymbol\sigma}}$ converges pointwise to $f_{\boldsymbol{\varsigma}}$. It then follows from Egoroff's theorem \cite[Th.~2.5.5]{AsDo00} that, for every $\varepsilon>0$, there exists a set $\mathcal{A}_{\varepsilon}$ of probability less than $\varepsilon$ such that $f_{\tilde{\boldsymbol\sigma}}$ converges uniformly to $f_{\boldsymbol{\varsigma}}$ on $\tilde{\mathcal{B}_{\varepsilon}} \triangleq \mathcal{B}_{\varepsilon} \setminus \mathcal{A}_{\varepsilon}$.\footnote{Without loss of generality, we can assume that $\mathcal{A}_{\varepsilon}$ decreases monotonically to a set of probability zero, i.e., $\mathcal{A}_{\varepsilon_1} \supseteq \mathcal{A}_{\varepsilon_2}$ for $\varepsilon_1 \geq \varepsilon_2$. Indeed, if $f_{\tilde{\boldsymbol\sigma}}$ converges uniformly to $f_{\boldsymbol{\varsigma}}$ on $\mathcal{B}_{\varepsilon}\setminus\mathcal{A}_{\varepsilon_1}$ and $\mathcal{B}_{\varepsilon}\setminus\mathcal{A}_{\varepsilon_2}$, then it also converges uniformly on $\mathcal{B}_{\varepsilon}\setminus\tilde{\mathcal{A}}_{\varepsilon_2}$, where $\tilde{\mathcal{A}}_{\varepsilon_2} = \mathcal{A}_{\varepsilon_1} \cap\mathcal{A}_{\varepsilon_2}$. The set $\tilde{\mathcal{A}}_{\varepsilon_2}$ satisfies $\mathcal{A}_{\varepsilon_1} \supseteq \tilde{\mathcal{A}}_{\varepsilon_2}$ and has probability less than $\varepsilon_2$ (because $\tilde{\mathcal{A}}_{\varepsilon_2} \subseteq \mathcal{A}_{\varepsilon_2}$), hence the claim follows.} Thus, there exists a sufficiently large $\rho_0>0$ such that
\begin{equation}
\left|f_{\tilde{\boldsymbol\sigma}}(\Sigma)-f_{\boldsymbol{\varsigma}}(\Sigma)\right| < \varepsilon^2, \quad \Sigma\in\tilde{\mathcal{B}}_{\varepsilon}, \, \rho\geq \rho_0 \label{eq:approx_fsigma}
\end{equation}
in which case $\left|\log f_{\tilde{\boldsymbol\sigma}}(\Sigma)-\log f_{\boldsymbol{\varsigma}}(\Sigma)\right| \leq \frac{\varepsilon}{1-\varepsilon}$. Similarly, it can be shown that there exists a sufficiently large $\rho_0>0$ such that
\begin{equation}
\left|\bar{J}(T,\Sigma)-\tilde{J}(T,\Sigma)\right| \leq \varepsilon^2, \quad \Sigma\in\tilde{\mathcal{B}}_{\varepsilon}, \, \rho\geq \rho_0 \label{eq:approx_J}
\end{equation}
and, hence, $|\log\bar{J}(T,\Sigma)-\log\tilde{J}(T,\Sigma)| \leq \frac{\varepsilon}{1-\varepsilon}$.

We next write $\hat{\underline{U}}(T,\rho,\mathcal{B}_{\varepsilon})$ as
\begin{equation}
\hat{\underline{U}}(T,\rho,\mathcal{B}_{\varepsilon}) = \hat{\underline{U}}(T,\rho,\tilde{\mathcal{B}}_{\varepsilon})+\hat{\underline{U}}(T,\rho,\mathcal{B}_{\varepsilon}\cap\mathcal{A}_{\varepsilon}). \label{eq:Btilde+Aeps}
\end{equation}
Using \eqref{eq:approx_fsigma}--\eqref{eq:approx_J} and the Cauchy-Schwarz inequality, we can approximate $\hat{\underline{U}}(T,\rho,\tilde{\mathcal{B}}_{\varepsilon})$ as
\begin{equation}
\tilde{\underline{U}}(T,\rho,\tilde{\mathcal{B}}_{\varepsilon})\left(1-\frac{4\varepsilon^2}{(1-\varepsilon)^2}\right) - \frac{4\varepsilon^2}{(1-\varepsilon)^2} \leq \hat{\underline{U}}(T,\rho,\tilde{\mathcal{B}}_{\varepsilon}) \leq \tilde{\underline{U}}(T,\rho,\tilde{\mathcal{B}}_{\varepsilon})\left(1+\frac{4\varepsilon^2}{(1-\varepsilon)^2}\right) + \frac{4\varepsilon^2}{(1-\varepsilon)^2} \label{eq:Utilde_approx}
\end{equation}
so the limit of $\hat{\underline{U}}(T,\rho,\tilde{\mathcal{B}}_{\varepsilon})$ as $\rho\to\infty$ follows from the limit of $\tilde{\underline{U}}(T,\rho,\tilde{\mathcal{B}}_{\varepsilon})$ upon letting $\varepsilon$ tend to zero from above.

We next apply the dominated convergence theorem to evaluate the limit of $\tilde{\underline{U}}(T,\rho,\tilde{\mathcal{B}}_{\varepsilon})$ as $\rho\to\infty$. Indeed, after algebraic manipulations similar to the ones that lead to \eqref{eq:logJ-logf-Xi}, it can be shown that
\begin{IEEEeqnarray}{lCl}
\IEEEeqnarraymulticol{3}{l}{n_t(T-n_t)\log \left(\frac{T\rho}{n_t}\right)-\log f_{\boldsymbol\varsigma}(\tilde{\boldsymbol\Sigma}_{\tilde{\mathbf{Y}}})+\log\tilde{J}(T,\tilde{\boldsymbol\Sigma}_{\tilde{\mathbf{Y}}})-\Xi(\mathbf{H},\mathbf{W})-I^*(T,\rho)} \nonumber\\
\quad & = & (T-n_t)\sum_{i=1}^{n_t} \log \tilde{\boldsymbol\sigma}_i^2 + \sum_{i=1}^{n_r} \tilde{\boldsymbol\sigma}_i^2 - \text{tr}(\mathbf{H}^\mathsf{H}\mathbf{H}) - \text{tr}(\mathbf{W}_{21}^\mathsf{H}\mathbf{W}_{21}) \nonumber\\
& & {} - \text{tr}(\mathbf{W}_{22}^\mathsf{H}\mathbf{W}_{22}) + n_t(T - n_t) - (T-n_t) \sum_{i=0}^{n_t-1} \varPsi(n_r-i) \label{eq:THIS}
\end{IEEEeqnarray}
which on the set
\begin{equation}
\tilde{\mathcal{B}}_{\varepsilon,1} \triangleq \left\{\Sigma\in\mathcal{D}_{n_r}^{+}\colon \varepsilon \leq \Sigma_{ii} \leq \frac{1}{\varepsilon}, i=1,\ldots,n_r\right\}
\end{equation}
is bounded by a random variable of finite second moment. Furthermore, by Lemma~\ref{Lem_singularY}, the RHS of \eqref{eq:THIS} converges almost surely to
\begin{equation}
 (T-n_t)\sum_{i=1}^{n_t} \log \boldsymbol{\varsigma}_i^2 - \text{tr}(\mathbf{W}_{21}^{\mathsf{H}}\mathbf{W}_{21}) + n_t(T-n_t) - (T-n_t) \sum_{i=0}^{n_t-1} \varPsi(n_r-i)
\end{equation}
as $\rho\to\infty$, which is equal to $i^*(\rho,\boldsymbol\varsigma,\mathbf{H},\mathbf{W}) - I^*(T,\rho)$ since $\sum_{i=1}^{n_t} \log \boldsymbol{\varsigma}_i^2 = \log\det(\mathbf{H}^{\mathsf{H}}\mathbf{H})$. Further using that, by \eqref{eq:approx_fsigma}--\eqref{eq:approx_J},
\begin{equation}
\tilde{\mathcal{B}}_{\varepsilon,2} \subseteq \tilde{\mathcal{B}}_{\varepsilon} \subseteq \tilde{\mathcal{B}}_{\varepsilon,1} \label{eq:B_inequalities}
\end{equation}
where
\begin{equation}
\tilde{\mathcal{B}}_{\varepsilon,2} \triangleq \left\{\Sigma\in \mathcal{D}_{n_r}^{+}\setminus \mathcal{A}_{\varepsilon}\colon \varepsilon \leq \Sigma_{ii} \leq \frac{1}{\varepsilon}, i=1,\ldots,n_r;\, f_{\boldsymbol\varsigma}(\Sigma) \geq \varepsilon-\varepsilon^2; \textnormal{ and } \tilde{J}(T,\Sigma) \geq \varepsilon-\varepsilon^2\right\}
\end{equation}
it follows from the dominated convergence theorem that
\begin{equation}
\mathsf{E}\left[\left(i^*(\rho,\boldsymbol\varsigma,\mathbf{H},\mathbf{W})-I^*(T,\rho)\right)^2\mathds{1}\{\tilde{\boldsymbol\Sigma}_{\boldsymbol\varsigma}\in\tilde{\mathcal{B}}_{\varepsilon,2}\}\right] \leq \lim_{\rho\to\infty} \tilde{\underline{U}}(T,\rho,\tilde{\mathcal{B}}_{\varepsilon}) \leq \mathsf{E}\left[\left(i^*(\rho,\boldsymbol\varsigma,\mathbf{H},\mathbf{W})-I^*(T,\rho)\right)^2\mathds{1}\{\tilde{\boldsymbol\Sigma}_{\boldsymbol\varsigma}\in\tilde{\mathcal{B}}_{\varepsilon,1}\}\right].
\end{equation}
Noting that both $\tilde{\mathcal{B}}_{\varepsilon,1}$ and $\tilde{\mathcal{B}}_{\varepsilon,2}$ increase monotonically to $\mathcal{D}_{n_r}^+$ as $\varepsilon\downarrow 0$, we then obtain from \eqref{eq:Utilde_approx} and the monotone convergence theorem that
\begin{equation}
\lim_{\varepsilon\downarrow 0} \lim_{\rho\to\infty} \hat{\underline{U}}(T,\rho,\tilde{\mathcal{B}}_{\varepsilon}) = \mathsf{E}\left[\left(i^*(\rho,\boldsymbol\varsigma,\mathbf{H},\mathbf{W})-I^*(T,\rho)\right)^2\right]. \label{eq:Btilde+}
\end{equation}

We conclude the analysis of $\hat{\underline{U}}(T,\rho,\mathcal{B}_{\varepsilon})$ by showing that
\begin{equation}
\lim_{\varepsilon\downarrow 0}\lim_{\rho\to\infty}\hat{\underline{U}}(T,\rho,\mathcal{B}_{\varepsilon}\cap\mathcal{A}_{\varepsilon}) = 0. \label{eq:Aeps}
\end{equation}
Indeed, by Part~\ref{Lem_13_2)}) of Lemma~\ref{Lem_13} and the definitions of $\mathcal{B}_{\varepsilon}$ and $\bar{J}(T,\Sigma)$, we have that
\begin{equation}
|\log f_{\tilde{\boldsymbol\sigma}}(\Sigma)|+|\log\bar{J}(T,\Sigma)| \leq \kappa \log\frac{1}{\varepsilon}, \quad \Sigma\in\mathcal{B}_{\varepsilon}
\end{equation}
for some $\kappa$ that only depends on $T$, $n_r$, and $n_t$. Using that $(a+b)^2 \leq 2a^2 + 2b^2$, we then obtain that
\begin{IEEEeqnarray}{lCl}
\IEEEeqnarraymulticol{3}{l}{\left(n_t(T-n_t)\log \left(\frac{T\rho}{n_t}\right)-\log f_{\tilde{\boldsymbol\sigma}}(\tilde{\boldsymbol\Sigma}_{\tilde{\mathbf{Y}}})+\log\bar{J}(T,\tilde{\boldsymbol\Sigma}_{\tilde{\mathbf{Y}}})-\Xi( \mathbf{H},\mathbf{W})-I^*(T,\rho)\right)^2} \nonumber\\
\quad & \leq & 2 \kappa^2 \log^2(\varepsilon) + 2 \left(\Xi(\mathbf{H}, \mathbf{W})+I^*(T,\rho) - n_t(T-n_t)\log \left(\frac{T\rho}{n_t}\right)\right)^2.
\end{IEEEeqnarray}
We can thus upper-bound $\hat{\underline{U}}(T,\rho,\mathcal{B}_{\varepsilon}\cap\mathcal{A}_{\varepsilon})$ as
\begin{IEEEeqnarray}{lCl}
\hat{\underline{U}}(T,\rho,\mathcal{B}_{\varepsilon}\cap\mathcal{A}_{\varepsilon}) & \leq & 2 \kappa^2 \log^2(\varepsilon)\Pr\left[\tilde{\boldsymbol\Sigma}_{\tilde{\mathbf{Y}}}\in \mathcal{B}_{\varepsilon}\cap\mathcal{A}_{\varepsilon}\right] \nonumber\\
& & {} + 2 \mathsf{E}\left[\left(\Xi(\mathbf{H},\mathbf{W})+I^*(T,\rho) - n_t(T-n_t)\log \left(\frac{T\rho}{n_t}\right)\right)^2 \mathds{1}\left\{\tilde{\boldsymbol\Sigma}_{\tilde{\mathbf{Y}}}\in \mathcal{B}_{\varepsilon}\cap\mathcal{A}_{\varepsilon}\right\}\right] \nonumber\\
& \leq & 2 \kappa^2 \log^2(\varepsilon)\varepsilon + 2 \mathsf{E}\left[\left(\Xi(\mathbf{H},\mathbf{W})+I^*(T,\rho) - n_t(T-n_t)\log \left(\frac{T\rho}{n_t}\right)\right)^4\right] \varepsilon \label{eq:egoroff}
\end{IEEEeqnarray}
where we used the Cauchy-Schwarz inequality and that $\Pr[\tilde{\boldsymbol\Sigma}_{\tilde{\mathbf{Y}}}\in \mathcal{B}_{\varepsilon}\cap\mathcal{A}_{\varepsilon}] \leq \Pr[\tilde{\boldsymbol\Sigma}_{\tilde{\mathbf{Y}}}\in\mathcal{A}_{\varepsilon}] \leq \varepsilon$. The term $\Xi(\mathbf{H},\mathbf{W})+I^*(T,\rho)-n_t(T-n_t)\log(T\rho/n_t)$ is independent of $\rho$ and has a finite fourth moment. Consequently, the RHS of \eqref{eq:egoroff} tends to zero as $\varepsilon\downarrow 0$, from which \eqref{eq:Aeps} follows. To conclude, \eqref{eq:Btilde+Aeps}, \eqref{eq:Btilde+}, and \eqref{eq:Aeps} demonstrate that
\begin{equation}
\lim_{\varepsilon\downarrow 0} \lim_{\rho\to\infty} \hat{\underline{U}}(T,\rho,\mathcal{B}_{\varepsilon}) = \mathsf{E}\left[\left(i^*(\rho,\boldsymbol\varsigma,\mathbf{H},\mathbf{W})-I^*(T,\rho)\right)^2\right]. \label{eq:UB_final}
\end{equation}

\subsubsection{Limit of $\hat{\underline{U}}(T,\rho,\mathcal{B}_{\varepsilon}^c)$}\label{sub:UBc}

We shall show that
\begin{equation}
\lim_{\varepsilon\downarrow 0}\lim_{\rho\to\infty} \hat{\underline{U}}(T,\rho,\mathcal{B}_{\varepsilon}^c) = 0. \label{eq:UBc_final}
\end{equation}
To this end, we use that $(a + b + c)^2\leq 9a^2 + 9b^2 + 9c^2$ for any real numbers $(a,b,c)$ to upper-bound $\hat{\underline{U}}(T,\rho,\mathcal{B}_{\varepsilon}^c)$ as
\begin{IEEEeqnarray}{lCl}
\hat{\underline{U}}(T,\rho,\mathcal{B}_{\varepsilon}^c) & \leq & 9\mathsf{E}\left[\left(\Xi(\mathbf{H},\mathbf{W})+I^*(T,\rho) - n_t(T-n_t)\log \left(\frac{T\rho}{n_t}\right)\right)^2 \mathds{1}\left\{\tilde{\boldsymbol\Sigma}_{\tilde{\mathbf{Y}}}\in \mathcal{B}_{\varepsilon}^c\right\}\right] \nonumber\\
& & {} + 9 \mathsf{E}\left[\left(\log f_{\tilde{\boldsymbol\sigma}}(\tilde{\boldsymbol\Sigma}_{\tilde{\mathbf{Y}}})\right)^2 \mathds{1}\left\{\tilde{\boldsymbol\Sigma}_{\tilde{\mathbf{Y}}}\in \mathcal{B}_{\varepsilon}^c\right\}\right]  + 9 \mathsf{E}\left[\left(\log\bar{J}(T,\tilde{\boldsymbol\Sigma}_{\tilde{\mathbf{Y}}})\right)^2 \mathds{1}\left\{\tilde{\boldsymbol\Sigma}_{\tilde{\mathbf{Y}}}\in \mathcal{B}_{\varepsilon}^c\right\}\right]. \label{eq:UBc_splitup}
\end{IEEEeqnarray}
We next analyze each term on the RHS of \eqref{eq:UBc_splitup} separately:
\begin{enumerate}[leftmargin=0pt, itemindent=15pt,
labelwidth=10pt, labelsep=5pt, listparindent=10pt,align=left]
\item Using the Cauchy-Schwarz inequality, the first term on the RHS of \eqref{eq:UBc_splitup} can be upper-bounded by
\begin{equation}
9\mathsf{E}\left[\left(\Xi(\mathbf{H},\mathbf{W})+I^*(T,\rho) - n_t(T-n_t)\log \left(\frac{T\rho}{n_t}\right)\right)^4\right] \Pr\left[\tilde{\boldsymbol\Sigma}_{\tilde{\mathbf{Y}}}\in \mathcal{B}_{\varepsilon}^c\right].\label{eq:UBc_first1}
\end{equation}
As noted before, the expected value in \eqref{eq:UBc_first1} is independent of $\rho$ and finite. Furthermore, by \eqref{eq:B_inequalities}, we have $\mathcal{B}_{\varepsilon}^c \subseteq \tilde{\mathcal{B}}_{\varepsilon,2}^c$ and, as noted before, $ \tilde{\mathcal{B}}_{\varepsilon,2}$ is independent of $\rho$ and increases monotonically to $\mathcal{D}_{n_r}^{+}$ as $\varepsilon\downarrow 0$. It thus follows from the continuity of measures that $\sup_{\rho>0}\Pr[\tilde{\boldsymbol\Sigma}_{\tilde{\mathbf{Y}}}\in \mathcal{B}_{\varepsilon}^c]\to 0$ as $\varepsilon\downarrow 0$. Consequently, the first term on the RHS of \eqref{eq:UBc_splitup} vanishes as $\rho\to\infty$ and then $\varepsilon\downarrow 0$.
\item The second expected value on the RHS of \eqref{eq:UBc_splitup} can be divided into the two terms
 \begin{IEEEeqnarray}{lCl}
 \mathsf{E}\left[\left(\log f_{\tilde{\boldsymbol\sigma}}(\tilde{\boldsymbol\Sigma}_{\tilde{\mathbf{Y}}})\right)^2 \mathds{1}\left\{\tilde{\boldsymbol\Sigma}_{\tilde{\mathbf{Y}}}\in \mathcal{B}_{\varepsilon}^c\right\}\right] & = & \mathsf{E}\left[\left(\log f_{\tilde{\boldsymbol\sigma}}(\tilde{\boldsymbol\Sigma}_{\tilde{\mathbf{Y}}})\right)^2 \mathds{1}\left\{\tilde{\boldsymbol\Sigma}_{\tilde{\mathbf{Y}}}\in \mathcal{B}_{\varepsilon}^c\cap \mathcal{A}_{\|\Sigma\|,r}\right\}\right] \nonumber\\
& & {} + \mathsf{E}\left[\left(\log f_{\tilde{\boldsymbol\sigma}}(\tilde{\boldsymbol\Sigma}_{\tilde{\mathbf{Y}}})\right)^2 \mathds{1}\left\{\tilde{\boldsymbol\Sigma}_{\tilde{\mathbf{Y}}}\in \mathcal{B}_{\varepsilon}^c\cap \mathcal{A}^c_{\|\Sigma\|,r}\right\}\right] \label{eq:UBc_second1}
 \end{IEEEeqnarray}
 where
 \begin{equation}
 \mathcal{A}_{\|\Sigma\|, r} \triangleq \left\{\Sigma\in\mathcal{D}_{n_r}^{+}\colon \|\Sigma\|_{\textnormal{F}} \leq r\right\}
 \end{equation}
for some $r>0$ that we shall let tend to infinity at the end of the proof. 

By Part~\ref{Lem_13_2)}) of Lemma~\ref{Lem_13}, the pdf $f_{\tilde{\boldsymbol\sigma}}$ is bounded by $\mathsf{K}_2$. Consequently, $f_{\tilde{\boldsymbol\sigma}}\log^2 f_{\tilde{\boldsymbol\sigma}}$ is bounded by $\max\{4/e^2,\mathsf{K}_2\log^2(\mathsf{K}_2)\}$ and
 \begin{IEEEeqnarray}{lCl}
 \mathsf{E}\left[\left(\log f_{\tilde{\boldsymbol\sigma}}(\tilde{\boldsymbol\Sigma}_{\tilde{\mathbf{Y}}})\right)^2 \mathds{1}\left\{\tilde{\boldsymbol\Sigma}_{\tilde{\mathbf{Y}}}\in \mathcal{B}_{\varepsilon}^c\cap \mathcal{A}_{\|\Sigma\|,r}\right\}\right] & = & \int_{\mathcal{B}_{\varepsilon}^c\cap \mathcal{A}_{\|\Sigma\|,r}} f_{\tilde{\boldsymbol\sigma}}(\Sigma) \log^2 f_{\tilde{\boldsymbol\sigma}}(\Sigma) \textnormal{d}\Sigma \nonumber\\
 & \leq & \max\{4/e^2,\mathsf{K}_2\log^2(\mathsf{K}_2)\} \lambda\left(\mathcal{B}_{\varepsilon}^c\cap \mathcal{A}_{\|\Sigma\|,r}\right) \label{eq:UBc_second1.5}
 \end{IEEEeqnarray}
 where $\lambda(\cdot)$ denotes the Lebesgue measure. By \eqref{eq:B_inequalities}, we have $\mathcal{B}_{\varepsilon}^c\cap \mathcal{A}_{\|\Sigma\|,r} \subseteq \tilde{\mathcal{B}}_{\varepsilon,2}^c\cap \mathcal{A}_{\|\Sigma\|,r}$ which, for a fixed $r>0$, is bounded, independent of $\rho$, and decreases to the empty set as $\varepsilon\downarrow 0$. It thus follows from the continuity of measures that the first expected value on the RHS of \eqref{eq:UBc_second1} vanishes as $\rho\to\infty$ and then $\varepsilon\downarrow 0$.

As for the second term on the RHS of \eqref{eq:UBc_second1}, we further divide the set $\mathcal{B}_{\varepsilon}^c\cap \mathcal{A}^c_{\|\Sigma\|,r}$ into $\mathcal{B}_{\varepsilon}^c\cap \mathcal{A}^c_{\|\Sigma\|,r}\cap \mathcal{A}_{f_{\tilde{\boldsymbol\sigma}}}$ and $\mathcal{B}_{\varepsilon}^c\cap \mathcal{A}^c_{\|\Sigma\|,r}\cap \mathcal{A}_{f_{\tilde{\boldsymbol\sigma}}}^c$, where
 \begin{equation}
 \mathcal{A}_{f_{\tilde{\boldsymbol\sigma}}} \triangleq \left\{\Sigma\in\mathcal{D}_{n_r}^+\colon f_{\tilde{\boldsymbol\sigma}}(\Sigma) \leq \|\Sigma\|_{\text{F}}^{-n_r-1}\right\}.
 \end{equation}
 When $r> e^{2/(n_r+1)}$, we have that
 \begin{equation}
 f_{\tilde{\boldsymbol\sigma}}(\Sigma)\log^2 f_{\tilde{\boldsymbol\sigma}}(\Sigma) \leq (n_r+1)^2\frac{\log^2(\|\Sigma\|_{\text{F}})}{\|\Sigma\|_{\text{F}}^{n_r+1}}, \quad \Sigma \in \mathcal{A}^c_{\|\Sigma\|,r}\cap \mathcal{A}_{f_{\tilde{\boldsymbol\sigma}}}.
 \end{equation}
 Consequently,
 \begin{IEEEeqnarray}{lCl}
 \mathsf{E}\left[\left(\log f_{\tilde{\boldsymbol\sigma}}(\tilde{\boldsymbol\Sigma}_{\tilde{\mathbf{Y}}})\right)^2 \mathds{1}\left\{\tilde{\boldsymbol\Sigma}_{\tilde{\mathbf{Y}}}\in \mathcal{B}_{\varepsilon}^c\cap \mathcal{A}^c_{\|\Sigma\|,r}\cap \mathcal{A}_{f_{\tilde{\boldsymbol\sigma}}}\right\}\right] & \leq & (n_r+1)^2\int_{\|\Sigma\|_{\text{F}}>r} \frac{\log^2(\|\Sigma\|_{\text{F}})}{\|\Sigma\|_{\text{F}}^{n_r+1}} \text{d} \Sigma \nonumber\\
 & \leq & (n_r+1)^2\frac{5{\pi}^{\frac{n_r}{2}}}{\Gamma(\frac{n_r}{2})} \frac{\log^2(r)}{r} \label{eq:UBc_second2}
 \end{IEEEeqnarray}
 where we bounded the integral by performing a change of variable from $\Sigma$ to $\|\Sigma\|_{\text{F}}$. Moreover, when \mbox{$r > \mathsf{K}_2^{1/(n_r+1)}$}, we have that
 \begin{equation}
\log^2 f_{\tilde{\boldsymbol\sigma}}(\Sigma)  \leq (n_r+1)^2 \log^2(\|\Sigma\|_{\text{F}}), \quad \Sigma\in \mathcal{A}^c_{\|\Sigma\|,r}\cap \mathcal{A}_{f_{\tilde{\boldsymbol\sigma}}}^c.
 \end{equation}
 Since $x\mapsto \log^2(x)$ is a concave function for $x\geq e$, it thus follows from Jensen's inequality that, for $r >\max\{\mathsf{K}_2^{1/(n_r+1)},e\}$,
 \begin{IEEEeqnarray}{lCl}
\IEEEeqnarraymulticol{3}{l}{ \mathsf{E}\left[\left(\log f_{\tilde{\boldsymbol\sigma}}(\tilde{\boldsymbol\Sigma}_{\tilde{\mathbf{Y}}})\right)^2 \mathds{1}\left\{\tilde{\boldsymbol\Sigma}_{\tilde{\mathbf{Y}}}\in \mathcal{B}_{\varepsilon}^c\cap \mathcal{A}^c_{\|\Sigma\|,r}\cap \mathcal{A}^c_{f_{\tilde{\boldsymbol\sigma}}}\right\}\right]}\nonumber\\
\qquad & \leq & (n_r+1)^2 \int_{\|\Sigma\|_{\text{F}}>r}  f_{\tilde{\boldsymbol\sigma}}(\Sigma) \log^2(\|\Sigma\|_{\text{F}}) \text{d}\Sigma \nonumber\\
 & \leq & (n_r+1)^2\Pr\left[\|\tilde{\boldsymbol\Sigma}_{\tilde{\mathbf{Y}}}\|_{\text{F}}>r\right] \log^2\mathsf{E}\Bigl[\|\tilde{\boldsymbol\Sigma}_{\tilde{\mathbf{Y}}}\|_{\text{F}} \Bigm| \|\tilde{\boldsymbol\Sigma}_{\tilde{\mathbf{Y}}}\|_{\text{F}} > r\Bigr] \nonumber\\
 & \leq & (n_r+1)^2\Pr\left[\|\tilde{\boldsymbol\Sigma}_{\tilde{\mathbf{Y}}}\|_{\text{F}}>r\right] \log^2\left(\frac{\sqrt{\mathsf{K}_1}}{\Pr\left[\|\tilde{\boldsymbol\Sigma}_{\tilde{\mathbf{Y}}}\|_{\text{F}}>r\right] }\right) \IEEEeqnarraynumspace \label{eq:UBc_second2.5}
 \end{IEEEeqnarray}
 where the last step follows by upper-bounding the conditional expectation by $\mathsf{E}[\|\tilde{\boldsymbol\Sigma}_{\tilde{\mathbf{Y}}}\|_{\text{F}}]/\Pr[\|\tilde{\boldsymbol\Sigma}_{\tilde{\mathbf{Y}}}\|_{\text{F}}>r]$, and by then applying Part~\ref{Lem_13_1}) of Lemma~\ref{Lem_13}. Applying Chebyshev's inequality together with Part~\ref{Lem_13_1}) of Lemma~\ref{Lem_13}, and noting that the function $x\mapsto \log^2(x/a)$ is monotonically increasing for $x>ae^2$, this can be further upper-bounded by
 \begin{equation}
 \mathsf{E}\left[\left(\log f_{\tilde{\boldsymbol\sigma}}(\tilde{\boldsymbol\Sigma}_{\tilde{\mathbf{Y}}})\right)^2 \mathds{1}\left\{\tilde{\boldsymbol\Sigma}_{\tilde{\mathbf{Y}}}\in \mathcal{B}_{\varepsilon}^c\cap \mathcal{A}^c_{\|\Sigma\|,r}\cap \mathcal{A}^c_{f_{\tilde{\boldsymbol\sigma}}}\right\}\right]   \leq (n_r+1)^2\frac{\mathsf{K}_1}{r^2} \log^2\left(\frac{r^2}{\sqrt{\mathsf{K}_1}}\right) \label{eq:UBc_second3}
 \end{equation}
 for $r > \max\{\mathsf{K}_2^{1/(n_r+1)},e,\sqrt{\mathsf{K}}_1e^2\}$. With \eqref{eq:UBc_second2} and \eqref{eq:UBc_second3}, we can upper-bound the second expected value on the RHS of \eqref{eq:UBc_second1} as
\begin{equation}
\mathsf{E}\left[\left(\log f_{\tilde{\boldsymbol\sigma}}(\tilde{\boldsymbol\Sigma}_{\tilde{\mathbf{Y}}})\right)^2 \mathds{1}\left\{\tilde{\boldsymbol\Sigma}_{\tilde{\mathbf{Y}}}\in \mathcal{B}_{\varepsilon}^c\cap \mathcal{A}^c_{\|\Sigma\|,r}\right\}\right] \leq (n_r+1)^2\frac{5{\pi}^{\frac{n_r}{2}}}{\Gamma(\frac{n_r}{2})} \frac{\log^2(r)}{r} + (n_r+1)^2\frac{\mathsf{K}_1}{r^2} \log^2\left(\frac{r^2}{\sqrt{\mathsf{K}_1}}\right) \label{eq:UBc_second4}
\end{equation}
which is independent of $\rho$ and $\varepsilon$, and vanishes as $r\to\infty$. We thus obtain from \eqref{eq:UBc_second1}, \eqref{eq:UBc_second1.5}, and \eqref{eq:UBc_second4} that the second term on the RHS of \eqref{eq:UBc_splitup} vanishes as we let first $\rho\to\infty$, then $\varepsilon\downarrow 0$, and then $r\to\infty$.
\item We upper-bound the third term on the RHS of \eqref{eq:UBc_splitup} using the definition of $\bar{J}(T,\bar\Sigma_{\mathsf{Y}})$ in \eqref{Def_J_sigma} and that
 \begin{equation}
	 |a_1+\ldots+a_{\mu}|^{\nu}\leq \mathsf{c}_{\mu,\nu}\big( |a_1|^{\nu}+\ldots+|a_{\mu}|^{\nu}\big), \quad \mu, \nu\in \mathbb{N} \label{Equ_cnv}
	 \end{equation}
	 for some positive constant $\mathsf{c}_{\mu,\nu}$ that only depends on $\mu$ and $\nu$. It follows that\footnote{For the sake of compactness, we omit the subscripts $\mu$ and $\nu$ of the constant $\mathsf{c}_{\mu,\nu}$.}
	 \begin{IEEEeqnarray}{lCl}
	 \frac{1}{\mathsf{c}}\left(\log\bar{J}(T,\tilde{\boldsymbol\Sigma}_{\tilde{\mathbf{Y}}})\right)^2 & \leq & \bigl(2(T-n_r)+1\bigr)^2 \sum_{i=1}^{n_r} \log^2(\tilde{\boldsymbol\sigma}_i) + 4\sum_{1\leq i < j \leq n_t} \log^2(\tilde{\boldsymbol\sigma}_i^2-\tilde{\boldsymbol\sigma}_j^2) \nonumber\\
	 & & {} + 4\sum_{n_t < i < j \leq n_t} \log^2(\tilde{\boldsymbol\sigma}_i^2-\tilde{\boldsymbol\sigma}_j^2) + 4 \sum_{i=1}^{n_t}\sum_{j=n_t+1}^{n_r}  \log^2\biggl(\tilde{\boldsymbol\sigma}_i^2-\frac{n_t}{T\rho}\tilde{\boldsymbol\sigma}_j^2\biggr). \label{eq:UBc_third1}
	 \end{IEEEeqnarray}
To show that the RHS of \eqref{eq:UBc_splitup} vanishes as we first let $\rho\to\infty$ and then $\varepsilon\downarrow 0$, it thus suffices to show that
\begin{subequations}
\label{eq:UBc_third2}
\begin{IEEEeqnarray}{rCl}
\lim_{\varepsilon\downarrow 0}\lim_{\rho\to\infty} \mathsf{E}\left[\log^2(\tilde{\boldsymbol\sigma}_i) \mathds{1}\left\{\tilde{\boldsymbol\Sigma}_{\tilde{\mathbf{Y}}}\in \mathcal{B}_{\varepsilon}^c\right\}\right] & = & 0 \label{eq:UBc_third2a}\\
\lim_{\varepsilon\downarrow 0}\lim_{\rho\to\infty} \mathsf{E}\left[\log^2(\tilde{\boldsymbol\sigma}_i^2-\tilde{\boldsymbol\sigma}_j^2) \mathds{1}\left\{\tilde{\boldsymbol\Sigma}_{\tilde{\mathbf{Y}}}\in \mathcal{B}_{\varepsilon}^c\right\}\right] & = & 0 \label{eq:UBc_third2b} \\
\lim_{\varepsilon\downarrow 0}\lim_{\rho\to\infty} \mathsf{E}\left[\log^2\biggl(\tilde{\boldsymbol\sigma}_i^2-\frac{n_t}{T\rho}\tilde{\boldsymbol\sigma}_j^2\biggr) \mathds{1}\left\{\tilde{\boldsymbol\Sigma}_{\tilde{\mathbf{Y}}}\in \mathcal{B}_{\varepsilon}^c\right\}\right] & = & 0 \label{eq:UBc_third2c}
\end{IEEEeqnarray}
\end{subequations}
for the combinations of indices $(i,j)$ indicated in \eqref{eq:UBc_third1}.

To prove \eqref{eq:UBc_third2a}, we bound the expected value as
\begin{IEEEeqnarray}{lCl}
 \mathsf{E}\left[\log^2(\tilde{\boldsymbol\sigma}_i) \mathds{1}\left\{\tilde{\boldsymbol\Sigma}_{\tilde{\mathbf{Y}}}\in \mathcal{B}_{\varepsilon}^c\right\}\right] & \leq &  \mathsf{E}\left[\log^2(\tilde{\boldsymbol\sigma}_i) \mathds{1}\left\{\tilde{\boldsymbol\Sigma}_{\tilde{\mathbf{Y}}}\in \mathcal{B}_{\varepsilon}^c\right\}\mathds{1}\left\{\delta\leq \tilde{\boldsymbol\sigma}_i \leq \frac{1}{\delta}\right\}\right] \nonumber\\
 & & {} + \mathsf{E}\left[\log^2(\tilde{\boldsymbol\sigma}_i) \mathds{1}\left\{\tilde{\boldsymbol\sigma}_i < \delta\right\}\right]  + \mathsf{E}\left[\log^2(\tilde{\boldsymbol\sigma}_i) \mathds{1}\left\{\tilde{\boldsymbol\sigma}_i > \frac{1}{\delta}\right\}\right] \label{eq:UBc_third3}
\end{IEEEeqnarray}
for some $\delta>0$ that we shall let tend to zero from above at the end of the proof. The first term on the RHS of \eqref{eq:UBc_third3} can be upper-bounded as
\begin{equation}
\mathsf{E}\left[\log^2(\tilde{\boldsymbol\sigma}_i) \mathds{1}\left\{\tilde{\boldsymbol\Sigma}_{\tilde{\mathbf{Y}}}\in \mathcal{B}_{\varepsilon}^c\right\}\mathds{1}\left\{\delta\leq \tilde{\boldsymbol\sigma}_i \leq \frac{1}{\delta}\right\}\right] \leq \log^2(\delta) \Pr\left[\tilde{\boldsymbol\Sigma}_{\tilde{\mathbf{Y}}}\in \mathcal{B}_{\varepsilon}^c\right]. \label{eq:UBc_third4}
\end{equation}
As noted before, $\sup_{\rho>0}\Pr[\tilde{\boldsymbol\Sigma}_{\tilde{\mathbf{Y}}}\in \mathcal{B}_{\varepsilon}^c]\to 0$ as $\varepsilon\downarrow 0$ by the continuity of measures. It follows that the first term on the RHS of \eqref{eq:UBc_third3} vanishes as we first let $\rho\to\infty$, then $\varepsilon\downarrow 0$, and then $\delta\downarrow 0$.

For $\delta \leq 1$, the second term on the RHS of \eqref{eq:UBc_third3} can be bounded as
\begin{IEEEeqnarray}{lCl}
\mathsf{E}\left[\log^2(\tilde{\boldsymbol\sigma}_i) \mathds{1}\left\{\tilde{\boldsymbol\sigma}_i < \delta\right\}\right] & \leq & \int_{\substack{\sigma < \delta,\\  f_{\tilde{\boldsymbol\sigma}_i}(\sigma) \leq \sigma^{-1/2}}} f_{\tilde{\boldsymbol\sigma}_i}(\sigma)\log^2(\sigma)\text{d}\sigma + \int_{\substack{\sigma < \delta,\\  f_{\tilde{\boldsymbol\sigma}_i}(\sigma) > \sigma^{-1/2}}} f_{\tilde{\boldsymbol\sigma}_i}(\sigma)\log^2(\sigma)\text{d}\sigma \nonumber\\
& \leq & \int_0^{\delta} \frac{\log^2(\sigma)}{\sqrt{\sigma}} \text{d}\sigma + 4\int_{0}^{\delta} f_{\tilde{\boldsymbol\sigma}_i}(\sigma)\log^2\bigl(f_{\tilde{\boldsymbol\sigma}_i}(\sigma)\bigr) \text{d}\sigma \label{eq:UBc_third5}
\end{IEEEeqnarray}
where in the last step we bounded the first integral by upper-bounding $f_{\tilde{\boldsymbol\sigma}_i}$ by $\sigma^{-1/2}$, and we bounded the second integral by noting that, for $\sigma<\delta<1$, $\log^2(\sigma) = \log^2(1/\sigma)$ and by upper-bounding $1/\sigma$ by $f_{\tilde{\boldsymbol\sigma}_i}^{2}$. The first integral on the RHS of \eqref{eq:UBc_third5} can be evaluated directly. For the second integral, we use that, by Part~\ref{Lem_13_2)}) of Lemma~\ref{Lem_13}, $f_{\tilde{\boldsymbol\sigma}}\log^2 f_{\tilde{\boldsymbol\sigma}}$ is bounded by $\max\{4/e^2,\mathsf{K}_2\log^2(\mathsf{K}_2)\}$. It follows that
\begin{equation}
\mathsf{E}\left[\log^2(\tilde{\boldsymbol\sigma}_i)\mathds{1}\left\{\tilde{\boldsymbol\sigma}_i < \delta\right\}\right] \leq 40\log^2(\delta) \sqrt{\delta} + \delta \max\{4/e^2,\mathsf{K}_2\log^2(\mathsf{K}_2)\} \label{eq:UBc_third6}
\end{equation}
from which we obtain that the second term on the RHS of \eqref{eq:UBc_third3} vanishes as we first let $\rho\to\infty$, then $\varepsilon\downarrow 0$, and then $\delta\downarrow 0$.

As for the third term on the RHS of \eqref{eq:UBc_third3}, we recall that $x\mapsto \log^2(x)$ is a concave function for $x\geq e$. Consequently, by following the same steps as in \eqref{eq:UBc_second2.5}--\eqref{eq:UBc_second3}, we obtain that, for $\delta\leq (\max\{e,\sqrt{\mathsf{K}}_1e^2\})^{-1}$,
\begin{equation}
\mathsf{E}\left[\log^2(\tilde{\boldsymbol\sigma}_i) \mathds{1}\left\{\tilde{\boldsymbol\sigma}_i > \frac{1}{\delta}\right\}\right] \leq \mathsf{K}_1 \delta^2\log^2\left(\sqrt{\mathsf{K}}_1 \delta^2\right). \label{eq:UBc_third7}
\end{equation}
Thus, the third term on the RHS of \eqref{eq:UBc_third3} also vanishes as we first let $\rho\to\infty$, then $\varepsilon\downarrow 0$, and then $\delta\downarrow 0$.

Combining \eqref{eq:UBc_third3}, \eqref{eq:UBc_third4}, \eqref{eq:UBc_third6}, and \eqref{eq:UBc_third7}, we obtain \eqref{eq:UBc_third2a}.

We next prove \eqref{eq:UBc_third2b}. To this end, we use that $\tilde{\boldsymbol\sigma}_i^2-\tilde{\boldsymbol\sigma}_j^2 = (\tilde{\boldsymbol\sigma}_i-\tilde{\boldsymbol\sigma}_j)(\tilde{\boldsymbol\sigma}_i + \tilde{\boldsymbol\sigma}_j)$ and the inequality $(a+b)^2\leq 2a^2+2b^2$ to upper-bounded the expected value in  \eqref{eq:UBc_third2b} as
\begin{IEEEeqnarray}{lCl}
\mathsf{E}\left[\log^2(\tilde{\boldsymbol\sigma}_i^2-\tilde{\boldsymbol\sigma}_j^2) \mathds{1}\left\{\tilde{\boldsymbol\Sigma}_{\tilde{\mathbf{Y}}}\in \mathcal{B}_{\varepsilon}^c\right\}\right] & \leq & 2\mathsf{E}\left[\log^2(\tilde{\boldsymbol\sigma}_i-\tilde{\boldsymbol\sigma}_j) \mathds{1}\left\{\tilde{\boldsymbol\Sigma}_{\tilde{\mathbf{Y}}}\in \mathcal{B}_{\varepsilon}^c\right\}\right] \nonumber\\
&& {} + 2 \mathsf{E}\left[\log^2(\tilde{\boldsymbol\sigma}_i + \tilde{\boldsymbol\sigma}_j) \mathds{1}\left\{\tilde{\boldsymbol\Sigma}_{\tilde{\mathbf{Y}}}\in \mathcal{B}_{\varepsilon}^c\right\}\right]. \label{eq:UBc_third8}
\end{IEEEeqnarray}
We continue similarly to the proof of \eqref{eq:UBc_third2a}. Indeed, we can bound the expected values on the RHS of \eqref{eq:UBc_third8} as
\begin{IEEEeqnarray}{lCl}
\mathsf{E}\left[\log^2(\tilde{\boldsymbol\sigma}_i-\tilde{\boldsymbol\sigma}_j) \mathds{1}\left\{\tilde{\boldsymbol\Sigma}_{\tilde{\mathbf{Y}}}\in \mathcal{B}_{\varepsilon}^c\right\}\right] & \leq & \mathsf{E}\left[\log^2(\tilde{\boldsymbol\sigma}_i-\tilde{\boldsymbol\sigma}_j) \mathds{1}\left\{\tilde{\boldsymbol\Sigma}_{\tilde{\mathbf{Y}}}\in \mathcal{B}_{\varepsilon}^c\right\}\mathds{1}\left\{\delta \leq \tilde{\boldsymbol\sigma}_i-\tilde{\boldsymbol\sigma}_j \leq \frac{1}{\delta}\right\}\right] \nonumber\\
& & {} + \mathsf{E}\left[\log^2(\tilde{\boldsymbol\sigma}_i-\tilde{\boldsymbol\sigma}_j) \mathds{1}\left\{\tilde{\boldsymbol\sigma}_i-\tilde{\boldsymbol\sigma}_j <\delta\right\}\right] \nonumber\\
& & {} + \mathsf{E}\left[\log^2(\tilde{\boldsymbol\sigma}_i-\tilde{\boldsymbol\sigma}_j) \mathds{1}\left\{ \tilde{\boldsymbol\sigma}_i-\tilde{\boldsymbol\sigma}_j > \frac{1}{\delta}\right\}\right] \label{eq:UBc_third9}
\end{IEEEeqnarray}
and
\begin{IEEEeqnarray}{lCl}
\mathsf{E}\left[\log^2(\tilde{\boldsymbol\sigma}_i+\tilde{\boldsymbol\sigma}_j) \mathds{1}\left\{\tilde{\boldsymbol\Sigma}_{\tilde{\mathbf{Y}}}\in \mathcal{B}_{\varepsilon}^c\right\}\right] & \leq & \mathsf{E}\left[\log^2(\tilde{\boldsymbol\sigma}_i+\tilde{\boldsymbol\sigma}_j) \mathds{1}\left\{\tilde{\boldsymbol\Sigma}_{\tilde{\mathbf{Y}}}\in \mathcal{B}_{\varepsilon}^c\right\}\mathds{1}\left\{\delta \leq \tilde{\boldsymbol\sigma}_i+\tilde{\boldsymbol\sigma}_j \leq \frac{1}{\delta}\right\}\right] \nonumber\\
& & {} + \mathsf{E}\left[\log^2(\tilde{\boldsymbol\sigma}_i+\tilde{\boldsymbol\sigma}_j) \mathds{1}\left\{\tilde{\boldsymbol\sigma}_i+\tilde{\boldsymbol\sigma}_j <\delta\right\}\right] \nonumber\\
& & {} + \mathsf{E}\left[\log^2(\tilde{\boldsymbol\sigma}_i+\tilde{\boldsymbol\sigma}_j) \mathds{1}\left\{ \tilde{\boldsymbol\sigma}_i+\tilde{\boldsymbol\sigma}_j > \frac{1}{\delta}\right\}\right] \label{eq:UBc_third10}.
\end{IEEEeqnarray}
The first terms on the RHSs of \eqref{eq:UBc_third9} and \eqref{eq:UBc_third10} are upper-bounded by $\log^2(\delta) \Pr[\tilde{\boldsymbol\Sigma}_{\tilde{\mathbf{Y}}}\in \mathcal{B}_{\varepsilon}^c]$, which tends to zero as $\varepsilon\downarrow 0$ by the continuity of measures. Hence, these terms vanish as we first let $\rho\to\infty$, then $\varepsilon\downarrow 0$, and then $\delta\downarrow 0$.

As for the remaining terms, we next note that the random variables $\boldsymbol\alpha_{ij} \triangleq \tilde{\boldsymbol\sigma}_i-\tilde{\boldsymbol\sigma}_j$ and $\boldsymbol\beta_{ij} \triangleq \tilde{\boldsymbol\sigma}_i+\tilde{\boldsymbol\sigma}_j$ have the joint pdf
\begin{equation}
f_{\boldsymbol\alpha_{ij},\boldsymbol\beta_{ij}}(\alpha,\beta) = \frac{1}{2}f_{\boldsymbol\sigma_i,\boldsymbol\sigma_j}\left(\frac{\alpha+\beta}{2},\frac{\beta-\alpha}{2}\right), \quad \alpha\geq 0, \,\beta > \alpha.\label{eq:UBc_third11}
\end{equation}
By marginalizing over $\alpha$ or $\beta$, and by bounding the corresponding integral, it can be shown that both $f_{\boldsymbol\alpha_{ij}}$ and $f_{\boldsymbol\beta_{ij}}$ are bounded. Furthermore, by Part~\ref{Lem_13_1}) of Lemma~\ref{Lem_13}, the second moments of $\boldsymbol\alpha_{ij}$ and $\boldsymbol\beta_{ij}$ are bounded, too. We can thus follow the steps \eqref{eq:UBc_third5}--\eqref{eq:UBc_third7} to show that  the second and third terms on the RHSs of \eqref{eq:UBc_third9} and \eqref{eq:UBc_third10} also vanish as we first let $\rho\to\infty$, then $\varepsilon\downarrow 0$, and then $\delta\downarrow 0$. Thus, \eqref{eq:UBc_third2b} follows.

Finally, to prove \eqref{eq:UBc_third2c}, we can follow along the lines of the proof of \eqref{eq:UBc_third2b}. Indeed, we have that
\begin{IEEEeqnarray}{lCl}
\mathsf{E}\left[\log^2\left(\tilde{\boldsymbol\sigma}_i^2-\frac{n_t}{T\rho}\tilde{\boldsymbol\sigma}_j^2\right) \mathds{1}\left\{\tilde{\boldsymbol\Sigma}_{\tilde{\mathbf{Y}}}\in \mathcal{B}_{\varepsilon}^c\right\}\right] & \leq & 2\mathsf{E}\left[\log^2\left(\tilde{\boldsymbol\sigma}_i-\frac{n_t}{T\rho}\tilde{\boldsymbol\sigma}_j\right) \mathds{1}\left\{\tilde{\boldsymbol\Sigma}_{\tilde{\mathbf{Y}}}\in \mathcal{B}_{\varepsilon}^c\right\}\right] \nonumber\\
&& {} + 2 \mathsf{E}\left[\log^2\left(\tilde{\boldsymbol\sigma}_i + \frac{n_t}{T\rho}\tilde{\boldsymbol\sigma}_j\right) \mathds{1}\left\{\tilde{\boldsymbol\Sigma}_{\tilde{\mathbf{Y}}}\in \mathcal{B}_{\varepsilon}^c\right\}\right]. \label{eq:UBc_third12}
\end{IEEEeqnarray}
We can then bound the expected values depending on whether $\boldsymbol\alpha_{\rho,ij}\triangleq \tilde{\boldsymbol\sigma}_i - \frac{n_t}{T\rho}\tilde{\boldsymbol\sigma}_j$ and $\boldsymbol\beta_{\rho,ij}\triangleq \tilde{\boldsymbol\sigma}_i + \frac{n_t}{T\rho}\tilde{\boldsymbol\sigma}_j$ lie in one of the intervals $[\delta,1/\delta]$, $(0,\delta)$, or $(1/\delta,\infty)$. In the first case, the expected values are bounded by $\log^2(\delta) \Pr[\tilde{\boldsymbol\Sigma}_{\tilde{\mathbf{Y}}}\in \mathcal{B}_{\varepsilon}^c]$, which vanishes as we first let $\rho\to\infty$, then $\varepsilon\downarrow 0$, and then $\delta\downarrow 0$. For the latter two cases, we note that the marginal pdfs $f_{\boldsymbol\alpha_{\rho,ij}}$ and $f_{\boldsymbol\beta_{\rho,ij}}$ as well as the second moments of $\boldsymbol\alpha_{\rho,ij}$ and $\boldsymbol\beta_{\rho,ij}$ are bounded. We can thus follow the steps \eqref{eq:UBc_third5}--\eqref{eq:UBc_third7} to show that these expected values vanish as we first let $\rho\to\infty$, then $\varepsilon\downarrow 0$, and then $\delta\downarrow 0$. Thus, \eqref{eq:UBc_third2c} follows.
\end{enumerate}

\subsubsection{Summary}

By \eqref{eq:U-splitup}, we can express $\hat{\underline{U}}(T,\rho)$ as $\hat{\underline{U}}(T,\rho,\mathcal{B}_{\varepsilon}) + \hat{\underline{U}}(T,\rho,\mathcal{B}_{\varepsilon}^c)$. As shown in Appendix~\ref{sub:UB}, the first term converges to $\mathsf{E}[\left(i^*(\rho,\boldsymbol\varsigma,\mathbf{H},\mathbf{W})-I^*(T,\rho)\right)^2]$ as we first let $\rho\to\infty$ and then $\varepsilon\downarrow 0$; cf.~\eqref{eq:UB_final}. As shown in Appendix~\ref{sub:UBc}, the second term converges to zero as we first let $\rho\to\infty$ and then $\varepsilon\downarrow 0$; cf.~\eqref{eq:UBc_final}. Since, by \eqref{eq:Uhat_approx}, $\hat{\underline{U}}(T,\rho)$ converges to $\underline{U}(T,\rho)$ as we let $\varepsilon$ tend to zero from above, we conclude that
\begin{equation}
\lim_{\rho\to\infty} \underline{U}(T,\rho) = \lim_{\varepsilon\downarrow 0}\lim_{\rho\to\infty} \hat{\underline{U}}(T,\rho) = \mathsf{E}\left[\left(i^*(\rho,\boldsymbol\varsigma,\mathbf{H},\mathbf{W})-I^*(T,\rho)\right)^2\right]
\end{equation}
which is \eqref{eq:i_to_i*}.

\section{Proof of Lemma \ref{DT_Lem_23order}} \label{Sec_Pf_DTLem2} 

The result that $\underline{U}(T,\rho)$ is uniformly bounded in $\rho\geq \rho_0$ for a sufficiently large $\rho_0$ (that only depends on $T$) follows directly from Lemma~\ref{DT_Lem_1}. Indeed, according to this lemma, $\underline{U}(T,\rho)$ can be approximated as
		\begin{IEEEeqnarray}{lCl}
		\underline{U}(T, \rho) & = & T^2\tilde{V}(T)+K_{\underline{U}}(T,\rho) \nonumber\\
		& = & n_t(T-n_t)+(T-n_t)^2  \sum_{i=0}^{n_t-1} \varPsi'(n_r-i)+K_{\underline{U}}(T,\rho) 
		\end{IEEEeqnarray}
		where $K_{\underline{U}}(T,\rho)$ is a function of $T$ and $\rho$ that satisfies $\lim_{\rho\to\infty}K_{\underline{U}}(T,\rho)=0$.
		Since $\tilde{V}(T)$ is independent of $\rho$ and finite, the claim thus follows.
		
We next show that $\mathsf{E}[|i^{(U)}(\mathbf{X};\mathbf{Y})-I(T,\rho)|^3]$ is uniformly bounded in $\rho\geq\rho_0$ for sufficiently large $\rho_0$. Indeed, by following similar steps as in the proof of Lemma~\ref{DT_Lem_1}, it can be shown that
		\begin{equation}
		\lim\limits_{\rho\rightarrow \infty}\mathsf{E}\left[\left|i^{(U)}(\mathbf{X};\mathbf{Y})-I(T,\rho)\right|^3\right] =\mathsf{E}\left[\left|i^{\ast}(\rho,\boldsymbol\varsigma,\mathbf{H},\mathbf{W})-I^{\ast}(T,\rho)\right|^3\right]. \label{eq:proof_Lemma2_asymp_i3}
		\end{equation}
		Furthermore, by the definitions of $i^{\ast}(\rho,\boldsymbol\varsigma,\mathbf{H},\mathbf{W})$ and $I^{\ast}(T,\rho)$ in \eqref{Def_i_ast} and \eqref{Def_I_ast}, we have that
		\begin{IEEEeqnarray}{lCl}
		\IEEEeqnarraymulticol{3}{l}{\mathsf{E}\left[ \left|i^{\ast}(\rho,\boldsymbol\varsigma,\mathbf{H},\mathbf{W})-I^{\ast}(T,\rho)\right|^3\right]} \nonumber\\
		\quad & =& \mathsf{E}\left[ \left|
		-\text{tr}(\mathsf{W}_{21}^\mathsf{H}\mathsf{W}_{21})+(T-n_t)n_t 
		+(T-n_t)\log \left(\prod_{i=1}^{n_t}\boldsymbol{\varsigma}_i^2\right) -(T-n_t)\mathbb{E}\left[\log \left(\prod_{i=1}^{n_t}\boldsymbol{\varsigma}_i^2\right)\right]  \right|^3\right] \nonumber  \\
		& \leq& \mathsf{c}_{2,3} \left( \mathbb{E}\left[\left|-\text{tr}(\mathsf{W}_{21}^\mathsf{H}\mathsf{W}_{21})\right|^3\right]+(T-n_t)^3n_t^3
		+2\left|T-n_t\right|^3 \mathsf{E}\left[\left|\log\left(\prod_{i=1}^{n_t}\boldsymbol{\varsigma}_i^2\right)\right|^3\right]   \right) \label{Equ_U3}  
		\end{IEEEeqnarray}
	where we use \eqref{Equ_cnv} in the last step.
	 
	 Note that the RHS of \eqref{Equ_U3} is independent of $\rho$. We next show that each term on the RHS of \eqref{Equ_U3} is bounded. Indeed, the first term on the RHS of \eqref{Equ_U3} is the third absolute moment of a gamma-distributed random variable with parameters $(T- n_t,1)$. As for the second term on the RHS of \eqref{Equ_U3}, by the definition of $\boldsymbol{\varsigma}_1,\ldots,\boldsymbol{\varsigma}_{n_t}$, we have
			\begin{equation}
			\left| \log\left(\prod_{i=1}^{n_t}\boldsymbol{\varsigma}_i^2\right) \right|
			= \left|\log \det(\mathbf{H}_{\ell}^\mathsf{H} \mathbf{H}_{\ell}) \right|.
			\end{equation}
			Since $\log\det(\mathbf{H}_{\ell}^\mathsf{H} \mathbf{H}_{\ell})$ is independent of $\rho$ and its third absolute moment is finite, it follows that the second term on the RHS of \eqref{Equ_U3} is bounded, too.
We conclude that $\mathsf{E}[|i^{\ast}(\rho,\boldsymbol\varsigma,\mathbf{H},\mathbf{W})-I^{\ast}(T,\rho)|^3]$ is bounded, which together with \eqref{eq:proof_Lemma2_asymp_i3} yields that, for sufficiently large $\rho_0>0$,
		\begin{equation}
		\sup_{\rho\geq \rho_0} \mathsf{E}\left[ \left|i^{(U)}(\mathbf{X};\mathbf{Y})-I(T,\rho)\right|^3\right]<\infty.
		\end{equation} 
This completes the proof of Lemma~\ref{DT_Lem_23order}.

\section{An Upper Bound on the Mismatched Information Density} \label{Sec_UB_j}
In this appendix, we demonstrate that, conditioned on $\mathbf{X}_{\ell}=\mathsf{X}_{\ell}$, the mismatched information density $j_{\ell}(\mathbf{X}_{\ell};\mathbf{Y}_{\ell})$ can be upper-bounded by a term that has the same distribution as $\bar{j}(\mathsf{D}_{\ell}, \mathbf{Z}_{\ell}',\mathbf{Z}_{\ell}'',\mathbf{H}_{\ell},\mathbf{Q}_{\ell})$ defined in \eqref{MC_Def_j_up}. To this end, we first use that $\sum_{i=1}^{n_t} \boldsymbol{\sigma}_{i,\ell}^2 = \text{tr}\left(\mathbf{Y}_{\ell}^{\mathsf{H}}\mathbf{Y}_{\ell}\right)-\sum_{i=n_t+1}^{n_r} \boldsymbol{\sigma}_{i,\ell}^2$ to write the third, fourth, and fifth terms on the RHS of \eqref{eq:def_j(D,s)} as
\begin{IEEEeqnarray}{lCl}
\frac{n_t}{T\rho} \sum_{i=1}^{n_t} \boldsymbol{\sigma}_{i,\ell}^2 + \sum_{i=n_t+1}^{n_r}\boldsymbol{\sigma}_{i,\ell}^2  - \text{tr}\left(\mathbf{Y}_{\ell}^{\mathsf{H}}\left(\mathsf{I}_{T}+\mathsf{X}_{\ell}\mathsf{X}_{\ell}^{\mathsf{H}}\right)^{-1}\mathbf{Y}_{\ell}\right) =   \text{tr}\left(\mathbf{Y}_{\ell}^{\mathsf{H}}\Upsilon_{\ell}\mathbf{Y}_{\ell}\right) + \left(1-\frac{n_t}{T\rho}\right) \sum_{i=n_t+1}^{n_r}\boldsymbol{\sigma}_{i,\ell}^2 \label{eq:app_UB_j_0}
\end{IEEEeqnarray}
where
\begin{equation}
\Upsilon_{\ell} \triangleq \frac{n_t}{T\rho}\mathsf{I}_{T} - \left(\mathsf{I}_{T}+\mathsf{X}_{\ell}\mathsf{X}_{\ell}^{\mathsf{H}}\right)^{-1}.
\end{equation}
We next use that we can assume without loss of optimality that $\mathsf{X}_{\ell}$ is a $T\times n_t$ rectangular diagonal matrix with diagonal entries $d_{1,\ell},\ldots,d_{n_t,\ell}$, so the channel output $\mathbf{Y}_{\ell}$ can be expressed as
\begin{equation}
\label{eq:app_UB_j_1}
\mathbf{Y}_{\ell} = \left(\begin{array}{c} \mathsf{D}_{\ell}\mathbf{H}_{\ell} + \mathbf{W}_{1,\ell} \\ \mathbf{W}_{2,\ell}\end{array}\right)
\end{equation}
where $\mathsf{D}_{\ell} \triangleq  \text{diag}\bigl\{d_{1,\ell},\ldots,d_{n_t,\ell}\bigr\}$, and $\mathbf{W}_{1,\ell}$ and $\mathbf{W}_{2,\ell}$ denote the first $n_t$ rows and the last $(T-n_t)$ rows of $\mathbf{W}_{\ell}$, respectively. It can then be shown that
\begin{IEEEeqnarray}{lCl}
\text{tr}\left(\mathbf{Y}_{\ell}^{\mathsf{H}}\Upsilon_{\ell}\mathbf{Y}_{\ell}\right) & = & \text{tr}\left(\left(\mathsf{D}_{\ell}\mathbf{H}_{\ell} + \mathbf{W}_{1,\ell}\right)^{\mathsf{H}}\Upsilon_{1,\ell}\left(\mathsf{D}_{\ell}\mathbf{H}_{\ell} + \mathbf{W}_{1,\ell}\right)\right) + \text{tr}\left(\mathbf{W}_{2,\ell}^{\mathsf{H}}\Upsilon_{2,\ell}\mathbf{W}_{2,\ell}\right) \nonumber\\
& = & \text{tr}\left(\left(\mathsf{D}_{\ell}\mathbf{H}_{\ell} + \mathbf{W}_{1,\ell}\right)^{\mathsf{H}}\Upsilon_{1,\ell}\left(\mathsf{D}_{\ell}\mathbf{H}_{\ell} + \mathbf{W}_{1,\ell}\right)\right) + \left(\frac{n_t}{T\rho}-1\right)\text{tr}\left(\mathbf{W}_{2,\ell}^{\mathsf{H}}\mathbf{W}_{2,\ell}\right) \label{eq:app_UB_j_1.5}
\end{IEEEeqnarray}
where the matrix $\Upsilon_{1,\ell}$ contains the first $n_t$ rows and columns of $\Upsilon_{\ell}$ and the matrix $\Upsilon_{2,\ell}$ contains the last $(T-n_t)$ rows and columns of $\Upsilon_{\ell}$, i.e.,
\begin{subequations}
\begin{IEEEeqnarray}{lCl}
\Upsilon_{1,\ell} & \triangleq & \frac{n_t}{T\rho}\mathsf{I}_{n_t} - \left(\mathsf{I}_{n_t}+\mathsf{D}_{\ell}^2\right)^{-1} \\
\Upsilon_{2,\ell} & \triangleq & \left(\frac{n_t}{T\rho}-1\right)\mathsf{I}_{T-n_t}.
\end{IEEEeqnarray}
\end{subequations}

We continue by upper-bounding $\sum_{i=n_t+1}^{n_r}\boldsymbol{\sigma}_{i,\ell}^2$ following the steps in \cite[p.~377]{zheng2002communication}. Specifically, we use that $\mathsf{D}_{\ell}\mathbf{H}_{\ell} + \mathbf{W}_{1,\ell}$ has rank $n_t$, so we can find a semi-unitary $(n_r-n_t) \times n_r$ matrix $\mathbf{Q}_0$\footnote{A semi-unitary $(n_r-n_t) \times n_r$ matrix is a $(n_r-n_t)\times n_r$ matrix $\mathsf{Q}$ satisfying $\mathsf{Q}\mathsf{Q}^{\mathsf{H}}=\mathsf{I}_{n_r-n_t}$.} (possibly depending on $\mathbf{H}_{\ell}$ and $\mathbf{W}_{1,\ell}$) such that $\mathbf{Q}_0 (\mathsf{D}_{\ell}\mathbf{H}_{\ell} + \mathbf{W}_{1,\ell})^{\mathsf{H}} = 0$. Together with \eqref{eq:app_UB_j_1}, this implies that $\text{tr}\left(\mathbf{Q}_0\mathbf{Y}_{\ell}^\mathsf{H}\mathbf{Y}_{\ell}\mathbf{Q}_0^\mathsf{H}\right) =  \text{tr}(\mathbf{Q}_0 \mathbf{W}_{2,\ell}^\mathsf{H} \mathbf{W}_{2,\ell} \mathbf{Q}_0^\mathsf{H})$. Furthermore, it follows from \cite[Corollary 4.3.39]{horn1990matrix} that
\begin{IEEEeqnarray}{lCl}
\sum_{i=n_t+1}^{n_r}\boldsymbol{\sigma}_{i,\ell}^2 & = & \min_{\substack{\mathbf{Q}\in\mathbb{C}^{(n_r-n_t)\times n_r}\colon \\ \mathbf{Q}\mathbf{Q}^{\mathsf{H}} = \mathsf{I}_{n_r-n_t}}} \text{tr}\left(\mathbf{Q}\mathbf{Y}_{\ell}^\mathsf{H}\mathbf{Y}_{\ell}\mathbf{Q}^\mathsf{H}\right)\nonumber\\
& \leq & \text{tr}\left(\tilde{\mathbf{Q}}\mathbf{Y}_{\ell}^\mathsf{H}\mathbf{Y}_{\ell}\tilde{\mathbf{Q}}^\mathsf{H}\right) \label{eq:app_UB_j_2}
\end{IEEEeqnarray}
for any $(n_r-n_t) \times n_r$ matrix $\tilde{\mathbf{Q}}$ satisfying $\tilde{\mathbf{Q}}\tilde{\mathbf{Q}}^{\mathsf{H}}=\mathsf{I}_{n_r-n_t}$. By applying \eqref{eq:app_UB_j_2} with $\tilde{\mathbf{Q}}=\mathbf{Q}_0$, we thus obtain that
\begin{equation}
\label{eq:app_UB_j_3}
\sum_{i=n_t+1}^{n_r}\boldsymbol{\sigma}_{i,\ell}^2 \leq \text{tr}\left(\mathbf{Q}_0 \mathbf{W}_{2,\ell}^\mathsf{H} \mathbf{W}_{2,\ell} \mathbf{Q}_0^\mathsf{H}\right).
\end{equation}

We next consider a unitary $n_r\times n_r$ matrix $\mathbf{Q}$ that contains $\mathbf{Q}_0$ as a submatrix. Let $\mathbf{Q}_1$ denote the submatrix composed of the remaining rows of $\mathbf{Q}$ so that
\begin{equation}
\mathbf{Q} = \left(\begin{array}{c} \mathbf{Q}_0 \\ \mathbf{Q}_1 \end{array}\right).
\end{equation}
It then follows that
\begin{IEEEeqnarray}{lCl}
\text{tr}\left(\mathbf{W}_{2,\ell}^{\mathsf{H}}\mathbf{W}_{2,\ell}\right) & = & \text{tr}\left(\mathbf{Q}\mathbf{W}_{2,\ell}^{\mathsf{H}}\mathbf{W}_{2,\ell}\mathbf{Q}^{\mathsf{H}}\right) \nonumber\\
& = & \text{tr}\left(\mathbf{Q}_0\mathbf{W}_{2,\ell}^{\mathsf{H}}\mathbf{W}_{2,\ell}\mathbf{Q}_0^{\mathsf{H}}\right) + \text{tr}\left(\mathbf{Q}_1\mathbf{W}_{2,\ell}^{\mathsf{H}}\mathbf{W}_{2,\ell}\mathbf{Q}_1^{\mathsf{H}}\right). \label{eq:app_UB_j_4}
\end{IEEEeqnarray}

Combining \eqref{eq:app_UB_j_1.5}, \eqref{eq:app_UB_j_3}, and \eqref{eq:app_UB_j_4}, we obtain
\begin{IEEEeqnarray}{lCl}
\IEEEeqnarraymulticol{3}{l}{\text{tr}\left(\mathbf{Y}_{\ell}^{\mathsf{H}}\Upsilon_{\ell}\mathbf{Y}_{\ell}\right) + \left(1-\frac{n_t}{T\rho}\right) \sum_{i=n_t+1}^{n_r}\boldsymbol{\sigma}_{i,\ell}^2} \nonumber\\
\quad & \leq & \text{tr}\left(\left(\mathsf{D}_{\ell}\mathbf{H}_{\ell} + \mathbf{W}_{1,\ell}\right)^{\mathsf{H}}\Upsilon_{1,\ell}\left(\mathsf{D}_{\ell}\mathbf{H}_{\ell} + \mathbf{W}_{1,\ell}\right)\right) + \left(\frac{n_t}{T\rho}-1\right)\text{tr}\left(\mathbf{Q}_1\mathbf{W}_{2,\ell}^{\mathsf{H}}\mathbf{W}_{2,\ell}\mathbf{Q}_1^{\mathsf{H}}\right). \label{eq:app_UB_j_5}
\end{IEEEeqnarray}
Using that the matrices $\mathbf{H}_{\ell}$, $\mathbf{W}_{1,\ell}$, and $\mathbf{W}_{2,\ell}$ have i.i.d. $\mathcal{CN}(0,1)$ entries, both traces on the RHS of \eqref{eq:app_UB_j_5} have Gamma distributions. Specifically, we have that, conditioned on $\mathbf{X}_{\ell}=\mathsf{D}_{\ell}$, $\mathsf{D}_{\ell}\mathbf{H}_{\ell}+\mathbf{W}_{1,\ell}$ has the same distribution as $(\mathsf{I}_{n_t}+\mathsf{D})^{1/2}\mathbf{H}_{\ell}$ \cite[p.~377]{zheng2002communication}. Furthermore, $\mathbf{Q}_1$ depends on $\mathbf{Q}_0$ (which in turn may depend on $\mathbf{H}_{\ell}$ and $\mathbf{W}_{1,\ell}$), but it is independent of $\mathbf{W}_{2,\ell}$. It can then be shown that
\begin{subequations}
\begin{IEEEeqnarray}{rCl}
\text{tr}\left(\left(\mathsf{D}_{\ell}\mathbf{H}_{\ell} + \mathbf{W}_{1,\ell}\right)^{\mathsf{H}}\Upsilon_{1,\ell}\left(\mathsf{D}_{\ell}\mathbf{H}_{\ell} + \mathbf{W}_{1,\ell}\right)\right) & \stackrel{d}{=} &\frac{n_t}{T\rho}\text{tr}\left(\mathbf{H}^{\mathsf{H}} (\mathsf{I}_{n_t}+\mathsf{D}_{\ell})\mathbf{H}\right) - \text{tr}\left(\mathbf{H}^{\mathsf{H}}\mathbf{H}\right) \nonumber\\
& \stackrel{d}{=} & \sum_{i=1}^{n_t}\left(\frac{n_t+n_t d^2_{i,\ell}}{T\rho}-1\right) Z_{i,\ell}' \label{eq:app_UB_j_z'}\\
\left(\frac{n_t}{T\rho}-1\right)\text{tr}\left(\mathbf{Q}_1\mathbf{W}_{2,\ell}^{\mathsf{H}}\mathbf{W}_{2,\ell}\mathbf{Q}_1^{\mathsf{H}}\right) & \stackrel{d}{=} & \frac{n_t-T\rho}{T\rho}\sum_{i=1}^{n_t} Z_{i,\ell}'' \label{eq:app_UB_j_z''}
\end{IEEEeqnarray}
\end{subequations}
where $Z_{i,\ell}'$ and $Z_{i,\ell}''$ were defined after \eqref{MC_Def_j_up}. In fact, it follows from \eqref{eq:app_UB_j_z'} that $Z_{i,\ell}'$ is given by the squared Euclidean norm of the $i$-th row of $\mathbf{H}_{\ell}$. Thus, \eqref{eq:app_UB_j_0} can be upper-bounded by a term that, conditioned on $\mathbf{X}_{\ell}=\mathsf{X}_{\ell}$, has the same distribution as
\begin{equation}
\label{eq:app_UB_j_5.5}
 \sum_{i=1}^{n_t}\left(\frac{n_t+n_t d^2_{i,\ell}}{T\rho}-1\right) Z_{i,\ell}' - \frac{T\rho-n_t}{T\rho}\sum_{i=1}^{n_t} Z_{i,\ell}''.
\end{equation}

It remains to upper-bound the last two terms on the RHS of \eqref{eq:def_j(D,s)}. To this end, we follow along the lines of \cite[Sec.~IV-A]{yang2013capacity}. Indeed, we begin by upper-bounding
\begin{equation}
(T-2n_r+n_t)\log \left(\prod_{i=1}^{n_t}{\boldsymbol{\sigma}}_{i,\ell}^2\right) + 2\log\left( \prod_{i=1}^{n_t}\prod_{j=n_t+1}^{n_r}({\boldsymbol{\sigma}}_{i,\ell}^2-{\boldsymbol{\sigma}}_{j,\ell}^2)\right) \leq (T-n_t)\log \left(\prod_{i=1}^{n_t}{\boldsymbol{\sigma}}_{i,\ell}^2\right) \label{eq:UB_j_6}
\end{equation}
using that $\boldsymbol{\sigma}_{j,\ell}^2 \geq 0$, $j=n_t+1,\ldots,n_r$. To bound the product of the singular values on the RHS of \eqref{eq:UB_j_6}, we next use that, conditioned on $\mathbf{X}_{\ell}=\mathsf{D}_{\ell}$, $\mathbf{Y}_{\ell}^{\mathsf{H}}\mathbf{Y}_{\ell}$ has the same distribution as $\mathbf{H}_{\ell}^{\mathsf{H}}(\mathsf{I}_{n_t}+\mathsf{D}_{\ell}^2)\mathbf{H}_{\ell}) + \mathbf{Q}_{\ell}^{\mathsf{H}}\mathbf{Q}_{\ell}$ \cite[p.~377]{zheng2002communication}, where $\mathbf{Q}_{\ell}$ was defined after \eqref{MC_Def_j_up}. It then follows from Weyl's theorem  \cite[Th.~4.3.1]{horn1990matrix} that (cf.~\cite[Sec.~IV-A]{yang2013capacity})
\begin{IEEEeqnarray}{lCl}
		\log \left(\prod_{i=1}^{n_t}{\boldsymbol{\sigma}}_{i,\ell}^2\right) 
				&\leq &\log \det \left(\big(\mathsf{I}_{n_t}+\mathsf{D}^2_{\ell}\big) \mathbf{H}_{\ell} \mathbf{H}_{\ell}^\mathsf{H}+\lambda_1(\mathbf{Q}_{\ell}^\mathsf{H}\mathbf{Q}_{\ell}) \mathsf{I}_{n_t} \right) \\
		&= & \log\det\bigl(\mathsf{I}_{n_t}+\mathsf{D}^2_{\ell}\bigr)
		+\log \det \Bigl( \mathbf{H}_{\ell} \mathbf{H}_{\ell}^\mathsf{H}+  \lambda_1(\mathbf{Q}_{\ell}^\mathsf{H}\mathbf{Q}_{\ell})\bigl(\mathsf{I}_{n_t}+\mathsf{D}^2_{\ell}\bigr)^{-1}  \Bigr) \label{eq:app_UB_j_7}
	\end{IEEEeqnarray}
	where the second step follows because $\mathbf{H}_{\ell}^{\mathsf{H}}\big(\mathsf{I}_{n_t}+\mathsf{D}_{\ell}^2\big)\mathbf{H}_{\ell}$ and $\big(\mathsf{I}_{n_t}+\mathsf{D}^2_{\ell}\big) \mathbf{H}_{\ell} \mathbf{H}_{\ell}^\mathsf{H}$ have the same $n_t$ nonzero eigenvalues \cite[Th.~1.3.20]{horn1990matrix}.

Combining \eqref{eq:app_UB_j_5.5}--\eqref{eq:app_UB_j_7} with \eqref{MC_Def_j_up}, we conclude that, conditioned on $\mathbf{X}_{\ell}=\mathsf{X}_{\ell}$, the mismatched information density $j_{\ell}(\mathsf{D}_{\ell},\boldsymbol{\sigma}_{\ell},T,\rho)$ can be upper-bounded by a term that has the same distribution as
\begin{IEEEeqnarray}{l}
n_tn_r\log \left(\frac{T\rho}{n_t}\right) +\log \left( \frac{\Gamma_{n_t}(n_t)}{\Gamma_{n_t}(T)} \right) +  \sum_{i=1}^{n_t}\left(\frac{n_t+n_t d^2_{i,\ell}}{T\rho}-1\right) Z_{i,\ell}' - \frac{T\rho-n_t}{T\rho}\sum_{i=1}^{n_t} Z_{i,\ell}'' \nonumber\\
{} +(T-n_t-n_r) \log\det\bigl(\mathsf{I}_{n_t}+ \mathsf{D}^2_{\ell}\bigr)  +(T-n_t)\log \det \Bigl( \mathbf{H}_{\ell} \mathbf{H}_{\ell}^\mathsf{H}+  \lambda_1(\mathbf{Q}_{\ell}^\mathsf{H}\mathbf{Q}_{\ell})\bigl(\mathsf{I}_{n_t}+\mathsf{D}^2_{\ell}\bigr)^{-1}  \Bigr)
\end{IEEEeqnarray}
which is $\bar{j}(\mathsf{D}_{\ell}, \mathbf{Z}_{\ell}',\mathbf{Z}_{\ell}'',\mathbf{H}_{\ell},\mathbf{Q}_{\ell})$.

\section{Proof of Lemma \ref{MC_Lemma4}} \label{Appx_Pf_MC_Lem4}

\subsection{Proof of \eqref{eq:Lemma3_first}}

	By the definitions of $\bar{j}(\mathsf{D}_{\ell}, \mathbf{Z}_{\ell}',\mathbf{Z}_{\ell}'',\mathbf{H}_{\ell},\mathbf{Q}_{\ell})$ and $\bar{J}(\mathsf{D}_{\ell},T,\rho)$ in \eqref{MC_Def_j_up} and \eqref{MC_Def_J_up}, we have that
	\begin{IEEEeqnarray}{lCl}
	\bar{U}(\mathsf{D},T,\rho) & = & \mathsf{E}\Biggl[ \sum_{i=1}^{n_t} \left(  \frac{n_t}{T\rho}+   \frac{n_t d_{i,\ell}^2 }{T\rho} -1 \right)(Z’_{i, \ell}-n_r)
			- \frac{T\rho-n_t}{T\rho} \sum_{i=1}^{n_t}\left(Z_{i,\ell}''-(T-n_t)\right) \nonumber\\
			& & \qquad {}  +(T-n_t)\log \det \left( \mathbf{H}_{\ell} \mathbf{H}_{\ell}^\mathsf{H}
			+\lambda_1(\mathbf{Q}_{\ell}^\mathsf{H}\mathbf{Q}_{\ell})(\mathsf{I}_{n_t}+\mathsf{D}^2)^{-1} \right) \nonumber\\
   			& & \qquad {} - (T-n_t)\mathsf{E}\left[ \log \det \left( \mathbf{H}_{\ell} \mathbf{H}_{\ell}^\mathsf{H} 
			+ \lambda_1(\mathbf{Q}_{\ell}^\mathsf{H}\mathbf{Q}_{\ell})(\mathsf{I}_{n_t}+\mathsf{D}^2)^{-1} \right) \right] \biggr)^2\Biggr] \nonumber\\
			& \leq & \mathsf{c}_{3,2} \Biggl(  \sum_{i=1}^{n_t} \left(  \frac{n_t}{T\rho}+   \frac{n_t d_{i,\ell}^2 }{T\rho} -1 \right)^2 \mathsf{E}\left[\left(Z_{i,\ell}'-n_r\right)^2\right] 
			 + \left(  \frac{T\rho-n_t}{T\rho}\right)^2 \sum_{i=1}^{n_t}\mathsf{E}\left[\left(Z_{i,\ell}''-(T-n_t)\right)^2\right] \nonumber\\
			& & \qquad\quad {} +  (T-n_t)^2 \mathsf{E}\left[\log^2 \det \left( \mathbf{H}_{\ell} \mathbf{H}_{\ell}^\mathsf{H} 
			+ \lambda_1(\mathbf{Q}_{\ell}^\mathsf{H}\mathbf{Q}_{\ell})(\mathsf{I}_{n_t}+\mathsf{D}^2)^{-1} \right)\right]\Biggr) \nonumber\\
			& \leq & \mathsf{c}_{3,2}  \left(  \frac{n_t}{T\rho_0}+   n_t \right)^2 n_r n_t + \mathsf{c}_{3,2} \left(1+\frac{n_t}{T\rho_0}\right)^2 n_t (T-n_t) \nonumber\\
			& & {} + \mathsf{c}_{3,2} (T-n_t)^2 \mathsf{E}\left[\log^2 \det \left( \mathbf{H}_{\ell} \mathbf{H}_{\ell}^\mathsf{H} 
			+ \lambda_1(\mathbf{Q}_{\ell}^\mathsf{H}\mathbf{Q}_{\ell})(\mathsf{I}_{n_t}+\mathsf{D}^2)^{-1} \right)\right] \IEEEeqnarraynumspace \label{MC_Equ_V}
	\end{IEEEeqnarray}
	where the second step follows from \eqref{Equ_cnv} and because $\mathsf{E}[(X-\mathsf{E}[X])^2] \leq \mathsf{E}[X^2]$ for every random variable $X$; the third step follows by bounding $d_{i,\ell}^2\leq T\rho$, $\rho\geq\rho_0$, and $|T\rho-n_t|/(T\rho) \leq 1+n_t/(T\rho_0)$, and because the variances of $Z_{i,\ell}'$ and $Z_{i,\ell}''$ are $n_r$ and $T-n_t$, respectively.
		
It remains to upper-bound the third term on the RHS of \eqref{MC_Equ_V}. To this end, we first note that
			\begin{IEEEeqnarray}{lCl}
			\log\det\left(\mathbf{H}_{\ell} \mathbf{H}_{\ell}^\mathsf{H}\right)
			& \leq & \log \det \left( \mathbf{H}_{\ell} \mathbf{H}_{\ell}^\mathsf{H} 
			+ \lambda_1(\mathbf{Q}_{\ell}^\mathsf{H}\mathbf{Q}_{\ell})(\mathsf{I}_{n_t}+\mathsf{D}^2)^{-1}\right) \nonumber\\
			& \leq &  \log\det\left(\mathbf{H}_{\ell} \mathbf{H}_{\ell}^\mathsf{H}+\lambda_1(\mathbf{Q}_{\ell}^\mathsf{H}\mathbf{Q}_{\ell}) \mathsf{I}_{n_t}  \right) \nonumber\\
			& \leq & n_t \log\left(1+\text{tr}\left(\mathbf{H}_{\ell} \mathbf{H}_{\ell}^\mathsf{H}\right)+\lambda_1(\mathbf{Q}_{\ell}^\mathsf{H}\mathbf{Q}_{\ell}) \right)\label{MC_Equ_log_b1n}
			\end{IEEEeqnarray}
			where the first two inequalities follow because $\det(\mathsf{A} + \mathsf{B}) \geq \det(\mathsf{A})$ for every pair of positive definite matrices $\mathsf{A}$ and $\mathsf{B}$ \cite[Th.~7.8.21]{horn1990matrix} and because $(\mathsf{I}_{n_t}+\mathsf{D}^2)^{-1} \preceq \mathsf{I}_{n_t}$; the last inequality follows from Hadamard's inequality \cite[Th.~7.8.1]{horn1990matrix} and the fact that every diagonal element of $\mathbf{H}_{\ell} \mathbf{H}_{\ell}^\mathsf{H}$ is upper-bounded by its trace. Therefore, 
			\begin{IEEEeqnarray}{lCl}
			\IEEEeqnarraymulticol{3}{l}{\mathsf{E}\left[\log^2 \det \left( \mathbf{H}_{\ell} \mathbf{H}_{\ell}^\mathsf{H} 
			+ \lambda_1(\mathbf{Q}_{\ell}^\mathsf{H}\mathbf{Q}_{\ell})(\mathsf{I}_{n_t}+\mathsf{D}^2)^{-1} \right)\right]} \nonumber\\
			\quad & \leq & \mathsf{c}_{2,2}  \mathsf{E} \left[\log^2\det\left( \mathbf{H}_{\ell} \mathbf{H}_{\ell}^\mathsf{H}\right) \right]
				+\mathsf{c}_{2,2} n_t^2 \mathsf{E} \left[\log^2\left(1+\text{tr}\left(\mathbf{H}_{\ell} \mathbf{H}_{\ell}^\mathsf{H}\right)+\lambda_1(\mathbf{Q}_{\ell}^\mathsf{H}\mathbf{Q}_{\ell}) \right) \right].  \label{MC_L4_T3o}
			\end{IEEEeqnarray}
			By Lemma~\ref{lem:grant_wishart}, the first expected value on the RHS of \eqref{MC_L4_T3o} is equal to
			\begin{equation}
			 \mathsf{E} \left[\log^2\det\left( \mathbf{H}_{\ell} \mathbf{H}_{\ell}^\mathsf{H}\right) \right]
					= \left(\sum_{i=0}^{n_t-1} \varPsi(n_r-i)\right)^2+\sum_{i=0}^{n_t-1} \varPsi'(n_r-i).
			\label{MC_L4_T3a}
			\end{equation} 
		To bound the second expected value on the RHS of \eqref{MC_L4_T3o}, we use that the function $x\mapsto \log^2(1+x)$ is concave for $x\geq e-1$. Consequently,
			\begin{IEEEeqnarray}{lCl}
			\IEEEeqnarraymulticol{3}{l}{\mathsf{E} \left[\log^2\left(1+\text{tr}\left(\mathbf{H}_{\ell} \mathbf{H}_{\ell}^\mathsf{H}\right)+\lambda_1(\mathbf{Q}_{\ell}^\mathsf{H}\mathbf{Q}_{\ell}) \right) \right]} \nonumber\\
			\quad & = & \mathsf{E} \left[\log^2\left(1+\text{tr}\left(\mathbf{H}_{\ell} \mathbf{H}_{\ell}^\mathsf{H}\right)+\lambda_1(\mathbf{Q}_{\ell}^\mathsf{H}\mathbf{Q}_{\ell}) \right) \mathds{1}\{\text{tr}\left(\mathbf{H}_{\ell} \mathbf{H}_{\ell}^\mathsf{H}\right)+\lambda_1(\mathbf{Q}_{\ell}^\mathsf{H}\mathbf{Q}_{\ell}) \leq 2\}\right] \nonumber\\
			& & {} + \mathsf{E} \left[\log^2\left(1+\text{tr}\left(\mathbf{H}_{\ell} \mathbf{H}_{\ell}^\mathsf{H}\right)+\lambda_1(\mathbf{Q}_{\ell}^\mathsf{H}\mathbf{Q}_{\ell}) \right) \mathds{1}\{\text{tr}\left(\mathbf{H}_{\ell} \mathbf{H}_{\ell}^\mathsf{H}\right)+\lambda_1(\mathbf{Q}_{\ell}^\mathsf{H}\mathbf{Q}_{\ell}) > 2\}\right] \nonumber\\
					& \leq &\log^2(3) \nonumber\\
					& & {} + \textnormal{Pr}\left[\text{tr}\left(\mathbf{H}_{\ell} \mathbf{H}_{\ell}^\mathsf{H}\right)+\lambda_1(\mathbf{Q}_{\ell}^\mathsf{H}\mathbf{Q}_{\ell}) > 2\right] \mathsf{E} \left[\left.\log^2\left(1+\text{tr}\left(\mathbf{H}_{\ell} \mathbf{H}_{\ell}^\mathsf{H}\right)+\lambda_1(\mathbf{Q}_{\ell}^\mathsf{H}\mathbf{Q}_{\ell}) \right)\right| \text{tr}\left(\mathbf{H}_{\ell} \mathbf{H}_{\ell}^\mathsf{H}\right)+\lambda_1(\mathbf{Q}_{\ell}^\mathsf{H}\mathbf{Q}_{\ell}) > 2\right] \nonumber\\
			& \leq &  \log^2(3) \nonumber\\
			& & {} + \textnormal{Pr}\left[\text{tr}\left(\mathbf{H}_{\ell} \mathbf{H}_{\ell}^\mathsf{H}\right)+\lambda_1(\mathbf{Q}_{\ell}^\mathsf{H}\mathbf{Q}_{\ell}) > 2\right]  \log^2\left(1+\mathsf{E}\left[\left.\text{tr}\left(\mathbf{H}_{\ell} \mathbf{H}_{\ell}^\mathsf{H}\right)+\lambda_1(\mathbf{Q}_{\ell}^\mathsf{H}\mathbf{Q}_{\ell})\right|\text{tr}\left(\mathbf{H}_{\ell} \mathbf{H}_{\ell}^\mathsf{H}\right)+\lambda_1(\mathbf{Q}_{\ell}^\mathsf{H}\mathbf{Q}_{\ell}) > 2\right]\right) \nonumber\\
			& \leq & \log^2(3) + \textnormal{Pr}\left[\text{tr}\left(\mathbf{H}_{\ell} \mathbf{H}_{\ell}^\mathsf{H}\right)+\lambda_1(\mathbf{Q}_{\ell}^\mathsf{H}\mathbf{Q}_{\ell}) > 2\right] \log^2\left(1+\frac{\mathsf{E}\left[\text{tr}\left(\mathbf{H}_{\ell} \mathbf{H}_{\ell}^\mathsf{H}\right)+\lambda_1(\mathbf{Q}_{\ell}^\mathsf{H}\mathbf{Q}_{\ell})\right]}{\textnormal{Pr}\left[\text{tr}\left(\mathbf{H}_{\ell} \mathbf{H}_{\ell}^\mathsf{H}\right)+\lambda_1(\mathbf{Q}_{\ell}^\mathsf{H}\mathbf{Q}_{\ell}) > 2\right]}\right) \nonumber\\
			& \leq & \log^2(3)  + \sup_{0<\delta\leq 1} \left\{\delta \log^2\left(1+\frac{n_t n_r (n_t n_r +1 ) + (T-n_r) n_r}{\delta}\right)\right\}\label{MC_L4_T3b} 
			\end{IEEEeqnarray}
			where the third step follows from Jensen's inequality; the fourth step follows by noting that
			\begin{IEEEeqnarray}{lCl}
			1 & \leq & \mathsf{E}\left[\left.\text{tr}\left(\mathbf{H}_{\ell} \mathbf{H}_{\ell}^\mathsf{H}\right)+\lambda_1(\mathbf{Q}_{\ell}^\mathsf{H}\mathbf{Q}_{\ell})\right|\text{tr}\left(\mathbf{H}_{\ell} \mathbf{H}_{\ell}^\mathsf{H}\right)+\lambda_1(\mathbf{Q}_{\ell}^\mathsf{H}\mathbf{Q}_{\ell}) > 2\right] \nonumber\\
					& \leq & \frac{\mathsf{E}\left[\text{tr}\left(\mathbf{H}_{\ell} \mathbf{H}_{\ell}^\mathsf{H}\right)+\lambda_1(\mathbf{Q}_{\ell}^\mathsf{H}\mathbf{Q}_{\ell})\right]}{\textnormal{Pr}\left[\text{tr}\left(\mathbf{H}_{\ell} \mathbf{H}_{\ell}^\mathsf{H}\right)+\lambda_1(\mathbf{Q}_{\ell}^\mathsf{H}\mathbf{Q}_{\ell}) > 2\right]};
			\end{IEEEeqnarray}
			the last step follows by evaluating $\mathsf{E}[(\text{tr}(\mathbf{H}_{\ell} \mathbf{H}_{\ell}^\mathsf{H} ))^2]=n_tn_r(n_tn_r + 1)$ \cite[Lemma 2.9]{tulino2004random} and by bounding $\mathsf{E}[\lambda_1(\mathbf{Q}_{\ell}^\mathsf{H}\mathbf{Q}_{\ell})]\leq \mathsf{E}[\text{tr}(\mathbf{Q}_{\ell}^\mathsf{H}\mathbf{Q}_{\ell})] = (T-n_t) n_r$ \cite[Lemma 2.9]{tulino2004random}, and by optimizing over $\delta = \textnormal{Pr}\left[\text{tr}\left(\mathbf{H}_{\ell} \mathbf{H}_{\ell}^\mathsf{H}\right)+\lambda_1(\mathbf{Q}_{\ell}^\mathsf{H}\mathbf{Q}_{\ell}) > 2\right]$.
		
		Since the function $\delta\mapsto \delta\log^2(1+(n_t n_r (n_t n_r +1 ) + (T-n_r) n_r)/\delta)$ is continuous on $0 < \delta \leq 1$ and tends to zero as $\delta\downarrow 0$, we conclude from \eqref{MC_Equ_V}--\eqref{MC_L4_T3b} that
		\begin{equation}
		\sup_{\rho \geq \rho_0} \sup_{\mathsf{D}\in \mathcal{D}_{n_t}} \bar{U}(\mathsf{D},T,\rho) < \infty.
		\end{equation}
		This is \eqref{eq:Lemma3_first}.
	
\subsection{Proof of \eqref{eq:Lemma3_second}}

By the definitions of $\bar{j}(\mathsf{D}_{\ell}, \mathbf{Z}_{\ell}',\mathbf{Z}_{\ell}'',\mathbf{H}_{\ell},\mathbf{Q}_{\ell})$ and $\bar{J}(\mathsf{D}_{\ell},T,\rho)$ in \eqref{MC_Def_j_up} and \eqref{MC_Def_J_up}, we have that
\begin{IEEEeqnarray}{lCl}
\bar{S}(\mathsf{D},T,\rho)  & = & \mathsf{E}\Biggl[\biggl|  \sum_{i=1}^{n_t} \left(  \frac{n_t}{T\rho}+   \frac{n_t d_{i,\ell}^2 }{T\rho} -1 \right) \left(Z_{i,\ell}'-\mathsf{E}[Z_{i,\ell}']\right)
			-  \frac{T\rho-n_t}{T\rho} \sum_{i=1}^{n_t}\left(Z_{i,\ell}''-\mathsf{E}[Z_{i,\ell}'']\right) \nonumber\\
			& & \qquad {}  +(T-n_t)\log \det \left( \mathbf{H}_{\ell} \mathbf{H}_{\ell}^\mathsf{H}
			+\lambda_1(\mathbf{Q}_{\ell}^\mathsf{H}\mathbf{Q}_{\ell})(\mathsf{I}_{n_t}+\mathsf{D}^2)^{-1} \right) \nonumber\\
   			& & \qquad {} - (T-n_t)\mathsf{E}\left[ \log \det \left( \mathbf{H}_{\ell} \mathbf{H}_{\ell}^\mathsf{H} 
			+ \lambda_1(\mathbf{Q}_{\ell}^\mathsf{H}\mathbf{Q}_{\ell})(\mathsf{I}_{n_t}+\mathsf{D}^2)^{-1} \right) \right] \biggr|^3\Biggr] \nonumber\\
			& \leq &2 \mathsf{c}_{2+4n_t,3}\Biggl(  
			\sum_{i=1}^{n_t} \left|  \frac{n_t}{T\rho}+   \frac{n_t d_{i,\ell}^2 }{T\rho} -1 \right|^3\mathsf{E}\left[\left|Z_{i,\ell}'\right|^3\right] 
			+  \left|\frac{T\rho-n_t}{T\rho} \right|^3 \sum_{i=1}^{n_t}\mathsf{E}\left[\left|Z_{i,\ell}''\right|^3\right] \nonumber\\
			& & \quad\qquad\qquad\qquad {} +  (T-n_t)^3 \mathsf{E}\left[\left|\log \det \left( \mathbf{H}_{\ell} \mathbf{H}_{\ell}^\mathsf{H} 
			+ \lambda_1(\mathbf{Q}_{\ell}^\mathsf{H}\mathbf{Q}_{\ell})(\mathsf{I}_{n_t}+\mathsf{D}^2)^{-1} \right)\right|^3\right]\Biggr) \nonumber\\
			& \leq & 2\mathsf{c}_{2+4n_t,3}\Biggl(n_t \left|  \frac{n_t}{T\rho_0}+  n_t  \right|^3 \frac{\Gamma(n_r+3)}{\Gamma(n_r)} + \left|1+\frac{n_t}{T\rho_0}\right|^3 n_t \frac{\Gamma(T-n_t+3)}{\Gamma(T-n_t+3)} \nonumber\\
			& & \quad\qquad\qquad\qquad {}  + (T-n_t)^3 \mathsf{E}\left[\left|\log \det \left( \mathbf{H}_{\ell} \mathbf{H}_{\ell}^\mathsf{H} 
			+ \lambda_1(\mathbf{Q}_{\ell}^\mathsf{H}\mathbf{Q}_{\ell})(\mathsf{I}_{n_t}+\mathsf{D}^2)^{-1} \right)\right|^3\right]\Biggr) \label{MC_Equ_V3}
\end{IEEEeqnarray}
where the second step follows from \eqref{Equ_cnv} and because $|\mathsf{E}[X]|^3 \leq \mathsf{E}[|X|^3]$ for every random variable $X$; the third step follows because by bounding $d_{i,\ell}^2\leq T\rho$, $\rho\geq\rho_0$, and $|T\rho-n_t|/(T\rho) \leq 1+n_t/(T\rho_0)$, and because the third moment of a random variable with Gamma distribution $\Gamma(k,\theta)$ is given by $\theta^3 \Gamma(k+3)/\Gamma(k)$.

It remains to upper-bound the third term on the RHS of \eqref{MC_Equ_V3}. By \eqref{MC_Equ_log_b1n} and \eqref{Equ_cnv}, we have that
\begin{IEEEeqnarray}{lCl}
\IEEEeqnarraymulticol{3}{l}{\left|\log \det \left( \mathbf{H}_{\ell} \mathbf{H}_{\ell}^\mathsf{H} 
			+ \lambda_1(\mathbf{Q}_{\ell}^\mathsf{H}\mathbf{Q}_{\ell})(\mathsf{I}_{n_t}+\mathsf{D}^2)^{-1} \right)\right|^3} \nonumber\\
			\quad & \leq & \mathsf{c}_{2,3} \left|\log\det\left(\mathbf{H}_{\ell} \mathbf{H}_{\ell}^\mathsf{H}\right)\right|^3 + \mathsf{c}_{2,3} n_t^3 \log^3\left(1+\text{tr}\left(\mathbf{H}_{\ell} \mathbf{H}_{\ell}^\mathsf{H}\right)+\lambda_1(\mathbf{Q}_{\ell}^\mathsf{H}\mathbf{Q}_{\ell}) \right).
\end{IEEEeqnarray}
The expected value of $|\log\det(\mathbf{H}_{\ell} \mathbf{H}_{\ell}^\mathsf{H})|^3$ is finite. Moreover, the function $x \mapsto \log^3(1+x)$ is concave for $x\geq e^2-1$. By following similar steps as in \eqref{MC_L4_T3b}, it can thus be shown that
\begin{equation}
\mathsf{E}\left[\log^3\left(1+\text{tr}\left(\mathbf{H}_{\ell} \mathbf{H}_{\ell}^\mathsf{H}\right)+\lambda_1(\mathbf{Q}_{\ell}^\mathsf{H}\mathbf{Q}_{\ell}) \right)\right] \leq \log^3(8) + \sup_{0<\delta\leq 1} \left\{\log^3\left(1+\frac{n_t n_r(n_t n_r+1) + (T-n_r)n_r}{\delta}\right)\right\}. \label{eq:pf(49)_last}
\end{equation}

Since the function $\delta\mapsto \log^3(1+(n_t n_r (n_t n_r +1 ) + (T-n_r) n_r)/\delta)$ is continuous on $0<\delta\leq 1$ and tends to zero as $\delta\downarrow 0$, we conclude from \eqref{MC_Equ_V3}--\eqref{eq:pf(49)_last} that
\begin{equation}
\sup_{\rho{\geq \rho_0}} \sup_{\mathsf{D}\in \mathcal{D}_{n_t}} \bar{S}(\mathsf{D},T,\rho) < \infty.
\end{equation}
This is \eqref{eq:Lemma3_second}. 

\section{Proof of Lemma \ref{MC_Lemma5}} \label{Appx_Pf_MC_Lem5} 
Recall that (cf.~\eqref{MC_Def_j_up} and \eqref{MC_Def_J_up})
\begin{IEEEeqnarray}{lCl} 
	\bar{U}(T,\mathsf{D},\rho) & =& \mathsf{E}\Biggl[\sum_{i=1}^{n_t} \left(  \frac{n_t}{T\rho}+   \frac{n_t d_{i,\ell}^2 }{T\rho} -1 \right)(Z'_{i, \ell}-n_r)
	-\frac{T\rho-n_t}{T\rho}  \sum_{i=1}^{n_t}\left(Z_{i,\ell}''-(T-n_t)\right) \nonumber\\	
	& &\quad {} +(T-n_t)\log \det \left( \mathbf{H}_{\ell} \mathbf{H}_{\ell}^\mathsf{H}
	+ \lambda_1(\mathbf{Q}_{\ell}^\mathsf{H}\mathbf{Q}_{\ell})(\mathsf{I}_{n_t}+\mathsf{D}^2)^{-1} \right) \nonumber \\
	&& \quad {} -(T-n_t)\mathsf{E}\left[ \log \det \big( \mathbf{H}_{\ell} \mathbf{H}_{\ell}^\mathsf{H} 
	+ \lambda_1(\mathbf{Q}_{\ell}^\mathsf{H}\mathbf{Q}_{\ell})(\mathsf{I}_{n_t}+\mathsf{D}^2)^{-1} \big) \right] 
	\biggr)^2 \Biggr]. \label{MC_Def_V}
	\end{IEEEeqnarray} 
To prove Lemma~\ref{MC_Lemma5}, we first show that
\begin{equation}
\label{L4_UU_ast}
\bar{U}(\mathsf{D},T,\rho) = \bar{U}^{\ast}(\mathsf{D},T,\rho)+K_{\bar{U}}(\mathsf{D},T,\rho) 
\end{equation}
where $\bar{U}^{\ast}(\mathsf{D},T,\rho) \triangleq \mathsf{E}[(\bar{j}^{\ast}(\mathsf{D},\mathbf{Z}_{\ell}',\mathbf{Z}_{\ell}'',\mathbf{H}_{\ell})-\bar{J}^{\ast}(\mathsf{D},T,\rho))^2]$;
\begin{IEEEeqnarray}{lCl}
 \bar{j}^{\ast}(\mathsf{D},\mathbf{Z}_{\ell}',\mathbf{Z}_{\ell}'',\mathbf{H}_{\ell}) & \triangleq & n_t n_r\log \left(\frac{T\rho}{n_t}\right) 
	+ \log \left( \frac{\Gamma_{n_t}(n_t)}{\Gamma_{n_t}(T)} \right)
	+\sum_{i=1}^{n_t} \left(  \frac{n_t d_{i,\ell}^2 }{T\rho} -1 \right)Z'_{i, \ell}  -\sum_{i=1}^{n_t}Z''_{i,\ell} \nonumber\\
	& & {} +(T-n_t-n_r) \log\det\left(\mathsf{I}_{n_t}+ \mathsf{D}^2\right) +(T-n_t)\log \det \left( \mathbf{H}_{\ell} \mathbf{H}_{\ell}^\mathsf{H}  \right); \label{Def_Highj}
\end{IEEEeqnarray}
$\bar{J}^{\ast}(\mathsf{D},T,\rho) \triangleq \mathsf{E}[\bar{j}^{\ast}(\mathsf{D},\mathbf{Z}_{\ell}',\mathbf{Z}_{\ell}'',\mathbf{H}_{\ell})]$ was given in \eqref{MC_Ap3_Def_Jast}; and $K_{\bar{U}}(\mathsf{D},T,\rho)$ is a function of $\mathsf{D}$, $T$, and $\rho$ that satisfies
\begin{equation}
\lim_{\rho\to\infty} \sup_{\mathsf{D}\in\mathcal{D}_1} \left| K_{\bar{U}}(\mathsf{D},T,\rho) \right|= 0.
\end{equation}
To obtain \eqref{MC_Ap3_Def_Jast} as the expected value of $\bar{j}^{\ast}(\mathsf{D},\mathbf{Z}_{\ell}',\mathbf{Z}_{\ell}'',\mathbf{H}_{\ell})$, we use that $\mathsf{E}[Z_{i,\ell}']=n_r$, $\mathsf{E}[Z_{i,\ell}''] = T-n_r$, and $d_{1,\ell}^2+\ldots+d_{n_t,\ell}^2 = T\alpha$. 
We prove \eqref{L4_UU_ast} by bounding $|\bar{U}(\mathsf{D},T,\rho) - \bar{U}^{\ast}(\mathsf{D},T,\rho)|$ as follows. For ease of exposition, let
\begin{subequations}
\begin{IEEEeqnarray}{rCl}
\Delta(\mathsf{D},\mathbf{Z}_{\ell}',\mathbf{Z}_{\ell}'',\mathbf{H}_{\ell},\mathbf{Q}_{\ell}) & \triangleq & \bar{j}(\mathsf{D},\mathbf{Z}_{\ell}',\mathbf{Z}_{\ell}'',\mathbf{H}_{\ell},\mathbf{Q}_{\ell})-\bar{J}(\mathsf{D},T,\rho) \label{eq:def_Delta}\\
\Delta^{\ast}(\mathsf{D},\mathbf{Z}_{\ell}',\mathbf{Z}_{\ell}'',\mathbf{H}_{\ell}) & \triangleq & \bar{j}^{\ast}(\mathsf{D},\mathbf{Z}_{\ell}',\mathbf{Z}_{\ell}'',\mathbf{H}_{\ell})-\bar{J}^{\ast}(\mathsf{D},T,\rho).
\end{IEEEeqnarray}
\end{subequations}
It follows that
\begin{IEEEeqnarray}{lCl}
\IEEEeqnarraymulticol{3}{l}{\left|\bar{U}(\mathsf{D},T,\rho) - \bar{U}^{\ast}(\mathsf{D},T,\rho)\right|} \nonumber\\
\quad  & = & \left|\mathsf{E}\left[ \left( \Delta(\mathsf{D},\mathbf{Z}_{\ell}',\mathbf{Z}_{\ell}'',\mathbf{H}_{\ell},\mathbf{Q}_{\ell})  \right)^2\right] - \mathsf{E}\left[\left(\Delta^{\ast}(\mathsf{D}, \mathbf{Z}_{\ell}',\mathbf{Z}_{\ell}'',\mathbf{H}_{\ell}) \right)^2\right]\right| \nonumber\\
& = & \mathsf{E}\left[\left(\Delta(\mathsf{D},\mathbf{Z}_{\ell}',\mathbf{Z}_{\ell}'',\mathbf{H}_{\ell},\mathbf{Q}_{\ell}) -\Delta^{\ast}(\mathsf{D}, \mathbf{Z}_{\ell}',\mathbf{Z}_{\ell}'',\mathbf{H}_{\ell})\right)\left(\Delta(\mathsf{D},\mathbf{Z}_{\ell}',\mathbf{Z}_{\ell}'',\mathbf{H}_{\ell},\mathbf{Q}_{\ell}) +\Delta^{\ast}(\mathsf{D}, \mathbf{Z}_{\ell}',\mathbf{Z}_{\ell}'',\mathbf{H}_{\ell})\right)\right]  \nonumber\\
& \leq & \mathsf{E}\left[\left(\Delta(\mathsf{D},\mathbf{Z}_{\ell}',\mathbf{Z}_{\ell}'',\mathbf{H}_{\ell},\mathbf{Q}_{\ell}) -\Delta^{\ast}(\mathsf{D}, \mathbf{Z}_{\ell}',\mathbf{Z}_{\ell}'',\mathbf{H}_{\ell})\right)^2\right] \mathsf{E}\left[\left(\Delta(\mathsf{D},\mathbf{Z}_{\ell}',\mathbf{Z}_{\ell}'',\mathbf{H}_{\ell},\mathbf{Q}_{\ell}) +\Delta^{\ast}(\mathsf{D}, \mathbf{Z}_{\ell}',\mathbf{Z}_{\ell}'',\mathbf{H}_{\ell})\right)^2\right] \nonumber\\
& \leq & \mathsf{E}\left[\left(\Delta(\mathsf{D},\mathbf{Z}_{\ell}',\mathbf{Z}_{\ell}'',\mathbf{H}_{\ell},\mathbf{Q}_{\ell})-\Delta^{\ast}(\mathsf{D}, \mathbf{Z}_{\ell}',\mathbf{Z}_{\ell}'',\mathbf{H}_{\ell})\right)^2\right] \mathsf{c}_{2,2}\left(\bar{U}(\mathsf{D},T,\rho) + \bar{U}^{\ast}(\mathsf{D},T,\rho)\right) \label{eq:L4_what_to_do}
\end{IEEEeqnarray}
where the second-to-last step follows from the Cauchy-Schwarz inequality, and the last step follows from \eqref{Equ_cnv}. By Lemma~\ref{MC_Lemma4}, $\bar{U}(\mathsf{D},T,\rho)$ is bounded in $\rho$ and $\mathsf{D}$. Similarly, using \eqref{Equ_cnv}, $\bar{U}^{\ast}(\mathsf{D},T,\rho)$ can be upper-bounded as
\begin{IEEEeqnarray}{lCl}
\bar{U}^{\ast}(\mathsf{D},T,\rho) & \leq & \mathsf{c}_{3,2}\Biggl(\frac{(n_tT\alpha-T\rho)^2}{(T\rho)^2}\mathsf{E}\left[\left(\sum_{i=1}^{n_t}\left(Z'_{i,\ell} - n_t\right)\right)^2\right] + \mathsf{E}\left[\left(\sum_{i=1}^{n_t} \left( Z''_{i,\ell} - (T - n_t)\right) \right)^2\right] \nonumber\\
& & \qquad {} + (T-n_t)^2\mathsf{E}\left[\left(\log \det \left( \mathbf{H}_{\ell} \mathbf{H}_{\ell}^\mathsf{H}  \right) - \mathsf{E}\left[\log \det \left( \mathbf{H}_{\ell} \mathbf{H}_{\ell}^\mathsf{H}  \right)\right]\right)^2\right]\Biggr) \nonumber\\
& = &  \mathsf{c}_{3,2}\left(\frac{(n_t T\alpha-T\rho)^2}{(T\rho)^2} n_r n_t + n_t (T-n_t) + (T-n_t)^2  \sum_{i=0}^{n_t-1} \varPsi'(n_r-i)\right)
\end{IEEEeqnarray}
where the second step follows because the variances of $Z_{t,\ell}'$, $Z_{t,\ell}''$, and $\log \det \left( \mathbf{H}_{\ell} \mathbf{H}_{\ell}^\mathsf{H}  \right)$ are $n_t$, $(T-n_t)$, and $\sum_{i=0}^{n_t-1} \varPsi'(n_r-i)$, respectively. Consequently, $\bar{U}^{\ast}(\mathsf{D},T,\rho)$ is bounded in $\rho$ and $\mathsf{D}$.

We thus obtain \eqref{L4_UU_ast} by showing that the first expected value on the RHS of \eqref{eq:L4_what_to_do} vanishes as $\rho\to\infty$ uniformly in $\mathsf{D}\in\mathcal{D}_1$. To this end, we first note that
\begin{IEEEeqnarray}{lCl}
\Delta(\mathsf{D},\mathbf{Z}_{\ell}',\mathbf{Z}_{\ell}'',\mathbf{H}_{\ell},\mathbf{Q}_{\ell}) -\Delta^{\ast}(\mathsf{D}, \mathbf{Z}_{\ell}',\mathbf{Z}_{\ell}'',\mathbf{H}_{\ell}) & = & \frac{n_t}{T\rho} \left(\sum_{i=1}^{n_t}  (Z'_{i, \ell}-n_r) +\sum_{i=1}^{n_t} \left(Z''_{i, \ell}-(T-n_t)\right)  \right)  \nonumber\\
& & {} + (T-n_t)\log \frac{\det \left( \mathbf{H}_{\ell} \mathbf{H}_{\ell}^\mathsf{H}
	+ \lambda_1(\mathbf{Q}_{\ell}^\mathsf{H}\mathbf{Q}_{\ell})(\mathsf{I}_{n_t}+\mathsf{D}^2)^{-1} \right)}{\det \left( \mathbf{H}_{\ell} \mathbf{H}_{\ell}^\mathsf{H}  \right)}\nonumber\\
	& & {} - (T-n_t)\mathsf{E}\left[ \log \frac{\det \big( \mathbf{H}_{\ell} \mathbf{H}_{\ell}^\mathsf{H} 
	+ \lambda_1(\mathbf{Q}_{\ell}^\mathsf{H}\mathbf{Q}_{\ell})(\mathsf{I}_{n_t}+\mathsf{D}^2)^{-1} \big)}{\det \left( \mathbf{H}_{\ell} \mathbf{H}_{\ell}^\mathsf{H} \right)} \right]. \IEEEeqnarraynumspace
	\end{IEEEeqnarray}
Using that $\mathsf{E}[(X-\mathsf{E}[X])^2] \leq \mathsf{E}[X^2]$, we can thus upper-bound the first expected value on the RHS of \eqref{eq:L4_what_to_do} as
\begin{IEEEeqnarray}{lCl}
\IEEEeqnarraymulticol{3}{l}{\mathsf{E}\left[\left(\Delta(\mathsf{D},\mathbf{Z}_{\ell}',\mathbf{Z}_{\ell}'',\mathbf{H}_{\ell},\mathbf{Q}_{\ell}) -\Delta^{\ast}(\mathsf{D}, \mathbf{Z}_{\ell}',\mathbf{Z}_{\ell}'',\mathbf{H}_{\ell})\right)^2\right] } \nonumber\\
\quad & \leq &  \mathsf{c}_{3,2} \left( \frac{n_t}{T\rho}  \right)^2 n_t n_r + \mathsf{c}_{3,2} \left( \frac{n_t}{T\rho}  \right)^2 n_t(T-n_t)  \nonumber\\
&&{}+ \mathsf{c}_{3,2} (T-n_t)^2 \mathsf{E}\left[\log^2\left(\frac{\det \left( \mathbf{H}_{\ell} \mathbf{H}_{\ell}^\mathsf{H}
	+ \lambda_1(\mathbf{Q}_{\ell}^\mathsf{H}\mathbf{Q}_{\ell})(\mathsf{I}_{n_t}+\mathsf{D}^2)^{-1} \right)}{\det \left( \mathbf{H}_{\ell} \mathbf{H}_{\ell}^\mathsf{H}  \right)}\right)\right]. \label{eq:Lemma4_intermed}
\end{IEEEeqnarray}
Since $\det(\mathsf{A} + \mathsf{B}) \geq \det(\mathsf{A})$ for every pair of positive definite matrices $\mathsf{A}$ and $\mathsf{B}$ \cite[Th.~7.8.21]{horn1990matrix}, the fraction inside the logarithm function is greater than or equal to one. Consequently, an upper bound on this fraction yields an upper bound on the squared logarithm. Using that $d_i^2 \geq \bar{\delta}\rho$, $i=1,\ldots,n_t$, for $\mathsf{D} \in \mathcal{D}_1$, we obtain $\big(\mathsf{I}_{n_t}+  \mathbf{D}^2\big)^{-1} \leq (1+\bar{\delta}\rho)^{-1} \mathsf{I}_{n_t}$, so for $\mathsf{D}\in\mathcal{D}_1$, we can further upper-bound \eqref{eq:Lemma4_intermed} as
	\begin{IEEEeqnarray}{lCl}
	\IEEEeqnarraymulticol{3}{l}{\mathsf{E}\left[\left(\Delta(\mathsf{D},\mathbf{Z}_{\ell}',\mathbf{Z}_{\ell}'',\mathbf{H}_{\ell},\mathbf{Q}_{\ell}) -\Delta^{\ast}(\mathsf{D}, \mathbf{Z}_{\ell}',\mathbf{Z}_{\ell}'',\mathbf{H}_{\ell})\right)^2\right] } \nonumber\\
\quad	 & \leq &  \mathsf{c}_{3,2} \left( \frac{n_t}{T\rho}  \right)^2 n_t n_r + \mathsf{c}_{3,2} \left( \frac{n_t}{T\rho}  \right)^2 n_t(T-n_t)  \nonumber\\
	 &&{}+ \mathsf{c}_{3,2}  (T-n_t)^2 \mathsf{E}\left[\log^2\left(\frac{\det \left( \mathbf{H}_{\ell} \mathbf{H}_{\ell}^\mathsf{H}
	+ \lambda_1(\mathbf{Q}_{\ell}^\mathsf{H}\mathbf{Q}_{\ell})(1+\bar{\delta}\rho)^{-1}\mathsf{I}_{n_t} \right)}{\det \left( \mathbf{H}_{\ell} \mathbf{H}_{\ell}^\mathsf{H}  \right)}\right)\right]. \label{eq:Lemma4_beforeDCT}
\end{IEEEeqnarray}
The RHS of \eqref{eq:Lemma4_beforeDCT} is independent of $\mathsf{D}$ and vanishes by the dominated convergence theorem. Indeed, we have 
\begin{equation*}
\lim_{\rho\to\infty} \log^2\det \left( \mathbf{H}_{\ell} \mathbf{H}_{\ell}^\mathsf{H}
	+ \lambda_1(\mathbf{Q}_{\ell}^\mathsf{H}\mathbf{Q}_{\ell})(1+\bar{\delta}\rho)^{-1}\mathsf{I}_{n_t} \right)  = \log^2\det\left( \mathbf{H}_{\ell} \mathbf{H}_{\ell}^\mathsf{H}  \right)
\end{equation*}
by the continuity of the logarithm function and the determinant. Furthermore, by \eqref{Equ_cnv},
\begin{equation}
\log^2\left(\frac{\det \left( \mathbf{H}_{\ell} \mathbf{H}_{\ell}^\mathsf{H}
	+ \lambda_1(\mathbf{Q}_{\ell}^\mathsf{H}\mathbf{Q}_{\ell})(1+\bar{\delta}\rho)^{-1}\mathsf{I}_{n_t} \right)}{\det \left( \mathbf{H}_{\ell} \mathbf{H}_{\ell}^\mathsf{H}  \right)}\right) \leq \mathsf{c}_{2,2} \log^2\det\left(\mathbf{H}_{\ell} \mathbf{H}_{\ell}^\mathsf{H}\right) + \mathsf{c}_{2,2} \log^2\det \left( \mathbf{H}_{\ell} \mathbf{H}_{\ell}^\mathsf{H}
	+ \lambda_1(\mathbf{Q}_{\ell}^\mathsf{H}\mathbf{Q}_{\ell})\mathsf{I}_{n_t} \right) \label{eq:Lemma4_beforeDCT_bla}
\end{equation}
because the numerator on the LHS of \eqref{eq:Lemma4_beforeDCT_bla} is monotonically decreasing in $\rho$. The expected value of the RHS of \eqref{eq:Lemma4_beforeDCT} is finite (cf.~\eqref{MC_L4_T3o}--\eqref{MC_L4_T3b}), hence the dominated convergence theorem applies and
\begin{equation}
\lim_{\rho\to\infty} (T-n_t)^2 \mathsf{E}\left[\log^2\left(\frac{\det \left( \mathbf{H}_{\ell} \mathbf{H}_{\ell}^\mathsf{H}
	+ \lambda_1(\mathbf{Q}_{\ell}^\mathsf{H}\mathbf{Q}_{\ell})(1+\bar{\delta}\rho)^{-1}\mathsf{I}_{n_t} \right)}{\det \left( \mathbf{H}_{\ell} \mathbf{H}_{\ell}^\mathsf{H}  \right)}\right)\right] = 0. \label{eq:Lemma4_DCT}
\end{equation}

From \eqref{eq:L4_what_to_do}--\eqref{eq:Lemma4_DCT}, we obtain that
\begin{equation}
\lim_{\rho\to\infty} \sup_{\mathsf{D}\in\mathcal{D}_1} \left|\bar{U}(\mathsf{D},T,\rho) - \bar{U}^{\ast}(\mathsf{D},T,\rho)\right| = 0
\end{equation}
which is \eqref{L4_UU_ast}.

We next lower-bound $\bar{U}^{\ast}(\mathsf{D},T,\rho)$ for every $\mathsf{D}\in\mathcal{D}_1$. Indeed, we have
\begin{IEEEeqnarray}{lCl}
\bar{U}^{\ast}(\mathsf{D},T, \rho) & = & \mathsf{E}\Biggl[\Biggl( \sum_{i=1}^{n_t} \left(   \frac{n_t d_{i,\ell}^2 }{T\rho} -1 \right)(Z'_{i, \ell}-n_r) - \sum_{i=1}^{n_t} \left(Z''_{i,\ell} - (T-n_t)\right) \nonumber\\
& & \qquad {} + (T-n_t) \log \det \left( \mathbf{H}_{\ell} \mathbf{H}_{\ell}^\mathsf{H}  \right) - (T-n_t) \mathsf{E}\left[\log \det \left( \mathbf{H}_{\ell} \mathbf{H}_{\ell}^\mathsf{H}  \right)\right]\Biggr)^2\Biggr] \nonumber\\
& = & \mathsf{E}\left[\left(\sum_{i=1}^{n_t} \left(Z''_{i,\ell} - (T-n_t)\right)\right)^2\right] \nonumber\\
& & {} + \mathsf{E}\left[\left( \sum_{i=1}^{n_t} \left(   \frac{n_t d_{i,\ell}^2 }{T\rho} -1 \right)(Z'_{i, \ell}-n_r) + (T-n_t) \log \det \left( \mathbf{H}_{\ell} \mathbf{H}_{\ell}^\mathsf{H}  \right) - (T-n_t) \mathsf{E}\left[\log \det \left( \mathbf{H}_{\ell} \mathbf{H}_{\ell}^\mathsf{H}  \right)\right]\right)^2\right] \nonumber\\
& \geq & \mathsf{E}\left[\left(\sum_{t=1}^{n_t} \left(Z''_{t,\ell} - (T-n_t)\right)\right)^2\right] \nonumber\\
& = & (T-n_t)n_t \label{L5_Uast_lowera}
\end{IEEEeqnarray}
where the second step follows because $\{Z_{1,1}'',\ldots, Z_{n_t,L}''\}$ and $\{\mathbf{H}^L,Z_{1,1}',\ldots, Z_{n_t,L}'\}$ are independent, and the last step follows because $\{Z_{1,1}'',\ldots, Z_{n_t,L}''\}$ is a sequence of i.i.d.\ random variables with variance $(T-n_t)$.

Combining \eqref{L5_Uast_lowera} with \eqref{L4_UU_ast}, we obtain that
\begin{equation}
\bar{U}(\mathsf{D},T,\rho) \geq (T-n_t) n_t + K_{\bar{U}}(\mathsf{D},T,\rho) 
\end{equation}
which proves Lemma~\ref{MC_Lemma5}. 

\section{Proof of Lemma \ref{MC_Lemma6}} \label{Appx_Pf_MC_Lem6} 
	Recall that $\bar{U}(\mathsf{D},T,\rho ) = \mathsf{E}\left[\Delta^2(\mathsf{D},\mathbf{Z}_{\ell}',\mathbf{Z}_{\ell}'',\mathbf{H}_{\ell},\mathbf{Q}_{\ell})\right]$, where $\Delta(\mathsf{D},\mathbf{Z}_{\ell}',\mathbf{Z}_{\ell}'',\mathbf{H}_{\ell},\mathbf{Q}_{\ell})$ was defined in \eqref{eq:def_Delta}. Using the Cauchy-Schwarz inequality, we can thus upper-bound $\bar{U}\left(T,\sqrt{\frac{\rho}{\alpha}}\mathsf{D},\rho\right)-\bar{U}(T,\mathsf{D},\rho)$ as
	\begin{IEEEeqnarray}{lCl}
	\IEEEeqnarraymulticol{3}{l}{\bar{U}\left(T,\sqrt{\frac{\rho}{\alpha}}\mathsf{D},\rho\right)-\bar{U}(T,\mathsf{D},\rho)} \nonumber\\
	\quad & = &  \mathsf{E} \left[ \left(\Delta\left(\sqrt{\frac{\rho}{\alpha}}\mathsf{D},\mathbf{Z}_{\ell}',\mathbf{Z}_{\ell}'',\mathbf{H}_{\ell},\mathbf{Q}_{\ell}\right)-\Delta(\mathsf{D},\mathbf{Z}_{\ell}',\mathbf{Z}_{\ell}'',\mathbf{H}_{\ell},\mathbf{Q}_{\ell})\right) \left(\Delta\left(\sqrt{\frac{\rho}{\alpha}}\mathsf{D},\mathbf{Z}_{\ell}',\mathbf{Z}_{\ell}'',\mathbf{H}_{\ell},\mathbf{Q}_{\ell}\right)+\Delta(\mathsf{D},\mathbf{Z}_{\ell}',\mathbf{Z}_{\ell}'',\mathbf{H}_{\ell},\mathbf{Q}_{\ell})\right)\right] \nonumber\\
	 & \leq & \sqrt{\mathsf{E} \left[ \left(\Delta\left(\sqrt{\frac{\rho}{\alpha}}\mathsf{D},\mathbf{Z}_{\ell}',\mathbf{Z}_{\ell}'',\mathbf{H}_{\ell},\mathbf{Q}_{\ell}\right)-\Delta(\mathsf{D},\mathbf{Z}_{\ell}',\mathbf{Z}_{\ell}'',\mathbf{H}_{\ell},\mathbf{Q}_{\ell})\right)^2\right]} \times \nonumber\\
	& & {} \times \sqrt{\mathsf{E} \left[ \left(\Delta\left(\sqrt{\frac{\rho}{\alpha}}\mathsf{D},\mathbf{Z}_{\ell}',\mathbf{Z}_{\ell}'',\mathbf{H}_{\ell},\mathbf{Q}_{\ell}\right)+\Delta(\mathsf{D},\mathbf{Z}_{\ell}',\mathbf{Z}_{\ell}'',\mathbf{H}_{\ell},\mathbf{Q}_{\ell})\right)^2\right]}. \label{Equ_MC_Omega}
		\end{IEEEeqnarray}
	We next note that   
	\begin{IEEEeqnarray}{lCl}
	\IEEEeqnarraymulticol{3}{l}{\sup_{\rho \geq \rho_0} \sup_{\mathsf{D}\in\mathcal{D}_{n_t}} \mathsf{E} \left[ \left(\Delta\left(\sqrt{\frac{\rho}{\alpha}}\mathsf{D},\mathbf{Z}_{\ell}',\mathbf{Z}_{\ell}'',\mathbf{H}_{\ell},\mathbf{Q}_{\ell}\right)+\Delta(\mathsf{D},\mathbf{Z}_{\ell}',\mathbf{Z}_{\ell}'',\mathbf{H}_{\ell},\mathbf{Q}_{\ell})\right)^2\right]} \nonumber\\
	\quad
	& \leq & 2\sup_{\rho \geq \rho_0}\sup_{\mathsf{D}\in\mathcal{D}_{n_t}}\bar{U}\left(\sqrt{\frac{\rho}{\alpha}}\mathsf{D},T,\rho\right)  
		+2\sup_{\rho \geq \rho_0}\sup_{\mathsf{D}\in\mathcal{D}_{n_t}}\bar{U}\left(\mathsf{D},T,\rho\right) \nonumber \\
	& \leq & 4\bar{U}\left(T\right)   \label{MC_Equ_V ower1}
	\end{IEEEeqnarray}
	for some constant $\bar{U}(T)$ that only depends on $T$ and any $\rho_0 \geq 0$. Here, the first inequality follows because $(a_1+a_2)^2\leq 2a_1^2+2a_2^2$ and the second inequality follows from Lemma \ref{MC_Lemma4}. We further have that
	\begin{IEEEeqnarray*}{lCl}
		\IEEEeqnarraymulticol{3}{l}{\mathsf{E} \left[ \left(\Delta\left(\sqrt{\frac{\rho}{\alpha}}\mathsf{D},\mathbf{Z}_{\ell}',\mathbf{Z}_{\ell}'',\mathbf{H}_{\ell},\mathbf{Q}_{\ell}\right)-\Delta(\mathsf{D},\mathbf{Z}_{\ell}',\mathbf{Z}_{\ell}'',\mathbf{H}_{\ell},\mathbf{Q}_{\ell}) \right)^2\right]} \nonumber\\
\quad & = &\mathsf{E} \Biggl[ \Biggl( 
		\sum_{i=1}^{n_t} \left(   \frac{\rho }{\alpha} -1 \right)\frac{n_t d_{i,\ell}^2 }{T\rho}(Z'_{i, \ell}-n_r)
		+(T-n_t)\log  \left(\frac{\det \left( \mathbf{H}_{\ell} \mathbf{H}_{\ell}^\mathsf{H}(\mathsf{I}_{n_t}+\frac{\rho}{\alpha}\mathsf{D}^2)
			+  \lambda_1(\mathbf{Q}_{\ell}^\mathsf{H}\mathbf{Q}_{\ell})\mathsf{I}_{n_t} \right) }{\det \left( \mathbf{H}_{\ell} \mathbf{H}_{\ell}^\mathsf{H}(\mathsf{I}_{n_t}+\mathsf{D}^2)
			+  \lambda_1(\mathbf{Q}_{\ell}^\mathsf{H}\mathbf{Q}_{\ell})\mathsf{I}_{n_t} \right)}\right) \nonumber \\
		& & \quad {} -(T-n_t)\mathsf{E} \left[ \log\left(\frac{\det \left( \mathbf{H}_{\ell} \mathbf{H}_{\ell}^\mathsf{H}(\mathsf{I}_{n_t}+\frac{\rho}{\alpha}\mathsf{D}^2)
			+  \lambda_1(\mathbf{Q}_{\ell}^\mathsf{H}\mathbf{Q}_{\ell})\mathsf{I}_{n_t} \right) }{\det \left( \mathbf{H}_{\ell} \mathbf{H}_{\ell}^\mathsf{H}(\mathsf{I}_{n_t}+\mathsf{D}^2)
			+  \lambda_1(\mathbf{Q}_{\ell}^\mathsf{H}\mathbf{Q}_{\ell})\mathsf{I}_{n_t} \right)}\right)\right] \Biggr)^2\Biggr]. \nonumber 
		\end{IEEEeqnarray*} 
	Consequently, using the inequalities $\mathsf{E}\left[\left(X-\mathsf{E}[X]\right)^2\right]\leq \mathsf{E}[X^2]$ and $(a_1+a_2)^2\leq 2a_1^2+2a_2^2$, we obtain
	\begin{IEEEeqnarray}{lCl}
		\IEEEeqnarraymulticol{3}{l}{\mathsf{E} \left[ \left(\Delta\left(\sqrt{\frac{\rho}{\alpha}}\mathsf{D},\mathbf{Z}_{\ell}',\mathbf{Z}_{\ell}'',\mathbf{H}_{\ell},\mathbf{Q}_{\ell}\right)-\Delta(\mathsf{D},\mathbf{Z}_{\ell}',\mathbf{Z}_{\ell}'',\mathbf{H}_{\ell},\mathbf{Q}_{\ell}) \right)^2\right]} \nonumber\\ 
		\quad & \leq &2 \left( \frac{\rho }{\alpha} -1\right)^2\mathsf{E} \left[ \left(\sum_{i=1}^{n_t}\frac{n_t d_{i,\ell}^2 }{T\rho}(Z'_{i, \ell}-n_r)\right)^2 \right]   \nonumber\\
 && {}		+2\mathsf{E} \left[(T-n_t)^2\log^2 \left(\frac{\det \left( \mathbf{H}_{\ell} \mathbf{H}_{\ell}^\mathsf{H}(\mathsf{I}_{n_t}+\frac{\rho}{\alpha}\mathsf{D}^2)
			+  \lambda_1(\mathbf{Q}_{\ell}^\mathsf{H}\mathbf{Q}_{\ell})\mathsf{I}_{n_t} \right)}{\det \left( \mathbf{H}_{\ell} \mathbf{H}_{\ell}^\mathsf{H}(\mathsf{I}_{n_t}+\mathsf{D}^2)
			+  \lambda_1(\mathbf{Q}_{\ell}^\mathsf{H}\mathbf{Q}_{\ell})\mathsf{I}_{n_t} \right)}\right) \right] \nonumber\\ 
		& \leq & 2 \left( \frac{\rho }{\alpha} -1\right)^2 n_t^3 n_r
		+ 2(T-n_t)^2 \mathsf{E} \left[ \log^2 \left(\frac{\det \left( \mathbf{H}_{\ell} \mathbf{H}_{\ell}^\mathsf{H}(\mathsf{I}_{n_t}+\frac{\rho}{\alpha}\mathsf{D}^2)
			+  \lambda_1(\mathbf{Q}_{\ell}^\mathsf{H}\mathbf{Q}_{\ell})\mathsf{I}_{n_t} \right) }{\det \left( \mathbf{H}_{\ell} \mathbf{H}_{\ell}^\mathsf{H}(\mathsf{I}_{n_t}+\mathsf{D}^2)
			+  \lambda_1(\mathbf{Q}_{\ell}^\mathsf{H}\mathbf{Q}_{\ell})\mathsf{I}_{n_t} \right)}\right)\right] \nonumber\\
		& \leq & 2n_t^2 \left( \frac{\rho }{\alpha} -1\right)^2 \left(n_tn_r+(T-n_t)^2\right) \label{MC_Equ_V ower2}
		\end{IEEEeqnarray} 
	where the second step follows by bounding $d_{i,\ell}^2 \leq T\rho$ and because $Z_{1,\ell}',\ldots,Z_{n_t,\ell}'$ are i.i.d.\ with variance $n_r$; and the third step follows by upper-bounding the second term as shown below (see \eqref{MC_Lem3_C1T2_first}--\eqref{MC_Lem3_C1T2_last}).
	
	Combining \eqref{Equ_MC_Omega}--\eqref{MC_Equ_V ower2}, and using that, by the lemma's assumptions, we have $T\alpha \geq T\rho(1-\delta)$ and $0<\delta<1/2$, so $(\rho/\alpha-1)^2 \leq \delta^2/(1-\delta)^2\leq 4\delta^2$, we obtain that $\bar{U}\left(\sqrt{\frac{\rho}{\alpha}}\mathsf{D},T,\rho\right)-\bar{U}(\mathsf{D},T,\rho)$  is upper-bounded by
	\begin{IEEEeqnarray}{lCl}
	\bar{U}\left(\sqrt{\frac{\rho}{\alpha}}\mathsf{D},T,\rho\right)-\bar{U}(\mathsf{D},T,\rho)
	& \leq & \sqrt{32\bar{U}(T)n_t^2 \delta^2\left(n_tn_r+(T-n_t)^2\right) } \nonumber\\
	& = & \delta\Upsilon(T)
	\end{IEEEeqnarray} 
	where $\Upsilon(T)\triangleq 4n_t\sqrt{2 \bar{U}(T)(n_t n_r + (T- n_t)^2)}$.
	
	It remains to prove the last step in \eqref{MC_Equ_V ower2}. To this end, we first note that, since $\frac{\rho}{\alpha}\geq 1$, 
	\begin{equation}
	\frac{\det \left( \mathbf{H}_{\ell} \mathbf{H}_{\ell}^\mathsf{H}(\mathsf{I}_{n_t}+\frac{\rho}{\alpha}\mathsf{D}^2)
		+ \lambda_1(\mathbf{Q}_{\ell}^\mathsf{H}\mathbf{Q}_{\ell})\mathsf{I}_{n_t} \right) }{\det \left( \mathbf{H}_{\ell} \mathbf{H}_{\ell}^\mathsf{H}(\mathsf{I}_{n_t}+\mathsf{D}^2)
		+ \lambda_1(\mathbf{Q}_{\ell}^\mathsf{H}\mathbf{Q}_{\ell})\mathsf{I}_{n_t} \right)}
	\geq 1. \label{MC_Lem3_C1T2_first}
	\end{equation}
	An upper bound on the expected value in the second-to-last line in \eqref{MC_Equ_V ower2} thus follows by upper-bounding the logarithm inside the expected value. We have
	\begin{IEEEeqnarray}{lCl} 
		\IEEEeqnarraymulticol{3}{l}{\mathsf{E} \left[ \log^2 \left(\frac{\det \left( \mathbf{H}_{\ell} \mathbf{H}_{\ell}^\mathsf{H}(\mathsf{I}_{n_t}+\frac{\rho}{\alpha}\mathsf{D}^2)
			+ \lambda_1(\mathbf{Q}_{\ell}^\mathsf{H}\mathbf{Q}_{\ell})\mathsf{I}_{n_t} \right) }{\det \left( \mathbf{H}_{\ell} \mathbf{H}_{\ell}^\mathsf{H}(\mathsf{I}_{n_t}+\mathsf{D}^2)
			+ \lambda_1(\mathbf{Q}_{\ell}^\mathsf{H}\mathbf{Q}_{\ell})\mathsf{I}_{n_t} \right)}\right)\right]} \nonumber\\
		\qquad & =& \mathsf{E} \left[ \log^2\det \left( \mathsf{I}_{n_t}
			+ \left( (\mathsf{I}_{n_t}+\mathsf{D}^2)
			+ \lambda_1(\mathbf{Q}_{\ell}^\mathsf{H}\mathbf{Q}_{\ell})(\mathbf{H}_{\ell} \mathbf{H}_{\ell}^\mathsf{H})^{-1} \right)^{-1}\left(\frac{\rho}{\alpha}-1\right)\mathsf{D}^2 \right)\right] \label{eq:Hey_Ho}
\end{IEEEeqnarray}
which follows by writing $\frac{\rho}{\alpha}\mathsf{D}^2 = \mathsf{D}^2 + \left(\frac{\rho}{\alpha}-1\right) \mathsf{D}^2$ and by multiplying the matrices inside the determinants in the numerator and denominator by $\left( (\mathsf{I}_{n_t}+\mathsf{D}^2)
			+ \lambda_1(\mathbf{Q}_{\ell}^\mathsf{H}\mathbf{Q}_{\ell})(\mathbf{H}_{\ell} \mathbf{H}_{\ell}^\mathsf{H})^{-1} \right)^{-1} (\mathbf{H}_{\ell} \mathbf{H}_{\ell}^\mathsf{H})^{-1}$.\footnote{The matrix $\mathbf{H}_{\ell} \mathbf{H}_{\ell}^\mathsf{H}$ has a complex Wishart distribution so, for $n_r\geq n_t$, it is invertible with probability one \cite[Th. 3.1.4]{muirhead2005aspects}.}
			
Using that $\log\det(\mathsf{A})\leq \text{tr}(\mathsf{A}-\mathsf{I}_{n_t})$ for every $n_t\times n_t$ positive-definite matrix $\mathsf{A}$, this can be further upper-bounded as
\begin{IEEEeqnarray}{lCl}
	\IEEEeqnarraymulticol{3}{l}{\mathsf{E} \left[ \log^2\det \left( \mathsf{I}_{n_t}
			+ \left( (\mathsf{I}_{n_t}+\mathsf{D}^2)
			+ \lambda_1(\mathbf{Q}_{\ell}^\mathsf{H}\mathbf{Q}_{\ell})(\mathbf{H}_{\ell} \mathbf{H}_{\ell}^\mathsf{H})^{-1} \right)^{-1}\left(\frac{\rho}{\alpha}-1\right)\mathsf{D}^2 \right)\right]} \nonumber\\	
			\quad & \leq &\mathsf{E} \left[  \left(\text{tr} \left(  \left( (\mathsf{I}_{n_t}+\mathsf{D}^2)
			+ \lambda_1(\mathbf{Q}_{\ell}^\mathsf{H}\mathbf{Q}_{\ell})(\mathbf{H}_{\ell} \mathbf{H}_{\ell}^\mathsf{H})^{-1} \right)^{-1}\left(\frac{\rho}{\alpha}-1\right)\mathsf{D}^2 \right)\right)^2\right]  \nonumber\\
		 & \leq & \mathsf{E} \left[  \left(\sum_{k=1}^{n_t} \lambda_k\left( \left( (\mathsf{I}_{n_t}+\mathsf{D}^2)
			+ \lambda_1(\mathbf{Q}_{\ell}^\mathsf{H}\mathbf{Q}_{\ell})(\mathbf{H}_{\ell} \mathbf{H}_{\ell}^\mathsf{H})^{-1} \right)^{-1}\right) 
		\lambda_k\left(\left(\frac{\rho}{\alpha}-1\right)\mathsf{D}^2 \right)   \right)^2 \right] \nonumber \\
		& \leq & \left(\sum_{k=1}^{n_t}  \frac{1}{\lambda_k\left(\mathsf{I}_{n_t}+\mathsf{D}^2\right) } 
		\lambda_k\left(\left(\frac{\rho}{\alpha}-1\right)\mathsf{D}^2 \right)   \right)^2  \nonumber\\
		& =& \left(\frac{\rho}{\alpha}-1\right)^2\left(\sum_{k=1}^{n_t}\frac{d_k^2}{1+d_k^2}\right)^2  \nonumber\\
		& \leq & \left(\frac{\rho}{\alpha}-1\right)^2 n_t^2 \label{MC_Lem3_C1T2}
		\end{IEEEeqnarray}
		where $\lambda_k(\mathsf{A})$ denotes the $k$-th largest eigenvalue of the matrix $\mathsf{A}$. Here, the second step is by von Neumann's trace inequality \cite[Th.~8.7.6]{horn1990matrix}; the third step follows because, by Weyl's theorem \cite[Th.~4.3.1]{horn1990matrix},
		\begin{equation}
		\lambda_k\left( \left( (\mathsf{I}_{n_t}+\mathsf{D}^2)
		+ \lambda_1(\mathbf{Q}_{\ell}^\mathsf{H}\mathbf{Q}_{\ell})(\mathbf{H}_{\ell} \mathbf{H}_{\ell}^\mathsf{H})^{-1} \right)^{-1}\right) 
		= \frac{1}{ \lambda_k \left( (\mathsf{I}_{n_t}+\mathsf{D}^2)
			+ \lambda_1(\mathbf{Q}_{\ell}^\mathsf{H}\mathbf{Q}_{\ell})(\mathbf{H}_{\ell} \mathbf{H}_{\ell}^\mathsf{H})^{-1} \right)  }
		\leq \frac{1}{ \lambda_k \left(\mathsf{I}_{n_t}+\mathsf{D}^2\right) }. \label{MC_Lem3_C1T2_last}
		\end{equation}
		Combining \eqref{eq:Hey_Ho} and \eqref{MC_Lem3_C1T2} yields \eqref{MC_Equ_V ower2} and concludes the proof of Lemma~\ref{MC_Lemma6}.

\section{Proof of Lemma \ref{MC_Lemma7b}} \label{Appx_Pf_MC_Lem7b} 
We begin by noting that, for every $\mathsf{D}\in\mathcal{D}_{n_t}$,
\begin{equation}
	\bar{J}(\mathsf{D},T,\rho)
	= \bar{J}^{\ast}(\mathsf{D},T,\rho)+ K_{\mathcal{D}_1}(\mathsf{D},T,\rho) \label{Def_J_rhoHq}
	\end{equation}
where  $\bar{J}^{\ast}(\mathsf{D},T,\rho)$ is defined in \eqref{MC_Ap3_Def_Jast}
	and
	\begin{equation}
	K_{\mathcal{D}_1}(\mathsf{D},T) \triangleq \frac{n_t}{T\rho}\left(n_tn_r+n_t(T-n_t)\right) + (T-n_t)\mathsf{E}\left[ \log \frac{ \det \Big( \mathbf{H}_{\ell} \mathbf{H}_{\ell}^\mathsf{H}
		+ \lambda_1(\mathbf{Q}_{\ell}^\mathsf{H}\mathbf{Q}_{\ell})\big(\mathsf{I}_{n_t}+  \mathsf{D}^2\big)^{-1}  \Big)}{\det \Big( \mathbf{H}_{\ell} \mathbf{H}_{\ell}^\mathsf{H} \Big) } \right].\label{MC_Ap3_Def_K}
		\end{equation}
		Since $\det(\mathsf{A} + \mathsf{B}) \geq \det(\mathsf{A})$ for every pair of positive definite matrices $\mathsf{A}$ and $\mathsf{B}$ \cite[Th.~7.8.21]{horn1990matrix}, it follows that $K_{\mathcal{D}_1}(\mathsf{D},T) \geq 0$. Furthermore, for $\mathsf{D} \in \mathcal{D}_1$, we have $d_i^2 \geq \bar{\delta}\rho$, $i=1,\ldots,n_t$, which in turn implies that $\big(\mathsf{I}_{n_t}+  \mathbf{D}^2\big)^{-1} \leq (1+\bar{\delta}\rho)^{-1} \mathsf{I}_{n_t}$. Consequently,
  \begin{IEEEeqnarray}{lCl}
			K_{\mathcal{D}_1}(\mathsf{D},T) 
			& \leq & \frac{n_t}{T\rho}\left(n_tn_r+n_t(T-n_t)\right) + (T-n_t)\mathsf{E}\left[ \log \frac{ \det \left( \mathbf{H}_{\ell} \mathbf{H}_{\ell}^\mathsf{H}
				+ \lambda_1(\mathbf{Q}_{\ell}^\mathsf{H}\mathbf{Q}_{\ell})\bigl(1+ \bar{\delta}\rho \bigr)^{-1} \mathsf{I}_{n_t} \right)}{\det \left( \mathbf{H}_{\ell} \mathbf{H}_{\ell}^\mathsf{H} \right) } \right] \nonumber\\
			& \triangleq & K_{\mathcal{D}_1}(T,\rho), \quad \mathsf{D} \in \mathcal{D}_1.\label{Equ_MC_supD1_diff}
			\end{IEEEeqnarray}
			We further have that
			\begin{equation} 
	\left| \log \det \left( \mathbf{H}_{\ell} \mathbf{H}_{\ell}^\mathsf{H}
				+ \lambda_1(\mathbf{Q}_{\ell}^\mathsf{H}\mathbf{Q}_{\ell})\bigl(1+ \bar{\delta}\rho \bigr)^{-1} \mathsf{I}_{n_t} \right)\right|
	\leq \left|\log\det\left(\mathbf{H}_{\ell} \mathbf{H}_{\ell}^\mathsf{H}\right)\right|+\log\left(1+\det\left(\mathbf{H}_{\ell} \mathbf{H}_{\ell}^\mathsf{H}+\lambda_1(\mathbf{Q}_{\ell}^\mathsf{H}\mathbf{Q}_{\ell}) \mathsf{I}_{n_t}  \right)\right) 
\end{equation} 
whose expected value is finite (cf.~\eqref{MC_Equ_log_b1n}--\eqref{MC_L4_T3b}).\footnote{Equations \eqref{MC_L4_T3a}--\eqref{MC_L4_T3b} demonstrate that the expected values of the squared logarithms are finite. This implies that the expected values of the absolute values of the logarithms are finite, too, since by Jensen's inequality $\mathsf{E}[|X|]\leq \sqrt{\mathsf{E}[X^2]}$ for any random variable $X$.} By the continuity of the logarithm function and the determinant, and by the dominated convergence theorem, we thus have that
	\begin{equation}
		\lim\limits_{\rho\to\infty} \sup_{\mathbf{D}\in {\mathcal{D}}_1}   \mathsf{E}\left[\log \det \left( \mathbf{H}_{\ell} \mathbf{H}_{\ell}^\mathsf{H}
				+ \lambda_1(\mathbf{Q}_{\ell}^\mathsf{H}\mathbf{Q}_{\ell})\bigl(1+ \bar{\delta}\rho \bigr)^{-1} \mathsf{I}_{n_t} \right)\right]  
		= \mathsf{E}\left[\log \det \left( \mathbf{H}_{\ell} \mathbf{H}_{\ell}^\mathsf{H}\right)\right].  \label{Equ_CL2_logUL}
	\end{equation} 
Combining \eqref{Equ_CL2_logUL} with \eqref{Equ_MC_supD1_diff} yields then that
\begin{equation}
\label{eq:SUI_sucks}
\bar{J}(\mathsf{D},T,\rho)
	\leq \bar{J}^{\ast}(\mathsf{D},T,\rho) + K_{\mathcal{D}_1}(T,\rho)
\end{equation}
for a nonnegative constant $K_{\mathcal{D}_1}(T,\rho)$ that only depends on $T$ and $\rho$ and that satisfies $\lim_{\rho\to \infty} K_{\mathcal{D}_1}(T,\rho)= 0$. Since $K_{\mathcal{D}_1}(T,\rho)$ does not depend on $\mathsf{D}\in\mathcal{D}_1$, this gives
\begin{IEEEeqnarray}{lCl}
\sup_{\mathsf{D}\in\mathcal{D}_1} \left\{\bar{J}(\mathsf{D},T,\rho) 
	-\sqrt{\frac{\bar{U}(\mathsf{D},T,\rho)}{L}}Q^{-1}(\epsilon)\right\} & \leq & \sup_{\mathsf{D}\in\mathcal{D}_1} \left\{\bar{J}^{\ast}(\mathsf{D},T,\rho)
	-\sqrt{\frac{\bar{U}(\mathsf{D},T,\rho)}{L}}Q^{-1}(\epsilon)\right\} + K_{\mathcal{D}_1}(T,\rho). \label{eq:app_lem7b_THIS}\IEEEeqnarraynumspace
\end{IEEEeqnarray}

We next show that, for $\rho \geq \rho_0$, $L\geq L_0$ and sufficiently large $\rho_0$ and $L_0$, the supremum over $\mathsf{D}\in\mathcal{D}_1$ can be replaced by a supremum over all matrices $\mathsf{D} \in \mathcal{D}_{n_t}$ that satisfy $\textnormal{tr}(\mathsf{D}^2) \geq T\rho\left(1-K(T)/L\right)$. To this end, we show first that, for $\rho\geq\rho^{(1)}$ and a sufficiently large $\rho^{(1)}$, we can assume without loss of optimality that
\begin{equation}
\textnormal{tr}(\mathsf{D}^2) \geq T\rho\left(1 - K^{(1)}(T,\rho^{(1)})/\sqrt{L}\right)
\end{equation}
for a constant $K^{(1)}(T,\rho^{(1)})$ that depends on $T$ and $\rho^{(1)}$. We then gradually improve this bound on the trace until we obtain the desired result.

\subsection{Lower Bound on $\textnormal{tr}(\mathsf{D}^2)$: First Round}

Recall that $\alpha = \textnormal{tr}(\mathsf{D}^2)/T$, and define $\bar{\mathsf{D}} \triangleq \mathsf{D}/\sqrt{\alpha}$. Clearly, $\textnormal{tr}(\bar{\mathsf{D}}^2) = T$ and, for $\mathsf{D}\in\mathcal{D}_1$, $n_t \bar{\delta} \rho/T \leq \alpha \leq \rho$. We then compare
\begin{equation}
\label{eq:app_lem7b_rhoD}
\bar{J}^{\ast}(\sqrt{\rho}\bar{\mathsf{D}},T,\rho) 
	-\sqrt{\frac{\bar{U}(\sqrt{\rho}\bar{\mathsf{D}},T,\rho)}{L}}Q^{-1}(\epsilon)
\end{equation}
with
\begin{equation}
\label{eq:app_lem7b_alphaD}
\bar{J}^{\ast}(\sqrt{\alpha}\bar{\mathsf{D}},T,\rho) 
	-\sqrt{\frac{\bar{U}(\sqrt{\alpha}\bar{\mathsf{D}},T,\rho)}{L}}Q^{-1}(\epsilon).
\end{equation}
Intuitively, every $\alpha$ for which \eqref{eq:app_lem7b_rhoD} is larger than \eqref{eq:app_lem7b_alphaD} is suboptimal and can be discarded without loss of optimality, since multiplying the matrix $\mathsf{D}$ by $\sqrt{\rho/\alpha}$ would increase the difference
\begin{equation}
\bar{J}^{\ast}(\mathsf{D},T,\rho) 
	-\sqrt{\frac{\bar{U}(\mathsf{D},T,\rho)}{L}}Q^{-1}(\epsilon).
\end{equation}

To formalize this argument, we define the function
\begin{equation}
f_{L,\rho}(\alpha,\bar{\mathsf{D}}) \triangleq \bar{J}^{\ast}(\sqrt{\rho}\bar{\mathsf{D}},T,\rho) 
	-\sqrt{\frac{\bar{U}(\sqrt{\rho}\bar{\mathsf{D}},T,\rho)}{L}}Q^{-1}(\epsilon) - \bar{J}^{\ast}(\sqrt{\alpha}\bar{\mathsf{D}},T,\rho) 
	 + \sqrt{\frac{\bar{U}(\sqrt{\alpha}\bar{\mathsf{D}},T,\rho)}{L}}Q^{-1}(\epsilon)
\end{equation}
and analyze for which values of $n_t \bar{\delta}\rho/T \leq \alpha \leq \rho$ it is nonnegative. Indeed, Lemma~\ref{MC_Lemma4} implies that there exists an $\bar{U}(T)$ that depends on $T$ and an arbitrary $\rho_0>0$ such that
\begin{equation}
0 \leq \bar{U}(\mathsf{D}, T,\rho) \leq   \bar{U}(T), \quad \mathsf{D} \in \mathcal{D}_{n_t},\,\rho\geq \rho_0.
\end{equation}
Consequently, when $T=n_t+n_r$, the function $f_{L,\rho}(\cdot)$ can be lower-bounded as
\begin{IEEEeqnarray}{lCl}
f_{L,\rho}(\alpha,\bar{\mathsf{D}}) & = &  
\frac{T\rho-T\alpha}{T\rho} n_t n_t
 - \left(\sqrt{\frac{\bar{U}(\sqrt{\rho}\bar{\mathsf{D}},T,\rho)}{L}} - \sqrt{\frac{\bar{U}(\sqrt{\alpha}\bar{\mathsf{D}},T,\rho)}{L}}\right) Q^{-1}(\epsilon) \nonumber\\
& \geq &   \frac{T\rho-T\alpha}{T\rho} n_t n_t- \sqrt{\frac{ \bar{U}(T)}{L}}Q^{-1}(\epsilon)
\end{IEEEeqnarray}
which is positive if
\begin{equation}
\alpha < \rho \left(1 - \frac{\sqrt{ \bar{U}(T)}\frac{Q^{-1}(\epsilon)}{n_t n_r}}{\sqrt{L}}\right).
\end{equation}
It follows that we can assume without loss of optimality that $\textnormal{tr}(\mathsf{D}^2) \geq T\rho\left(1 - K^{(1)}(T,\rho_0)/\sqrt{L}\right)$, where
\begin{equation}
K^{(1)}(T,\rho_0) = \sqrt{ \bar{U}(T)}\frac{Q^{-1}(\epsilon)}{n_t n_r}.
\end{equation}

Similarly, when $T>n_t + n_r$, the function $f_{L,\rho}(\cdot)$ can be lower-bounded as
\begin{IEEEeqnarray}{lCl}
f_{L,\rho}(\alpha,\bar{\mathsf{D}}) & = & \frac{T\rho-T\alpha}{T\rho} n_t n_t + (T- n_t - n_r) \log\frac{\det(\mathsf{I}_{n_t} + \rho\bar{\mathsf{D}}^2)}{\det(\mathsf{I}_{n_t} + \alpha \bar{\mathsf{D}}^2)} \nonumber\\
& & {} - \left(\sqrt{\frac{\bar{U}(\sqrt{\rho}\bar{\mathsf{D}},T,\rho)}{L}} - \sqrt{\frac{\bar{U}(\sqrt{\alpha}\bar{\mathsf{D}},T,\rho)}{L}}\right) Q^{-1}(\epsilon) \nonumber\\
& \geq & (T- n_t - n_r) \log\frac{\det(\mathsf{I}_{n_t} + \rho\bar{\mathsf{D}}^2)}{\det(\mathsf{I}_{n_t} + \alpha \bar{\mathsf{D}}^2)} - \sqrt{\frac{ \bar{U}(T)}{L}}Q^{-1}(\epsilon) \nonumber\\
& \triangleq & \underline{f}^{(1)}_{L,\rho}(\alpha,\bar{\mathsf{D}}).
\end{IEEEeqnarray}
Using that $\bar{\mathsf{D}}$ is a diagonal matrix, the first derivative of $\underline{f}^{(1)}(\cdot)$ with respect to $\alpha$ can be computed as
\begin{equation}
\label{eq:app_lem7b_deriv_f1}
\frac{\partial}{\partial \alpha} \underline{f}^{(1)}_{L,\rho}(\alpha,\bar{\mathsf{D}}) = - (T-n_t-n_r) \sum_{i=1}^{n_t} \frac{\bar{d}_i^2}{1+\alpha \bar{d}_i^2}
\end{equation}
where $\bar{d}_i$ denotes the $i$-th diagonal element of $\bar{\mathsf{D}}$. Furthermore, by definition, we have $\bar{d}_i = d_i/\sqrt{\alpha}$, so for every $\mathsf{D} \in\mathcal{D}_1$
\begin{equation}
\bar{d}_i^2 \geq \frac{\bar{\delta}\rho}{\alpha} \geq \bar{\delta}, \quad i=1,\ldots,n_t
\end{equation}
since, for such $\mathsf{D}$, we have that $d_i^2 \geq \bar{\delta}\rho$ and $n_t\bar{\delta}\rho/T \leq \alpha \leq \rho$. It follows that
\begin{equation}
\frac{\partial}{\partial \alpha} \underline{f}^{(1)}_{L,\rho}(\alpha,\bar{\mathsf{D}}) \leq -(T-n_t - n_r) n_t \frac{\bar{\delta}}{1+\alpha \bar{\delta}}
\end{equation}
which is strictly negative and bounded away from zero for every $n_t\bar{\delta}\rho/T \leq \alpha \leq \rho$. Thus, $\underline{f}^{(1)}_{L,\rho}(\cdot)$ is a strictly decreasing function of $n_t\bar{\delta}\rho/T \leq \alpha \leq \rho$.

We next note that
\begin{equation}
\label{eq:app_lem7b_f1_rho}
\underline{f}^{(1)}_{L,\rho}(\rho,\bar{\mathsf{D}}) = - \sqrt{\frac{ \bar{U}(T)}{L}}Q^{-1}(\epsilon)
\end{equation}
and
\begin{IEEEeqnarray}{lCl}
\underline{f}^{(1)}_{L,\rho}\left(\frac{n_t\bar{\delta}\rho}{T},\bar{\mathsf{D}}\right) & = & (T-n_t-n_r) \log\frac{\det(\mathsf{I}_{n_t} + \rho\bar{\mathsf{D}}^2)}{\det(\mathsf{I}_{n_t} + \frac{n_t\bar{\delta}\rho}{T} \bar{\mathsf{D}}^2)}  - \sqrt{\frac{ \bar{U}(T)}{L}}Q^{-1}(\epsilon) \nonumber\\
& \geq & (T-n_t - n_r) n_t \log\frac{1+\rho\bar{\delta}}{1+\frac{n_t \bar{\delta}^2}{T}\rho} - \sqrt{\frac{ \bar{U}(T)}{L}}Q^{-1}(\epsilon) \label{eq:app_lem7b_f1_LLB}
\end{IEEEeqnarray}
where the inequality follows because the first term is monotonically increasing in the diagonal elements of $\bar{\mathsf{D}}$ and $\bar{d}^2_i \geq \bar{\delta}$, $i=1,\ldots,n_t$. The first term on the RHS of \eqref{eq:app_lem7b_f1_LLB} depends on $\rho$ but not on $L$, and it converges to
\begin{equation}
\label{eq:app_lem7b_a}
(T-n_t - n_r) n_t \log\frac{T}{n_t \bar{\delta}}
\end{equation}
as $\rho\to\infty$. Since $\bar{\delta}$, as defined in \eqref{eq:delta_bar}, is strictly smaller than $T/n_t$, this expression is strictly positive. Similarly, the second term on the RHS of \eqref{eq:app_lem7b_f1_LLB} depends on $L$ but not on $\rho$. Thus, we can find a $\rho^{(1)}$ such that, for $\rho\geq\rho^{(1)}$ and $T>n_t+n_r$,
\begin{equation*}
 (T-n_t - n_r) n_t \log\frac{1+\rho\bar{\delta}}{1+\frac{n_t \bar{\delta}^2}{T}\rho} \geq \frac{1}{2} \log\frac{T}{n_t \bar{\delta}}.
\end{equation*}
This in turn implies that
\begin{equation}
\label{eq:app_lem7b_b}
\underline{f}^{(1)}_{L,\rho}\left(\frac{n_t\bar{\delta}\rho}{T},\bar{\mathsf{D}}\right) \geq \frac{1}{4} \log\frac{T}{n_t \bar{\delta}} > 0
\end{equation}
for $\rho \geq\rho^{(1)}$ and
\begin{equation}
L \geq L^{(1)} \triangleq 16  \bar{U}(T)\left(\frac{Q^{-1}(\epsilon)}{\log\frac{T}{n_t\bar{\delta}}}\right)^2.
\end{equation}

We conclude that, when $T>n_t+n_r$, $\rho\geq\rho^{(1)}$, and $L \geq L^{(1)}$, $\alpha\mapsto\underline{f}^{(1)}_{L,\rho}(\alpha,\bar{\mathsf{D}})$ is a strictly decreasing function that is positive at $\alpha=n_t\bar{\delta}\rho/T$ and negative at $\alpha=\rho$. We can thus find an $\alpha_0^{(1)}(\bar{\mathsf{D}},L,\rho)$ such that
\begin{subequations}
\begin{IEEEeqnarray}{lCll}
\underline{f}^{(1)}_{L,\rho}(\alpha,\bar{\mathsf{D}}) & > & 0, \quad & \alpha < \alpha_0^{(1)}(\bar{\mathsf{D}},L,\rho) \\
\underline{f}^{(1)}_{L,\rho}(\alpha,\bar{\mathsf{D}}) & \leq & 0, \quad & \alpha \geq \alpha_0^{(1)}(\bar{\mathsf{D}},L,\rho).
\end{IEEEeqnarray}
\end{subequations}
Defining
\begin{equation}
\delta_0^{(1)}(\bar{\mathsf{D}},L,\rho) \triangleq 1-\frac{\alpha_0^{(1)}(\bar{\mathsf{D}},L,\rho)}{\rho}
\end{equation}
this implies that we can assume without loss of optimality that $\alpha \geq \rho(1-\delta_0^{(1)}(\bar{\mathsf{D}},L,\rho))$. We next show that, for $\rho\geq \rho^{(1)}$ and $L \geq L^{(1)}$, we have that $\delta_0^{(1)}(\bar{\mathsf{D}},L,\rho) \leq K^{(1)}(T,\rho^{(1)})/\sqrt{L}$ for a constant $K^{(1)}(T,\rho_0)$ that depends on $T$ and $\rho^{(1)}$. It then follows that, for such $\rho$ and $L$, we can assume without loss of optimality that $\textnormal{tr}(\mathsf{D}^2) \geq T\rho\left(1-K^{(1)}(T,\rho_0)/\sqrt{L}\right)$.

To upper-bound $\delta_0^{(1)}(\bar{\mathsf{D}},L,\rho)$, we apply the mean value theorem \cite[Th.~5.10]{rudin-principles} to express $\underline{f}^{(1)}_{L,\rho}(\cdot)$ as
\begin{IEEEeqnarray}{lCl}
\underline{f}^{(1)}_{L,\rho}(\rho,\bar{\mathsf{D}}) & = & \underline{f}^{(1)}_{L,\rho}(\rho,\bar{\mathsf{D}}) - \underline{f}^{(1)}_{L,\rho}(\alpha_0^{(1)}(\bar{\mathsf{D}},L,\rho),\bar{\mathsf{D}}) \nonumber\\
& = & \int_{\alpha_0^{(1)}(\bar{\mathsf{D}},L,\rho)}^{\rho} \dot{\underline{f}}^{(1)}_{L,\rho}(\alpha,\bar{\mathsf{D}}) \mathrm{d} \alpha \nonumber\\
& = & \left(\rho - \alpha_0^{(1)}(\bar{\mathsf{D}},L,\rho)\right)  \dot{\underline{f}}^{(1)}_{L,\rho}(\tilde{\alpha},\bar{\mathsf{D}}) \nonumber\\
& = & \rho \delta_0^{(1)}(\bar{\mathsf{D}},L,\rho) \dot{\underline{f}}^{(1)}_{L,\rho}(\tilde{\alpha},\bar{\mathsf{D}}) \label{eq:app_lem7b_MVT}
\end{IEEEeqnarray}
for some $\tilde{\alpha}$ between $\alpha_0^{(1)}(\bar{\mathsf{D}},L,\rho)$ and $\rho$. Here, we denote by $ \dot{\underline{f}}^{(1)}_{L,\rho}(\cdot)$ the derivative of $\alpha\mapsto\underline{f}^{(1)}_{L,\rho}(\alpha,\bar{\mathsf{D}})$ with respect to $\alpha$. In \eqref{eq:app_lem7b_MVT}, the first step follows because, by definition, $\underline{f}^{(1)}_{L,\rho}(\alpha_0^{(1)}(\bar{\mathsf{D}},L,\rho),\bar{\mathsf{D}}) = 0$; the last step follows by the definition of $\delta_0^{(1)}(\bar{\mathsf{D}},L,\rho)$. Together with \eqref{eq:app_lem7b_deriv_f1}--\eqref{eq:app_lem7b_f1_rho}, this yields that, for every $\rho \geq \rho^{(1)}$,
\begin{IEEEeqnarray}{lCl}
\delta_0^{(1)}(\bar{\mathsf{D}},L,\rho) & = & \frac{\underline{f}^{(1)}_{L,\rho}(\rho,\bar{\mathsf{D}})}{\rho\dot{\underline{f}}^{(1)}_{L,\rho}(\tilde{\alpha},\bar{\mathsf{D}})} \nonumber\\
& = & \frac{\sqrt{\frac{ \bar{U}(T)}{L}}Q^{-1}(\epsilon)}{\rho(T-n_t-n_r) \sum_{i=1}^{n_t} \frac{\bar{d}_i^2}{1+\tilde{\alpha} \bar{d}_i^2}} \nonumber\\
& \leq & \left(1+\frac{1}{\bar{\delta}\rho}\right) \frac{\sqrt{\frac{ \bar{U}(T)}{L}}Q^{-1}(\epsilon)}{(T-n_t-n_r) n_t}  \nonumber\\
& \leq & \frac{K^{(1)}(T,\rho^{(1)})}{\sqrt{L}}  \label{eq:app_lem7b_R1_almost}
\end{IEEEeqnarray}
where
\begin{equation}
K^{(1)}(T,\rho^{(1)}) \triangleq \left(1+\frac{1}{\bar{\delta}\rho^{(1)}}\right)\frac{\sqrt{ \bar{U}(T)}Q^{-1}(\epsilon)}{(T-n_t-n_r) n_t}.
\end{equation}
In \eqref{eq:app_lem7b_R1_almost}, the third step follows by bounding $\bar{d}^2_i \geq \bar{\delta}$ and $\tilde{\alpha} \leq \rho$.

\subsection{Lower Bound on $\textnormal{tr}(\mathsf{D}^2)$: $k$-th Round}

Suppose that, for $\rho\geq\rho^{(k-1)}$ and $L \geq L^{(k-1)}$ (for some $\rho^{(k-1)}$ and $L^{(k-1)}$), we can assume without loss of optimality that
\begin{equation}
\textnormal{tr}(\mathsf{D}^2) \geq T\rho\left(1-\delta^{(k-1)}\right).
\end{equation}
We next show that, in this case, we can also assume without loss of optimality that
\begin{equation}
\label{eq:app_lem7b_LBk}
\textnormal{tr}(\mathsf{D}^2) \geq T\rho\left(1-\sqrt{\frac{K(T)\delta^{(k-1)}}{L}}\right)
\end{equation}
for $\rho\geq\rho_0$, $L\geq L_0$, and sufficiently large $\rho_0$ and $L_0$ that depend on $(T,\rho^{(1)},L^{(1)})$ but not on $(\rho,L,k)$, where $K(T)$ is a constant that depends on $T$. Indeed, using Lemma~\ref{MC_Lemma6}, we can lower-bound $\bar{U}(\sqrt{\alpha}\bar{\mathsf{D}},T,\rho)$ as
\begin{IEEEeqnarray}{lCl}
\bar{U}(\sqrt{\alpha}\bar{\mathsf{D}},T,\rho) \geq \bar{U}(\sqrt{\rho}\bar{\mathsf{D}},T,\rho)-\Upsilon(T) \delta^{(k-1)}
\end{IEEEeqnarray}
where $\Upsilon(T)$ is a constant that depends on $T$. Together with the bound $\sqrt{a -b} \geq \sqrt{a} - \sqrt{b}$, $a\geq b$, this yields that
\begin{equation}
\sqrt{\frac{\bar{U}(\sqrt{\alpha}\bar{\mathsf{D}},T,\rho)}{L}} \geq \sqrt{\frac{\bar{U}(\sqrt{\rho}\bar{\mathsf{D}},T,\rho)}{L}} - \sqrt{\frac{\Upsilon(T)\delta^{(k-1)}}{L}}. \label{eq:this_is_220}
\end{equation}
Note that the condition $\bar{U}(\sqrt{\rho}\bar{\mathsf{D}},T,\rho) \geq \Upsilon(T) \delta^{(k-1)}$, required to apply the inequality $\sqrt{a -b} \geq \sqrt{a} - \sqrt{b}$, is satisfied for sufficiently large $\rho$ and $L$. Indeed, by Lemma~\ref{MC_Lem_Vhigh}, we have that $\bar{U}(\sqrt{\rho}\bar{\mathsf{D}},T,\rho) \geq T^2\tilde{V}(T)/2$ for $\rho \geq \rho'$ and some sufficiently large $\rho'$, where $\tilde{V}(T)$ was defined in \eqref{Def_Vt}. Furthermore, $\delta^{(1)} = K^{(1)}(T,\rho^{(1)})/\sqrt{L}$ and, as we shall show in \eqref{eq:this_is_232}--\eqref{eq:this_is_235} below, we have $\delta^{(k)} \leq \delta^{(k-1)}$, $k=2,3,\ldots$ for $L\geq  L^{\circ}$ with $ L^{\circ}$ defined in \eqref{eq:app_lem7b_Lcirc}. We thus obtain that $\bar{U}(\sqrt{\rho}\bar{\mathsf{D}},T,\rho) \geq \Upsilon(T) \delta^{(k-1)}$, $k=2,3,\ldots$ when $\rho\geq\rho'$ and
\begin{equation}
L \geq L' \triangleq \max\left\{ L^{\circ},\left(\frac{2\Upsilon(T) K^{(1)}(T,\rho^{(1)})}{T^2\tilde{V}(T)}\right)^2\right\}. \label{eq:this_is_a}
\end{equation}
We next use \eqref{eq:this_is_220} to lower-bound $f_{L,\rho}(\cdot)$. Indeed, when $T=n_t  + n_r$, we have
\begin{IEEEeqnarray}{lCl}
f_{L,\rho}(\alpha,\bar{\mathsf{D}}) & = & \frac{T\rho-T\alpha}{T\rho} n_t n_t- \left(\sqrt{\frac{\bar{U}(\sqrt{\rho}\bar{\mathsf{D}},T,\rho)}{L}} - \sqrt{\frac{\bar{U}(\sqrt{\alpha}\bar{\mathsf{D}},T,\rho)}{L}}\right) Q^{-1}(\epsilon) \nonumber\\
& \geq & \frac{T\rho-T\alpha}{T\rho} n_t n_t- \sqrt{\frac{\Upsilon(T)\delta^{(k-1)}}{L}} Q^{-1}(\epsilon)
\end{IEEEeqnarray}
which is positive if
\begin{equation}
\alpha < \rho\left(1 - \sqrt{\frac{\left(\frac{Q^{-1}(\epsilon)}{n_t n_r}\right)^2 \Upsilon(T) \delta^{(k-1)}}{L}}\right).
\end{equation}
It follows that we can assume without loss of optimality that \eqref{eq:app_lem7b_LBk} holds for $L_0 = L'$ and
\begin{equation}
K(T) = \Upsilon(T)\left(\frac{Q^{-1}(\epsilon)}{n_t n_r}\right)^2.
\end{equation}

We next consider the case where $T>n_t+n_r$. In this case, we can lower-bound $f_{L,\rho}(\cdot)$ as
\begin{IEEEeqnarray}{lCl}
f_{L,\rho}(\alpha,\bar{\mathsf{D}}) & = &\frac{T\rho-T\alpha}{T\rho} n_t n_t+ (T- n_t - n_r) \log\frac{\det(\mathsf{I}_{n_t} + \rho\bar{\mathsf{D}}^2)}{\det(\mathsf{I}_{n_t} + \alpha \bar{\mathsf{D}}^2)} \nonumber\\
& & {} - \left(\sqrt{\frac{\bar{U}(\sqrt{\rho}\bar{\mathsf{D}},T,\rho)}{L}} - \sqrt{\frac{\bar{U}(\sqrt{\alpha}\bar{\mathsf{D}},T,\rho)}{L}}\right) Q^{-1}(\epsilon) \nonumber\\
& \geq & (T- n_t - n_r) \log\frac{\det(\mathsf{I}_{n_t} + \rho\bar{\mathsf{D}}^2)}{\det(\mathsf{I}_{n_t} + \alpha \bar{\mathsf{D}}^2)} - \sqrt{\frac{\Upsilon(T)\delta^{(k-1)}}{L}} Q^{-1}(\epsilon) \nonumber\\
& \triangleq & \underline{f}^{(k)}_{L,\rho}(\alpha,\bar{\mathsf{D}}).
\end{IEEEeqnarray}
Since $\underline{f}^{(k)}_{L,\rho}(\cdot)$ differs from $\underline{f}^{(1)}_{L,\rho}(\cdot)$ only in terms that are independent of $\alpha$, it follows that
\begin{equation}
\label{eq:app_lem7b_der_k}
\frac{\partial}{\partial \alpha}  \underline{f}^{(k)}_{L,\rho}(\alpha,\bar{\mathsf{D}}) = \dot{\underline{f}}^{(1)}_{L,\rho}(\alpha,\bar{\mathsf{D}}).
\end{equation}
Consequently, $\alpha\mapsto\underline{f}^{(k)}_{L,\rho}(\alpha)$ is a strictly decreasing function on $n_t\bar{\delta}\rho/T \leq \alpha \leq \rho$. Furthermore,
\begin{equation}
 \underline{f}^{(k)}_{L,\rho}(\rho,\bar{\mathsf{D}}) = - \sqrt{\frac{\Upsilon(T)\delta^{(k-1)}}{L}} Q^{-1}(\epsilon)
\end{equation}
and
\begin{IEEEeqnarray}{lCl}
\underline{f}^{(k)}_{L,\rho}\left(\frac{n_t\bar{\delta}\rho}{T},\bar{\mathsf{D}}\right) & = & (T-n_t-n_r) \log\frac{\det(\mathsf{I}_{n_t} + \rho\bar{\mathsf{D}}^2)}{\det(\mathsf{I}_{n_t} + \frac{n_t\bar{\delta}\rho}{T} \bar{\mathsf{D}}^2)}  - \sqrt{\frac{\Upsilon(T)\delta^{(k-1)}}{L}} Q^{-1}(\epsilon) \nonumber\\
& \geq & (T-n_t - n_r) n_t \log\frac{1+\rho\bar{\delta}}{1+\frac{n_t \bar{\delta}^2}{T}\rho} - \sqrt{\frac{\Upsilon(T)\delta^{(k-1)}}{L}} Q^{-1}(\epsilon)
\end{IEEEeqnarray}
similar to \eqref{eq:app_lem7b_f1_LLB}. Following the same steps as in \eqref{eq:app_lem7b_a}--\eqref{eq:app_lem7b_b}, we can lower-bound the function $\underline{f}^{(k)}_{L,\rho}(\cdot)$ as
\begin{equation}
\underline{f}^{(k)}_{L,\rho}\left(\frac{n_t\bar{\delta}\rho}{T},\bar{\mathsf{D}}\right) \geq \frac{1}{4}\log\frac{T}{n_t\bar{\delta}} > 0
\end{equation}
for $\rho\geq \rho^{(1)}$ and
\begin{equation}
L \geq L^{(k)} \triangleq 16 \Upsilon(T) \delta^{(k-1)} \left(\frac{Q^{-1}(\epsilon)}{\log\frac{T}{n_t\bar{\delta}}}\right)^2, \quad k=2,3,\ldots
\end{equation}
Since $\delta^{(1)} = K^{(1)}(T,\rho^{(1)})/\sqrt{L}$, we have that
\begin{equation}
\frac{L^{(2)}}{L^{(1)}} = \frac{\Upsilon(T) K^{(1)}(T,\rho^{(1)})}{\sqrt{L}\bar{U}(T)} 
\end{equation}
and
\begin{equation}
\frac{L^{(k)}}{L^{(k-1)}} = \frac{\delta^{(k-1)}}{\delta^{(k-2)}}, \quad k=3,4,\ldots
\end{equation}
The former ratio is less than or equal to $1$ if
\begin{equation}
L \geq L'' \triangleq \left(\frac{\Upsilon(T) K^{(1)}(T,\rho^{(1)})}{\bar{U}(T)}\right)^2. \label{eq:this_is_b}
\end{equation}
The latter ratio is less than or equal to $1$ since $\delta^{(k)} \leq \delta^{(k-1)}$, $k=2,3,\ldots$ for $L\geq L^{\circ}$, as we shall show in \eqref{eq:this_is_232}--\eqref{eq:this_is_235}. Comparing \eqref{eq:this_is_a} with \eqref{eq:this_is_b}, we further note that $L'' \leq L'$ since $\tilde{V}(T)T^2/2 \leq \bar{U}(T)$ for $\rho\geq\rho'$. Consequently,
\begin{equation}
f_{L,\rho}\left(\frac{n_t\bar{\delta}\rho}{T},\bar{\mathsf{D}}\right) \geq \frac{1}{4}\log\frac{T}{n_t\bar{\delta}} > 0, \quad k = 2,3,\ldots
\end{equation}
for $\rho\geq\max\{\rho^{(1)},\rho'\}$ and $L\geq \max\{L^{(1)},L'\}$.

We conclude that, when $T>n_t+n_r$, $\rho\geq\max\{\rho^{(1)},\rho'\}$, and $L \geq \max\{L^{(1)},L'\}$, $\alpha\mapsto \underline{f}^{(k)}_{L,\rho}(\alpha,\bar{\mathsf{D}})$ is a strictly decreasing function that is positive at $\alpha = n_t\bar{\delta}\rho/T$ and negative at $\alpha=\rho$. We can thus find an $\alpha_0^{(k)}(\bar{\mathsf{D}},L,\rho)$ such that
\begin{subequations}
\begin{IEEEeqnarray}{lCll}
\underline{f}^{(k)}_{L,\rho}(\alpha,\bar{\mathsf{D}}) & > & 0, \quad & \alpha < \alpha_0^{(k)}(\bar{\mathsf{D}},L,\rho) \\
\underline{f}^{(k)}_{L,\rho}(\alpha,\bar{\mathsf{D}}) & \leq & 0, \quad & \alpha \geq \alpha_0^{(k)}(\bar{\mathsf{D}},L,\rho).
\end{IEEEeqnarray}
\end{subequations}
Defining
\begin{equation}
\delta_0^{(k)}(\bar{\mathsf{D}},L,\rho) \triangleq 1-\frac{\alpha_0^{(k)}(\bar{\mathsf{D}},L,\rho)}{\rho}
\end{equation}
this implies that we can assume without loss of optimality that $\alpha \geq \rho(1-\delta_0^{(k)}(\bar{\mathsf{D}},L,\rho))$. Following the steps in \eqref{eq:app_lem7b_MVT}--\eqref{eq:app_lem7b_R1_almost}, and using \eqref{eq:app_lem7b_der_k}, it can be shown that, for $\rho\geq\max\{\rho^{(1)},\rho'\}$ and some $\tilde{\alpha}$ between $n_t\bar{\delta}\rho/T$ and $\rho$,
\begin{IEEEeqnarray}{lCl}
\delta_0^{(k)}(\bar{\mathsf{D}},L,\rho) &  = & \frac{\underline{f}^{(k)}_{L,\rho}(\rho,\bar{\mathsf{D}})}{\rho\dot{\underline{f}}^{(1)}_{L,\rho}(\tilde{\alpha},\bar{\mathsf{D}})} \nonumber\\
& = & \frac{\sqrt{\frac{\Upsilon(T)\delta^{(k-1)}}{L}} Q^{-1}(\epsilon)}{\rho(T-n_t-n_r) \sum_{i=1}^{n_t} \frac{\bar{d}_i^2}{1+\tilde{\alpha} \bar{d}_i^2}} \nonumber\\
& \leq & \sqrt{\frac{K(T) \delta^{(k-1)}}{L}}
\end{IEEEeqnarray}
for
\begin{equation}
K(T) = \Upsilon(T)\left(1+\frac{1}{\bar{\delta}\rho^{(1)}}\right)^2 \left(\frac{Q^{-1}(\epsilon)}{(T-n_t-n_r) n_t}\right)^2.
\end{equation}
Thus, we can assume without loss of optimality that \eqref{eq:app_lem7b_LBk} holds with $\rho_0=\max\{\rho^{(1)},\rho'\}$ and $L_0=\max\{L^{(1)},L'\}$.

It remains to show that $\delta^{(k)} \leq \delta^{(k-1)}$, $k=2,3,\ldots$ Indeed, by \eqref{eq:app_lem7b_LBk},
\begin{equation}
\delta^{(k)} = \sqrt{\frac{K(T) \delta^{(k-1)}}{L}}, \quad k=2,3,\ldots \label{eq:this_is_232}
\end{equation}
Since $\delta^{(1)} = K^{(1)}(T,\rho^{(1)})/\sqrt{L}$, we obtain for $k=2$ that
\begin{equation}
\frac{\delta^{(2)}}{\delta^{(1)}} = \sqrt{\frac{K(T)}{K^{(1)}(T,\rho^{(1)})L^{1/2}}}
\end{equation}
which is less than or equal to $1$ if
\begin{equation}
\label{eq:app_lem7b_Lcirc}
L \geq L^{\circ} \triangleq \left(\frac{K(T)}{K^{(1)}(T,\rho^{(1)})}\right)^2.
\end{equation}
Furthermore, if $\delta^{(k-1)} \leq \delta^{(k-2)}$, then we also have $\delta^{(k)} \leq \delta^{(k-1)}$, since the square-root function is a monotonically increasing:
\begin{IEEEeqnarray}{lCl}
\delta^{(k)} & = & \sqrt{\frac{K(T) \delta^{(k-1)}}{L}} \nonumber\\
& \leq & \sqrt{\frac{K(T) \delta^{(k-2)}}{L}} \nonumber\\
& = & \delta^{(k-1)}.\label{eq:this_is_235}
\end{IEEEeqnarray}
We conclude that $\delta^{(k)} \leq \delta^{(k-1)}$, $k=2,3,\ldots$ for $L\geq L^{\circ}$.

\subsection{Lower Bound on $\textnormal{tr}(\mathsf{D}^2)$: Limit as $k\to\infty$}
In the previous section, we have demonstrated that, for $\rho\geq\rho_0$ and $L\geq L_0$, we can assume without loss of optimality that
\begin{equation}
\textnormal{tr}(\mathsf{D}^2) \geq T\rho\left(1-\sqrt{\frac{K(T)\delta^{(k-1)}}{L}}\right), \quad k=1,2,\ldots
\end{equation}
where
\begin{subequations}
\begin{IEEEeqnarray}{lCl}
\delta^{(1)} & = & \frac{K^{(1)}(T,\rho^{(1)})}{\sqrt{L}} \\
\delta^{(k)} & = & \sqrt{\frac{K(T) \delta^{(k-1)}}{L}}, \quad k=2,3,\ldots \label{eq:app_lem7b_deltak}
\end{IEEEeqnarray}
\end{subequations}
Since $\rho_0$ and $L_0$ do not depend on $k$, this implies that, for $\rho\geq\rho_0$ and $L\geq L_0$, we can assume without loss of optimality that
\begin{equation}
\label{eq:app_lem7b_delta_bound}
\textnormal{tr}(\mathsf{D}^2) \geq T\rho\left(1-\sqrt{\frac{K(T)\delta}{L}}\right)
\end{equation}
where $\delta \triangleq \lim_{k\to\infty} \delta^{(k)}$ is the limit of the sequence $\{\delta^{(k)}\}$, which exists since the sequence is nonnegative and decreasing. We next determine $\delta$. By the continuity of the square-root function, we obtain from \eqref{eq:app_lem7b_deltak} that $\delta$ must satisfy
\begin{equation}
\delta = \sqrt{\frac{K(T) \delta}{L}}.
\end{equation}
This equation has the two solutions
\begin{equation}
\delta= 0 \quad \textnormal{and} \quad \delta = \frac{K(T)}{L}.
\end{equation}
It follows that $\delta \leq K(T)/L$, which together with \eqref{eq:app_lem7b_delta_bound} demonstrates that,  for $\rho \geq \rho_0$ and $L\geq L_0$, we can assume without loss of optimality that
\begin{equation}
\textnormal{tr}(\mathsf{D}^2) \geq T\rho\left(1-\frac{K(T)}{L}\right).
\end{equation}
Applied to \eqref{eq:app_lem7b_THIS}, this in turn demonstrates that, for $\rho \geq \rho_0$ and $L\geq L_0$,
\begin{IEEEeqnarray}{lCl}
		\IEEEeqnarraymulticol{3}{l}{\sup_{\mathsf{D}\in\mathcal{D}_1} \left\{\bar{J}(\mathsf{D},T,\rho) 
	-\sqrt{\frac{\bar{U}(\mathsf{D},T,\rho)}{L}}Q^{-1}(\epsilon)\right\}} \nonumber\\
	\qquad & \leq & \sup_{\mathsf{D}\in\mathcal{D}_{n_t}\colon \textnormal{tr}(\mathsf{D}^2) \geq T\rho\left(1-\frac{K(T)}{L}\right)} \left\{\bar{J}^{\ast}(\mathsf{D},T,\rho)
	-\sqrt{\frac{\bar{U}(\mathsf{D},T,\rho)}{L}}Q^{-1}(\epsilon)\right\} + K_{\mathcal{D}_1}(T,\rho)
		\end{IEEEeqnarray}
		which is Lemma~\ref{MC_Lemma7b}.

\section{Proof of Lemma \ref{MC_Lemma7}} \label{Appx_Pf_MC_Lem7} 
To maximize $\bar{J}^{\ast}(\mathsf{D},T,\rho)$ over $\mathsf{D}\in\mathcal{D}_1$, we note that the term in $\bar{J}^{\ast}(\mathsf{D},T,\rho)$ that depends on $\mathsf{D}$ is  $(T-n_t-n_r)\log\det(\mathsf{I}_{n_t}+\mathsf{D}^2)$. 
 By the concavity of the logarithm function and Jensen's inequality, we have for all diagonal matrices $\mathsf{D}$ satisfying $\text{tr}(\mathsf{D}^2) \leq T\rho$
		\begin{IEEEeqnarray}{lCl}
		\log\det(\mathsf{I}_{n_t}+\mathsf{D}^2) & = & \sum_{i=1}^{n_t} \log(1+d_i^2) \nonumber\\
		& \leq & n_t \log\left(1+\frac{1}{n_t} \sum_{i=1}^{n_t} d_i^2\right) \nonumber\\
		& \leq & n_t \log\left(1+\frac{T\rho}{n_t}\right)
		\end{IEEEeqnarray}
		which holds with equality if $d_i^2 = T\rho/n_t$, $i=1,\ldots,n_t$. Since $\bar{\delta}$, as defined in \eqref{eq:delta_bar}, does not exceed $T/n_t$, this choice of $d_i^2$ satisfies $d_i^2 \geq \bar{\delta}\rho$ and, hence, the corresponding $\mathsf{D}$ lies in $\mathcal{D}_1$. Thus,
		\begin{equation}
		\sup_{\mathsf{D} \in \mathcal{D}_1} \log\det(\mathsf{I}_{n_t}+\mathsf{D}^2) = n_t  \log\left(1+\frac{T\rho}{n_t}\right)
		\end{equation}
		which together with the definition of $\bar{J}^{\ast}(\mathsf{D},T,\rho)$ in \eqref{MC_Ap3_Def_Jast} proves Lemma~\ref{MC_Lemma7}.

\section{Proof of Lemma \ref{MC_Lem_Vhigh}} \label{Appx_Pf_MC_Lem8} 

	By replacing in \eqref{L4_UU_ast} $\mathsf{D}$ by $\sqrt{\rho/\alpha}\mathsf{D}$, we obtain that
	\begin{equation}
	\bar{U}\left(\sqrt{\frac{\rho}{\alpha}} \mathsf{D},T,\rho\right) = \bar{U}^{\ast}\left(\sqrt{\frac{\rho}{\alpha}} \mathsf{D},T,\rho\right) + K_{\bar{U}}\left(\sqrt{\frac{\rho}{\alpha}}\mathsf{D},T,\rho\right)
	\end{equation}
	where $K_{\bar{U}}(\mathsf{D},T,\rho)$ is a function of $\mathsf{D}$, $T$, and $\rho$ that satisfies $\lim_{\rho\to\infty}\sup_{\mathsf{D}\in\mathcal{D}_1} \left|K_{\bar{U}}(\mathsf{D},T,\rho)\right| = 0$.
	
	We next evaluate $\bar{U}^{\ast}(\tilde{\mathsf{D}},T,\rho)$ for $\tilde{\mathsf{D}} = \sqrt{\rho/\alpha}\mathsf{D}$. Note that the trace of $\tilde{\mathsf{D}}$ is equal to $T\rho$, so the corresponding $\tilde{\alpha} = \textnormal{tr}(\tilde{\mathsf{D}}^2)/T$ is equal to $\rho$. It follows that
	\begin{IEEEeqnarray}{lCl}
		\IEEEeqnarraymulticol{3}{l}{\bar{U}^{\ast}\left(\sqrt{\frac{\rho}{\alpha}} \mathsf{D},T,\rho\right)} \nonumber\\
		& = &  \mathsf{E}\left[ \left(\sum_{i=1}^{n_t} \left( \frac{n_t d_{i,\ell}^2 }{T\alpha}  -1 \right)(Z’_{i, \ell}-n_r)
		-\sum_{i=1}^{n_t}\bigl(Z''_{i,\ell}-(T-n_t)\bigr) 
		+(T-n_t)\left(\log \det \bigl( \mathbf{H}_{\ell} \mathbf{H}_{\ell}^\mathsf{H} \bigr) 
		-\mathsf{E}\left[ \log \det \left( \mathbf{H}_{\ell} \mathbf{H}_{\ell}^\mathsf{H}  \right) \right] \right) \right)^2 \right] \nonumber\\
		& = &   \mathsf{E}\left[\left(\sum_{i=1}^{n_t} \left( \frac{n_t d_{i,\ell}^2 }{T\alpha}  -1 \right)(Z’_{i, \ell}-n_r)\right)^2\right] + \mathsf{E}\left[\left(\sum_{i=1}^{n_t}\bigl(Z''_{i,\ell}-(T-n_t)\bigr) \right)^2 \right] \nonumber \\
		&&{} +(T-n_t)^2 \mathsf{E}\left[\left(\log \det \left( \mathbf{H}_{\ell} \mathbf{H}_{\ell}^\mathsf{H} \right)	
		-\mathsf{E}\left[\log \det \left( \mathbf{H} \mathbf{H}^\mathsf{H} \right)\right] \right)^2\right] \nonumber\\
		& & {} + 2(T-n_t)\sum_{i=1}^{n_t} \left( \frac{n_t d_{i,\ell}^2 }{T\alpha}  -1 \right)\mathsf{E}\left[(Z’_{i, \ell}-n_r)\log \det \left( \mathbf{H}_{\ell} \mathbf{H}_{\ell}^\mathsf{H} \right)\right] \nonumber \\
		& \geq & \mathsf{E}\left[\left(\sum_{i=1}^{n_t}\bigl(Z''_{i,\ell}-(T-n_t)\bigr) \right)^2 \right] + (T-n_t)^2 \mathsf{E}\left[\left(\log \det \left( \mathbf{H}_{\ell} \mathbf{H}_{\ell}^\mathsf{H} \right)	
		-\mathsf{E}\left[\log \det \left( \mathbf{H} \mathbf{H}^\mathsf{H} \right)\right] \right)^2\right] \nonumber\\
		& = & n_t (T-n_t)+(T-n_t)^2  \sum_{i=0}^{n_t-1} \varPsi'(n_r-i) \nonumber\\
		& = & T^2 \tilde{V}(T)
		\end{IEEEeqnarray} 
	where the second step follows because $\{Z_{1,1}'',\ldots, Z_{n_t,L}''\}$ and $\{\mathbf{H}^L,Z_{1,1}',\ldots, Z_{n_t,L}'\}$ are independent, so the term depending on $\{Z_{1,1}'',\ldots, Z_{n_t,L}''\}$ is uncorrelated with the other terms; the third step follows because the first term is nonnegative and the last term is zero, since $\mathsf{E}[(Z’_{i, \ell}-n_r)\log \det \left( \mathbf{H}_{\ell} \mathbf{H}_{\ell}^\mathsf{H} \right)]$ does not depend on $i$ and the sum over $i$ is zero because $d_{1,\ell}^2+\ldots+d_{n_t,\ell}^2=T\alpha$; the fourth step follows because the variance of $\log \det (\mathbf{H}_{\ell} \mathbf{H}_{\ell}^\mathsf{H} )$ is $\sum_{i=0}^{n_t-1} \varPsi'(n_r-i)$ (Lemma~\ref{lem:grant_wishart}); and the last step follows by the definition of $\tilde{V}(T)$ in \eqref{Def_Vt}. We conclude that
	\begin{equation}
	\label{eq:qpp_lem8_K}
	\bar{U}\left(\sqrt{\frac{\rho}{\alpha}} \mathsf{D},T,\rho\right) \geq T^2 \tilde{V}(T) + K_{\bar{U}}(\mathsf{D},T,\rho)
	\end{equation}
where, with a slight abuse of notation, we replace $K_{\bar{U}}(\sqrt{\rho/\alpha}\mathsf{D},T,\rho)$ by $K_{\bar{U}}(\mathsf{D},T,\rho)$. Since the set of matrices $\tilde{\mathcal{D}}_1 \triangleq \{\tilde{\mathsf{D}}\in\mathcal{D}_{n_t}\colon \tilde{\mathsf{D}} = \sqrt{\rho/\alpha}\mathsf{D}, \mathsf{D} \in\mathcal{D}_1\}$ is contained in $\mathcal{D}_1$, we have that
\begin{equation}
\sup_{\mathsf{D}\in\mathcal{D}_1} \left|K_{\bar{U}}\left(\sqrt{\frac{\rho}{\alpha}}\mathsf{D},T,\rho\right)\right| = \sup_{\mathsf{D}\in\tilde{\mathcal{D}}_1}\left|K_{\bar{U}}(\mathsf{D},T,\rho)\right| \leq \sup_{\mathsf{D}\in\mathcal{D}_1}\left|K_{\bar{U}}(\mathsf{D},T,\rho)\right| 
\end{equation}
so the constant $K_{\bar{U}}(\mathsf{D},T,\rho)$ in \eqref{eq:qpp_lem8_K} satisfies $\lim_{\rho\to\infty} \sup_{\mathsf{D}\in\mathcal{D}_1} \left|K_{\bar{U}}(\mathsf{D},T,\rho)\right|  = 0$. This proves the lemma.

\section{Proof of Lemma \ref{MC_Lem8}} \label{Appx_Pf_MC_Lem8b} 
We begin with \eqref{Def_J_rhoHq} and optimize over $\mathsf{D}\in\mathcal{D}_2$. A lower bound on $\bar{J}(\mathsf{D},T,\rho)$ follows by noting that $K_{\mathcal{D}_1}(\mathsf{D},T) \geq 0$, so for every $\mathsf{D}\in\mathcal{D}_{n_t}$
\begin{equation}
\label{eq:app_mc_lem8b_LB}
\bar{J}(\mathsf{D},T,\rho) \geq \bar{J}^{\ast}(\mathsf{D},T,\rho).
\end{equation}
To obtain an upper bound, and to optimize $\bar{J}^{\ast}(\mathsf{D},T,\rho)$ over $\mathsf{D} \in \mathcal{D}_2$, we have to distinguish between the different cases of $d_i^2$, $i=1,\ldots,n_t$ that can occur for $\mathsf{D}\in\mathcal{D}_2$. Indeed, without loss of optimality, assume that the diagonal elements of $\mathsf{D}$ are ordered so that $d_1 \leq d_2 \leq \ldots \leq d_{n_t}$. Then, for every $\mathsf{D}\in\mathcal{D}_2$, we have that
\begin{IEEEeqnarray*}{lCll}
d_i^2 & \leq & \bar{\delta}\rho, \quad &  i=1,\ldots,n_k\\
{d}_i^2 & > & \bar{\delta} \rho, \quad & i=n_k+1,\ldots n_t
\end{IEEEeqnarray*}
for some $n_k=1,\ldots,n_t$. We need to distinguish between the following two cases:
\begin{enumerate}
\item $d_i$ tends to infinity as $\rho\to\infty$ for all $i=1,\ldots,n_t$;
\item there exists a finite $\bar{\rho}$ such that $d_i \leq \bar{\rho}$ for $i=1,\ldots,n_s$ and some $n_s=1,\ldots,n_k$.
\end{enumerate}
We analyze the former case in Appendix~\ref{subsub:lem7b_unbounded} and the latter case in Appendix~\ref{subsub:lem7b_bounded}. We then combine both cases in Appendix~\ref{subsub:lem7b_combination}.

\subsection{Unbounded $d_i$}
\label{subsub:lem7b_unbounded}
When $\lim_{\rho\to\infty} d_i = \infty$ for every $i=1,\ldots,n_t$, we can find a $\xi(\rho)$ satisfying $\lim_{\rho\to\infty}\xi(\rho)=\infty$ such that $d_i^2 \geq \xi(\rho)$, $i=1,\ldots,n_t$. It follows from \eqref{Def_J_rhoHq} and \eqref{MC_Ap3_Def_K} that
\begin{IEEEeqnarray}{lCl}
\bar{J}(\mathsf{D},T,\rho) & = &  \bar{J}^{\ast}(\mathsf{D},T,\rho) 
		+\frac{n_t}{T\rho}\left(n_tn_r+n_t(T-n_t)\right)
		+ (T-n_t)\mathsf{E}\left[ \log \frac{ \det \Big( \mathbf{H}_{\ell} \mathbf{H}_{\ell}^\mathsf{H}
		+ \lambda_1(\mathbf{Q}_{\ell}^\mathsf{H}\mathbf{Q}_{\ell})\big(\mathsf{I}_{n_t}+  \mathsf{D}^2\big)^{-1}  \Big)}{\det \Big( \mathbf{H}_{\ell} \mathbf{H}_{\ell}^\mathsf{H} \Big) } \right] \nonumber\\
		& \leq & \bar{J}^{\ast}(\mathsf{D},T,\rho) +\frac{n_t}{T\rho}\left(n_tn_r+n_t(T-n_t)\right) + (T-n_t)\mathsf{E}\left[ \log \frac{ \det \Big( \mathbf{H}_{\ell} \mathbf{H}_{\ell}^\mathsf{H}
		+ \lambda_1(\mathbf{Q}_{\ell}^\mathsf{H}\mathbf{Q}_{\ell})\big(1+ \xi(\rho)\big)^{-1} \mathsf{I}_{n_t} \Big)}{\det \Big( \mathbf{H}_{\ell} \mathbf{H}_{\ell}^\mathsf{H} \Big) } \right] \nonumber\\
		& \triangleq & \bar{J}^{\ast}(\mathsf{D},T,\rho) + K_{\mathcal{D}_2}(T,\rho)
		\end{IEEEeqnarray}
		where the last step should be viewed as the definition of $K_{\mathcal{D}_2}(T,\rho)$. Consequently,
		\begin{equation}
		\bar{J}(\mathsf{D},T,\rho) \leq \bar{J}^{\ast}(\mathsf{D},T,\rho) + K_{\mathcal{D}_2}(T,\rho) \label{eq:lemma7b_UB_Jbar}
		\end{equation}
		and, by the dominated convergence theorem, $\lim_{\rho\to\infty} K_{\mathcal{D}_2}(T,\rho) = 0$.

We next optimize $\bar{J}^{\ast}(\mathsf{D},T,\rho)$ over $\mathsf{D}\in\mathcal{D}_2$. To this end, we recall that the only term in $\bar{J}^{\ast}(\mathsf{D},T,\rho)$ that depends on $\mathsf{D}$ is $(T-n_t-n_r)\log\det(\mathsf{I}_{n_t}+\mathsf{D}^2)$. Defining $\sum_{i=1}^{n_k} d_i^2\triangleq \bar{d}_{n_k}$, and using the concavity of the logarithm function and Jensen's inequality, we have that, for all diagonal matrices $\mathsf{D}$ in $\mathcal{D}_2$,
\begin{IEEEeqnarray}{lCl}
		\log\det(\mathsf{I}_{n_t}+\mathsf{D}^2)
		& = &\sum_{i=1}^{n_k}  \log(1+d_i^2) + \sum_{i=n_k+1}^{n_t} \log(1+d_i^2) \nonumber\\
		& \leq & n_k \log\left(1+\frac{\bar{d}_{n_k}}{n_k}\right) + (n_t-n_k)\log\left(1+\frac{T\rho - \bar{d}_{n_k}}{n_t-n_k}\right)\nonumber\\
		& \leq & \log\left(1+\bar{\delta}\rho\right) + (n_t-1) \log\left(1+\frac{T\rho - \bar{\delta}\rho}{n_t-1}\right) \label{eq:lemma7b_unbounded_max}
		\end{IEEEeqnarray}
		where the second step follows from Jensen's inequality and because $\sum_{i=1}^{n_t} d_i^2 \leq T\rho$, so $\sum_{i=n_k+1}^{n_t} d_i^2 \leq T\rho - \bar{d}_{n_k}$; the third step follows because the expression is monotonically increasing in $\bar{d}_{n_k} \leq n_k\bar{\delta}\rho$ and monotonically decreasing in $n_k=1,\ldots,n_t$.
		
		The inequality \eqref{eq:lemma7b_unbounded_max} holds with equality if $d_1^2 = \bar{\delta}\rho$ and $d_i^2 = (T\rho-\bar{\delta}\rho)/(n_t-1)$, $i=2,\ldots,n_t$. It follows that
	\begin{IEEEeqnarray}{lCl}
	\sup_{\mathsf{D}\in\mathcal{D}_2} \bar{J}^{\ast}(\mathsf{D},T,\rho) &= & n_tn_r\log \left(\frac{T\rho}{n_t}\right)
	+\log \left(\frac{\Gamma_{n_t}(n_t)}{\Gamma_{n_t}(T)} \right)
	-n_t(T-n_t) +(T-n_t)\mathsf{E}\left[ \log \det \left( \mathbf{H}_{\ell} \mathbf{H}_{\ell}^\mathsf{H} \right) \right] \nonumber\\
	& & {}  + (T-n_t-n_r)\log\left((1+\bar{\delta}\rho)\left(1+\frac{T-\bar{\delta}}{n_t-1}\rho\right)^{n_t-1}\right) \nonumber\\
	& \triangleq & \bar{J}_{\mathcal{D}_2}(T,\rho). \label{eq:lemma7b_max_logdet}
	\end{IEEEeqnarray}
	We conclude from \eqref{eq:lemma7b_UB_Jbar} and \eqref{eq:lemma7b_max_logdet} that, when $d_i$ tends to infinity as $\rho\to\infty$ for every $i=1,\ldots,n_t$, we have that
	\begin{equation}
	\sup_{\mathsf{D}\in\mathcal{D}_2}\bar{J}(\mathsf{D},T,\rho) \leq \bar{J}_{\mathcal{D}_2}(T,\rho) + K_{\mathcal{D}_2}(T,\rho). \label{eq:lemma7b_unbounded}
	\end{equation}
	
\subsection{Bounded $d_i$}
\label{subsub:lem7b_bounded}
	For every $\mathsf{D} \in \mathcal{D}_{n_t}$, the term $\bar{J}(\mathsf{D},T,\rho)$ can be upper-bounded as 
	\begin{IEEEeqnarray}{lCl} 
		\bar{J}(\mathsf{D},T,\rho)
		& =& n_tn_r\log \left(\frac{T\rho}{n_t}\right)
			+\log \left(\frac{\Gamma_{n_t}(n_t)}{\Gamma_{n_t}(T)} \right)
			+\left(\frac{T\alpha-T\rho}{T\rho}\right)n_rn_t-n_t(T-n_t) 
			+\frac{n_t}{T\rho}\left(n_tn_r+n_t(T-n_t)\right)\nonumber\\
			&& {} +(T-n_t-n_r)\log\det\left(\mathsf{I}_{n_t}+  \mathsf{D}^2\right)
			 +(T-n_t)\mathsf{E}\left[ \log \det \left( \mathbf{H}_{\ell} \mathbf{H}_{\ell}^\mathsf{H}
			+ \lambda_1(\mathbf{Q}_{\ell}^\mathsf{H}\mathbf{Q}_{\ell})\big(\mathsf{I}_{n_t}+  \mathsf{D}^2\big)^{-1}   \right) \right] \nonumber \\
		& \leq &  n_tn_r\log \left(\frac{T\rho}{n_t}\right)
			+\log \left(\frac{\Gamma_{n_t}(n_t)}{\Gamma_{n_t}(T)} \right) -n_t(T-n_t)+ \frac{n_t}{T\rho}\left(n_tn_r+n_t(T-n_t)\right) \nonumber\\
			& & {} +(T-n_t-n_r)\log\det\left(\mathsf{I}_{n_t}+  \mathsf{D}^2\right) +(T-n_t)\mathsf{E}\left[ \log \det \left( \mathbf{H}_{\ell} \mathbf{H}_{\ell}^\mathsf{H}
			+ \lambda_1(\mathbf{Q}_{\ell}^\mathsf{H}\mathbf{Q}_{\ell})\mathsf{I}_{n_t}   \right) \right]. \label{Equ_sup_D2C2} 
		\end{IEEEeqnarray}
		To optimize the RHS of \eqref{Equ_sup_D2C2} over $\mathsf{D}$, we note that the only term that depends on $\mathsf{D}$ is again $(T-n_t-n_r)\log\det\left(\mathsf{I}_{n_t}+  \mathsf{D}^2\right)$. Following similar steps as in \eqref{eq:lemma7b_unbounded_max}, it can be shown that, for all matrices $\mathsf{D}$ satisfying
		\begin{subequations}
		\label{eq:lemma7b_all}
	\begin{IEEEeqnarray}{lCll}
	d_i^2 & \leq & \bar{\rho}, \quad & i=1,\ldots,n_s \label{eq:lemma7b_cond1}\\
	d_i^2 & \leq & \bar{\delta}\rho, \quad & i=n_s+1,\ldots,n_k \label{eq:lemma7b_cond2}\\
	d_i^2 & > & \bar{\delta}\rho, \quad & i= n_k+1,\ldots,n_t \label{eq:lemma7b_cond3}
	\end{IEEEeqnarray}	
	\end{subequations}
	for some $n_s=1,\ldots,n_k$ and $n_k=1,\ldots,n_t$, and for $\rho \geq \bar{\rho}/\bar{\delta}$, we have that
	\begin{IEEEeqnarray}{lCl}
	\log\det\left(\mathsf{I}_{n_t}+  \mathsf{D}^2\right) & \leq & n_s \log\left(1+\bar{\rho}\right) + (n_k - n_s)\log\left(1+\bar{\delta}\rho\right) + (n_t-n_k) \log\left(1+\frac{T\rho - n_s\bar{\rho}-(n_k-n_s)\bar{\delta}\rho}{n_t-n_k}\right) \nonumber\\
	& \leq & \log\left(1+\bar{\rho}\right) + (n_t-1) \log\left(1+\frac{T\rho-\bar{\rho}}{n_t-1}\right)
	\end{IEEEeqnarray}
	where the second inequality follows because, for $\rho\geq \bar{\rho}/\bar{\delta}$, the expression is monotonically decreasing in $n_s$ and $n_k$. Thus, for all matrices $\mathsf{D}$ satisfying \eqref{eq:lemma7b_all} and $\rho\geq \bar{\rho}/\bar{\delta}$,
	\begin{IEEEeqnarray}{lCl}
	\bar{J}(\mathsf{D},T,\rho) & \leq & n_tn_r\log \left(\frac{T\rho}{n_t}\right)
			+\log \left(\frac{\Gamma_{n_t}(n_t)}{\Gamma_{n_t}(T)} \right)+\frac{n_t}{T\rho}\left(n_tn_r+n_t(T-n_t)\right) -n_t(T-n_t) \nonumber\\			
		&& {}	+(T-n_t-n_r)\log\left((1+\bar{\rho})\left(1+\frac{T\rho-\bar{\rho}}{n_t-1}\right)^{n_t-1}\right) 
		 +(T-n_t)\mathsf{E}\left[ \log \det \left( \mathbf{H}_{\ell} \mathbf{H}_{\ell}^\mathsf{H}
			+ \lambda_1(\mathbf{Q}_{\ell}^\mathsf{H}\mathbf{Q}_{\ell})\mathsf{I}_{n_t}   \right) \right] \nonumber\\
			& \triangleq &  \bar{J}'_{\mathcal{D}_2}(T,\rho). \label{eq:lemma7b_bounded}
	\end{IEEEeqnarray}
		
\subsection{Combining both cases}
\label{subsub:lem7b_combination}
We obtain from \eqref{eq:app_mc_lem8b_LB} and \eqref{eq:lemma7b_max_logdet} that
\begin{equation}
\label{eq:lemma7b_LB}
\sup_{\mathsf{D}\in\mathcal{D}_2} \bar{J}(\mathsf{D},T,\rho) \geq \bar{J}_{\mathcal{D}_2}(T,\rho).
\end{equation}
Similarly, we obtain from \eqref{eq:lemma7b_unbounded} and \eqref{eq:lemma7b_bounded} that, for $\rho\geq\bar{\rho}/\bar{\delta}$,
\begin{equation}
\sup_{\mathsf{D}\in\mathcal{D}_2} \bar{J}(\mathsf{D},T,\rho) \leq \max\left\{\bar{J}_{\mathcal{D}_2}(T,\rho) + K_{\mathcal{D}_2}(T,\rho),\bar{J}'_{\mathcal{D}_2}(T,\rho)\right\}.
\end{equation}
We next show that, for sufficiently large $\rho$, $\bar{J}'_{\mathcal{D}_2}(T,\rho) \leq \bar{J}_{\mathcal{D}_2}(T,\rho) + K_{\mathcal{D}_2}(T,\rho)$. Hence, for such $\rho$,
\begin{equation}
\sup_{\mathsf{D}\in\mathcal{D}_2} \bar{J}(\mathsf{D},T,\rho) \leq \bar{J}_{\mathcal{D}_2}(T,\rho) + K_{\mathcal{D}_2}(T,\rho). \label{eq:lemma7b_UB}
\end{equation}
Indeed,
	\begin{IEEEeqnarray}{lCl}
	\IEEEeqnarraymulticol{3}{l}{\bar{J}_{\mathcal{D}_2}(T,\rho) + K_{\mathcal{D}_2}(T,\rho) -\bar{J}'_{\mathcal{D}_2}(T,\rho)} \nonumber \\
	\quad & =& \log \left(\left(1+\bar{\delta}\rho\right)\left(1+\frac{T-\bar{\delta}}{n_t-1}\rho\right)^{n_t-1}\right) 
	-\log \left( \left(1+\bar{\rho}\right)\left(1+\frac{T\rho-\bar{\rho}}{n_t-1}\right)^{n_t-1}\right) \nonumber \\
	& & {}+ (T-n_t)\mathsf{E}\left[ \log \frac{\det \left( \mathbf{H}_{\ell} \mathbf{H}_{\ell}^\mathsf{H} \right)}{\det \left( \mathbf{H}_{\ell} \mathbf{H}_{\ell}^\mathsf{H} + \lambda_1(\mathbf{Q}_{\ell}^\mathsf{H}\mathbf{Q}_{\ell})\mathsf{I}_{n_t} \right)} \right]  + K_{\mathcal{D}_2}(T,\rho)  \nonumber \\
	& =& \log  \left( \frac{1+\bar{\delta}\rho}{1+\bar{\rho}}\right)
	+ (n_t-1)\log \left( \frac{n_t-1+T\rho-\bar{\delta}\rho}{n_t-1+T\rho-\bar{\rho}} \right) \nonumber\\
	&& {}+ (T-n_t)\mathsf{E}\left[ \log \frac{\det \left( \mathbf{H}_{\ell} \mathbf{H}_{\ell}^\mathsf{H} \right)}{\det \left( \mathbf{H}_{\ell} \mathbf{H}_{\ell}^\mathsf{H} + \lambda_1(\mathbf{Q}_{\ell}^\mathsf{H}\mathbf{Q}_{\ell})\mathsf{I}_{n_t} \right)} \right]  + K_{\mathcal{D}_2}(T,\rho). \label{Equ_sup_Com} 
	\end{IEEEeqnarray}
	The first term on the RHS of \eqref{Equ_sup_Com} tends to infinity as $\rho\to\infty$. The remaining terms are bounded in $\rho\geq\bar{\rho}/\bar{\delta}$. Thus, for $\rho\geq\rho_0$ and a sufficiently large $\rho_0$, the RHS of \eqref{Equ_sup_Com} is nonnegative and \eqref{eq:lemma7b_UB} follows.
	
Combining \eqref{eq:lemma7b_UB} with \eqref{eq:lemma7b_LB}, we conclude that
\begin{equation}
\bar{J}_{\mathcal{D}_2}(T,\rho) \leq \sup_{\mathsf{D}\in\mathcal{D}_2} \bar{J}(\mathsf{D},T,\rho) \leq \bar{J}_{\mathcal{D}_2}(T,\rho) + K_{\mathcal{D}_2}(T,\rho), \quad \rho\geq \rho_0
\end{equation}
for a sufficiently large $\rho_0$, which is Lemma~\ref{MC_Lem8}.

\section*{Acknowledgment}
The authors gratefully acknowledge fruitful discussions with W.~Yang. They further wish to thank the Associate Editor E. Jorswieck and the anonymous referees for their valuable comments.
 
\bibliographystyle{IEEEtran}
\bibliography{MIMOFading}

\begin{thebibliography}{10}
\providecommand{\url}[1]{#1}
\csname url@samestyle\endcsname
\providecommand{\newblock}{\relax}
\providecommand{\bibinfo}[2]{#2}
\providecommand{\BIBentrySTDinterwordspacing}{\spaceskip=0pt\relax}
\providecommand{\BIBentryALTinterwordstretchfactor}{4}
\providecommand{\BIBentryALTinterwordspacing}{\spaceskip=\fontdimen2\font plus
\BIBentryALTinterwordstretchfactor\fontdimen3\font minus
  \fontdimen4\font\relax}
\providecommand{\BIBforeignlanguage}[2]{{%
\expandafter\ifx\csname l@#1\endcsname\relax
\typeout{** WARNING: IEEEtran.bst: No hyphenation pattern has been}%
\typeout{** loaded for the language `#1'. Using the pattern for}%
\typeout{** the default language instead.}%
\else
\language=\csname l@#1\endcsname
\fi
#2}}
\providecommand{\BIBdecl}{\relax}
\BIBdecl

\bibitem{durisi2016toward}
G.~Durisi, T.~Koch, and P.~Popovski, ``Toward massive, ultrareliable, and
  low-latency wireless communication with short packets,'' \emph{Proc. IEEE},
  vol. 104, no.~9, pp. 1711--1726, Sep. 2016.

\bibitem{hayashi09}
M.~Hayashi, ``Information spectrum approach to second-order coding rate in
  channel coding,'' \emph{IEEE Trans. Inf. Theory}, vol.~55, no.~11, pp.
  4947--4966, Nov. 2009.

\bibitem{polyanskiy2010channel}
Y.~Polyanskiy, H.~V. Poor, and S.~Verd{\'u}, ``Channel coding rate in the
  finite blocklength regime,'' \emph{IEEE Trans. Inf. Theory}, vol.~56, no.~5,
  pp. 2307--2359, May 2010.

\bibitem{Shirvanimoghaddam19}
M.~Shirvanimoghaddam, M.~S. Mohammadi, R.~Abbas, A.~Minja, C.~Yue, B.~Matuz,
  G.~Han, Z.~Lin, W.~Liu, Y.~Li, S.~Johnson, and B.~Vucetic, ``Short
  block-length codes for ultra-reliable low latency communications,''
  \emph{IEEE Communications Magazine}, vol.~57, no.~2, pp. 130--137, Feb. 2019.

\bibitem{coskun19}
M.~C. Co\c{s}kun, G.~Durisi, T.~Jerkovits, G.~Liva, W.~Ryan, B.~Stein, and
  F.~Steiner, ``Efficient error-correcting codes in the short blocklength
  regime,'' \emph{Physical Communication}, vol.~34, pp. 66--79, Jun. 2019.

\bibitem{Makki15}
B.~Makki, T.~Svensson, and M.~Zorzi, ``Finite block-length analysis of spectrum
  sharing networks using rate adaptation,'' \emph{IEEE Trans. Commun.},
  vol.~63, no.~8, pp. 2823--2835, Aug. 2015.

\bibitem{Makki16}
------, ``Wireless energy and information transmission using feedback: Infinite
  and finite block-length analysis,'' \emph{IEEE Trans. Commun.}, vol.~64,
  no.~12, pp. 5304--5318, Dec. 2016.

\bibitem{Zhang18}
L.~Zhang and Y.-C. Liang, ``Average throughput analysis and optimization in
  cooperative {IoT} networks with short packet communication,'' \emph{IEEE
  Trans. Veh. Technol.}, vol.~67, no.~12, pp. 11\,549--11\,562, Dec. 2018.

\bibitem{Li19}
C.~Li, N.~Yang, and S.~Yan, ``Optimal transmission of short-packet
  communications in multiple-input single-output systems,'' \emph{IEEE Trans.
  Veh. Technol.}, vol.~68, no.~7, pp. 7199--7203, Jul. 2019.

\bibitem{Mahmood21}
N.~H. Mahmood, O.~A. López, H.~Alves, and M.~Latva-Aho, ``A predictive
  interference management algorithm for {URLLC} in beyond {5G} networks,''
  \emph{IEEE Commun. Lett.}, vol.~25, no.~3, pp. 995--999, Mar. 2021.

\bibitem{Munari21}
A.~Munari and F.~Clazzer, ``Spectral coexistence of {QoS}-constrained and {IoT}
  traffic in satellite systems,'' \emph{Sensors}, vol.~21, no.~14, Jul. 2021.

\bibitem{Feng23}
J.~Feng, H.~Q. Ngo, and M.~Matthaiou, ``{(Non)}-coherent {MU-MIMO} block fading
  channels with finite blocklength and linear processing,'' \emph{IEEE Trans.
  Wireless Commun.}, vol.~22, no.~4, pp. 2441--2461, Apr. 2023.

\bibitem{Xie23}
M.~Xie, J.~Gong, X.~Jia, Q.~Wang, and X.~Ma, ``Age and energy analysis for
  $l$th best relay enabled cooperative status update systems with short packet
  communications,'' \emph{IEEE Trans. Veh. Technol.}, vol.~72, no.~5, pp.
  6294--6308, May 2023.

\bibitem{Zheng24}
C.~Zheng, F.-C. Zheng, J.~Luo, P.~Zhu, X.~You, and D.~Feng, ``Differential
  modulation for short packet transmission in {URLLC},'' \emph{IEEE Trans.
  Wireless Commun.}, vol.~23, no.~7, pp. 7230--7245, Jul. 2024.

\bibitem{Li25}
L.~Li, W.~Chen, P.~Popovski, and K.~B. Letaief, ``Reliability-latency-rate
  tradeoff in low-latency communications with finite-blocklength coding,''
  \emph{IEEE Trans. Inf. Theory}, vol.~71, no.~1, pp. 360--389, Jan. 2025.

\bibitem{polyanskiy2011scalar}
Y.~Polyanskiy and S.~Verd{\'u}, ``Scalar coherent fading channel: Dispersion
  analysis,'' in \emph{Proc. IEEE Int. Symp. on Inf. Theory (ISIT)},
  St.~Petersburg, Russia, Jul.~31--Aug.~5, 2011, pp. 2959--2963.

\bibitem{collins2014orthogonal}
A.~Collins and Y.~Polyanskiy, ``Orthogonal designs optimize achievable
  dispersion for coherent {MISO} channels,'' in \emph{Proc. IEEE Int. Symp. on
  Inf. Theory (ISIT)}, Honolulu, HI, USA, Jun.~29--Jul.~4, 2014, pp.
  2524--2528.

\bibitem{collins2016dispersion}
------, ``Dispersion of the coherent {MIMO} block-fading channel,'' in
  \emph{Proc. IEEE Int. Symp. on Inf. Theory (ISIT)}, Barcelona, Spain,
  Jul.~10--15, 2016, pp. 1068--1072.

\bibitem{Collins2019coherent}
------, ``Coherent multiple-antenna block-fading channels at finite
  blocklength,'' \emph{IEEE Trans. Inf. Theory}, vol.~65, no.~1, pp. 380--405,
  Jan. 2019.

\bibitem{hoydis2015second}
J.~Hoydis, R.~Couillet, and P.~Piantanida, ``The second-order coding rate of
  the {MIMO} quasi-static {R}ayleigh fading channel,'' \emph{IEEE Trans. Inf.
  Theory}, vol.~61, no.~12, pp. 6591--6622, Dec. 2015.

\bibitem{Yang15}
W.~Yang, G.~Caire, G.~Durisi, and Y.~Polyanskiy, ``Optimum power control at
  finite blocklength,'' \emph{IEEE Trans. Inf. Theory}, vol.~61, no.~9, pp.
  4598--4615, Nov. 2015.

\bibitem{yang2014quasi}
W.~Yang, G.~Durisi, T.~Koch, and Y.~Polyanskiy, ``Quasi-static multiple-antenna
  fading channels at finite blocklength,'' \emph{IEEE Trans. Inf. Theory},
  vol.~60, no.~7, pp. 4232--4265, 2014.

\bibitem{yang2012diversity}
------, ``Diversity versus channel knowledge at finite block-length,'' in
  \emph{Proc. IEEE Inf. Theory Workshops (ITW)}, Lausanne, Switzerland,
  Sep.~3--7, 2012, pp. 572--576.

\bibitem{durisi2015short}
G.~Durisi, T.~Koch, J.~{\"O}stman, Y.~Polyanskiy, and W.~Yang, ``Short-packet
  communications over multiple-antenna {R}ayleigh-fading channels,'' \emph{IEEE
  Trans. Commun.}, vol.~64, no.~2, pp. 618--629, Feb. 2016.

\bibitem{lancho2019saddlepoint}
A.~Lancho, J.~{\"O}stman, G.~Durisi, T.~Koch, and G.~Vazquez-Vilar,
  ``Saddlepoint approximations for noncoherent single-antenna {Rayleigh}
  block-fading channels,'' in \emph{Proc. IEEE Int. Symp. on Inf. Theory
  (ISIT)}, Paris, France, Jul.~7--12 2019, pp. 612--616.

\bibitem{LanchoSaddlepoint2020}
------, ``Saddlepoint approximations for short-packet wireless
  communications,'' \emph{IEEE Trans. Wireless Commun.}, vol.~19, no.~7, pp.
  4831--4846, Jul. 2021.

\bibitem{lancho2019single}
A.~Lancho, T.~Koch, and G.~Durisi, ``On single-antenna {R}ayleigh block-fading
  channels at finite blocklength,'' \emph{IEEE Trans. Inf. Theory}, vol.~66,
  no.~1, pp. 496--519, Jan. 2019.

\bibitem{zheng2002communication}
L.~Zheng and D.~N.~C. Tse, ``Communication on the {G}rassmann manifold: A
  geometric approach to the noncoherent multiple-antenna channel,'' \emph{IEEE
  Trans. Inf. Theory}, vol.~48, no.~2, pp. 359--383, Feb. 2002.

\bibitem{yang2013capacity}
W.~Yang, G.~Durisi, and E.~Riegler, ``On the capacity of large-{MIMO}
  block-fading channels,'' \emph{IEEE J. Sel. Areas Commun.}, vol.~31, no.~2,
  pp. 117--132, Feb. 2013.

\bibitem{telatar1999capacity}
E.~Telatar, ``Capacity of multi-antenna {G}aussian channels,'' \emph{European
  Trans. Telecom.}, vol.~10, no.~6, pp. 585--595, 1999.

\bibitem{grant2002rayleigh}
A.~Grant, ``Rayleigh fading multi-antenna channels,'' \emph{EURASIP Journal on
  Advances in Signal Processing}, vol. 2002, no.~3, pp. 316--329, Mar. 2002.

\bibitem{ostman2019}
J.~Östman, G.~Durisi, E.~G. Ström, M.~C. Coşkun, and G.~Liva, ``Short
  packets over block-memoryless fading channels: Pilot-assisted or noncoherent
  transmission?'' \emph{IEEE Trans. Commun.}, vol.~67, no.~2, pp. 1521--1536,
  Feb. 2019.

\bibitem{collins2016spectre}
Spectre, \emph{SPECTRE: short packet communication toolbox}, 2015, GitHub
  repository, [Online]. Available: \url{https://github.com/yp-mit/spectre}.

\bibitem{zhengtse2003DMT}
L.~Zheng and D.~Tse, ``Diversity and multiplexing: A fundamental tradeoff in
  multiple-antenna channels,'' \emph{IEEE Trans. Inf. Theory}, vol.~49, no.~5,
  pp. 1073--1096, May 2003.

\bibitem{zhengtse2002DMT_noncoherent}
------, ``The diversity-multiplexing tradeoff for non-coherent multiple antenna
  channels,'' in \emph{Proc. Allerton Conf. Commun., Contr., Comput.},
  Monticello, IL, USA, 2002, pp. 1011--1020.

\bibitem{feller71}
W.~Feller, \emph{An Introduction to Probability Theory and Its Applications},
  2nd~ed.\hskip 1em plus 0.5em minus 0.4em\relax New York: John Wiley \& Sons,
  1971, vol.~II.

\bibitem{polyanskiy2013saddlepoint}
Y.~Polyanskiy, ``Saddle point in the minimax converse for channel coding,''
  \emph{IEEE Trans. Inf. Theory}, vol.~59, no.~5, pp. 2576--2595, May 2013.

\bibitem{tulino2004random}
A.~M. Tulino and S.~Verd{\'u}, \emph{Random Matrix Theory and Wireless
  Communications}.\hskip 1em plus 0.5em minus 0.4em\relax Foundations and
  Trends in Communications and Information Theory, 2004, vol.~1, no.~1.

\bibitem{horn1990matrix}
R.~A. Horn and C.~R. Johnson, \emph{Matrix Analysis}.\hskip 1em plus 0.5em
  minus 0.4em\relax Cambridge University Press, 1990.

\bibitem{Tse_Viswanath_2005}
D.~Tse and P.~Viswanath, \emph{Fundamentals of Wireless Communication}.\hskip
  1em plus 0.5em minus 0.4em\relax Cambridge University Press, 2005.

\bibitem{AsDo00}
R.~B. Ash and C.~A. Dol\'eans-Dade, \emph{Probability and Measure Theory},
  2nd~ed.\hskip 1em plus 0.5em minus 0.4em\relax Elsevier/Academic Press, 2000.

\bibitem{muirhead2005aspects}
R.~J. Muirhead, \emph{Aspects of Multivariate Statistical Theory},
  2nd~ed.\hskip 1em plus 0.5em minus 0.4em\relax John Wiley \& Sons, 2005.

\bibitem{rudin-principles}
W.~Rudin, \emph{Principles of Mathematical Analysis}, 3rd~ed.\hskip 1em plus
  0.5em minus 0.4em\relax New York: McGraw-Hill Book Co., 1976.

\end{thebibliography}

\end{document}